\documentclass{lmcs} %%% last changed 2014-08-20 
\pdfoutput=1

\usepackage{lastpage}
\lmcsdoi{15}{1}{36}
\lmcsheading{}{\pageref{LastPage}}{}{}%
{Aug.~09,~2017}{Mar.~29,~2019}{}

\keywords{Abstract Hidden Markov Models, Giry Monad, Quantitative Information Flow.}

\usepackage{hyperref}
\theoremstyle{plain} %\crefname{satz}{Satz}{S\"atze}
\newcommand{\cal}{\mathcal}
\usepackage{caption}
\usepackage[normalem]{ulem} % For strikethroughs, as in \Cr.
\usepackage{stmaryrd} % For \rightarrowtriangle.
\usepackage{bbding} % For \PencilLeftDown.
\usepackage{amssymb} % \mathbb
\usepackage{units} % \nicefrac
\usepackage{color} % For coloured markups.
\usepackage{endnotes} % For endnotes, like meta-footnotes but longer.
\usepackage{graphicx} % For including PDF's etc.
\usepackage{amsmath} % For substack.
\usepackage{picinpar} % To get windows in paragraphs.
\usepackage[noadjust]{cite} % To get runs of citations in the IEEE style.

\newcommand\AppFrom[1] {\hfill{\rm[#1]}}
\newcommand\Atimes {\times} % Arithmetic multipication of scalars,
\newcommand\Same {\textsf{\small same}}
\newcommand\Diff {\textsf{\small diff}}
\newcommand\NullChannel {{\sf nc}}
\newcommand\MH {\CalX{\MFun}\CalY{\times}\CalX} % Matrix-H.
\newcommand\AHSpc {\Dist\CalX{\Fun}\Dist^2\CalX} % The "raw" denotational space of forward HMM's.
\newcommand\AHS {\HS\CalX} % The subset of \AHSpc satisfying our two conditions.
\newcommand\UT {\Unc\CalX{\Fun}\Unc\CalX} % The "raw" space of transformers.
\newcommand\TS {{\mathbb T}\CalX} % The subset of \UT satisfying the Riesz-like conditions.
\newcommand\CalD {{\cal D}}
\newcommand\CalX {{\cal X}}
\newcommand\CalY {{\cal Y}}
\newcommand\CalZ {{\cal Z}}
\newcommand\Leak {leak}
\newcommand\Imp {\Rightarrow}
\newcommand\Iff {\Leftrightarrow}
\newcommand\Fun {\mathbin{\rightarrow}}
\newcommand\LiftBox {\vspace{-1em}} % Avoid extra space when display ends environment.
\newcommand\MFun {\mathbin{\rightarrowtriangle}}
\newcommand\VType[1] {\stackrel{\raisebox{-1.5pt}[0pt][0pt]{$\rightarrowtriangle$}}{#1}}
\newcommand\Dist {{\mathbb D}}
\newcommand\HS {{\mathbb H}}
\newcommand\TT {{\mathbb T}}
\newcommand\Size[1] {{\##1}}
\newcommand\MDist[2] {d_M(#1,#2)}
\newcommand\Pow {{\mathbb P}}
\newcommand\ChPar[2] {#1{\parallel}#2}
\newcommand\WLOG {\textit{wlog}}
\newcommand\CM[2] {(#1{;}#2)}
\newcommand\SubDist {\underline{{\mathbb D}}}
\newcommand\SubHyp {\underline{\Dist^2}}
\newcommand\Unc {{\mathbb U}}
\newcommand\In {{:}\,}
\newcommand\Spot {\mbox{\boldmath$\mathrel{\cdot}$}}

\newcommand\OpenSem {[\![}
\newcommand\CloseSem {]\!]}
\newcommand\Hyp[1] {\OpenSem#1\CloseSem}
\newcommand\AbCh[1] {\OpenSem#1\CloseSem}
\newcommand\MM[1] {\OpenSem#1\CloseSem}
\newcommand\HMMsem[2] {\OpenSem#1{:}#2\CloseSem}
\newcommand\HMMoneDal[2] {\OpenSem#1\CloseSem_{{\times}\kern-.2em#2}}
\newcommand\Dal[2] {#1_{{\times}\kern-.2em#2}}
\newcommand\HMMoneD[1] {\OpenSem#1\CloseSem_\times}
\newcommand\HMMone[1] {\OpenSem#1\CloseSem}
\newcommand\Point[1] {[#1]}
\newcommand\SubPt[1] {\underline{[#1]}}
\newcommand\Ref {\mathrel\sqsubseteq}
\newcommand\RRef {\mathrel\sqsupseteq}
\newcommand\StrictRef {\mathrel\sqsubset}
\newcommand\NotRef {\mathrel{\not\sqsubseteq}}
\newcommand\Avg {{\sf avg}}
\newcommand\Nrm[1] {\lfloor#1\rfloor}

\newcommand\Wide[1] {\hspace{1em}#1\hspace{1em}}
\newcommand\WideRM[1] {\hspace{1em}\textrm{#1}\hspace{1em}}
\newcommand\Defs {{:=}\,}
\newcommand\Gets {{:=}\,}
\newcommand\Min {\textsf{min}}

\newcommand\MMult {\mathbin{\cdot}}
\newcommand\Comp {\mathbin{\circ}}
\newcommand\OProb {\{\!\{}
\newcommand\CProb {\}\!\}}
\newcommand\DDist[1] {\OProb#1\CProb}
\newcommand\DSet[1] {\{#1\}}
\newcommand\AtP {@} %Something defining \At; can't figure out what.
\newcommand\LMat[3] {\makebox[0pt]{\raisebox{#2}[0pt][0pt]{$\left(\rule{0pt}{#3}\right.\hspace{#1}$}}}
\newcommand\RMat[3] {\makebox[0pt]{\raisebox{#2}[0pt][0pt]{$\hspace{#1}\left.\rule{0pt}{#3}\right)$}}}
\newcommand\RVec[1] {(#1)}
\newcommand\CVec[1] {(#1)}
\newcommand\ID {{\sf id}}
\newcommand\SSum{\textrm{$\footnotesize\Sigma$}}
\newcommand\Weight[1] {\SSum#1}
\newcommand\VecV {{\mathbb V}}

\newcommand\Wp[1] {\textrm{wp}.#1} % Wp  of a program P.
\newcommand\WpE[1] {\overline{\textrm{wp}}.#1} % Wp  of a program P.
\newcommand\Real {{\mathbb R}}
\newcommand\NNReal {{\mathbb R}^\geq}
\newcommand\PC[1] {{\mathbin{{}_#1\kern-.05em\oplus}}}
\newcommand\WS[1] {{\mathbin{{}_#1\kern-.05em+}}}
\newcommand\Expt[2]{{\cal E}_{#1}\,#2}
\newcommand{\Skip}{\textit{Skip}}
\newcommand\Pnt {\mathop{\textsf{pnt}}}
\newcommand\LEntan[2]{#1\! \triangleleft\! #2}
\newcommand\REntan[2]{#2\! \triangleright\! #1}

\newcommand\RightMar[1]{\stackrel{\rightarrow}{#1}}
\newcommand\LeftMar[1]{\stackrel{\leftarrow}{#1}}

\newcommand\AND{\mathbin{\&}}

\newcommand\HMM {\textit{HMM}}
\newcommand\UM {\textit{UM}}
\newcommand\Dx[1] {\,\textrm{d}#1}
\newcommand\Supp[1] {\lceil#1\rceil}
\newcommand\ChApp[2] {#1{\rhd}#2}

\newcommand\LossApp[2] {#2{\lhd}#1}
\newcommand\XS {\texttt{xs}}
\newcommand\Sm[1] {{\mbox{\small$#1$}}}

\newcommand\Natural{\mathbb N}

\newcommand\Cont{{\mathbb C}}
\newcommand\UNorm[1]{\|#1\|_\infty}
\newcommand\UMet[2]{\UNorm{#1{-}#2}}
\newcommand\KDist{d_K}
\newcommand\KMet[2]{\KDist(#1,#2)}

\newcommand\Integral[2]{\int{#1}\mathrm{d}{#2}}
\newcommand\FunctOne{\mathbf 1}
\newcommand\FunctZero{\mathbf 0}
\newcommand\Lip{\mathbb L}
\newcommand\rhs{\textit{rhs} }

\newcommand\wrt{wrt. }
\newcommand\CalA{\mathcal{A}}

\newcommand\CalS{\mathcal{S}}

\newcommand\CT{\Cont\CalX{\Fun}\Cont\CalX}
\newcommand\Span[1]{{\left\langle#1\right\rangle}}
\newcommand{\LAbs}[2]{\lambda {#1}\ \!\Spot\ \! {#2}}

\newenvironment{ReTheorem}[2]{\par\textit{Theorem \ref{#2}:~}\textit{#1}\rm\quad}{\possiblyCloseProof}
\newenvironment{ReCorollary}[2]{\par\textit{Corollary \ref{#2}:~}\textit{#1}\rm\quad}{\possiblyCloseProof}
\newenvironment{ReLemma}[2]{\par\textit{Lemma \ref{#2}:~}\textit{#1}\rm\quad}{\possiblyCloseProof}
\newtheorem{theorem}{Theorem}

\newtheorem{definition}[theorem]{Definition}
\usepackage{bbding}				% For the small pencil.
\usepackage{listings}			% 0923: For programs and ASCII: see environments for that.
\usepackage{units}				% For \nicefrac.
\usepackage{wasysym}			% For \smiley and \hexagon.
\def\Vhrulefill{\leavevmode\leaders\hrule height 0.7ex depth \dimexpr0.4pt-0.7ex\hfill\kern0pt}

\newcommand\ImageInText[4]%
{\makebox[0pt][l]{%
{\def\Up{#1}\def\Right{#2}% Use \def in case they're already defined.
\hfill%
\raisebox{\Up}[0pt][0pt]{% No height, no depth.
\makebox[0pt][l]{\hspace{\Right}% No width.
\includegraphics[scale=#3]{#4}%
}}}}}

\newcommand\ImageInTextBlock[7]%
{\def\Height{#1}% Use \def in case they're already defined.
 \def\Depth{#2}
 \def\Width{#3}
 \def\Up{#4}
 \def\Right{#5}
\raisebox{\Up}[\Height][\Depth]{\makebox[\Width][l]{\hspace{\Right}\includegraphics[scale=#6]{#7}
}}\\} % Without the \\ the following line is not separated by \Depth.

\newcommand\FirstLineOfTheorem[1][\empty] {\ifx#1\empty\quad\else~\hrulefill~#1\\\fi}
\newcommand\MakeFact[3]{%
  \newenvironment{#2}[2]{%
    \begin{#1}[##1]\label{##2}
    \def\possiblyCloseProof{}
  }{%
    \possiblyCloseProof%
    \end{#1}%
    \def\possiblyCloseProof{}
  }%
}
\newcommand\Proof{%
  \begin{proof}
  \def\possiblyCloseProof{\end{proof}}
}
\MakeFact{defi}{Definition}{definition}
\MakeFact{cor}{Corollary}{corollary}
\MakeFact{lem}{Lemma}{lemma}
\MakeFact{thm}{Theorem}{theorem}
\MakeFact{exa}{Example}{example}

\newenvironment{Figure}[2] {\begin{figure}\def\z{#1}\def\zz{#2}\small}{\caption{\zz\quad}\label{\z}\end{figure}}

\newcommand\Section[1]{\section{#1}}

\newcommand\Cor[1] {Cor.~\ref{#1}}

\newcommand\Def[1] {Def.~\ref{#1}}

\newcommand\Eqn[1] {(\ref{#1})}
\newcommand\Fig[2][\empty] {\ifx#1\empty Fig.~\ref{#2}\else Fig.~\ref{#2}(#1)\fi}

\newcommand\Lem[1] {Lem.~\ref{#1}}
\newcommand\Note[1] {\FnSym\ref{#1}}

\newcommand\Sec[1] {\S\ref{#1}}
\newcommand\App[1] {\S\ref{#1}}
\newcommand\Thm[1] {Thm.~\ref{#1}}

\lstnewenvironment{Prog} {
	\lstset{
		language=csh,
		basicstyle=\ttfamily,	% 1350: Should agree with \PF, i.e. program font.
  		breaklines=false,	% 0922: Don't wrap lines in Prog.
  		showspaces=false,
  		keywords={vis,hid,var,while,if,end,begin,to,then,fi,skip},
  		columns=flexible,
  		escapechar=\%,	% 0920: Use "%" for escape to LaTeX, since it doesn't work as comment in verbatim anyway.
  		mathescape=true
	}
}{}

\lstnewenvironment{ASCII} {
	\lstset{
		language=csh,
		basicstyle=\ttfamily,
  		showspaces=false,
  		breaklines=true,	% 0922: Wrap lines in ASCII.
  		breakindent=0pt,
  		keywords={},
  		columns=flexible,
  		escapechar=\%,	% 0920.
  		mathescape=true
	}
}{}

\newenvironment{Reason}{\begin{tabbing}\hspace{4em}\= \hspace{1cm} \= \kill}
    {\end{tabbing}\vspace{-1em}}
\newcommand\Step[2] {#1 \> $\begin{array}[t]{@{}llll}#2\end{array}$ \\}
\newcommand\StepR[3] {#1 \> $\begin{array}[t]{@{}llll}#3\end{array}$
    \` {\RF \makebox[0pt][r]{\begin{tabular}[t]{r}``#2''\end{tabular}}} \\}
\newcommand\WideStepR[3] {#1 \>
    $\begin{array}[t]{@{}ll}~\\#3\end{array}$ \`
    {\RF \makebox[0pt][r]{\begin{tabular}[t]{r}``#2''\end{tabular}}} \\}
\newcommand\Space {~ \\}
\newcommand\RF {\small}

\usepackage{amssymb}			% For \boxtimes.
\usepackage[normalem]{ulem}		% For strikethroughs.
\usepackage{bbding}				% For the small pencil.
\usepackage{endnotes}			% For endnotes.
\usepackage{hyperref}			% For hyperlinks to and from endnotes.

\definecolor{PurplePlum}{rgb}{0.1,0,0.55} 
\definecolor{Brown}{rgb}{0.5,.25,0}
\definecolor{Orange}{rgb}{1,.6,0}
\definecolor{Gray}{rgb}{.7,.7,.7}
\definecolor{DarkGreen}{rgb}{0,.6,0}

\newif\ifBleck
\newcommand\Bleck {\Blecktrue} % Carry out all deletes and replacements; remove all footnotes/endnotes; make everything black.
\newcommand\Colour[1] {\color{#1}}
\newcommand\ToCLinks {
 \ifBleck\else % 1031: Put hyperlink to ToC  and GR at bottom of every page.
   \expandafter\def\csname@oddfoot\endcsname{
    {\Colour{blue}\mbox{\hyperlink{w1619}{\sf$\rightarrow$~top}\quad
    				    \hyperlink{w1031}{\sf$\rightarrow$~toc}\quad
    				    \hyperlink{w1148}{\sf$\rightarrow$~lof}\quad
                                      \hyperlink{GreenRoom}{\sf$\rightarrow$~gr}\quad
                                      \hyperlink{EndNotes}{\sf$\rightarrow$~en}\quad
                                      \hyperlink{Sargasso}{\sf$\rightarrow$~sg}\quad
                                      \hyperlink{GlossaryOfMacros}{\sf$\rightarrow$~gm}}\hfill
   }}
   \expandafter\def\csname@evenfoot\endcsname{
    {\Colour{blue}\mbox{\hyperlink{w1619}{\sf$\rightarrow$~top}\quad
                                      \hyperlink{w1031}{\sf$\rightarrow$~toc}\quad
    				    \hyperlink{w1148}{\sf$\rightarrow$~lof}\quad
                                      \hyperlink{GreenRoom}{\sf$\rightarrow$~gr}\quad
                                      \hyperlink{EndNotes}{\sf$\rightarrow$~en}\quad
                                      \hyperlink{Sargasso}{\sf$\rightarrow$~sg}\quad
                                      \hyperlink{GlossaryOfMacros}{\sf$\rightarrow$~gm}}\hfill
   }}
 \fi
}

\newif\ifEndNotes % 0831: Workaround for endnotes-bug: missing ".ent" file when there are no endnotes.

\newcommand\FnSym{{\scriptsize\PencilLeftDown\kern.1em}} % Needs bbding.
\newcommand\EnSym {{$\bigtriangledown$}}
\newcommand\Fn[1] {[\FnSym\ref{#1}]} % For referring to footnotes.
\newcommand\En[1] {[\EnSym\ref{#1}]} % For referring to EndNotes.
\newcommand\ENSection[1] {\par\noindent\textbf{\large #1}\par\medskip}
\newcommand\ENSubsection[1] {\par\noindent\textbf{#1}\par\medskip}

\newif\ifMakeMarkupsCalled %1121: Print help text only once.
\newif\ifSuppress % Whether to suppress these annotations.
\newcommand\MakeMarkups[3][.]{%1119
 \Suppressfalse % Suppress these markups if (Bleck OR [0]) and not [1].
 \ifBleck\Suppresstrue\fi
 \ifx0#1\Suppresstrue\fi
 \ifx1#1\Suppressfalse\fi
 \expandafter\providecommand\csname#2x\endcsname {} % 1613: This trick allows MakeMarkups to be called more than once for a particular person.
 \ifSuppress\expandafter\renewcommand\csname#2x\endcsname{\relax}\else
                   \expandafter\renewcommand\csname#2x\endcsname{#3}\fi % \Xx: Change to colour.
 \expandafter\providecommand\csname#2\endcsname {} % 1613.
 \ifSuppress\expandafter\renewcommand\csname#2\endcsname[1]{##1}\else
                   \expandafter\renewcommand\csname#2\endcsname[1]{{\csname#2x\endcsname##1}}\fi % \X{#1}: Put #1 in colour.
 \expandafter\providecommand\csname#2d\endcsname {} % 1613.
 \ifSuppress\expandafter\renewcommand\csname#2d\endcsname[1]{\relax}\else
                   \expandafter\renewcommand\csname#2d\endcsname[1]{{\csname#2x\endcsname\sout{##1}}}\fi % \Xd{#1}: Delete (strikeout) #1; needs ulem package.
 \expandafter\providecommand\csname#2r\endcsname {} % 1613.
 \ifSuppress\expandafter\renewcommand\csname#2r\endcsname[2]{{##2}}\else
                   \expandafter\renewcommand\csname#2r\endcsname[2]{\csname#2d\endcsname{##1} \csname#2\endcsname{##2}}\fi % \Xr{#1}{#2}: Replace #1 with #2.
 \expandafter\providecommand\csname#2t\endcsname {} % 1613.
 \ifSuppress\expandafter\renewcommand\csname#2t\endcsname[1]{\relax}\else
                   \expandafter\renewcommand\csname#2t\endcsname[1]{{\csname#2x\endcsname{$\langle\!\langle$##1$\rangle\!\rangle$}}}\fi % \Xt{#1}: Like \Xd{#1} but without strikeout; use where \Xf{#1} not allowed because e.g. "not in outer par mode"..
 \expandafter\providecommand\csname#2b\endcsname {} % 1613.
 \ifSuppress\expandafter\renewcommand\csname#2b\endcsname[1][empty]{\relax}\else % \Xb[#1]: Put bookmark in the margin, with optional parameter.
                   \expandafter\renewcommand\csname#2b\endcsname[1][\empty]{\ifx\empty##1\empty
                   	\label{#2-bookmark} % With hyperlinks you can jump directly to your bookmark by clicking on an appropriate \pageref.
                          \marginpar [\raggedleft\csname#2\endcsname{{\footnotesize\fbox{#2 working here}}~$\Rightarrow$}]
                                            {\csname#2\endcsname{$\Leftarrow$~{\footnotesize\fbox{#2 working here}}}}
                   \else % No label if there's an explicit parameter.
                   	\marginpar [\raggedleft\csname#2\endcsname{\ifx\empty##1\empty\else\fbox{\tiny\parbox{6em}{\raggedright##1}}~\fi$\Rightarrow$}]
                                            {\csname#2\endcsname{$\Leftarrow$\ifx\empty##1\empty\else~{\tiny\fbox{\parbox{6em}{\raggedright##1}}}\fi}}\fi}\fi
 \expandafter\providecommand\csname#2TD\endcsname {} % 1613.
 \ifSuppress\expandafter\renewcommand\csname#2TD\endcsname{\relax}\else
                   \expandafter\renewcommand\csname#2TD\endcsname{\csname#2\endcsname{\fbox{#2 to do}}}\fi % \XTD: This person to do.
 \expandafter\providecommand\csname#2Bar\endcsname {} % 1613.
 \ifSuppress\expandafter\renewcommand\csname#2Bar\endcsname{\relax}\else
                   \expandafter\renewcommand\csname#2Bar\endcsname{\csname#2\endcsname{\scriptsize\XSolidBrush}}\fi % \XBar: Left by other people: delete when you're happy.
 \expandafter\providecommand\csname#2f\endcsname {} % 1613.
 \ifSuppress\expandafter\renewcommand\csname#2f\endcsname[2][]{\relax}\else
  \expandafter\renewcommand\csname#2f\endcsname[2][\empty]{ % Have forgotten what the optional parameter was for.
    {\mbox{\csname#2x\endcsname\tiny$\boxtimes$}\marginpar{\csname#2x\endcsname\fbox{\FnSym\footnotemark}}\relax 
    \footnotetext{\csname#2x\endcsname##2}}}\fi
 \expandafter\providecommand\csname#2e\endcsname {} % 1613.
 \ifSuppress\expandafter\renewcommand\csname#2e\endcsname[1]{\relax}\else
  \expandafter\renewcommand\csname#2e\endcsname[1]{
   \global\EndNotestrue %0831
   \mbox{\scriptsize\csname#2x\endcsname$\boxtimes$}\relax
   \marginpar{\csname#2x\endcsname\fbox{\EnSym\endnotemark
                      \hypertarget{ENmark\thepage-\theendnote}{}~\hyperlink{ENtext\thepage-\theendnote}{{\Colour{blue}$\downarrow$}}}
   }
   { % 1641 Make \z,\zz local.
    \def\zz{\noexpand#3}
    \edef\z{~{\zz[Endnote \theendnote\ on p.\noexpand\hypertarget{ENtext\thepage-\theendnote}{}\thepage
                ~\noexpand\hyperlink{ENmark\thepage-\theendnote}{{\noexpand\Colour{blue}$\uparrow$}}]}
    }
    \expandafter\endnotetext\expandafter{\z\vspace{2ex}\\ ##1\newpage}
   } %1641
  }\fi
 \expandafter\providecommand\csname#2fe\endcsname {} % 1613.
 \ifSuppress\expandafter\renewcommand\csname#2fe\endcsname[2][]{\relax}\else %1538
  \expandafter\renewcommand\csname#2fe\endcsname[2][]{ % 1540: \File is local.
   \def\File{##1}\relax
   \ifx\File\empty\csname#2f\endcsname{##2}\else % 1536: Without file, behave like \Xf.
    \global\EndNotestrue %0831
    \mbox{\scriptsize\csname#2x\endcsname$\boxtimes$}
    \marginpar{\csname#2x\endcsname\fbox{\FnSym\footnotemark}}\relax
    \footnotetext{~\csname#2x\endcsname##2\
                         --- See [\EnSym\endnotemark\hypertarget{ENmark\thepage-\theendnote}{}
                         \kern-.2em\hyperlink{ENtext\thepage-\theendnote}{{\Colour{blue}$\downarrow$}}].}\relax
   { % 1641 Make \z,\zz local.
     \def\zz{\noexpand#3}
     \edef\z{~{\zz[Footnote~\thefootnote~on~p.\noexpand\hypertarget{ENtext\thepage-\theendnote}{}\thepage
                 ~\noexpand\hyperlink{ENmark\thepage-\theendnote}
                 {{\noexpand\Colour{blue}\kern-0.1em$\uparrow$}]}}
                 {\noexpand\footnotesize\noexpand\newline\noexpand\hspace*{2em} (~from file {\noexpand\tt\File.tex}~)}
     }    
     \expandafter\endnotetext\expandafter{\z~\par\input{##1}\newpage}
    } %1641
   \fi % 1536
  } % 1540
 \fi % 1538
 \ifSuppress\relax\else%0749
  \csname#2f\endcsname{%1009
   $\backslash$\texttt{#2}$\cdots$\ markups are in \textbf{this} colour\ifx#1..\else\ifx1#1.\else, for #1.\fi\fi
   \ifMakeMarkupsCalled\relax\else %1121
    \begin{quote}\begin{tabular}{l@{\hspace{2em}}p{.8\linewidth}}
     \multicolumn{2}{l}{\texttt{$\backslash$MakeMarkups\ifx#1.\relax\else[#1]\fi\{#2\}\{{\it$\langle$colour command\/$\rangle$}\}}
     				 --- Defines the macros below:}\\
         & see comments at \texttt{$\backslash$MakeMarkups} definition. \\[1ex]
     \texttt{$\backslash$#2\{$\langle$text$\rangle$\}} & Sets \texttt{$\langle$text$\rangle$} in \texttt{#2}'s colour. \\
     \texttt{$\backslash$#2x} & Changes to \texttt{#2}'s colour (until end of context). \\
     \texttt{$\backslash$#2d\{$\langle$text$\rangle$\}} & Sets \texttt{$\langle$text$\rangle$} in \texttt{#2}'s colour with a strikethrough (i.e.\ delete). \\
     \texttt{$\backslash$#2r\{$\langle$this$\rangle$\}\{$\langle$that$\rangle$\}} &
      Strikes through \texttt{$\langle$this$\rangle$} and inserts \texttt{$\langle$that$\rangle$} (i.e.\ replace). \\
     \texttt{$\backslash$#2f\{$\langle$text$\rangle$\}} & Meta-comment: puts \texttt{$\langle$text$\rangle$} in a \texttt{#2}-footnote with a {\tiny$\boxtimes$} in the main text. \\
     \texttt{$\backslash$#2t\{$\langle$text$\rangle$\}} & Use for meta when  \texttt{$\backslash$#2f} isn't allowed (``Not in outer-par mode.'') \\
     \texttt{$\backslash$#2b[$\langle$optional$\rangle$]} & Marginal pointer, with label for hyper-linking directly there. \\
     \texttt{$\backslash$#2e\{$\langle$text$\rangle$\}} & Puts \texttt{$\langle$text$\rangle$} in a \texttt{#2}-endnote with a (big) $\boxtimes$ in the main text. \\[.5ex]
     \texttt{$\backslash$#2fe[$\langle$this$\rangle$]\{$\langle$that$\rangle$\}} & Makes a  \texttt{$\backslash$#2f\{$\langle$that$\rangle$\}} that refers to a \\
       & \texttt{$\backslash$#2e\{$\langle$contents of file this.tex$\rangle$\}}. \\ 
       & Without the optional argument, acts as \texttt{$\backslash$#2f\{$\langle$that$\rangle$\}}. \\[.5ex]
     \texttt{$\backslash$#2Bar} & Inserts ``burn after reading'' symbol \csname#2Bar\endcsname, meaning
      \begin{quote}\begin{itemize}\setlength\itemsep{0pt}
       \item If yours is the only \csname#2Bar\endcsname\ in this (presumably someone else's) footnote, and you are happy that the footnote has been addressed,
       go ahead and comment-out the whole footnote. (The \csname#2Bar\endcsname\ is their request for you to ``approve and remove''.)
       \item If you are not happy, delete only your \csname#2Bar\endcsname\ and follow-on in the footnote
        (in your colour, i.e.\ with \texttt{$\backslash$#2x}) saying why you are not happy.
       \item If you are happy, but there are others' burn-after-reading symbols as well as yours, just delete yours; the other people have not yet responded.
      \end{itemize}
      \end{quote}
      The idea is that when everyone's happy, the last person will comment-out the meta-text. \\[0.5ex]
     \texttt{$\backslash$#2TD} & Inserts {\csname#2TD\endcsname}\ . \\
    \end{tabular}\end{quote}
   \fi
  }%1009
  \MakeMarkupsCalledtrue %1121
 \fi%0749
}%1119

\newif\ifNoGreenRoom

\newif\ifNoEndNotes

\newif\ifNoSargasso
\newcommand\MakeSargasso {
 \hypertarget{Sargasso}{}
 \newcommand\NewLabel[1] {\OldLabel{Sargasso-##1}} % All Sargasso labels "branded".
 \newcommand\NewRef[1] % All Sargasso references take branded version if possible.
 {\expandafter\ifx\csname r@Sargasso-##1\endcsname\relax\OldRef{##1}\else\OldRef{Sargasso-##1}\fi}
 \let\OldLabel\label \let\label\NewLabel
 \let\OldRef\ref \let\ref\NewRef
\ifBleck\end{document}\else\ifNoSargasso
\begin{center}\huge \textit{Sargasso suppressed in this run}\end{center}
\else
  \hrule
  ~\\\begin{center}\Huge Sargasso
  \end{center}~\\
  \hrule
 \fi\fi
}

\newcommand\Cite[2][\empty] {{\color{red}\ifx#1\empty[#2]\else[#2,~#1]\fi}}

\Bleck % Enable to suppress all markups etc.

\begin{document}
\MakeMarkups[Carroll]{C}{\Colour{blue}}
\MakeMarkups[Annabelle]{A}{\Colour{red}}
\MakeMarkups[Ghost]{X}{\Colour{Gray}}
\MakeMarkups[Tahiry]{T}{\Colour{Orange}}
\MakeMarkups[Black]{B}{\Colour{black}}
% For TeXshop: make it LaTeX the root file if you cmt-T here.
% !TEX root = LMCS18FromLiCS15.tex

%%% Label hints --- 1720: Has to be called after the (Springer) \begin{document}, which resets \label.
%%% Also it doesn't seem to work inside of \begin{equation}, which also resets \label.
%%% Carroll Morgan 3 August 2016
%

% Detect whether in a float.
% https://tex.stackexchange.com/questions/27172/how-can-i-detect-if-im-inside-or-outside-of-a-float-environment
\makeatletter   
\newcommand\IfFloating[2]{\ifnum\@floatpenalty<0\relax #1 \else #2 \fi}
\makeatother

\newif\ifShowLabels
\newcommand\SL[1] {\mbox{\tt\scriptsize[#1]}}
\newcommand\LabelWithShow[1] {\ifShowLabels\ifinner\SL{#1}\else\ifmmode\SL{#1}\else\IfFloating{\SL{#1}}{\marginpar{\SL{#1}}}\fi\fi\fi\Label{#1}}
\let\Label\label\let\label\LabelWithShow

\newcommand\ShowLabels {\ShowLabelstrue} % Put labels in the margin for convenience.
\newcommand\NoShowLabels {\ShowLabelsfalse} % Don't do that.

% Kludge to get around Springer's re-redefining \label inside \Begin{Equation}.
% Later could replace equation with Equation "behind the scenes"; but for now just use \begin{Equation} directly if you want this feature.
% See http://latex-community.org/forum/viewtopic.php?f=45&t=23165
\newif\ifShowLabelsWas
\newenvironment{Equation}{\ShowLabelsWasfalse\ifShowLabels\ShowLabelsWastrue\begin{equation}\let\LabelWas\Label\let\Label\label\let\label\LabelWithShow\else\begin{equation}\fi}{\end{equation}\ifShowLabelsWas\let\label\Label\let\Label\LabelWas\fi\ShowLabelsWasfalse}
%%% Local Variables:
%%% mode: latex
%%% TeX-master: "LMCS18.tex"
%%% End:

\title[Abstract Hidden Markov Models]{ Abstract Hidden Markov Models: \\
	a monadic account of quantitative information flow}
%\titlecomment{{\lsuper*}OPTIONAL comment concerning the title, \eg, 
%	if a variant or an extended abstract of the paper has appeared elsewhere.}

\author[A.~McIver]{Annabelle McIver\rsuper{a}}	%required
\address{\lsuper{a}Dept.~Computing\\
	Macquarie University.\\
	Sydney, Australia}	%required
\email{\{annabelle.mciver,tahiry.rabehaja\}@mq.edu.au}  %optional
%\thanks{thanks 1, optional.}	%optional

\author[C.~Morgan]{Carroll Morgan\rsuper{b}}	%optional
\address{\lsuper{b}School of Comp.\ Sci.\ and Eng.\\
	Univ.~New South Wales,
	and Data61.\\
	Sydney, Australia}	%optional
\email{carroll.morgan@unsw.edu.au}  %optional
%\thanks{thanks 2, optional.}	%optional

\author[T.~Rabehaja]{Tahiry Rabehaja\rsuper{a}}	%optional
%\urladdr{name3@url3\quad\rm{(optionally, a web-page can be specified)}}  %optional
%\thanks{thanks 3, optional.}	%optional

%% etc.

%% required for running head on odd and even pages, use suitable
%% abbreviations in case of long titles and many authors:

%%%%%%%%%%%%%%%%%%%%%%%%%%%%%%%%%%%%%%%%%%%%%%%%%%%%%%%%%%%%%%%%%%%%%%%%%%%

%% the abstract has to PRECEDE the command \maketitle:
%% be sure not to issue the \maketitle command twice!
	
\begin{abstract}
	Hidden Markov Models, \HMM's, are mathematical models of Markov processes with state that is hidden, but from which information can leak. They are typically represented as 3-way joint-probability distributions.
	
	We use \HMM's as denotations of probabilistic hidden-state sequential programs: for that, we recast them as ``abstract'' \HMM's, computations in the Giry monad $\Dist$, and we equip them with a partial order of increasing security. However to encode the monadic type \emph{with hiding} over some state $\CalX$ we use $\AHSpc$ rather than the conventional $\CalX{\Fun}\Dist\CalX$ that suffices for Markov models whose state is not hidden. We illustrate the $\AHSpc$ construction with a small Haskell prototype.
	
	We then present \emph{uncertainty measures} as a generalisation of the extant diversity of probabilistic entropies, with characteristic analytic properties for them, and show how the new entropies interact with the order of increasing security. Furthermore, we give a ``backwards'' uncertainty-\emph{transformer} semantics for \HMM's that is dual to the ``forwards'' abstract \HMM's --- it is an analogue of the duality between forwards, relational semantics and backwards, predicate-transformer semantics for imperative programs with demonic choice.
	
	Finally, we argue that, from this new denotational-semantic viewpoint, one can see that the Dalenius desideratum for statistical databases is actually an issue in compositionality. We propose a means for taking it into account.
\end{abstract}

\maketitle

\section{Introduction}\label{s1656}

\subsection{Setting and overview}

We can represent probabilistic sequential programs with hidden state as Hidden Markov Models, i.e.\ \HMM's\,%
\footnote{We use apostrophe uniformly between acronyms and suffixes, even when they are not possessive.}
formulated as probabilistic mechanisms that take prior, input probability distributions and give posterior distributions over (leaked) observations and final state. Here, however, we recast \HMM's as computations over the Giry monad, making them more suitable for denotational semantics. Indeed the monadic view of simple Markov processes in particular is well established \cite{Moggi:89,Giry:81}, using $\CalX{\Fun}\Dist\CalX$ where type-constructor $\Dist$ makes distributions on its base type $\CalX$; \label{g1320}the Kleisli extension is then of type $\Dist\CalX{\Fun}\Dist\CalX$, representing the action of multiplying an initial-state-distribution vector by a Markov matrix. But that simplicity cannot account for hidden state and information flow.

We treat hidden state by beginning with $\Dist\CalX$ (not $\CalX$): the computation type we obtain is then ``one level up'', of type $\Dist\CalX{\Fun}\Dist^2\CalX$, the Kleisli extension is $\Dist^2\CalX{\Fun}\Dist^2\CalX$; and we call the double-distribution type $\Dist^2\CalX$ \emph{hyper-distributions}, or ``hypers'' for short.

Although the Giry monad is formulated in terms of general measures \cite{Giry:81}, we will need only discrete distributions for matrix-based \HMM's. Nevertheless, we give our constructions and results in more general terms, anticipating e.g.\ infinite sequences of \HMM's, nondeterminism, and iterations for which proper measures will be necessary \cite{McIver:2014ab}.

In earlier work, we have used the hypers $\Dist^2\CalX$, equipped with a partial order of increasing security, to establish compositionality results \cite{mcivermeinicke10a}, to explore the effect of including demonic nondeterminism \cite{McIver:12} and to give an abstract treatment of probabilistic channels \cite{McIver:2014aa,Alvim:2014aa}. A second earlier theme has been the generalisation of entropies (such as Shannon) to a more abstract setting where only their essential properties are preserved \cite{mcivermeinicke10a,Alvim:2012aa,McIver:12,Alvim:2014aa}. Here we use monads to bring all those separate strands together and to go further.

One major further step is to show that there is a dual, backwards view for abstract \HMM's, based on ``uncertainty'' \emph{transformers} that transform post- uncertainty measures into pre- uncertainty measures where, in turn, \emph{uncertainty measures} generalise probabilistic entropies.

We and others have argued that specific entropies (e.g.\ Shannon) have limitations in security work generally \cite{Smith:2009aa,mcivermeinicke10a}. Therefore we focus here on their essential properties: continuity and concavity. That view is supported by powerful theorems that such a generalisation supports, and a methodological criterion that uncertainty measures capture contexts in a way that individual styles of entropy cannot.

A second further step is to extend our recent treatment \cite{Alvim:2014aa} of the \emph{Dalenius Desideratum}, the ``collateral'' leakage of information due to unknown correlations with third-party data, from merely channels (a ``read only'' scenario  \cite{Dalenius:1977aa,Dwork:2006aa}, such as access to a statistical database) to programs that might alter the database (thus ``read/write'' as well). The Dalenius perspective here is the fact that care must be taken wrt.\ compositionality in a context containing extra variables even when a program fragment does not explicitly refer to them\cite{mcivermeinicke10a}.

\emph{To remain accessible to a broader security community}, we do not begin from Giry: rather we first work in elementary terms. In \Sec{s1250} the monadic structures will be seen to have informed our earlier definitions and theorems.

%References to the appendix can be resolved at \cite{McIver:2015aa}.
\Cf{Check this --- maybe include them?}
\subsection{Principal contributions and aims: summary}\label{s0429}
\noindent Our principal contributions are these, in which the \textbf{new constructions and results} are given in bold:
\begin{itemize}

\item We note that (finite) classical \HMM's are a model for straight-line sequential probabilistic programs with hidden state
\item We formulate \emph{abstract \HMM's} over a state as a monadic model for \HMM's over that same state, and give their characteristic properties.

\item We formulate \emph{uncertainty measures} as a generalisation of diverse entropies (top centre), and give their characteristic properties.\,%
\footnote{They were studied, but less extensively, as ``disorders'', in \cite{McIver:12}.}

\item We note that uncertainty measures have a complete representation based on real-valued functions of state and adversarial strategy.

\item We give a dual, uncertainty-transformer semantics of \HMM's and prove the duality.

\item We show how all of the above is an instance of the general Giry monad as a computation, of which (finite) \HMM's use a discrete portion.

\item We explain how the ``Dalenius effect'' is manifested as a compositional issue in this framework, and how it can be treated.
\end{itemize}
In other sections we review abstract channels (\Sec{s0737}), hyper-distributions (\Sec{s1129}) and the security order (\Sec{s1432}) on hypers.

We believe that \Thm{t1005}, in particular its assumptions and proof, is a significant new result.

\medskip
\noindent Our principal \underline{aims} are these:

\begin{itemize}
\item (More abstract) To construct forward- and dual backward semantic spaces for probabilistic sequential computations over hidden state, using monadic computations and partial (refinement) orders in this new context, and we formulate and prove the general properties that make them suitable for embedding finite (for the moment) \HMM's.

\item (More concrete) To provide the basis for a source-level reasoning method, analogous to Hoare logic or weakest preconditions, for quantitative non-interference in sequential programs. For this, the dual, transformer semantics for \HMM's seems to be a necessary first step, together with a link between the social aspects of security and the mathematical behaviour of a program (\Sec{s0914}).
\end{itemize}

The conclusion \Sec{s1323} discusses the \underline{benefits} of doing this.

\subsection{General notations --- see also \App{s1241}}
\label{g1054}Application of function $f$ to argument $x$ is written $f.x$ to reduce parentheses. It associates to the left.

\label{g1042}Although \label{g1036}a matrix $M$ with rows, columns indexed by $R,C$ is a function $R{\times}C\Fun\Real$, we avoid constant reference to the reals $\Real$ by writing just $R{\MFun}C$ for that type; \label{g1043}similarly we write the type of a vector over $X$ as $\VType{X}$. We write $M_{r,c}$ for the element of matrix $M$ indexed by row $r$ and column $c$; then the $r$-th row of $M$ is $M_{r,-}$; and the $c$-th column is $M_{-,c}$, of types $\VType{Y},\VType{X}$ resp. For row- or column vector $v{\In}{\VType{I}}$ we write $v_i$ for its $i$-th element. Thus e.g.\ we have ${(M_{-,c})}_{r}{=}M_{r,c}$.

\label{g1034}When multiplying vectors and matrices we assume without comment that the vector has been oriented properly, i.e.\ as a row or column as required. Thus $v$ acts as a row in $v{\MMult}M$ but as a column in $M{\MMult}v$.\,%
Thus for $v{\In}\VType{X}$ and $M{\In}X{\MFun}Y$ the matrix product $v{\MMult}M$ is in $\VType{Y}$, where here we are using dot $(\cdot)$ for matrix multiplication. Multiplication of scalars will usually be juxtaposition, but occasionally $\times$ when we are avoiding ambiguity.

\label{g1052}We write for example $x{\In}X$, i.e.\ with a colon, when we are introducing a fresh variable $x$ into the discussion at that point; with $x{\in}X$ we are instead stating a property of some $x$ and $X$ that have been already introduced at some earlier point.\,%
\footnote{For example we could write ``Because we have already established that $s{\in}\Pow X$, we know  that for any $x{\In}s$ we have $x{\in}X$.'' Both $s,X$ are defined in the surrounding text, but $x$ here is a local (i.e.\ bound) variable just used temporarily.}
That means in the former case that one need not search backwards to see what $x$ is being referred to (and in the latter case, one might).

Other specific notations are explained at first use, and (as noted above) a full glossary in occurrence order is given in \App{s1241}.

\section{Abstract channels and hyper-distributions}\label{s0918}
We now review abstract channels as a conceptual stepping-stone to hyper-distributions --- recall they are ``hypers'' for short. (Channels are the special case of \HMM's where the state is not updated.)

\subsection{Channels and distributions as matrices and vectors}
\label{g1022} A \emph{channel} is a (stochastic) matrix of non-negative reals with 1-summing rows; we use upper-case Roman letters like $C$ for them. \label{g1024}The rows are labelled with elements from some set $\CalX$; and the columns from some set $\CalY$. Thus a channel typically has type $\CalX{\MFun}\CalY$; here, both $\CalX$ and $\CalY$ will be finite.

\label{g1026}A distribution in $\Dist\CalX$ can be presented as a 1-summing vector in $\VType{\CalX}$, usually lower-case Greek: generally $\delta$ for ``distribution'', but especially $\pi$ for prior and sometimes $\rho$ for posterior.

\begin{Definition}{Weight}{d1358B}
	\label{g1038}Let $M$ or $v$ be a matrix or vector resp. Then $\Weight{M}$ or $\Weight{v}$ is its \emph{weight}, the sum $\SSum_{x,y} M_{x,y}$ or $\SSum_x v_x$ taken over all its indices.
\end{Definition}
Thus e.g.\ we have $\Weight{M_{x,-}}= \SSum_y M_{x,y}$ and that $M$ is stochastic (i.e.\ represents a channel) just when $\Weight{M_{x,-}}$ is 1 for all $x$.

Each row $C_{x,-}$ of a channel $C$ is a conditional probability distribution over $\CalY$ given that particular $x{\In}\CalX$. That is, the $y$-th element $C_{x,y}$ of $C_{x,-}$ is the probability that $C$ takes input $x$ to output $y$.

\subsection{Informal channel semantics: \emph{abstract} channels}\label{s0737}
A (1-summing) prior $\pi$ and (stochastic) channel $C$ together determine a joint distribution as follows.
\begin{Definition}{Channel applied to prior}{d1124B}
	\label{g1039}Given a prior $\pi{\In}{\VType{\CalX}}$ and channel $C{\In}\CalX{\MFun}\CalY$ we write $\ChApp{\pi}{C}$ for the joint-distribution matrix of type $\CalX{\MFun}\CalY$ resulting from \emph{applying} the channel to the prior, defined $(\ChApp{\pi}{C})_{x,y}\Defs \pi_x C_{x,y}$.
(Here juxtaposition is ordinary multiplication of reals.)
\end{Definition}
Note that matrix $\ChApp{\pi}{C}$ is not stochastic: rather because $C$ itself is stochastic we have
$\SSum(\SSum(\ChApp{\pi}{C})_{x,-})=\SSum\pi_x =1$.

A non-zero vector is normalised as follows.
\begin{Definition}{Normalisation}{d1357B}
	\label{g1049}Let $\delta{\In}{\VType{\CalX}}$ be such that $0{\neq}\SSum\delta$. Then the \emph{normalisation} $\Nrm{\delta}$ of $\delta$ is given by $\Nrm{\delta}_x\Defs \delta_x/\SSum\delta$ for each $x{\In}\CalX$.
\end{Definition}

Now for some $\pi{\In}\VType{\CalX}$ and channel $C{\In}\CalX{\MFun}\CalY$ define joint distribution $J{\In}\CalX{\MFun}\CalY$ by $J{=}\ChApp{\pi}{C}$. The (marginal) probability of each output $y{\In}\CalY$ is $\SSum J_{-,y}$ and, associated with each $J$, there is a posterior distribution $\Nrm{J_{-,y}}$ on $\CalX$.

Abstracting from the $y$-values, but retaining the link between the marginal probabilities and the posterior distributions, gives an informal description of our intended ``abstract channel'' semantics \cite{McIver:2014aa}. We make this precise in \Sec{s1956}.

\subsection{Hypers abstract from joint distributions}\label{s1129}
The joint-distribution matrix $J{=}\ChApp{\pi}{C}$ contains ``too much'' information if we do not need the actual value of $y$ that led to a particular posterior. Abstracting from those output values leads us to a representation of the possible posteriors on their own, retaining however the probability with which they occur (in fact the marginal probability of the value $y$ that produced each one). The advantage we take from that is that \HMM's acquire a monadic structure, acting as Kleisli maps, and furthermore can express other probabilistic notions in a way more suited to calculation: for example, conditional entropies become expected values (of the entropies) over the distribution of posteriors.

More intuitive reasons for the abstraction include that it is appropriate in security to consider the information leakage of a channel $C$ wrt.\ a prior $\pi$ to concern only what an adversary can discover about $\pi$, and not the actual observations that led to that discovery: whether a spy's vocabulary is ``da/nyet'' or ``yes/no'', or indeed whether ``yes'' means ``it's zero'' or ``it's one'', does not affect the information-theoretic threat that spy represents, provided of course that the spy and her controller have agreed on the vocabulary beforehand.

We can abstract from the observations in $\ChApp{\pi}{C}$ as follows.
If column $y$ of $J=\ChApp{\pi}{C}$ is all zero, then that $y$ will never occur (for any prior); thus we can omit that column.

And if two columns $y_{1,2}$ of $J$ are proportional to each other, i.e.\ are \label{g2002}\emph{similar} (as for triangles), then we can add them together, since for a given prior the same posterior will be inferred for $y_1$ as for $y_2$ and the overall probability of inferring that posterior will be the sum of the marginal probabilities for $y_{1,2}$.\,%
\footnote{\Label{g1056}For brevity we write $y_{1,2}$ rather than $y_1,y_2$.}

Finally, a 1-1 renaming of the $y$-values has no effect on the posteriors and their respective probabilities; so we can remove those names as long as we retain the distinction between separate (non-zero, non-similar) columns.

Abstracting from all that arguably inessential information (about $y$) leaves only a distribution of posteriors on $\CalX$ and, for us, this is the semantic view. \label{g1058}Writing in general $\Dist\CalX$ for 1-summing functions of type $\CalX{\Fun}\NNReal$, a discrete distribution over $\CalX$ has type $\Dist\CalX$ and so a discrete distribution of such distributions has type $\Dist(\Dist\CalX)$ that is $\Dist^2\CalX$.\,%
%\footnote{Thus 1-summing vectors $\delta$ in $\VType{\CalX}$ describe distributions in $\Dist\CalX$.}
Those latter are our hypers, and they are our abstraction of joint distributions $\CalX{\MFun}\CalY$.

The values of type $\Dist\CalX$ are called the \emph{inners} of a hyper, and the \emph{outer} distribution of a hyper is its distribution over those inners: that is, a hyper on $\CalX$ is a (single) outer distribution over (possibly many) inners, and each inner is a (single) distribution over $\CalX$ itself.
\label{g1121} 

As an example, recall the famous puzzle of \emph{Bertrand's Boxes}. Three identical boxes contain two balls each: one has two white balls; one has two black balls; and the remaining box has one of each. It is not known which box is which; and one of them is chosen randomly. A ball is drawn at random from it, and it is white. What is the probability that the other ball in that box is also white? We reason as follows.

The state space is $\CalX{=}\{0,1,2\}$, referring to the number of white balls in each box. The prior distribution in $\Dist\CalX$ is uniform, which we can write $(\nicefrac{1}{3},\nicefrac{1}{3},\nicefrac{1}{3})$. The \HMM\ is a channel that takes input $x$ to the distribution $(\textit{white}\mapsto \nicefrac{x}{2},\textit{black}\mapsto 1{-}\nicefrac{x}{2})$. The joint distribution $p$ say, of type $\Dist(\CalX{\times}\{\textit{white},\textit{black}\})$, would be such that $p(1,\textit{white}) = \nicefrac{1}{3}{\times}\nicefrac{1}{2} = \nicefrac{1}{6}$, the probability that Box 1 was chosen and that the ball taken from it was white. The overall probability that a white ball is taken (from whichever box) is the \textit{white}-marginal $0{+}\nicefrac{1}{6}{+}\nicefrac{1}{3}=\nicefrac{1}{2}$ (which is obvious from symmetry anyway), and the posterior distribution on $\CalX$ is in that case $(0,\nicefrac{1}{3},\nicefrac{2}{3})$ --- which nicely solves the puzzle. The posterior probability that $x{=}2$ is $\nicefrac{2}{3}$ given that a white is taken and, by the way, a white is taken with overall probability $\nicefrac{1}{2}$ (the marginal). And for that reasoning of course the value of the observation, the colour of the drawn ball, is used.

But now suppose instead you wanted to know only the decrease in Shannon entropy resulting from that experiment. Beforehand the entropy is $H(\nicefrac{1}{3},\nicefrac{1}{3},\nicefrac{1}{3})=\log_23=1.58$ (approximately). Afterwards, it will be the \emph{conditional} Shannon entropy of the distribution of posteriors, calculated by taking the expected value of $H()$ over the distribution of posteriors: and that is approximately
\begin{equation}\label{e1635}
	\nicefrac{1}{2}{\times}H(0,\nicefrac{1}{3},\nicefrac{2}{3})
	+ \nicefrac{1}{2}{\times}H(\nicefrac{2}{3},\nicefrac{1}{3},0)
	\Wide{=} \nicefrac{1}{2}{\times}0.92+\nicefrac{1}{2}{\times}0.92
	\Wide{=} 0.92\quad,
\end{equation}
so that $1.58{-}0.92 = \nicefrac{2}{3}$ (exactly) of a bit has been leaked. And we did not need colours for that: the calculation is done entirely with the hyper-distribution, that is with the distribution
\[
\textrm{the \textit{outer}}\left\{~
\begin{array}{rcl}
	\nicefrac{1}{2} &\mapsto& (0,\nicefrac{1}{3},\nicefrac{2}{3}) \\
	\nicefrac{1}{2} &\mapsto& (\nicefrac{2}{3},\nicefrac{1}{3},0)
\end{array}
~\right\}\textrm{the \textit{inners}}
\]
\emph{of distributions}, i.e.\ of posteriors: a hyper-distribution. The $\nicefrac{1}{2}$'s are the marginals, and the $({\cdots})$ are the posteriors associated with each. We call the channel-output marginal the \emph{outer}, and the posteriors the \emph{inners}.

Seeing this example as a security leak, we might imagine that the adversary is trying to guess the colour of the other ball in the box: in that case she would look at the colour she took and then guess that same colour. To describe that we use a different entropy $V_1$, called \emph{Bayes Vulnerability}, which is the probability the secret can be guessed in one try by an optimal adversary. Obviously she will guess the $x$-value with the largest probability in the posterior (the inner), and her conditional probability of guessing correctly is
\[
	\nicefrac{1}{2}{\times}V_1(0,\nicefrac{1}{3},\nicefrac{2}{3})
	+ \nicefrac{1}{2}{\times}V_1(\nicefrac{2}{3},\nicefrac{1}{3},0)
	\Wide{=} \nicefrac{1}{2}{\times}\nicefrac{2}{3}+\nicefrac{1}{2}{\times}\nicefrac{2}{3}
	\Wide{=} \nicefrac{2}{3}\quad.
\]
That's no surprise --- but what is worth noting is that we used the \emph{same} hyper-distribution for the $V_1$ calculation just above as for the $H$ calculation at \Eqn{e1635}. That is the utility of the abstraction: that the hyper contains enough information to handle many entropies one might use to measure leakage.

{\Cx %1506
\subsection{The semantic function from joints to hypers}\label{s1956}
\Cf{This is new; original in Sargasso.}
\label{g1102}
In this section we define precisely the denotation $\Hyp{J}$ in $\Dist^2\CalX$ of a joint-distribution matrix $J{\In}\CalX{\MFun}\CalY$.

\begin{Definition}{Sub-distribution, sup-hyper}{d1404}
	\label{g1401}
	A discrete sub-distribution over a set $\CalX$ is a function of type $\CalX{\Fun}[0,1]$ that sums to \emph{no more than 1}; we write that type as $\SubDist\CalX$. (Recall that a \emph{proper} distribution in $\Dist\CalX$ sums to exactly 1, and thus $\Dist\CalX\subseteq\SubDist\CalX$.)
		
	Similarly a discrete sub-\emph{hyper} over a set $\CalX$ is a sub-distribution over the (proper, inner) distributions $\Dist\CalX$, thus of type $\SubDist(\Dist\CalX)$; only the outer of a sub-hyper can sum to less than 1. We write that type as $\SubHyp\CalX$.
(Note that the inners of a sub-hyper are proper distributions.)
\end{Definition}
\begin{Definition}{One- and two-point distributions}{d1954}
	\label{g1011}\label{g1212}For $z,z'{\In}\CalZ$ in general we write $\Point{z}$ for the \emph{point distribution} on $z$, viz.\ assigning probability 1 to $z$ and 0 to all other elements of $\CalZ$.\,%
	\footnote{Function $\Point{\cdot}$ is the unit $\eta$ of the $\Dist$-monad: see \Sec{s1250}. \Ct{What about $\SubPt{\cdot}$. Is it a unit too?}}
	We write $z\PC{p}z'$ for the two-point distribution that assigns $p$ to $z$ and $1{-}p$ to $z'$ and 0 to everything else in $\CalZ$.
\end{Definition}
Thus $z\,\PC{1}\,z'=\Point{z}$ and $z\,\PC{0}\,z'=\Point{z'}$.

\begin{Definition}{Point sub-hyper}{d1411}
	\label{g1404}
	For sub-distribution $\delta{\In}\SubDist\CalX$ the \emph{point sub-hyper} $\SubPt{\delta}$ in $\SubHyp\CalX$ has weight $\Weight{\delta}$ concentrated on the single (inner) distribution $\Nrm{\delta}$,
	provided of course that $\Weight{\delta}{\neq}0$. If $\Weight{\delta}{=}0$ then $\SubPt{\delta}$ is the (unique) weight-zero sub-hyper.
	
	That is, the argument $\delta$ is normalised to make the inner, and its weight becomes the (one-point) sub-outer on that inner.
\end{Definition}
Note that when $\Weight{\delta}=1$ we have $\SubPt{\delta}=\Point{\delta}$.

We now define the semantic function itself:
\begin{Definition}{Joint-distribution denotes hyper}{d1426HB}
	Let $J{\In}\CalX{\MFun}\CalY$ satisfy $1{=}\SSum J$ so that it describes a discrete (proper) joint distribution in $\Dist(\CalX{\times}\CalY)$. Then its abstraction \Hyp{J} to a hyper in $\Dist^2\CalX$ is given by
	\[
	\Hyp{J} \Wide{=} \sum_{y{\in}\CalY} \SubPt{J_{-,y}}\quad,
	\]
	with summation $\SSum_{y{\in}\CalY}$ therefore being an addition of sub-hypers, i.e.\  sub-distributions on $\Dist\CalX$. Each column $J_{-,y}$ is regarded as a sub-distribution in $\SubDist\CalX$, and then $\SubPt{-}$ converts it to a sub-point hyper. %Then they are all added up.
	\end{Definition}
	
	Note that in \Def{d1426HB} any all-zero columns in $J$ are automatically ignored, since they become zero-weight sub-hypers in the sum and drop out automatically.
%	 provided some column was non-zero.
	 If however all columns of $J$ are zero, then its denotation $\Hyp{J}$ becomes automatically the weight-zero sub-point hyper.
	%\Cf{It's nice that we don't have to treat the weight-zero case specially.}
	
	\subsection{Abstract channels --- review}\label{s1011}
	In earlier work \cite{McIver:2014aa} we described an ``abstract channel'' as a function from prior distributions to hypers. We restate that here in our current denotational style:
	\begin{Definition}{Denotation of channel}{d1416CB}
		Let $C{\In}\CalX{\MFun}\CalY$ be a channel matrix. Its denotation, of type $\AHSpc$, is called an \emph{abstract channel} and is defined for $\pi{\In}\Dist\CalX$ by
		\[
		\AbCh{C}.\pi
		\Defs
		\Hyp{\ChApp{\pi}{C}}~,
		\]
		where the $\AbCh{\cdot}$ on the left is the denotational function for channels, being defined here; and on the right it is the denotational function on joint distributions that we have already from \Def{d1426HB}.\,%% just above.
(We use $\AbCh{-}$ uniformly for denotation functions, relying on context instead of e.g.\ using subscripts like $\Hyp{-}_{\rm chan}$ on the left and $\Hyp{-}_{\rm joint}$ on the right.)
	\end{Definition}
	
	In fact the prior $\pi$ can be recovered from $\ChApp{\pi}{C}$, as this definition shows:% using the average function:
	\begin{Definition}{Support of a distribution}{d1510}
		\label{g1041}Given discrete distribution $\delta{\In}\Dist\CalZ$, we write $\Supp{\delta}$ for the \emph{support} of $\delta$, the set of elements $z{\In}\CalZ$ for which $\delta.z$, the probability assigned by $\delta$ to $z$, is not zero. Obviously $\delta{\in}\Dist\CalZ$ implies $\Supp{\delta}{\subseteq}\CalZ$; if in fact $\Supp{\delta}{=}\CalZ$ then we say that $\delta$ is \emph{full support}.
	\end{Definition}
	\begin{Definition}{Average of a hyper}{d1154}\label{g1109}
		For hyper $\Delta{\In}\Dist^2\CalX$ define its average $\Avg.\Delta$ in $\Dist\CalX$ by
		\[
		\Avg.\Delta.x \Wide{\Defs} \sum_{\delta{\In}\Supp{\Delta}} (\Delta.\delta)(\delta.x) ~ \textrm{\quad\quad for all} ~x{\In}\CalX , \quad\footnotemark
		\]
		\label{g1140}where we use upper-case Greek for hypers.
	\end{Definition}
	\footnotetext{This $\Avg$ is multiplication $\mu$ from the Giry monad: see \Sec{s1250}.}
	We then have $\Avg.(\AbCh{C}.\pi) = \pi$, because
	{\small\[
		(\Avg.(\AbCh{C}.\pi))_x~~=~~(\Avg.\Hyp{\ChApp{\pi}{C}})_x~~=~~\SSum (\ChApp{\pi}{C})_{x,-}~~=~~\pi_x\quad.
		\]}%
	In fact $\Avg.\Hyp{J}$ for any $J$ in $\CalX{\MFun}\CalY$ is $J$'s $\CalX$ marginal in $\Dist\CalX$.
}%1506

\section{Classical- vs.\ abstract \HMM's}\label{s0939}
\subsection{Classical \HMM's, and single \HMM-steps as matrices}\label{s1403B}
Classically a Hidden Markov Model comprises a set $\CalX$ of states, a set $\CalY$ of observations and two stochastic matrices $C,M$ that give resp.\ the \emph{emission} probabilities $C_{x,y}$ that $x$ will emit observation $y$ and the \emph{transition} probabilities $M_{x,x'}$ that $x$ will change to $x'$ \cite{Jurafsky:00}. Usually, the \textit{homogeneous} case, computation evolves in (probabilistic) steps each determined by the same $C,M$, with each output state $x'$ becoming the following input $x$ and with the emissions $y$ accumulating. In our case however, \textit{heterogeneous}, we can vary the matrices from step to step, each standing for various (different) program fragments.

We show two computations in \Fig{f1420B}. If $\pi$ is the distribution of incoming $x$, the distribution $\pi''$ of intermediate $x''$ is $\pi{\MMult}M^1$. The distribution of observations $y^1$ is $\pi{\MMult}C^1$.  The second step's input $x''$ is the output of the first step.

A classical \HMM\ hides all of three of $x,x'',x'$, but still the observations $y^{1,2}$ tell us something about each of them provided we know $\pi,M^{1,2},C^{1,2}$. (This is analogous to knowing the source code of a program, but not being able to observe its variables as it executes.)

\begin{Figure}{f1420B}{Two successive steps $H^1$ and $H^2$ of a heterogeneous \HMM.}
	\begin{center}
	\includegraphics[scale=.6]{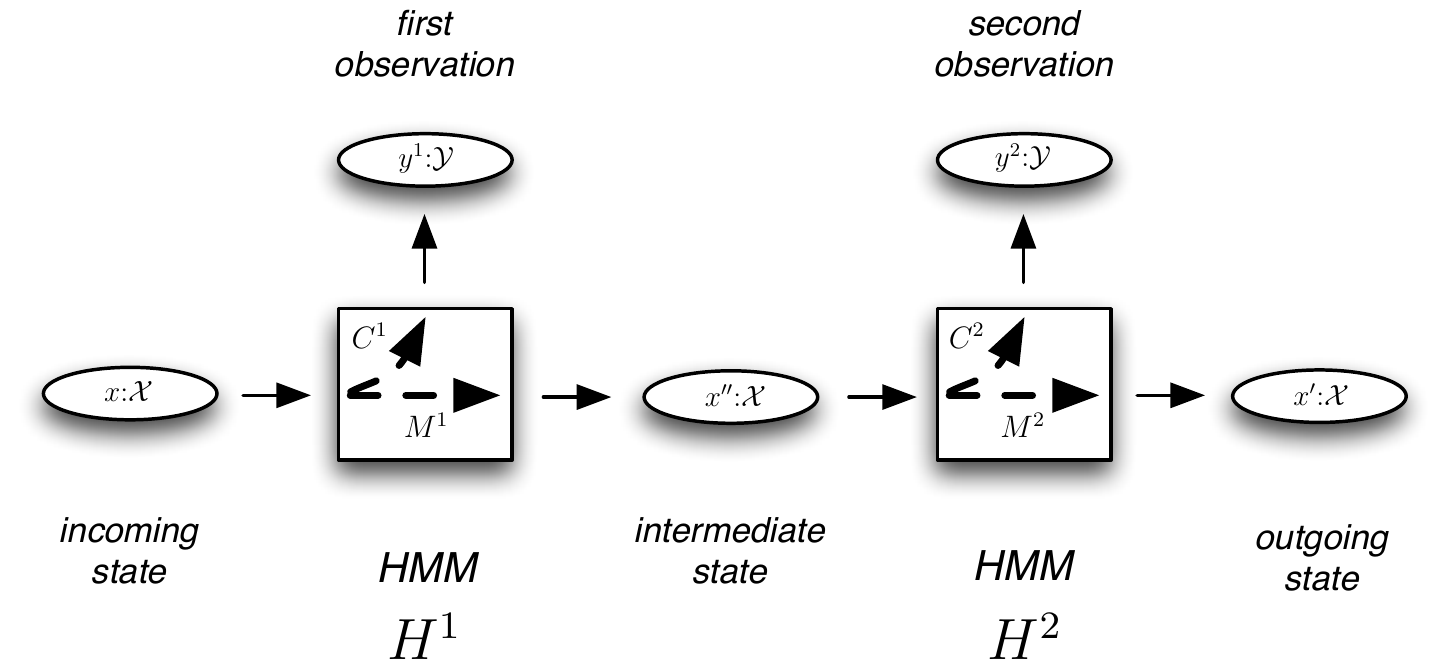}
	\end{center}
	\medskip
	\captionsetup{singlelinecheck=off,font=footnotesize}
	\caption*{Each step $H^{1,2}$ takes an input- to an output state in $\CalX$; the observations $y^{1,2}{\In}\CalY$ are accumulated. In each step $H^{1,2}$ the output state is determined by a markov $M^{1,2}$ on the input to that step, and the observation is determined independently by a channel $C^{1,2}$ on the same input, i.e.\ \emph{before} application of the markov.}
	\captionsetup{singlelinecheck=on,font=normalsize}
\end{Figure}

From now on we call the emission part of an \HMM\ the \emph{channel} and the transition part the \label{g0928}\emph{markov} (lower case).
\begin{Definition}{Single \HMM-step}{d1728}
	\label{g1117}Given channel $C{\In}\CalX{\MFun}\CalY$ and markov $M{\In}\CalX{\MFun}\CalX$,
	define the \HMM-matrix $\CM{C}{M}$ of type $\MH$ by 
	\[
	\CM{C}{M}_{x,y,x'}\Wide{\Defs}C_{x,y}\times M_{x,x'}~.
	\]
	This (row-1-summing) matrix $\CM{C}{M}$ produces a joint distribution of type $\Dist(\CalX{\times}\CalY{\times}\CalX)$, as top-left in \Fig{f1843}, once applied to a prior (\Def{d1124B}).\,%
\end{Definition}
Note that in $ \CM{C}{M}$ the probabilistic choices in $C$ (of $y$) and $M$ (of $x'$) are made independently;
although indeed $\CM{C}{M}$ has the property that for each $x{\In}\CalX$ the (remaining) joint distribution $\CM{C}{M}_{x,-,-}$ is independent in $y,x'$, this property is not preserved once steps are composed (\Sec{s1203}).

%\Cf{%0719
%Do we need to say somewhere that
%\begin{quote}
%Our ``a priori'' estimate of the output distribution is calculated by applying the markov to the (known) prior distribution on the input: that is, it is ``prior'' in the sense that it is our before-execution belief of what the final distribution will be. Our after-execution estimate of the final distribution is determined by the observations emitted by the \HMM's channel component, which we know only once the \HMM\ has run: therefore it is called ``a posteriori.'' This is consistent with the use of these terms for channels, because for a pure channel the final- and initial distributions on the state are the same: we use the channel's emissions (output) to revise our belief about both.
%\end{quote}
%}%0719

\subsection{Abstract \HMM's represent classical \HMM's}\label{s0913}
For abstract channels (\Sec{s1011}) we focussed on the hyper of posteriors on the \emph{input}; for \HMM's we focus on the hyper of posteriors on the \emph{output}, because \HMM's are computations and so it is over their outputs we wish to reason.\,
(The \emph{prior} on the output would be our calculation from the input prior and the markov of what the output distribution would be, but \emph{before} running the program and making observations in the type $\CalY$.)

\begin{Definition}{Matrix \HMM\ denotes abstract \HMM}{d1022}
	Let $H{\In}\MH$ be an \HMM\ presented as a matrix (stochastic in $y,x'$). Its denotation, of type $\AHSpc$, is called an \emph{abstract \HMM} and is defined $\HMMone{H}.\pi \Defs\Hyp{J}$, where $\pi{\In}\Dist\CalX$ and the joint-distribution matrix $J{\In}\CalX{\MFun}\CalY$ is given by $J_{x',y}\Defs\,
	\SSum_x\pi_x H_{x,y,x'}$.
\end{Definition}

In \Sec{s0917} we discuss the (Dalenius) implications of having abstracted from the \HMM's input (with the $\SSum_x$ just above) --- it is not always appropriate.

\subsection{Special cases of \HMM-steps: pure markovs}\label{s1407m}\label{s0652}
Markovs are the special case of \HMM\ where the channel-part effectively outputs nothing.
\label{g1124}If an \HMM-step $\CM{C}{M}$ has  for its channel $C$ an all-one column vector \NullChannel, 
where  \NullChannel\ stands for ``null channel''.
Then $\CalY$ is a singleton and $J$ becomes a column vector: i.e.\ $J_{x'}=\SSum_x\pi_xM_{x,x'}$, so that in fact $J$ is the usual matrix product $\pi{\MMult}M$.

Taking $\NullChannel$ as the default channel gives $\HMMsem{}{M}.\pi = \HMMsem{\NullChannel}{M}.\pi = \Point{\pi{\MMult}M}$, the point hyper on $\pi{\MMult}M$. This more general framework simply ``wraps'' a $\Point{-}$ around the final distributions; but it's that wrapping that enables treating markovs and channels within the same type.
A general $H$ is a markov just when $\SSum_{x'} H_{x,y,x'}$ is $\NullChannel$.

\begin{Figure}{f0544m}{Pure-markov \HMM\ program}
	{\tt\small
		\begin{tabbing}
			// \textit{\XS\ is initialised uniformly at random.} \\
			xs:= xs $\PC{{\nicefrac{1}{2}}}$ -xs \\
			// \textit{What does an attacker guess for \XS\ finally?}
	\end{tabbing}}
	\captionsetup{singlelinecheck=off,font=footnotesize}
	\caption*{\small The secret two-bit bit-string \XS\ is set initially from $\DSet{00,01,10,11}$ with equal probability $\nicefrac{1}{4}$ for each; the following assignment either leaves \XS\ unchanged (probability $\nicefrac{1}{2}$) or bit-wise inverts both components.}
	\captionsetup{singlelinecheck=on,font=normalsize}
\end{Figure}
Consider the program of \Fig{f0544m} whose single variable is a two-bit string \texttt{xs}.
We model it with $\CalX{=}\DSet{00,01,10,11}$; prior $\pi{\In}\Dist\CalX$ is uniform, and its markov $M$ is as just below:
\[
 \begin{array}{c@{\hspace{1.5em}}cccc}
  &\mbox{\Sm{00}}&\Sm{01}&\Sm{10}&\Sm{11} \\[.3em]
  \Sm{00{:}} & \LMat{1em}{-1.8em}{2.8em}\nicefrac{1}{2} & 0 & 0 & \nicefrac{1}{2}\RMat{1.1em}{-1.8em}{2.8em} \\
  \Sm{01{:}} & 0 & \nicefrac{1}{2} & \nicefrac{1}{2} & 0 \\
  \Sm{10{:}} & 0 & \nicefrac{1}{2} & \nicefrac{1}{2} & 0 \\
  \Sm{11{:}} & \nicefrac{1}{2} & 0 & 0 & \nicefrac{1}{2} 
 \end{array}
\]
The output distribution is of course $\pi'{=}\pi{\MMult}M{=}\pi$, and so the attacker's guess of the final state is optimally any of the four values in $\CalX$: they are equally good.

%\smallskip\begin{window}[0,r,%
%	$
%	\begin{array}{c@{\hspace{1.5em}}cccc}
%	&\mbox{\Sm{00}}&\Sm{01}&\Sm{10}&\Sm{11} \\[.3em]
%	\Sm{00{:}} & \LMat{1em}{-1.8em}{2.8em}\nicefrac{1}{2} & 0 & 0 & \nicefrac{1}{2}\RMat{1.1em}{-1.8em}{2.8em} \\
%	\Sm{01{:}} & 0 & \nicefrac{1}{2} & \nicefrac{1}{2} & 0 \\
%	\Sm{10{:}} & 0 & \nicefrac{1}{2} & \nicefrac{1}{2} & 0 \\
%	\Sm{11{:}} & \nicefrac{1}{2} & 0 & 0 & \nicefrac{1}{2} ~.
%	\end{array}
%	$, {}]
%	\noindent The output distribution is of course $\pi'{=}\pi{\MMult}M{=}\pi$, and so the attacker's guess is optimally any of the four values in $\CalX$: they are equally good.
%\end{window}

\bigskip
This system viewed as an abstract \HMM\ would give output \emph{hyper} $\Delta'=\HMMsem{}{M}.\pi=\Point{\pi}$, in fact the \emph{point} hyper on $\pi$ indicating that the attacker is certain (point-probability 1) that the posterior distribution $\pi'$ on the final value of \XS\ is equal to the prior $\pi$ in this case, i.e.\ it is still uniform.

\subsection{Special cases of \HMM-steps: pure channels}\label{s1407c}
Channels are the special case where the input- and the output state are the same. If  $\CM{C}{M}$ has markov $M$ as the identity \ID, then it is a ``pure channel'' with output the same as its input. In that case \Def{d1022} gives $J_{x',y}=\sum_x\pi_xC_{x,y}\ID_{x,x'}=(\ChApp{\pi}{C})_{x',y}$, and so $\HMMsem{C}{\ID}$ from \Def{d1022} is just $\AbCh{C}$ from \Def{d1416CB}.

\label{g1123}With $\ID$ as the default markov, we have $\HMMsem{C}{}=\AbCh{C}$.
%++\A{A general \HMM\ is a channel just when $\sum_y H_{x,y,x'}$ is $\ID$.}
%++\Cf{True? This is stronger than saying it's the identity on distributions ``of course''. {\Ax I have removed this last line which was queried "True?" for safety reasons.}}

\begin{Figure}{f0544c}{Pure-channel \HMM\ program.}
	{\tt\small
		\begin{tabbing}
			// \textit{\XS\ is initialised uniformly at random.} \\
			{\Leak}~~xs[0]$\PC{{\nicefrac{1}{2}}}$\,xs[1] \\
	\end{tabbing}}
	\captionsetup{singlelinecheck=off,font=footnotesize}
	\caption*{\small The value of either bit 0 or bit 1 of \XS\ is revealed; the attacker learns that value, but does not know which bit it came from. 
		
	What should he guess for \XS\ after execution in this case?}
	\captionsetup{singlelinecheck=on,font=normalsize}
\end{Figure}
Now consider \Fig{f0544c} where some of \XS\ is leaked, but \XS\ itself is not changed. Thus our state $\CalX$ and prior $\pi$ are as before, the observation space is $\CalY{=}\DSet{0,1}$ and the channel $C$ representing this program is here at left:
\[
C=
\begin{array}{c@{\hspace{1.7em}}cc}
&\Sm{0}&\Sm{1} \\
\Sm{00{:}} & \LMat{1em}{-1.8em}{2.8em}1 & 0\RMat{1.1em}{-1.8em}{2.8em} \\
\Sm{01{:}} & \nicefrac{1}{2} & \nicefrac{1}{2}\\
\Sm{10{:}} & \nicefrac{1}{2} & \nicefrac{1}{2} \\
\Sm{11{:}} & 0 & 1
\end{array}
\hspace{3em}
J=
\begin{array}{c@{\hspace{1.7em}}cc}
&\Sm{0}&\Sm{1} \\
\Sm{00{:}} & \LMat{1em}{-1.8em}{2.8em}\nicefrac{1}{4} &0\RMat{1.1em}{-1.8em}{2.8em} \\
\Sm{01{:}} & \nicefrac{1}{8} & \nicefrac{1}{8}\\
\Sm{10{:}} & \nicefrac{1}{8} & \nicefrac{1}{8} \\
\Sm{11{:}} & 0 & \nicefrac{1}{4}
\end{array}
\]
The joint distribution in $x',y$ is $J{=}\ChApp{\pi}{C}$. The construction of \Def{d1426HB} gives us a hyper $\Delta'$ as
\label{g1130}
\begin{equation}\label{e0852n}
\begin{array}{r@{\hspace{3em}}l@{\,}l}
\textrm{inner distributions} & \multicolumn{2}{@{}l}{\textrm{outer distribution}} \\
\CVec{\nicefrac{1}{2},\nicefrac{1}{4},\nicefrac{1}{4},0} &\AtP& \nicefrac{1}{2} \\
\CVec{0,\nicefrac{1}{4},\nicefrac{1}{4},\nicefrac{1}{2}} &\AtP& \nicefrac{1}{2}~,
\end{array}
\end{equation}%
\label{g0924}%
where in general we write $z_1\AtP p_1,~z_2 \AtP p_2,\cdots$ for the discrete distribution that assigns probability $p_1$ to $z_1$ etc. In \Eqn{e0852n} the values $z_1,z_2$ are themselves (inner, posterior) distributions. This hyper shows that with probability $\nicefrac{1}{2}$ an optimal attacker will guess \verb+00+ (because she saw a \verb+0+ {\Leak}ed, and deduces \emph{a posteriori} that \verb+00+ now has the highest probability, twice either of the others); and with probability $\nicefrac{1}{2}$ the attacker will guess \verb+11+ (because she saw a~\verb+1+).

%We have abstracted from the {\Leak}ed-out $\CalY$-values, i.e.\ what he saw, concentrating simply on what he deduces.

\section{Beyond steps: \HMM\ programming and sequential composition}\label{s1203}
\subsection{Classical \HMM\ composition: matrices}\label{s1243}
\label{g1134}Let $H^1,H^2{\In}\MH$ be two \HMM's. Their sequential composition $H=H^1;H^2$ describes the distribution on $x$, and $y_{1,2}$ together, and $x'$ as
\begin{equation}\label{e0946}
(H^1;H^2)_{x,(y_1,y_2),x'} \Wide{\Defs} \sum_{x''} H^1_{x,y_1,x''}H^2_{x'',y_2,x'} ~.
\end{equation}
%This can also be seen as rewriting $H^1$ as type $\CalX{\times}\CalY{\MFun}\CalX$, then matrix-multiplying by $H^2$, and then re-converting the resulting $\CalX{\times}\CalY{\MFun}\CalY{\times}\CalX$ back to $\CalX{\MFun}{(\CalY{\times}\CalY)}{\times}\CalX$.\,%
%\footnote{Lambert Meertens pointed out this nice formulation.}
Note how the set of observables is now $\CalY{\times}\CalY$, compounding the observations $\CalY$ from each component. (This is why infinite composition of \HMM's cannot easily be represented as a finite matrix.)

Remarkably, the action of \HMM-composition on pure-markovs \HMM's is effectively their matrix multiplication, yet its action on pure channels is effectively their ``parallel composition'': thus a single general definition of composition specialises automatically to the two principal sub-cases, as we now show.
%%%%%
First, we give the details for classical \HMM's; then \Thm{t1008} shows that the same holds for abstract \HMM's.

\subsubsection{Composition of pure markovs}
The usual composition of Markov matrices $M^{1,2}{\In}\CalX{\MFun}\CalX$ is via matrix multiplication $M^1{\MMult}M^2$, and the result is of the same type $\CalX{\MFun}\CalX$. If we do it at the \HMM-level, we find
\begin{Reason}
	\Step{}{
		\CM{}{M^1}{;}\CM{}{M^2}~_{x,(y_1,y_2),x'}
	}
	\Step{$=$}{
		\sum_{x''} \CM{}{M^1}_{x,y_1,x''}\CM{}{M_2}_{x'',y_2,x'}
	}
	\StepR{$=$}{Recall from \Sec{s0652} that channel $\NullChannel$ reveals nothing.}{
		\NullChannel_{x,(y_1,y_2)}\sum_{x''}M^1_{x,x''}M^2_{x'',x'}
	}
	\Step{$=$}{
		\CM{}{~M^1{\MMult}M^2}\quad _{x,(y_1,y_2),x'} ~,
	}
\end{Reason}
so that indeed $\CM{}{M^1}{;}\CM{}{M^2} = \CM{}{~M^1{\MMult}M^2}$.

\subsubsection{Composition of pure channels}
\label{g1136}Parallel composition of channels, which we write $\ChPar{C^1}{C^2}$, models applying \emph{both} channels to the same input and observing both outputs. Thus
\[
\ChPar{C^1}{C^2}\quad_{x,(y_1,y_2)}
\Wide{=}
C^1_{x,y_1}\Atimes C^2_{x,y_2} ~.
\]
This is different from channel \emph{cascading}, which applies the second channel $C^2$ to the observations of the first channel $C^1$ via matrix multiplication. A striking distinction is that the cascade of $C^1$ into $C^2$ releases no more information that $C^1$ alone (the Data-Processing Inequality \cite{Cover:2006aa}), whereas $\ChPar{C^1}{C^2}$ releases no \emph{less} information that either of $C^{1,2}$ alone. In this latter case we find
\begin{Reason}
	\Step{}{
		\CM{C^1}{}{;}\CM{C^2}{}~_{x,(y_1,y_2),x'}
	}
	\Step{$=$}{
		\sum_{x''} \CM{C^1}{}_{x,y_1,x''}\CM{C^2}{}_{x'',y_2,x'}
	}
	\StepR{$=$}{{Recall from \Sec{s1407c} markov $\ID$ is the identity.}}{
		\sum_{x''} C^1_{x,y_1}\ID_{x,x''}C^2_{x'',y_2}\ID_{x'',x'}
	}
	\Step{$=$}{
		C^1_{x,y_1}C^2_{x,y_2}\ID_{x,x'}
	}
	\Step{$=$}{
		(\ChPar{C^1}{C^2})_{x,(y_1,y_2)}\ID_{x,x'}
	}
	\Step{$=$}{
		\CM{\ChPar{C^1}{C^2}~}{}~ _{x,(y_1,y_2),x'} ~,
	}
\end{Reason}
so that indeed again $\CM{C^1}{}{;}\CM{C^2}{} = \CM{\ChPar{C^1}{C^2}~}{~}$.

\subsubsection{Pure channel followed by pure markov}\label{a1530}
Finally, note that a general \HMM-step (\Sec{s1403B}) is a pure channel followed by a pure markov. Let  $\NullChannel_{x,y_2}$ be $(\textrm{$1$ if $y_2{=}\hat{y}$ else $0$})$ for some fixed $\hat{y}$ in $\CalY$, and calculate
\begin{Reason}
	\Step{}{
		\CM{C}{}{;}\CM{}{M}\quad_{x,(y_1,y_2),x'}
	}
	\Step{$=$}{
		\sum_{x''} \CM{C}{}_{x,y_1,x''}\CM{}{M}_{x'',y_2,x'}
	}
	\Step{$=$}{
		\sum_{x''} C_{x,y_1}\ID_{x,x''}\NullChannel_{x'',y_2}M_{x'',x'}
	}
	\StepR{$=$}{$\ID_{x'',x'}$, 1-point rule}{
		C_{x,y_1}M_{x,x'}\NullChannel_{x,y_2}
	}
	\StepR{$=$}{above}{
		\textrm{$C_{x,y_1}M_{x,x'}$ if $y_2{=}\hat{y}$ else $0$}
	}
	\Step{$=$}{
		\CM{C}{M}_{x,(y_1,y_2),x'}~,
	}
\end{Reason}
so that $\CM{C}{}{\,;\,}\CM{}{M} = \CM{C}{M}$.\,%

The reason that $\CM{C}{}{;}\CM{}{M}$ and $\CM{}{M}{;}\CM{C}{}$ differ in general is that in the (mathematical) definition of an \HMM-step (e.g.\ \Fig{f1420B}) the emissions are determined by the input, initial state (rather than the output, final state). Had that original definition been the other way around, then we'd have had $\CM{}{M}{;}\CM{C}{}$ as an \HMM-step.

\subsubsection{Pure markov followed by pure channel}
This cannot, in general, be reduced to a single \HMM-step. In $\CM{}{M};\CM{C}{}$ let both $C,M$ be the identity. Then the observations and final state will be perfectly correlated, something that is not possible for single \HMM-step $\CM{C'}{M'}$.

\subsection{Abstract \HMM's: Kleisli composition}\label{s1105}
Now we consider $h_1;h_2$ where $h_{1,2}$ are abstract \HMM's.\,%
(We use upper-case for matrices and lower-case for denotations.)
Because the components' types $\AHSpc$ do not match directly, i.e.\ the co-domain $\Dist^2\CalX$ from the left is not the domain $\Dist\CalX$ required on the right, we use Kleisli composition for that.\,%
\footnote{This is the usual composition in a Kleisli category. See \Sec{s1250}.}
\begin{Definition}{Push-forward of a function}{d1341B}\label{g1108}
	Given sets $\CalZ,\CalZ'$ and function $f{\In}\CalZ{\Fun}{\CalZ'}$, we write $\Dist f$ for the \emph{push-forward} of $f$, a ``lifted'' function of type $\Dist\CalZ{\Fun}\Dist\CalZ'$ \cite{Fremlin:00}. For $z'{\In}\CalZ'$ and $\delta{\In}\Dist\CalZ$ we have
	\footnote{Lifting, as in $\Dist f$, binds tightest: the conventional notation for $\Dist f.\delta.z'$ would be $(\Dist f)(\delta)(z')$, so that $(\Dist f)(\delta){\in}\Dist\CalZ'$.}
	\[
	\Dist f.\delta.z' \Wide{\Defs} \sum_{\substack{z{\In}\CalZ\\f.z=z'}} \delta.z ~.\quad\footnotemark
	\LiftBox\]
	\footnotetext{$\Dist f$ is the action of functor $\Dist$ on arrow $f$: see \Sec{s1250}.}%In general $f$ must be measurable.}
\end{Definition}

\begin{Definition}{Kleisli composition of abstract \HMM's}{d1210}
	~\\Given two abstract \HMM's
	%\Cf{Measurable?}
	$h_{1,2}{\In}\AHSpc$, their Kleisli composition is defined
	\[ 
		(h_1;h_2).\pi\Wide{\Defs}\Avg.(\Dist h_2.(h_1.\pi))
	\]
for $\pi{\In}\Dist\CalX$, where $\Dist h_2$ is as above the push-forward of $h_2$.
Using functional composition, equivalently $h_1;h_2\Defs\Avg\Comp\Dist h_2\Comp h_1$.
\end{Definition}
That is, the lifting inherent in Kleisli-composition applies the right-hand abstract \HMM\ $h_2$ to each inner (i.e.\ posterior) produced by the left-hand $h_1$ from prior $\pi$, preserving the way in which they are all combined together by the outer distribution. Then the intermediate result, of type $\Dist^3\CalX$, is averaged to bring it back to the required type $\Dist^2\CalX$. 

\subsection{Proof that composition is faithfully denoted}\label{s1008}
It is important (though unsurprising) for our interpretation that composition of \HMM's as matrices \Eqn{e0946} is correctly mapped by $\HMMone{\cdot}$ to their Kleisli composition as abstract \HMM's (\Def{d1210}). That is, we expect

\begin{Theorem}{Composition faithfully denoted}{t1008}
	Let $H^{1,2}{\In}\MH$ be \HMM's as matrices. Then we have
	\[
	\HMMone{H^1;H^2} \Wide{=} \HMMone{H^1};\HMMone{H^2}~,
	\]
	where \Eqn{e0946} is used on the left and \Def{d1210} on the right.
	\Proof
	%Given in \App{a1008}.
	We reason as follows for any $\pi$. 
	\begin{Reason}
		\Step{}{(\HMMone{H^1};\HMMone{H^2}).\pi}
		\Step{$=$}{\Avg.(\Dist\HMMone{H^2}.(\HMMone{H^1}.\pi))}
		\Step{$=$}{\Avg.(\Dist\HMMone{H^2}.\Hyp{\HMMone{\pi}{H^1}})}   
		\Step{$=$}{\Avg.(\Dist\HMMone{H^2}.(\sum_{y^1} \SubPt{(\HMMone{\pi}{H^1})_{y^1,-}} ))}
		\StepR{$=$}{$\Dist f$ distributes $f$ through inners.}{(\sum_{y^1} \HMMone{H^2}.(\HMMone{\pi}{H^1})_{y^1,-})}
		\Step{$=$}{(\sum_{y^1} \Hyp{\HMMone{(\HMMone{\pi}{H^1})_{y^1,-}\,}{\,H^2}})}
		\Step{$=$}{(\sum_{y^1}(\sum_{y^2} \SubPt{(\HMMone{(\HMMone{\pi}{H^1})_{y^1,-}\,}{\,H^2})_{y^2,-}}))}
		\StepR{$=$}{\Lem{l1257}}{(\sum_{y^1,y^2}\SubPt{(\HMMone{\pi}{(H^1;H^2)})_{(y^1,y^2),-}})}
		\Step{$=$}{\HMMone{H^1;H^2}.\pi~,}
	\end{Reason}
	as required.
	\Cf{Probably the general lemma we're looking for at ``justify this'' is that for any linear $f$ from $\VecV\CalX$ to $\VecV\CalZ$ we have $(\Avg\Comp\Dist f).\Hyp{J} = (\sum_y f.(J_{-,y})).$}
\end{Theorem}

\begin{Lemma}{Double application of \HMM\ matrix}{l1257} 
	We have
	\[
	(\HMMone{\pi}{(H^1;H^2)})_{(y^1,y^2),-} \Wide{=}  (\HMMone{(\HMMone{\pi}{H^1})_{y^1,-}}{H^2})_{y^2,-}
	\]
	from this calculation for any $x'$ that
	\begin{Reason}
		\Step{}{(\HMMone{\pi}{(H^1;H^2)})_{(y^1,y^2),{x'}}}
		\Step{$=$}{(\sum_x \pi_x\,(H^1;H^2)_{x,(y^1,y^2),{x'}})}
		\Step{$=$}{(\sum_x \pi_x\,(\sum_{x''}H^1_{x,y^1,x''}H^2_{x'',y^2,x'}))}
		\Step{$=$}{(\sum_{x,x''} \pi_x\,H^1_{x,y^1,x''}H^2_{x'',y^2,x'})}
		\Step{$=$}{(\sum_{x''} (\sum_x \pi_x\,H^1_{x,y^1,x''})H^2_{x'',y^2,x'})}
		\Step{$=$}{(\sum_{x''} (\HMMone{\pi}{H^1})_{y^1,x''}H^2_{x'',y^2,x'})}
		\Step{$=$}{(\HMMone{(\HMMone{\pi}{H^1})_{y^1,-}}{H^2})_{y^2,x'}~,}
	\end{Reason}
	as required.
\end{Lemma}

\begin{Figure}{f0544cm}{\HMM\ program as sequential composition.}
	{\tt\small
		\begin{tabbing}
			// \textit{\XS\ is set uniformly at random.} \\
			{\Leak} xs[0] $\PC{{\nicefrac{1}{2}}}$ xs[1]~; \\
			xs:= xs $\PC{{\nicefrac{1}{2}}}$ -xs \\
	\end{tabbing}}
	\captionsetup{singlelinecheck=off,font=footnotesize}
	\caption*{    \small The value of either bit 0 or bit 1 of \XS\ is revealed; the attacker learns that value, but does not know which bit it is. Then \XS\ is either unchanged or inverted, but the attacker does not know which.
		
	What's his best guess now for the final value of \XS?}
	\captionsetup{singlelinecheck=on,font=normalsize}
\end{Figure}

\subsection{Channel/markov together: two examples of composition}\label{s0825}
\label{a0407}

For an example of sequential composition we return to \XS\ and consider \Fig{f0544cm} where the state is both leaked \emph{and} (possibly) changed.
The final hyper $\Delta'$ in this case is obtained by applying the markov $M$ to the \emph{inners} generated by $C$ in \Sec{s0652} while retaining their outers: that gives
%\pagebreak
\[%\begin{equation}\label{e0654}
	\hspace{-.5em}
	\begin{array}[t]{r@{~}r@{\hspace{1em}}l@{}l@{\,}l}
	%   \multicolumn{2}{r}{\textrm{``inner'' distributions}} & \multicolumn{2}{@{}l}{\textrm{``outer'' distribution}} \\
	(& \nicefrac{1}{2}{\times}\nicefrac{1}{2}+\nicefrac{1}{2}{\times}0, \\
	& \nicefrac{1}{2}{\times}\nicefrac{1}{4}+\nicefrac{1}{2}{\times}\nicefrac{1}{4}, \\
	& \nicefrac{1}{2}{\times}\nicefrac{1}{4}+\nicefrac{1}{2}{\times}\nicefrac{1}{4}, \\
	& \nicefrac{1}{2}{\times}0+\nicefrac{1}{2}{\times}\nicefrac{1}{2}\,&) &\AtP& \nicefrac{1}{2} \\[.5em]
	\end{array}\hspace{1em}\begin{array}[t]{{r@{~}r@{\hspace{1em}}l@{}l@{\,}l}}
	(& \nicefrac{1}{2}{\times}0+\nicefrac{1}{2}{\times}\nicefrac{1}{2}, \\
	& \nicefrac{1}{2}{\times}\nicefrac{1}{4}+\nicefrac{1}{2}{\times}\nicefrac{1}{4}, \\
	& \nicefrac{1}{2}{\times}\nicefrac{1}{4}+\nicefrac{1}{2}{\times}\nicefrac{1}{4}, \\
	& \nicefrac{1}{2}{\times}\nicefrac{1}{2}+\nicefrac{1}{2}{\times}0\,&) &\AtP& \nicefrac{1}{2}
	\end{array}
\]
\noindent which is simplified first to this 
\[
	\begin{array}{r@{\hspace{1.2em}}l@{\,}l}
	\CVec{\nicefrac{1}{4},\nicefrac{1}{4},\nicefrac{1}{4},\nicefrac{1}{4}} &\AtP& \nicefrac{1}{2} \\
	\CVec{\nicefrac{1}{4},\nicefrac{1}{4},\nicefrac{1}{4},\nicefrac{1}{4}} &\AtP& \nicefrac{1}{2}
	\end{array}	
\]
and then, since the two inners are the same, as a hyper-distribution is collapsed to just the singleton hyper $\Point{\pi}$,
where we are using an explicit $(\times)$ for multiplication of specific numbers.

Thus the program of \Fig{f0544cm} reveals nothing about the final value of \XS\ when the initial distribution was uniform.
Informally we would explain this by noting that the information about \XS\ released by the \texttt{{\Leak}} becomes ``stale'', irrelevant once we do not know whether \XS\ has subsequently been inverted or not. (See \Sec{s0917} however for a discussion of why the initial value of \XS\ might in some cases still be important.)

\smallskip % \end{window} loses some space.
It would be wrong however to conclude, from
$\Delta'{=}\Point{\pi}$
in this specific case, that the program is secure for \XS\ in general --- for when the initial distribution is \emph{not} uniform, the final value of \XS\ \emph{can} be less secure than the initial. This illustrates the danger in assuming something is uniformly distributed simply because we know nothing about it. (See \App{a0407}.)

We now reconsider \Fig{f0544cm} but with a non-uniform prior, showing that indeed the conclusion that the program was (wrt.\ the final state) ``leak free'' is unjustified. In \Fig{f0544cmnu} the initial hyper is ``skewed'', i.e.\ it is not uniform over the whole type \textit{XS} of \texttt{xs}, but rather is concentrated on only three of its values:
\begin{Figure}{f0544cmnu}{Simple-channel program excluding \XS=00 initially.}
	{\tt\small
		\begin{tabbing}
			// \textit{\XS\ is set uniformly from \{01,10,11\}.} \\
			{\Leak} xs[0] $\PC{{\nicefrac{1}{2}}}$ xs[1]~; \\
			xs:= xs $\PC{{\nicefrac{1}{2}}}$ -xs \\
	\end{tabbing}}
	\captionsetup{singlelinecheck=off,font=footnotesize}
	\caption*{\small This system is as in \Fig{f0544cm} except that the prior initial distribution differs: at least one bit of \XS\ is known to be 1.}
	\captionsetup{singlelinecheck=on,font=normalsize}
\end{Figure}
\begin{equation}\label{e0651a}
\begin{array}{r@{\hspace{3em}}l@{\,}l}
\CVec{0,\nicefrac{1}{3},\nicefrac{1}{3},\nicefrac{1}{3}} &\AtP& 1~,
\end{array}
\end{equation}
so that with certainty ($\AtP1$) it is known that the initial distribution is $\CVec{0,\nicefrac{1}{3},\nicefrac{1}{3},\nicefrac{1}{3}}$. Via the first statement \texttt{{\Leak} xs[0] $\PC{{\nicefrac{1}{2}}}$ xs[1]} an attacker will with probability $\nicefrac{1}{3}$ (resp.\ $\nicefrac{2}{3}$) observe \verb+0+ (resp.\ \verb+1+) and revise his belief of \XS's distribution as in the first (resp.\ second) row here:
\[
\begin{array}{r@{\hspace{3em}}l@{\,}l}
\CVec{0,\nicefrac{1}{2},\nicefrac{1}{2},0} &\AtP& \nicefrac{1}{3} \\
\CVec{0,\nicefrac{1}{4},\nicefrac{1}{4},\nicefrac{1}{2}} &\AtP& \nicefrac{2}{3}
\end{array}
\]
And after the second statement \texttt{xs:= xs $\PC{{\nicefrac{1}{2}}}$ -xs} the hyper for the \emph{current} (and final) distribution of $\XS$ will have become
\begin{equation}\label{e0651b}
\begin{array}{r@{\hspace{3em}}l@{\,}l}
\CVec{0,\nicefrac{1}{2},\nicefrac{1}{2},0} &\AtP& \nicefrac{1}{3} \\
\CVec{\nicefrac{1}{4},\nicefrac{1}{4},\nicefrac{1}{4},\nicefrac{1}{4}} &\AtP& \nicefrac{2}{3} ~,
\end{array}
\end{equation}
where in the $\nicefrac{1}{3}$-case he is better off finally than initially (since he knows \XS\ cannot be 00 or 11), but in the other case he is worse off (since \XS=00 has become possible). Thus if the attacker's choice is either to guess \XS's initial value or to run the program and guess \XS's final value, he can use these hypers to help make up his mind depending on his own criteria for the utility of his planned theft, that is the social context in which he is operating.\,%
Compare for example a thief's two alternatives for stealing a credit card: she might \emph{``Steal it now, since the wallet is just sitting there.''} or she might \emph{``Steal it after the card is used at an ATM where she can see some digit of the PIN.''} But in the second case there is a risk her victim will notice her, and choose a new PIN.

For example, the Shannon entropy of \XS\ is initially $\lg(3){\sim}1.6$, but finally, it is conditionally $\nicefrac{1}{3}{\Atimes}1+\nicefrac{2}{3}{\Atimes}2=\nicefrac{2}{3}{>}1.6$: if the attacker is using Shannon entropy to make his decision, he should act sooner rather than later.

On the other hand, the one-guess probability (R{\'e}nyi min-entropy) of \XS\ is initially $\nicefrac{1}{3}$; and finally it is the same, at $\nicefrac{1}{3}{\Atimes}\nicefrac{1}{2}+\nicefrac{2}{3}{\Atimes}\nicefrac{1}{4}=\nicefrac{1}{3}$. 
If the attacker is using this criterion, it does not matter when he acts.
\Cf{Can we find one where in the second case he is better off acting later?}

In either case, the hypers \Eqn{e0651a} and \Eqn{e0651b} contain all the information necessary for his decision: the bit-values printed are themselves not important \emph{for his decision}, which is why we can quotient our semantics by abstracting from them. (He does, however, need those values when he \emph{makes} his attack if indeed he decides ``later''.)

These calculations are confirmed in the next section.

\Section{Overview of Haskell-monadic prototype}% \AppFrom{\Sec{s0825},\Sec{s1418}}}
\label{a1409}

A Haskell prototype of our hyper-based monadic model has been constructed for discrete, finite \HMM's, and it has been applied to our examples of Figs.~\ref{f0544m}--\ref{f0544cmnu} \cite{Gibbons:2018aa}. We give a brief summary here.\,%
%\footnote{We have altered some of Schrijvers' variable names and types, in order better to correspond with our presentation above.}

A discrete probability distribution on a set $\CalX$ is modelled as a monadic type \verb+Dist x+ that is effectively a list \verb+[(x,Rational)]+ of elements from $\CalX$ and their associated probabilities. The type of (discrete) hypers $\Dist^2\CalX$ is then \verb+Dist(Dist x)+.

A Markov ``matrix'' on $\CalX$ is of type \verb+x->Dist x+, in fact encoding the matrix as a function from row-indices to distributions $\Dist\CalX$; a channel matrix is of type \verb+x->Dist y+ for any type $\CalY$ of observations whatever.

The mini- programming language has two elementary statements: to use a markov \verb+mm+ we have an \verb+update mm+ that updates the state according to \verb+mm+.
Note that \verb+mm+ is a \emph{Markov matrix}, but \verb+update mm+ is a \emph{markov \HMM} that is constructed from \verb+mm+, i.e.\ implements it.

To use a channel \verb+cm+ we have a \verb+reveal cm+ that emits (e.g.\ prints) the channel's output wrt.\ the hidden state at that point: the state is not changed and, in particular, the output is not assigned to anything. It is merely observed. Both of these statements are of type \verb+Dist x->Dist(Dist x)+, modelling our $\AHSpc$. In fact they are in $\AHS$, as Lemmas~\ref{l1432},\ref{l1433} show.

Sequential composition $({;})$ of programs is the Kleisli composition \verb+>=>+ provided by Haskell's conventions for monads, in this case the monad \verb+Dist+. Using that, and relying on \Sec{s1008} and \App{a1530}, we can define an elementary \HMM-step (\Sec{s1403B}) as
\begin{quote}\tt
	\begin{tabular}{cl}
		& hmmStep cm mm  \\
		= & reveal cm >=> update mm ~{\rm.}
	\end{tabular}
\end{quote}
Thus \verb+hmmStep+ does not have to be primitive.

Our example space $\CalX$ is \verb+(Bool,Bool)+, representing the bit-pair \XS, and our two example (input) priors are
\begin{quote}\tt
	\begin{tabular}{lcll}
		uniform &=& [ & ((False,False),1\%4), \\
		& &   & ((False,True),1\%4), \\
		& &   & ((True,False),1\%4), \\
		& &   & ((True,True),1\%4) \\
		& & ] \\[1ex]
		{\it and from \App{a1105}} \\[1ex]
		skewed &=& [ & ((False,False),0), \\
		& &   & ((False,True),1\%3), \\
		& &   & ((True,False),1\%3), \\
		& &   & ((True,True),1\%3) \\
		& & ] \\[1ex]
		
	\end{tabular}
\end{quote}
Our example observation space $\CalY$ is \verb+Bool+.

With suitable definitions typed as above for channel \verb+oneBit+ that outputs one of \verb+xs+'s two bits, uniformly at random, and \verb+invert+ that either inverts \verb+xs+ or does not, again uniformly at random, the four programs are then
\begin{quote}\tt
	\begin{tabular}{lclll}
		fig\ref{f0544m}    &=&     & update invert \\
		fig\ref{f0544c}    &=&     & reveal oneBit \\[1ex]
		fig\ref{f0544cm}   &=&     & -- Uniform prior. \\
		& &     & reveal oneBit \\
		& & >=> & update invert \\[1ex]
		fig\ref{f0544cmnu} &=&     & -- Skewed prior. \\
		& &     & reveal oneBit \\
		& & >=> & update invert \\[1ex]
	\end{tabular}
\end{quote}

The programs are run using the function
\begin{quote}\tt
	\begin{tabular}{cl}
		& runOn prior prog \\
		= & pretty (prog prior)
	\end{tabular}
\end{quote}
where \verb+pretty+ is an output-formatting function that prints hypers in a readable way.\,%
%\footnote{Ironically, the pretty-printing function's definition is the longest of any individual function in the prototype.}

The results of running the programs are as follows, where the third column gives the outer probabilities of the resulting hyper, and the first two columns give the corresponding inner distributions. We print \verb+True+,\verb+False+ as \verb+1+,\verb+0+ respectively:\,%
(The prototype prints probabilities as fractions; but here they are printed as reals, for neatness.)
\begin{quote}\tt
	runOn uniform fig\ref{f0544m} =
	\begin{tabular}[t]{lll@{\hspace{5ex}}p{20ex}}
		00 & 0.25 & 1.0 & \it point hyper \\
		01 &  0.25 \\
		10 & 0.25 \\   
		11 & 0.25
	\end{tabular}
\end{quote}

\bigskip
\begin{quote}\tt
	runOn uniform fig\ref{f0544c} =
	\begin{tabular}[t]{lll@{\hspace{5ex}}p{25ex}}
		01 & 0.25 & 0.5 & \it Half the time\ldots \\
		10 &  0.25 \\
		11 & 0.5 && \it {\tt 11} is most likely, and \\[1ex]   
		00 & 0.5 & 0.5 & \it the other half it's {\tt 00}. \\
		01 &  0.25 \\
		10 & 0.25 
	\end{tabular}
\end{quote}

\bigskip
\begin{quote}\tt
	runOn uniform fig\ref{f0544cm} =
	\begin{tabular}[t]{lll}
		00 & 0.25 & 1.0 \\
		01 &  0.25 \\
		10 & 0.25 \\   
		11 & 0.25
	\end{tabular}
\end{quote}

\bigskip
\begin{quote}\tt
	runOn skewed fig\ref{f0544cmnu} =
	\begin{tabular}[t]{lll}
		00 & 0.25 & 0.67 \\
		01 & 0.25 \\
		10 & 0.25 \\
		11 & 0.25 \\[1ex]   
		01 & 0.5 & 0.33 \\
		10 & 0.5
	\end{tabular}
\end{quote}

The prototype contains also a \verb+repeat+ feature: for example \verb+repeat 10 (reveal oneBit)+ is a program that reveals a random bit of \verb+xs+ 10 times independently.
(Such an iteration of parallel compositions is sometimes called ``repeated independent runs.'')
With the uniform prior we would expect that the resulting hyper would have three inners: one of them, occurring with probability approximately $\nicefrac{1}{2}$, would correspond to the case where the input bits of \XS\ differed, in which case with overall probability $\nicefrac{1023}{1024}$ there would be two different revelations among the 10 instances --- thus showing that indeed the bits differed. But we would still have no (more) information about whether the input was \verb+01+ or \verb+10+.

The remaining $\nicefrac{1}{1024}$ would be split between two cases: bit \verb+xs[0]+ was revealed every time, or \verb+xs[1]+ was; and those outcomes would contribute to the other two inners.

Those other two inners would have probability approximately $\nicefrac{1}{4}$ each, corresponding to input \verb+00+ or \verb+11+ where the two bits are the same. The program confirms this, giving
\begin{quote}\tt
	runOn uniform \\ (repeat 10 (reveal oneBit)) =\\[1ex]
	\begin{tabular}[t]{lll@{\hspace{3ex}}p{20ex}}
		01 & 1/1026 & 513/2048 & \it outer is about $\nicefrac{1}{4}$ \\
		10 & 1/1026 \\
		11 & 512/513  & \multicolumn{2}{l}{\it inner is ``almost certainly {\tt 11}''} \\[1ex]
		01 & 1/2 & 511/1024 & \it about $\nicefrac{1}{2}$ \\
		10 & 1/2 \\[1ex]
		00 & 512/513 & 513/2048 & \it about $\nicefrac{1}{4}$ \\
		10 & 1/1026 \\
		11 & 1/1026
	\end{tabular}
\end{quote}
(where this time we preserve the fractions).
The small perturbations away from $\nicefrac{1}{4}$ etc.\ reflect the small chance, mentioned above, that even when the inputs differ the random \verb+oneBit+ reveals the same bit 10 times in a row.

Finally, if we run the same program but with the final probabilistic inversion included, we get
\begin{quote}\tt
	runOn uniform \\ (repeat 10 (reveal oneBit) \\ >=> update invert) = \\[1ex]
	\begin{tabular}[t]{lll@{\hspace{3ex}}p{20ex}}
		00 & 256/513 & 513/1024 & \it two inners merged \\
		01 & 1/1026 \\
		10 & 1/1026 \\
		11 & 256/513 & \multicolumn{2}{l}{\it about $\nicefrac{1}{2}$} \\[1ex]
		01 & 1/2 & 511/1024 & \it about $\nicefrac{1}{2}$ \\
		10 & 1/2 \\[1ex]
	\end{tabular}
\end{quote}
in which the two ``bits equal'' inners from just above have merged: although the probabilistic inversion preserves the information concerning whether the bits are equal, it conceals in the equals case whether they were both \verb+00+ or both \verb+11+.

\section{The structure of hyper-space}\label{s1432}%\label{s1432G}
\newcommand\DELTA {\underline{\Delta}}

Our hyper-space $\Dist^2\CalX$ has been synthesised by abstraction from the classical ``matrix style'' description of \HMM's. We now recall that there is a partial order $(\Ref)$ of refinement on hypers, where for two hypers $\Delta_{S,I}{\In}\Dist^2\CalX$ we say that $\Delta_S$ (a specification) is ``refined by'' $\Delta_I$ (implementation) when, in a sense we make precise below, the implementation $\Delta_I$ releases no more information than the specification $\Delta_S$ does \cite{mcivermeinicke10a,McIver:12,Alvim:2012aa,McIver:2014ab}. That order lifts pointwise to $\AHSpc$, i.e.\ that $h_S{\Ref}h_I$ just when $h_S.\pi\Ref h_I.\pi$ for all $\pi{\In}\Dist\CalX$, thus giving a new \emph{refinement} order for (abstract) \HMM's. We write $\Delta_S{\Ref}\Delta_I$, and call it ``uncertainty refinement'' if we need to distinguish it from other kinds of refinement. Its ultimate antecedent is the lattice of information \cite{Landauer:1993aa} --- but it generalises those seminal ideas significantly.
%\Cf{In fact our quantitative $(\Ref)$ does not give a lattice \cite{FCS}.}

\begin{Definition}{Uncertainty refinement \cite{McIver:12,McIver:2014ab}}{d1449}\label{g1143}
	\label{g1142}Let $\Delta_{S,I}{\In}\Dist^2\CalX$ be two hypers on $\CalX$. We say that $\Delta_S$ is refined by $\Delta_I$ just when there is a distribution $\DELTA{\In}\Dist^3\CalX$, that is a distribution of \emph{hypers}, such that
	\[
	\Delta_S ~=~ \Avg.\DELTA \WideRM{and} (\Dist\Avg).\DELTA ~=~ \Delta_I~.
	\]
\end{Definition}
Recall that $\Dist\Avg$ is the push-forward of $\Avg$ (Defs.~\ref{d1341B},\ref{d1154}).

The advantage of the abstract formulation in \Def{d1449} is that it is defined on hypers directly, and can be generalised to proper measures, thus extending discrete distributions~\cite{McIver:12}. 
But in the case (as here) where we remain discrete, there is an equivalent matrix-style characterisation:
\begin{Lemma}{Refinement of joint-distributions \cite{mcivermeinicke10a,Alvim:2012aa}}{l0827}
	Let $J_S{\In}\CalX{\MFun}\CalY_S$ and $J_I{\In}\CalX{\MFun}\CalY_I$ be joint-distribution matrices
	%, both of them \emph{reduced}
	%\Cf{Do they have to be reduced?}
	%in the sense of \Def{d1426HB},
	such that $\Hyp{J_{S,I}}{=}\Delta_{S,I}$ resp.\,%
	\footnote{Recall that the $\CalX$ here in type $\CalX{\MFun}\CalY_S$ is the final-, not the initial state.}
	Then
	\begin{equation}\label{e1005}
	\Delta_S \Ref \Delta_I
	\WideRM{iff}
	J_S\MMult R = J_I
	\end{equation}
	for some stochastic \emph{refinement matrix} $R{\In}\CalY_S{\MFun}\CalY_I$. Note that the state-spaces of $\Delta_{S,I}$ are the same, but their observation spaces $\CalY_{S,I}$ can differ.
	\Proof
	Illustrated in \App{s0851}; sketch proof in \App{s0851A}.
\end{Lemma}

With \Lem{l0827} the reflexivity and transitivity of relation $(\Ref)$ is clear from elementary matrix properties.
For anti-symmetry we refer to \cite[Thm~6]{McIver:2014aa}, whose supporting Lemma 1 there is adapted to suit our purposes here:

\begin{Definition}{Expected value}{d1655}
	\label{g1659}For distribution $\delta{\In}\Dist\CalZ$ and function $f{\In}\CalZ{\Fun}V$ for vector space $V$, the expected value of $f$ on $\delta$ is $\Expt{\delta}f\Defs\sum_{z{\In}\CalZ} ~\delta_z{\times}f.z$, where $\sum$ and $(\times)$ are taken in the vector space.\,%
	\footnote{More generally it is $\int f \Dx{\delta}$ and requires measurability of $f$. One reason we do not use the standard notation $E(X)$ for the expected value of random variable $X$ is that the distribution over which $X$ is taken is implicit. In the calculations our \HMM-semantics entails, we often need to make it explicit.}
\end{Definition}
We will be using $\Expt{}{}$ principally over hypers, i.e. the case $\CalZ=\Dist\CalX$ in the definition.

\begin{Lemma}{(Strict) monotonicity}{l1228}
	Given are two hypers $\Delta_{S,I}{\In}\Dist^2\CalX$ and a strictly concave function $f{\In}\Dist\CalX{\Fun}\NNReal$.
	
	If $\Delta_S{\StrictRef}\Delta_I$ then $\Expt{\Delta_S}f < \Expt{\Delta_I} f$. And if $f$ is (non-strictly) concave, then $\Delta_S{\Ref}\Delta_I$ implies $\Expt{\Delta_S}f \leq \Expt{\Delta_I} f$.
	\Proof Proved for abstract channels in \cite[Lem~1]{McIver:2014aa}; the proof for hypers is essentially identical.
	%++\Cf{Did we prove it in \cite{McIver:2014ab} too?}
\end{Lemma}

We now have antisymmetry, because $\Delta_S{\Ref}\Delta_I{\Ref}\Delta_S$ and $\Delta_S{\neq}\Delta_I$ implies $\Delta_S{\StrictRef}\Delta_I{\StrictRef}\Delta_S$ whence we have from \Lem{l1228} the contradiction $\Expt{\Delta_S}f<\Expt{\Delta_I}f<\Expt{\Delta_S}f$ for any strictly concave $f{\In}\Dist\CalX{\Fun}\NNReal$ of our choice (for example Shannon entropy).% \cite[Thm~6]{McIver:2014aa}.

%\label{g1230}\label{g1124}
Hyper-space $\Dist^2\CalX$ also admits a metric, the Kantorovich metric \cite{Breugel:05} based on the Manhattan metric on $\Dist\CalX$ (\Sec{s1250}). It is used for continuity properties (as we will see in \Sec{s0842A}), and is chosen because of its hierarchical properties, i.e.\ that the Kantorovich metric on say $\CalX$ induces a metric on $\Dist\CalX$ and $\Dist^2\CalX$ etc. \cite{Breugel:05}.

\section{Monads: Giry, Kleisli and Kantorovich}\label{s1250}

With $\AHSpc$ we have given a discrete model of abstract \HMM's, suitable for interpreting probabilistic sequential programs with hidden state, together with concrete programming examples (Figs.~\ref{f0544m}--\ref{f0544cm}). We now provide a brief overview about how our setup embeds into structures based on a Giry monad.

\label{g1210}The Giry monad over the category of Polish spaces and continuous functions comprises an endofunctor $\Pi$ and two natural transformations $\eta$ (unit) and $\mu$ (multiply) \cite{Giry:81}; following \cite{Moggi:89} we take that as a basis for the denotation of computations. More precisely, we restrict ourselves to the category \textit{Comp} of compact metric spaces and continuous functions. We have been using $\Dist$ as a specialisation of $\Pi$ to this case. The object $\Dist\CalS$ is the set of Borel probability measures over the compact metric space $\CalS$ which is indeed a compact metric space~\cite[Thm 6.4]{Parthasarathy:1967}. To form a monad on $\textit{Comp}$, we have provided the unit-function $\Point{\cdot}$ specialising $\eta$ that makes a point measure, and multiply-function $\Avg$ specialising $\mu$ that takes the average of a distribution (of distributions). Typically we have $\Point{\cdot}_{\Dist\CalX}{\in}\AHSpc$ and $\Avg_\CalX{\in}\Dist^2\CalX{\Fun}\Dist\CalX$ where the subscripts are left implicit when they are clear from the context.

From Giry's construction, the arrows $\Point{\cdot}_\CalS$ and $\Avg_\CalS$ are continuous with respect to the weak topology on $\Dist\CalS$ but, in this paper, we are dealing with compact metric spaces. Fortunately, for compact metric spaces, the Kantorovich distance metrizes the weak topology~\cite{Gibbs:02}. This implies that the triple $(\Dist,\Point{\cdot},\Avg)$ is indeed a monad on the category $\textit{Comp}$.

\label{g1214}Monadic constructions based on the Kantorovich metric are not new. In~\cite{Breugel:05}, Van Breugel construct monads on the category \textit{Comp*} of compact metric spaces and 1-Lipschitz functions. His functor $\cal B$ coincides with our $\Dist$ on objects and he shows that $\cal B f$, $\Point{\cdot}_\CalS$ and $\Avg_\CalS$ are 1-Lipschitz; whenever $f$ is 1-Lipschitz and $\CalS$ is a compact metric space. Thus $(\cal B, \Point{\cdot}, \Avg)$ is a monad on the category \textit{Comp*}~\footnote{Note that Van Breugel's construction is more general than this since he also considers complete metric spaces and tight probability measures over them.}. In that work, the metric is crucial since the notion of 1-Lipschitzness is not a topological property. In fact, Van Breugel shows that the Kantorovich metric is \emph{the right} metric to construct probabilistic monads out of metric spaces. Such a construction does not necessarily work with other metrics that metrizes the weak topology (e.g. Prohorov metric which is equivalent to the Kantorovich metric from compact spaces). However, arrows in the category \textit{Comp*} are insufficient to denote probabilistic programs with hidden states because $\HMMone{C}$ is not necessarily 1-Lipschitz for some channel matrix $C$. This drives our choice of the Giry monad $(\Dist,\Point{\cdot},\Avg)$ on the category \textit{Comp} of compact metric spaces and continuous functions.

\label{g1233}Our construction of the Kantorovich metric begins with a finite set $\CalX$ endowed with the discrete metric $d_1$ (i.e. $d_1(x,x') \Defs 0 \textrm{ if } x {=} x' \textrm{ else } 1$). This is trivially a compact metric space. The space $\Dist\CalX$ of discrete distributions on $\CalX$ is endowed with the Kantorovich metric based on $d_1$ which coincides with the total variation metric on $\Dist\CalX$\label{g1234}. At this level, the Kantorovich metric reduces to $
\KMet{\delta^1}{\delta^2} = \frac{1}{2}\sum_x|\delta^1_x{-}\delta^2_x|$. At the next level, our hyper-space $\Dist^2\CalX$ has the Borel algebra generated by $\Dist$ from $\Dist\CalX$, which is in turn determined by the Kantorovich metric derived from $\KDist$\label{g1504} which we will also denote by $d_K$. These metrics are distinguished in terms of the arguments they are applied on, i.e.\ $\KMet{\delta^1}{\delta^2}$ is the Kantorovich metric on $\Dist\CalX$ while $\KMet{\Delta^1}{\Delta^2}$ is the Kantorovich metric on $\Dist^2\CalX$.

{\color{red}
}	
	\section{Characteristics of $\AHS$, the abstract \HMM's}\label{s0842}
	
	\subsection{Continuity and super-linearity}\label{s0842A}
	The semantic function $\HMMone{\cdot}$ (\Def{d1022}) takes \HMM\ matrices in $\MH$ to functions in $\AHSpc$; but not all of those functions are denotations $\HMMone{H}$ of some $H$. We now describe two important characteristics satisfied by $\HMMone{H}$ as $H$ ranges over \HMM's: they are continuity and super-linearity. We will define abstract \HMM's $\AHS$ to be just the functions satisfying those conditions.
	
	Sometimes called ``healthiness conditions'', these are essentially technical results giving general properties that are to hold for the denotation of any program. They are used for the proof of other, more specific properties of programs. Here we use them to define a subset $\AHS$ of $\AHSpc$, and we prove that $\HMMone{H}{\in}\AHS$ for all classical HMM's $H$.
	
	Our first condition concerns continuity wrt.\ the Kantorovich metrics on $\Dist\CalX$ and $\Dist^2\CalX$.
	\begin{Lemma}{Denotations of \HMM's are continuous}{l1432}
		For all $H{\In}\MH$ we have that $\HMMone{H}$ is a continuous function in $\AHSpc$ wrt.\ the Kantorovich metrics.

%		\begin{Figure}{f1705}{Illustration of the proof of \Lem{l1432}.}
%			\includegraphics[scale=0.4]{continuity.pdf}
%			
%			\bigskip
%			\captionsetup{singlelinecheck=off,font=footnotesize}
%			\caption*{We move from $\Delta^1$ to $\Delta^2$ in two steps. The first-step distance $\KMet{\Delta^1}{\Delta}$ is no more than the distance between corresponding inners times the total weight moved: the former is $<\delta$ and the latter is exactly 1. The second-step distance $\KMet{\Delta}{\Delta^2}$ is no more than the maximum distance between any two inners times the total weight moved: the former is $\leq1$ and the latter is $<\delta'$.}
%			\captionsetup{singlelinecheck=on,font=normalsize}
%		\end{Figure}
		
		%For all $H{\In}\MH$ we have that $\HMMone{H}$ is a continuous function in $\AHSpc$ wrt.\ the Kantorovich metric on both sides.
		\Proof
		We recall from \Def{d1022} that $\HMMone{H}.\pi = \Hyp{\ChApp{\pi}{J}}$ where $J_{x',y}=\sum_x H_{x,y,x'}$. We consider this as the composition of the two functions $\pi{\mapsto}(\ChApp{\pi}{J})$ and $\Hyp{\cdot}$, with the metric on the intermediate space $\CalX{\MFun}\CalY$ (of matrices) being the Kantorovich metric $\KDist$ on $\Dist(\CalX{\times}\CalY)$ which is $\KMet{J^1}{J^2} = \frac{1}{2}\sum_{(x,y)}|J^1_{x,y} - J^2_{x,y}|$. Furthermore, since $\pi{\mapsto}(\ChApp{\pi}{J})$ comprises only elementary arithmetic operations, and $\KDist$ is topologically equivalent to the Euclidean distance, the continuity is clear. Thus we concentrate on the continuity of $\Hyp{\cdot}$ at an arbitrary joint distribution $J{\In}\Dist(\CalX{\times}\CalY)$. 
		
		Let $\varepsilon{>} 0$. We denote $\Hyp{J} = \Delta$ and let $J'{\In}\Dist(\CalX{\times}\CalY)$ with $\Hyp{J'} = \Delta'$. Since $J$ and $J'$ are matrices, we can write $\Delta = \sum_{y} a_y\Point{\delta^y}$ and $\Delta' = \sum_{y} a_y'\Point{\delta'^y}$ with $\delta^y,\delta'^y{\In}\Dist\CalX$. These are sums over the full set $\CalY$ so if $a_y = 0$ (resp. $a'_y = 0$) then we define $\delta^y\Defs \delta'^y$ (resp. $\delta'^y\Defs \delta^y$). Let us define $\Delta'' = \sum_{y}a_y\Point{\delta'^y}$ which combines the coefficients of $\Delta$ with the inners of $\Delta'$. The triangular inequality tells us that 
\[
	\KMet{\Delta}{\Delta'} \leq \KMet{\Delta}{\Delta''} + \KMet{\Delta''}{\Delta'}~.
\]

On the one hand, 
\begin{Reason}
			\Step{}{
				\KMet{\Delta''}{\Delta'}
			}
			\StepR{$=$}{$\Delta''$ and $\Delta'$ have finite supports}{
				\frac{1}{2}\sum_y\left|a_y - a'_y\right|
			}
			\StepR{$=$}{Defn. $a_y$ and $a_y'$}{
				\frac{1}{2}\sum_{y}\left|\sum_xJ_{x,y} - \sum_xJ_{x,y}'\right|
			}
			\StepR{$\leq$}{$\left|\sum_xf.x\right|\leq\sum_x\left|f.x\right|$}{
				\frac{1}{2}\sum_{y,x}\left|J_{x,y} - J_{x,y}'\right|
			}
			\StepR{$=$}{Defn. $\KMet{J}{J'}$}{
				\KMet{J}{J'}
			}
		\end{Reason}

On the other hand, for every $y$, the function which maps $J$ to $\delta_y$ is continuous at $J$ since it is a composition of a $y$-projection and normalisation. Therefore, there exists $\alpha_y{>}0$ such that for every $J'{\In}\Dist(\CalX{\times}\CalY)$ with $\KMet{J}{J'}{<}\alpha_y$, we have $\KMet{\delta^y}{\delta'^y}{<}\frac{\varepsilon}{2}$. 
But we have
\begin{Reason}
			\Step{}{
				\KMet{\Delta}{\Delta''} 
			}
			\StepR{$\leq$}{Kantorovich-Rubeinstein Theorem~\cite[Pg. 8]{Gibbs:02}}{
				\sum_{y}a_y\KMet{\delta^y}{\delta'^y}
			}
			\StepR{$\leq$}{$\sum_ya_y = 1$ and $a_y{\geq}0$ for all $y$}{
				\max_y\KMet{\delta^y}{\delta'^y}
			}
\end{Reason}

Therefore, we choose $\beta = \min(\min_y\alpha_y,\frac{\varepsilon}{2})$ and for every $J'$ such that $\KMet{J}{J'} < \beta$, we have
\[
	\KMet{\Delta}{\Delta'} {<} \frac{\varepsilon}{2} +  \frac{\varepsilon}{2}= \varepsilon~.
\]
Hence, $\Hyp{\cdot}$ is continuous at $J$.
	\end{Lemma}
	
	Our second condition concerns linear combinations.
	\begin{Definition}{Weighted sum}{d1955}
		\label{g1621}For $\delta_{1,2}{\In}\Dist\CalX$ we write $\delta_1\WS{p} \delta_2$ for the weighted sum of the two distributions, so that $(\delta\WS{p} \delta')_x = p\delta_x+(1{-}p)\delta'_x$.
Note that $\delta\WS{p}\delta'$ defined here and the $\delta\PC{p}\delta'$ of \Def{d1154} differ: the former is a single distribution made from $p$-merging $\delta,\delta'$; the latter is a hyper whose support is just the two elements $\delta,\delta'$.
	\end{Definition}
	
	\begin{Lemma}{Denotations of \HMM's are super-linear}{l1433}
		For all $H{\In}\MH$ we have
		\begin{equation}\label{e1506}
		\Hyp{H}.\pi_1 ~\WS{p}~ \Hyp{H}.\pi_2 \Wide{\Ref} \Hyp{H}.(\pi_1 ~\WS{p}~ \pi_2)~,
		\end{equation}
		where $(\Ref)$ is refinement as defined in \Def{d1449}.\,%
		\footnote{Super-linearity can also be seen as a form of monotonicity. See \App{a1113}.}
%		\Proof %Given in \App{a0842}.
%		For all $H{\In}\MH$ we have
%		\begin{equation}
%		\Hyp{H}.\pi_1 ~\WS{p}~ \Hyp{H}.\pi_2 \Wide{\Ref} \Hyp{H}.(\pi_1 ~\WS{p}~ \pi_2)~,
%		\end{equation}
%		where $(\Ref)$ is refinement as defined in \Def{d1449}.
		\Proof
		Take any reduced $J^{1,2}\In\CalX{\MFun}\CalY$, and argue first that for any $0{\leq}p{\leq}1$ we have
		\begin{equation}\label{e1157}
		\Hyp{J^1} ~\WS{p}~ \Hyp{J^2}
		\Wide{\Ref}
		\Hyp{J^1 ~\WS{p}~ J^2}~,
		\end{equation}
		since the horizontal concatenation $J$ of the two (scaled) matrices $p{\Atimes}J^1$ and $(1{-}p){\Atimes}J^2$ satisfies $\Hyp{J}=\Hyp{J^1} \WS{p} \Hyp{J^2}$, and $J$ itself is refined to $J_1 ~\WS{p}~ J_2$ (in the sense of \Lem{l0827}) by the refinement matrix
		\[
		R \Wide{\Defs}
		\left(
		\begin{array}{ccc}
		1 & 0 & \cdots \makebox[0pt][l]{\hspace{1.5em}$\longleftarrow$\small\ corresp.\ to first col.\ of $J^1$} \\
		0 & 1 & \cdots \makebox[0pt][l]{\hspace{1.5em}$\longleftarrow$\small\ corresp.\ to snd.\ col.\ of $J^1$} \\
		\vdots & \vdots & \vdots \\
		1 & 0 & \cdots \makebox[0pt][l]{\hspace{1.5em}$\longleftarrow$\small\ corresp.\ to first col.\ of $J^2$} \\
		0 & 1 & \cdots \makebox[0pt][l]{\hspace{1.5em}$\longleftarrow$\small\ corresp.\ to snd.\ col.\ of $J^2$} \\
		\vdots & \vdots & \vdots
		\end{array}
		\right)~,
		\hspace{10em}
		\]
		that simply sums corresponding columns.	Now we observe that
		\begin{Reason}
			\Step{}{
				\HMMone{H}(\pi_1 ~\WS{p}~ \pi_2)
			}
			\Step{$=$}{
				\Hyp{\ChApp{(\pi_1 ~\WS{p}~ \pi_2)}{H})}
			}
			\Step{$=$}{
				\Hyp{\ChApp{\pi_1}{H}~\WS{p}~\ChApp{\pi_2}{H}}
			}
			\StepR{$\RRef$}{\Eqn{e1157} just above}{
				\Hyp{\ChApp{\pi_1}{H}}~\WS{p}~\Hyp{\ChApp{\pi_2}{H}}
			}
			\Step{$=$}{
				\Hyp{H}.\pi_1~\WS{p}~\Hyp{H}.\pi_2 ~,
			}
		\end{Reason}
		as required.
	\end{Lemma}
%}%1040

Motivated by those two lemmas, we now define
\begin{Definition}{The space $\AHS$ of abstract \HMM's}{d0943}
	\label{g1628}We write $\AHS$ for those $h$ in $\AHSpc$ satisfying Lemmas~\ref{l1432},\ref{l1433}, i.e.\ that are Kantorovich-continuous and super-linear.
\end{Definition}
Thus our two lemmas above establish that $\HMMone{H}{\in}\AHS$ for any classical \HMM\ $H$. 

Since we will therefore be restricting our denotations to $\AHS$, a subset of the arrows in the category \textit{Comp}, we expect $\AHS$ to be closed under composition.
\begin{Lemma}{Abstract \HMM's closed under composition}{l0903}
	For any two $h_{1,2}{\In}\AHS$ we have $h_1;h_2\in\AHS$ as well, where $({;})$ is as in \Def{d1210}.
	\Proof
	Although a direct proof is possible, the result is much easier once we have introduced ``uncertainty'' transformers (\Sec{s1244A}), because it is then a consequence of \Thm{t1005} and in particular its \Cor{c1020}, which depends crucially on the dual view we develop in \Sec{s1244A}.
%This follows from the fact that $\Avg, \Dist h_2$ and $h_1$ are all continuous.
\end{Lemma}
It is shown in \App{a1025} that composition in $\AHS$ is monotonic with respect to the refinement order $(\Ref)$. This completes our construction of our forward, abstract semantics for \HMM's. We now propose a dual view.

\section{A dual view: uncertainty measures, and  their transformers}\label{s1244A}
\subsection{Uncertainty measures, and their relation to refinement}\label{s0731}
``Uncertainty measures'' generalise the diversity of entropy measures (including e.g.\ Shannon), the functions from distributions to reals that measure increasing disorder.\,%
%Said this in the intro.
%\footnote{We introduced them earlier as ``disorder tests'' \cite{McIver:12}.}
\begin{Definition}{Uncertainty measure}{d1709A}
	An \emph{uncertainty measure} over $\CalX$ is a Kantorovich-continuous- and concave function in $\Dist\CalX\Fun\NNReal$\kern-.1em, i.e.\ one taking distributions (on $\CalX$ in this case) to non-negative reals. It is intended that a distribution's greater uncertainty indicates more resilience (less vulnerability) to the distribution's being exploited by an adversary.\,%
	\footnote{Smith's ``vulnerability measure'' based on Bayes Risk \cite{Smith:2009aa} is an uncertainty measure except that it goes in the opposite direction.}
	
	\label{g1147}We write $\Unc\CalX$ for the uncertainty measures over $\CalX$, and call them ``\UM's'' in the text for brevity.
\end{Definition}

A typical example of a \UM\ applied to a hyper is as follows. Given prior $\pi{\In}\Dist\CalX$ and channel $C{\In}\CalX{\MFun}\CalY$, the resulting hyper is $\Delta{\Defs}\HMMone{\ChApp{\pi}{C}}$ and the ``conditional $u$ uncertainty'' of that (compare conditional Shannon entropy) would be $\Expt{\Delta}{u}$.
We write $\Expt{\Delta}{u}$ because it makes explicit that the conditional uncertainty is an expected value and, as such, we can calculate with it. (More conventional notations such as $H(y|x)$ --in the Shannon case-- make those calculations more difficult.)
This could be compared to the uncertainty $u.\pi$ of the prior, to give a ``$u$-leakage'' of the channel on that prior.

There is a compelling connection between \UM's (\Def{d1709A}) and refinement (\Def{d1449}, \Lem{l0827}): we have
\begin{Lemma}{Soundness and completeness of uncertainty measures \cite{McIver:2014aa}}{l0755}
	For any hypers $\Delta_{1,2}{\In}\Dist^2\CalX$ we have  
	\[
	\Delta_1\Ref\Delta_2
	\WideRM{iff}
	\Expt{\Delta_1}{\!u}\leq \Expt{\Delta_2}{\!u}
	~\textrm{\quad for all $u{\In}\Unc\CalX$.}
	\]
\end{Lemma}
%\Proof
We regard ``only if'' as \emph{soundness} in the sense that if we have a witness to the refinement relation $\Delta_1{\Ref}\Delta_2$, i.e.\ either $\DELTA$ (\Def{d1449}) or $R$ (\Lem{l0827}), then no \UM\ can show $\Delta_2$ to be less uncertain than $\Delta_1$. It is related to the \emph{Data-Processing Inequality}, as explained in \cite{McIver:2014aa}.
%++\Cf{Or was it \Cite{CSF14}? {\Ax \Cite{CSF12}?}}

We regard ``if'' as \emph{completeness} in the sense that if refinement fails, that is if $\Delta_1{\NotRef}\Delta_2$, then there is a \UM\ demonstrating the failure \cite{mcivermeinicke10a,McIver:12,Alvim:2012aa,McIver:2014aa}.

In \App{a0810} is background on the proof of \Lem{l0755}, whose completeness part was originally called ``Coriaceous'' because it was hard to prove \cite{Alvim:2012aa}.

\subsection{Abstract \HMM's to \UM-transformers}\label{s1031}
In \Sec{s0939} we introduced a ``forward'' denotational view of \HMM's that takes initial distributions to final hypers. Here we take the dual view, where an \HMM\ takes a ``post-uncertainty'' to a ``pre-uncertainty''.

\begin{Definition}{Uncertainty-measure transformers}{d1701A}\label{g1151}
	Take $h{\In}\AHS$ and $u{\In}\Unc\CalX$.\,%
	%\footnote{\Cx In \Cor{c1020} below we briefly apply this definition of $\Wp{()}$ in the more general case that $h$ is in $\AHSpc$ and is measurable.}
	Define the \emph{uncertainty transformer} $\Wp{h}$ of type $\UT$ so that for any $u{\In}\Unc\CalX$ and $\pi{\In}\Dist\CalX$ we have
	\[
	\Wp{h}.u.\pi \Wide{\Defs} \Expt{h.\pi}{u}~,
	\]
	where on the right we are taking the expected value of $u$ on the hyper $h.\pi$. (Because $u$ is continuous, it is measurable.)
\end{Definition}
By analogy with weakest preconditions for ordinary sequential programming \cite{Dijkstra:76}, a \UM-transformer $\Wp{h}$ takes a \UM\ to be applied \emph{after} $h$ and produces a \UM\ that equivalently can be applied \emph{before} $h$. (Compare also \cite{Kozen:83,Morgan:96d,Jones:90} for probabilistic/demonic sequential programs.) The idea (and utility) is in goal-directed reasoning: if one knows the program, and knows also the uncertainty that it must achieve, with the uncertainty transformer one determines the minimum uncdertainty that is necessary \emph{before} the program begins its execution.

In \Lem{l1719A} we show well definedness of \Def{d1701A}, that is that $\Wp{h}.u$ is indeed in $\Unc\CalX$.

\begin{Lemma}{Well-definedness of \Def{d1701A}}{l1719A}
	If $h{\In}\HS\CalX$ is an abstract \HMM\ and $u{\In}\Unc\CalX$ is a \UM, then $\Wp{h}.u$ is in $\Unc\CalX$.
	\Proof See \App{a1613}.
\end{Lemma}

\subsection{Characteristic properties of $\Wp{h}$}\label{s0935}
For $h{\In}\AHS$ the \UM-transformer $\Wp{h}$ has a number of characteristic properties.

\begin{Lemma}{$\Wp{h}$ is linear and total}{l0846A}
	For every $h{\In}\AHS$ and $t=\Wp{h}$ we have  that $t$ is:
	\begin{enumerate}
		\item \emph{linear} so that for $a_{1,2}{\In}\NNReal$ and $u_{1,2}{\In}\Unc\CalX$ we have
		\[
		t.(a_1u_1+a_2u_2) \Wide{=} a_1t.u_1 + a_2t.u_2~;
		\]
		\item\label{i1243} \emph{monotonic}, so that $t.u_1.\delta\geq t.u_2.\delta$ for every $u_1{\geq}u_2$ with $u_{1,2}{\In}\Unc\CalX$ and $\delta{\In}\Dist\CalX$, where %
		we lift $(\geq)$ pointwise;
		and
		\item \emph{total}, so that $t.\FunctOne{=}\FunctOne$ where \label{g1827}$\FunctOne.\delta\Defs1$ for all $\delta{\In}\Dist\CalX$.
	\end{enumerate}
	\Proof
	These properties are immediate from  \Def{d1701A}.
\end{Lemma}
A further property of \UM-transformers is that they are 1-Lipshitz in a certain sense:

\begin{Lemma}{$\Wp{h}$ is 1-Lipschitz}{l0846B}
	Take $h{\In}\AHS$ and define $t{\Defs}\Wp h$. Let $|{\cdot}|$ be absolute value. Then $t$ is $1$-Lipschitz in the sense that
	\[
	\sup_{\delta{\In}\Dist\CalX}\left|t.{u_1}.\delta -t.{u_2}.\delta\right|
	\Wide{\leq}
	\sup_{\delta{\In}\Dist\CalX}\left|{u_1}.\delta - {u_2}.\delta\right|~.
	\]
	\Proof %See \App{a1620}.
	Consider arbitrary $u_{1,2}{\In}\Unc\CalX$. We reason
	\begin{Reason}
		\Step{}{
			\sup_{\delta{\In}\Dist\CalX}\left|\Wp{h}.u_1.\delta - \Wp{h}.u_2.\delta\right|
		}
		\Step{$=$}%{Definition of $\UMet{\cdot}{\cdot}$}
		{
			\sup_{\delta{\In}\Dist\CalX}\left|\Expt{h.\delta}{u_1} - \Expt{h.\delta}{u_2}\right|
		}
		\StepR{$\leq$}{property of $|\cdot|$}{
			\sup_{\delta{\In}\Dist\CalX}~\Expt{h.\delta}{|u_1{-}u_2|}
		}
		\StepR{$\leq$}{$\Expt{}{}$ monotonic}{%{$|u_1{-}u_2|\leq\UMet{u_1}{u_2}$}{
			\sup_{\delta{\In}\Dist\CalX}\quad\Expt{h.\delta}{(\sup_{\delta'{\In}\Dist\CalX} \left|u_1.{\delta'}- u_2.{\delta'}\right|)}
		}
		\StepR{$=$}{$\Expt{h.\delta}\FunctOne = 1$ and rename $\delta'$ to $\delta$}{%{$|u_1{-}u_2|\leq\UMet{u_1}{u_2}$}{
			\sup_{\delta{\In}\Dist\CalX} \left|u_1.{\delta}- u_2.{\delta}\right| ~.
		}
	\end{Reason}
\end{Lemma}

Motivated by those lemmas, we define uncertainty transformers to be exactly the functions in $\Unc\CalX{\Fun}\Unc\CalX$ that satisfy the properties listed.
\label{g1254}
\begin{Definition}{The uncertainty transformers $\TT\CalX$}{d1447}
	The uncertainty transformers $\TT\CalX$ are the functions in $\Unc\CalX{\Fun}\Unc\CalX$ that satisfy Lems.~\ref{l0846A},\ref{l0846B}.
\end{Definition}
We note that transformers $\TT\CalX$ are closed under composition. In \App{a1139} we show that refinement for $\TT\CalX$ is pointwise $(\leq)$.

\subsection{\UM-transformers back to abstract \HMM's}\label{s1408}
The function $\Wp{(\cdot)}$ has been shown to be of type $\AHS{\Fun}\TT\CalX$. Here we show that this correspondence is exact, i.e.\ that for every $t{\In}\TT\CalX$ there is an $h{\In}\AHS$ such that $t{=}\Wp{h}$ and, moreover, that the $h$ is unique.

The following theorem thus establishes the exact correspondence between $\AHS$ and $\TT\CalX$, giving an analogue for hidden-state probabilistic programs to the well-known correspondence between demonic relations and conjunctive predicate transformers \cite{Dijkstra:76} that the former correspond exactly to those functions from predicates to predicates that distribution conjunction.
(A further example, generalising that, is the correspondence between demonic/probabilistic programs and super-linear expectation transformers \cite{Morgan:96d,McIver:05a}.)

\begin{Theorem}{Characterisation of transformers}{t1005}
	For any $t{\In}\TT\CalX$ there is a unique $h{\In}\AHS$ such that $t{=}\Wp{h}$.
	\Proof
	% Move to appendix to save space.
	%The proof first establishes a non-trivial version of the Riesz representation for the current setting, with the main argument showing that our definitions extend more generally (i.e.\ outside of $\Unc\CalX$) thus allowing an appeal to established mathematical results on dualities. The constructed $h$ is then shown to have the characteristic properties set out at \Lem{l1432} and \Lem{l1433}.
	%See \App{s1805}.
	Let $t{\In}\TT$. The construction of $h$ starts by showing that the transformer $t$ can be extended into a linear function over the space $\Cont\CalX$ of continuous functions from $\Dist\CalX$ to itself. This extension is executed in two phases. Firstly, we show that the set $\Unc\CalX$ of continuous, concave and non-negative function over $\Dist\CalX$ generates a sub-vector space of $\Cont\CalX$. Thus, the first stage is an algebraic extension of $t$ to that generated sub-space. This extension is necessarily unique by  linearity (\Lem{l0846A}). The second stage is a topological extension where $\Cont\CalX$ is endowed with the norm uniform making it a Banach space. In fact, we show that the sub-vector space generated by $\Unc\CalX$ is a dense sub-algebra of $\Cont\CalX$ using the Stone-Weierstrass Theorem~\cite[Thm.~5]{Stone:1948}. Thus the second stage of the extension is also unique by continuity. Full technical details of these extensions are given in \App{s1805}.
	
We now have a unique continuous linear function $\hat{t}$ from $\Cont\CalX$ to itself which coincides with $t$ on $\Unc\CalX$. We shall construct an $h$ such that $\Wp{h} = t$.

Fix $\delta{\In}\Dist\CalX$. The function
\[
f \Wide{\mapsto} \tilde{t}.f.\delta
\]
maps  each continuous function $f{\In}\Cont\CalX$ to $\tilde{t}.f.\delta$ is a positive linear functional on $\Cont\CalX$;  moreover  $\tilde{t}.\FunctOne.\delta = t.\FunctOne.\delta = 1$, thus $\FunctOne \mapsto 1$. Therefore, the Riesz Representation Theorem for linear functionals~\cite[Ch.~2 Thm.~5.8]{Parthasarathy:1967} implies that there exists a unique Borel probability measure $\Delta_\delta$ on $\Dist\CalX$ such that $\tilde{t}.f.\delta = \Expt{\Delta_\delta}{f}$, for every $f{\In}\Cont\CalX$.

Define $h.\delta\Defs\Delta_\delta$ for each $\delta{\In}\Dist\CalX$. We now check that $h$ has the required properties demanded by Lemmas \ref{l1432},\ref{l1433}. 

\medskip

\underline{Continuity:}
For the continuity assumption in \Lem{l1432}, we let $\delta_n$ be a sequence of distributions in $\Dist\CalX$ converging to $\delta{\In}\Dist\CalX$ with respect to the Kantorovich metric on $\Dist\CalX$. It suffices to show that the limit of  $\KMet{h.\delta_n}{h.\delta}$ is $0$, as $n$ goes to infinity. Since $\Dist\CalX$ is compact, the Kantorivich metric metrizes the weak topology and it suffices to show that $h.\delta_n$ converges weakly to the Borel measure $h.\delta$. Let $f{\In}\Cont\CalX$, we have 
\[
	\Expt{h.\delta_n}f = \tilde{t}.f.\delta_n
\]
Since $\tilde{t}.f$ is also continuous, the sequence $\tilde{t}.f.\delta_n$ converges to $\tilde{t}.f.\delta$ and thus, $h.\delta_n$ coverges weakly to $h.\delta$.~\footnote{This proof crucially depends on the compactness of $\Dist\CalX$. For Polish spaces, we can achieve the same result but using a more general result by Rao \cite[Thm.~3.1]{Ranga:1962}.}

\medskip

\underline{Super-linear:}
For the super-linearity assumption in \Lem{l1433}, suppose that $t{=}\Wp{h}$ is in $\TT\CalX$ and take arbitrary $\delta_{1,2}{\In}\Dist\CalX$. Then we reason
\begin{Reason}
	\Step{}{
		h.\delta_1 \WS{p} h.\delta_2 ~\Ref~ h.(\delta_1 \WS{p} \delta_2)
	}
	\StepR{if}{for all $u{\In}\Unc\CalX$ \Lem{l0755} Coriaceous}{
		\Expt{(h.\delta_1 \WS{p} h.\delta_2)}{u} ~\leq~ \Expt{h.(\delta_1 \WS{p} \delta_2)} u
	}
	\StepR{if}{Defn.\ \Wp{()}}{
		 \Wp{h}.u.\delta_1\WS{p}\Wp{h}.u.\delta_2 \leq & \Wp{h}.u.(\delta_1 \WS{p} \delta_2)
	}
	\StepR{if}{Defn.~$\Unc\CalX$}{
		\Wp{h}.u \in \Unc\CalX ~,
	}
\end{Reason}
which was our assumption.
\end{Theorem}

With these characterisations, we now can prove two technical facts. In the discrete case (as earlier) they seem self-evident. In the more general setting, however, the work is mainly in ensuring well definedness (e.g.\ that only measurable functions are integrated, etc.) The first establishes the usual connection between composition, this time between the forward- and backward semantics; the second confirms that $\AHS$ is closed under composition (i.e.\ preserves continuity and super-linearity, as claimed in \Lem{l0903}).

%{\Cx%1627
\begin{Corollary}{Transformer composition}{c1020} 
	For any $h_{1,2}{\In}\AHS$ we have that also $h_1;h_2{\in}\AHS$, and furthermore that $\Wp{(h_1;h_2)} = \Wp{h_1}\Comp\Wp{h_2}$.
	\Proof Direct calculation shows that $\Wp{(h_1;h_2)} = \Wp{h_1}\Comp\Wp{h_2}$, although the working is intricate in the general (Giry) case. Well definedness of $h_1;h_2$ itself uses the simpler properties of (functional) composition on the transformer side. See \App{a1625}.
\end{Corollary}
%}%1627

Also, transformer composition respects refinement (\App{a1306}).

\section{Gain- and loss functions define uncertainty measures}\label{s1636}

\subsection{Gain- and loss functions}\label{s0534}
Although \Def{d1709A} of uncertainty measures is abstract, they can be made concrete via ``gain functions'' \cite{Alvim:2012aa} or equivalently ``loss functions'' \cite[Eqn.~(5)]{mcivermeinicke10a} that encode an attacker's (e.g.)\ economic interest in the secrets and the cost of obtaining them.
We use loss functions here.

\begin{Definition}{Loss function determines uncertainty measure}{d1321A}\label{g1156}
	%++\Cf{Again \Cite{Kalnishkan}?}
	A \emph{loss function} ${\ell}$ is of type $I{\Fun}\CalX{\Fun}\NNReal$ for some index set $I$, 
	with the intuitive meaning that ${\ell}.i.x$ is the cost to the attacker of using ``attack strategy'' $i$ when the hidden value turns out actually to be $x$. Her expected cost for an attack planned but not yet carried out is then $\Expt{\delta}{({\ell}.i)}$ if $\delta$ is the distribution in $\Dist\CalX$ she believes to be governing $x$ currently.
	
	From such an ${\ell}$ we define an uncertainty measure 
	\begin{equation}\label{e0843}
	U_{\ell}.\rho \Wide{\Defs} \inf_{i{\In} I}~\Expt{\rho}{({\ell}.i)} ~.
	\end{equation}
	When $I$ is finite, the $\inf$ can be replaced by $\min$.
\end{Definition}
The $\inf$ represents a rational strategy of minimising cost or risk, and a typical attacker will act as follows: she chooses the attack strategy (i.e.\ he chooses $i$) whose expected cost $\Expt{\rho}{({\ell}.i)}$ to her, where $\rho$ is the posterior in $\Dist\CalX$ she infers from her observations in $\CalY$, will be the least.

\begin{Lemma}{Well-definedness for \Def{d1321A}}{l0928}
	For any loss function ${\ell}{\In} I{\Fun}\CalX{\Fun}\NNReal$ the function $U_{\ell}$ in \Def{d1321A} is continuous and concave.
	\Proof
	We give here the proof for the finite-$I$ case. (The infinite case is considered in \cite[Sec III-B]{Alvim:2014aa}; it might require further assumptions on $I$.)
	Let ${\ell}$ be a loss function and $U_{\ell}$ be the associated uncertainty measure.
	
	\medskip
	\underline{$U_{\ell}$ is concave}: Take $\rho_{1,2}{\In}\Dist\CalX$ and $p{\In}[0,1]$. We have 
	\begin{Reason}
		\Step{}{
			U_{\ell}.(\rho_1\WS{p}\rho_2)
		}
		\StepR{$=$}{definition $U_{\ell}$}{
			\min_{i{\In}I}\,(\Expt{\rho_1\WS{p}\rho_2}{{\ell}.i})
		}
		\StepR{$=$}{$\LAbs{\delta}{\Expt{\delta}{u}}$ is linear}{
			\min_{i{\In}I}\,(\Expt{\rho_1}{{\ell}.i}\WS{p}\Expt{\rho_2}{{\ell}.i})
		}
		\StepR{$\geq$}{$(\min f){\WS{p}}(\min g)$ $\leq \min (f{\mathbin{\WS{p}}} g)$}{
			(\min_{i{\In}I}\Expt{\rho_1}{{\ell}.i})\mathbin{\WS{p}}(\min_{i{\In}I}\Expt{\rho_2}{{\ell}.i})
		}
		\StepR{$=$}{definition $U_{\ell}$}{
			U_{\ell}.\rho_1\WS{p}U_{\ell}.\rho_2 ~.
		}
	\end{Reason}
	
	\underline{$U_l$ is continuous}: 
	Since $I$ is finite and each function $(\LAbs{\rho}{\Expt{\rho}{{\ell}.i}}) = (\LAbs{\rho}{\sum_{x{\In}\CalX}\rho_x{\times}{\ell}.i.x})$ is continuous, the function $U_{\ell}$ is also continuous.
\end{Lemma}

\medskip
Remarkably, loss functions are \emph{complete} for uncertainty measures: any uncertainty measure in $\Unc\CalX$ can be expressed as $U_{\ell}$ for some loss function ${\ell}$ in $I{\Fun}\CalX{\Fun}\NNReal$, but possibly requiring $I$ to be infinite \cite{Chatzikokolakis:2014aa}. Roughly speaking, this is because of the way concave functions can be expressed as the envelope of their tangential hyperplanes: the coefficients of the hyperplanes' normals are the loss functions.\,%
\footnote{For example, Shannon entropy requires infinite $I$, and the encoding is then related to minimising the Kullback-Leibler divergence.}

It is compellingly shown elsewhere how versatile loss (equiv.\ gain) functions are \cite{Alvim:2012aa}. Of particular interest is that \Lem{l0755} applies, both in the discrete \cite{mcivermeinicke10a} and the continuous cases \cite{McIver:12}, even when uncertainties are restricted to those generated by loss functions: the ``distinguishing witness'' constructed for completeness is in fact a loss function \cite{mcivermeinicke10a}.

\section{A \UM-transformer example for \Sec{s0825}}\label{s0914}
\subsection{Profiling an attacker with a loss function}\label{s0903}
In the context of \Fig{f0544cm} we imagine an attacker whose livelihood depends on her guessing whether $\XS\verb+[0]+{=}\XS\verb+[1]+$ or not, finally. If he guesses incorrectly he loses \$1; if correctly, he breaks even (loses \$0). This is as much a mathematical- as a \emph{social} issue: attacks will be discouraged if they are not worthwhile for the attacker in terms of her own criteria. (See also \App{a0407} for this social aspect.)

In this example, following \Sec{s1636}, we express the attacker's criteria as two strategies ``guess same'' and ``guess different'' (thus $I=\DSet{\Same,\Diff}$) and a loss function ${\ell}$ therefore defined
\[
\begin{array}{rcl}
{\ell}.\Same.(\verb+00+) &=& 0 \\
\makebox[0pt][r]{$\dagger$\quad}
{\ell}.\Same.(\verb+01+) &=& 1 \\
{\ell}.\Same.(\verb+10+) &=& 1 \\
{\ell}.\Same.(\verb+11+) &=& 0 \\
\end{array}
\hspace{3em}
\begin{array}{rcl}
{\ell}.\Diff.(\verb+00+) &=& 1 \\
{\ell}.\Diff.(\verb+01+) &=& 0   \makebox[0pt][l]{\quad$\ddagger$} \\
{\ell}.\Diff.(\verb+10+) &=& 0 \\
{\ell}.\Diff.(\verb+11+) &=& 1~, \\
\end{array}
\]
based on the informal description just above: for example if $\XS{=}\verb+01+$ but he guesses \Same, the case indicated by $\dagger$, then he loses \$1; but if he guesses \Diff, he breaks even $\ddagger$. Using \Eqn{e0843} we define our \UM\ as $u.\delta=U_{\ell}.\delta=$
\[
\begin{array}{crcl}
& {\ell}.\Same.\delta &\min& {\ell}.\Diff.\delta \\
= & \Expt{\delta}{({\ell}.\Same)} &\min& \Expt{\delta}{({\ell}.\Diff)} \\
= &(\delta_{00}{+}\delta_{11}) &\min & (\delta_{01}{+}\delta_{10})~.
\end{array}
\]

\subsection{Using \UM's and transformers to plan an attack}\label{s1224}
We can use our transformer semantics to answer $u$-dependent questions about \Fig{f0544cm} over \emph{all} priors: we use the two we chose earlier in \Sec{s0825} as examples.

Writing $\HMMone{P}$ for the abstract \HMM\ denoted by the two lines of code in \Fig{f0544cm}, we have for any $\pi$ that
\begin{equation}\label{e0810}
\Wp{\HMMone{P}}.u.\pi \Wide{=}
\begin{array}[t]{@{}ll}
& \pi_{00}~\min~(\pi_{01}{+}\pi_{10})/2 \\
+ & \pi_{11}~\min~(\pi_{01}{+}\pi_{10})/2~.
\end{array}
\end{equation}
(See \App{a0751} below for how this $\Wp{(\cdot)}$ is calculated.)

Now let $\pi^{\ref{f0544cm}}$ be the prior described by the initial comment in \Fig{f0544cm}. The attacker's (expected) uncertainty wrt.\ the \emph{final} hyper $\HMMone{P}.\pi^{\ref{f0544cm}}$ is given by $\Wp{\HMMone{P}}.u$ applied to that \emph{initial} (uniform) prior $\pi^{\ref{f0544cm}}$, that is $\Wp{\HMMone{P}}.u.\pi^{\ref{f0544cm}}=\nicefrac{1}{2}$ directly from \Eqn{e0810}. Since $u.\pi^{\ref{f0544cm}}$ is also $\nicefrac{1}{2}$, he is indifferent wrt.\ whether he should attack before or after $P$ has been allowed to run.

Now suppose that $\XS[0]{=}1$ is known initially, thus with prior $\pi$ being $\CVec{0,0,\nicefrac{1}{2},\nicefrac{1}{2}}$ so that $u$ applied initially gives $u.\pi{=}\nicefrac{1}{2}$. But $u$ applied finally would give $\Wp{\HMMone{P}}.u.\pi=(0\min\nicefrac{1}{4})+(\nicefrac{1}{2}\min\nicefrac{1}{4})=\nicefrac{1}{4}<\nicefrac{1}{2}$, so that it is better to attack later even though \XS\ might have been altered by $P$.
This scenario confirms that in fact for some priors, the program in \Fig{f0544cm} cannot be regarded as secure.

\section{\HMM's and the Dalenius Desideratum}\label{s0917}
\Ct{This section has changed wrt.\ my printout. Check why.}
Our abstracting from initial-state correlations allows a semantics for programs' \emph{final} states alone. Sometimes, however, leakage from the \emph{initial} state is important, even if that state is overwritten by the markov part of the \HMM: what the initial state \emph{was} might reveal information about what some other correlated state still \emph{is}, even if that other state is not mentioned in the program at all. This general concern was raised wrt.\ statistical databases by Dalenius \cite{Dalenius:1977aa} who argued that it is inescapable; Dwork later gave a proof of this \cite{Dwork:2006aa}. Here is an (edited) extract from her paper:
\begin{quote}\small
Suppose we have a statistical database that [records] average heights of population subgroups, and suppose further that it is infeasible to learn this information (perhaps for financial reasons) in any other way (say, by conducting a new study). Finally, suppose that one's true height is considered sensitive. \par [An adversary having the] auxiliary information ``Turing is two inches taller than the average Lithuanian woman'' [would, with access to the database, learn] Turing's height. In contrast, anyone without access to the database, knowing only the auxiliary information, learns much less about Turing's height.
\end{quote}

%{\Cx%1018
With our constructions here, we are able to see the Dalenius effect in programming terms. The program that allows access to Dwork's database of Lithuanian heights (as above) might itself, in isolation, have been analysed for security leaks. But if that program is run in a larger programming context in which there is a (to be kept secret) variable \texttt{tHeight} containing Turing's height \emph{and} there is program code (external to the database-access code) that establishes a correlation between the two, then running the database-access program reveals information about \texttt{tHeight} even though that variable is not mentioned anywhere in the database program.

In more austere terms, we would explain the effect as follows. A ``classical'' sequential program does not affect variables to which it does not refer; for example \texttt{x:= E} does not affect some other variable \texttt{y} in any way. But the program \texttt{ {\Leak} x } (recalling the notation of \Fig{f0544c}) can affect \emph{what we know} about variable \texttt{y} even though the program \texttt{ {\Leak} x } does not refer to \texttt{y} at all.

Consider for example an input distribution $(0,0)\AtP\nicefrac{1}{2}, (1,1)\AtP\nicefrac{1}{2}$ on two variables \texttt{(x,y)}. Its \texttt{y}-marginal distribution is uniform on $\{0,1\}$. But the output hyper of that program, projected onto \texttt{y}, is $\Point{0}\AtP\nicefrac{1}{2}, \Point{1}\AtP\nicefrac{1}{2}$, showing that the distribution on \texttt{y} is now a point, no longer uniform:\,%
\footnote{Since the output is a hyper, if knowledge of \texttt{y} were unaffected we would have the point hyper on the uniform distribution, that is $\Point{0\AtP\nicefrac{1}{2}, 1\AtP\nicefrac{1}{2}}$.}
with probability $\nicefrac{1}{2}$ that point will be $\Point{0}$, and with probability $\nicefrac{1}{2}$ that point will be $\Point{1}$. Reviewing the leaks of \texttt{x} tells us which point distribution on \texttt{y} we have, and we see essentially the Dalenius effect between ``database'' \texttt{x}, ``query'' \texttt{ {\Leak} x } and ``third-party data'' \texttt{y}.

This effect is exacerbated when we include state updates, as we have done with our abstract \HMM's here. (Updates were not considered originally by Dalenius or by Dwork.) For then the program \texttt{ {\Leak} x; x:= 0 } and the program \texttt{ x:= 0 } have the same abstract-\HMM\ semantics on state-space (just) \texttt{x}, but different semantics on state-space \texttt{x,y}.\,%
\footnote{On state-space \texttt{x}, both programs produce the output hyper $\Point{\Point{0}}$ that denotes ``\texttt{x} is certainly 0.'' On \texttt{x,y} however, the first might reveal something about \texttt{y} while the second cannot.}
The Dalenius effect has become, in programming terms, a failure of compositionality wrt.\ unreferenced global variables.

We show in this section how to deal with that: in brief, we include both the initial- and the final values of the state in our semantics. The crucial point is that we do not have to do more than that, in particular that we do not have to consider ``all possible third-party data \texttt{y} of any type''.
%}%1018
%\Cf{%1019
%\emph{The text here was}\begin{quote}
%then its meaning is independent of that variable: the variable is just ``carried through''. In terms of weakest preconditions for terminating programs, this is a combination of conjunctivity and the fact that $\Wp\mathit{prog}.\Phi$ is just $\Phi$ whenever $\mathit{prog}$ does not refer to $\Phi$.
%
%We believe that the Dalenius effect is, in programming terms, the question of compositionality wrt.\ unreferenced global variables, and we show here how to deal with it.
%\end{quote}
%}%1019

%In fact whether a semantics should capture that Dalenius effect depends on whether you care about it. Here we assume that we do care, and show that the general machinery we have set out in this paper is able to model it.

We now address the details. Consider a ``constant'' overwrite-by-{\sf x} markov $M^{\sf x}_{x,x'} = \textit{1 if $x'{=}{\sf x}$ else 0}$ for some fixed ${\sf x}{\In} \CalX$. Then $\HMMsem{C}{M^{\sf x}} = \HMMsem{}{M^{\sf x}}$ for any channel $C$, because $C$ has no effect on our knowedge of the final state. We know already what it is going to be.

We now adjust the semantics so that leakage from the initial state is accounted for, even if it is subsequently overwritten.
Let $C{\In}\CalX{\MFun}\CalY$ be a channel and $M{\In}\CalX{\MFun}\CalX$ a markov, as usual, and let $\CalZ$ be fresh. \label{g1235}Write $\Dal{C}{\CalZ}$ in $(\CalX{\times}\CalZ){\MFun}\CalY$ for the expanded channel
\[
(\Dal{C}{\CalZ})_{(x,z),y}\Wide{\Defs}C_{x,y}~,
\]
i.e.\ that $C$ ignores $z$.
Similarly $\Dal{M}{\CalZ}{\In}(\CalX{\times}\CalZ){\MFun}(\CalX{\times}\CalZ)$ is given by
\[
(\Dal{M}{\CalZ})_{(x,z),(x',z')}\Wide{\Defs}\textrm{\quad$M_{x,x'}$ if $z{=}z'$ else $0$}~,
\]
i.e.\ so that $M$ does not change $x$.
Thus these definitions ensure that for any $\pi{\In}\Dist(\CalX{\times}\CalZ)$ neither $\ChApp{\pi}{(\Dal{C}{\CalZ})}$ nor $\ChApp{\pi}{(\Dal{M}{\CalZ})}$ depends on the $\CalZ$ component. Take for example $\CalZ \Defs \{z_0, z_1\}$ consider $C,M$ as below:
{\small%0629
	\[
	C~=~ 
	\begin{array}{ccc}
	& y_0 & y_1 \\[.5em]
	x_0{:} & \LMat{1.1em}{-.6em}{1.5em}1 & 0\RMat{1.1em}{-.6em}{1.5em} \\
	x_1{:} & \nicefrac{1}{4} & \nicefrac{3}{4}
	\end{array}
	\hspace{3em}
	M~=~ 
	\begin{array}{ccc}
	& x_0 & x_1 \\[.5em]
	x_0{:} & \LMat{.9em}{-.5em}{1.5em}\nicefrac{1}{2} & \nicefrac{1}{2}\RMat{.9em}{-.51em}{1.5em} \\
	x_1{:} & \nicefrac{1}{2} & \nicefrac{1}{2}
	\end{array}
	\]
	\medskip
	\[
	\Dal{C}{\CalZ}\Wide{=} 
	\begin{array}{c@{\quad\quad}cc}
	& y_0 & y_1 \\[.5em]
	(x_0, z_0){:} & \LMat{1.1em}{-1.8em}{2.8em} 1 & 0\RMat{1.4em}{-1.8em}{2.8em} \\
	(x_0, z_1){:} & 1 & 0 \\
	(x_1, z_0){:} & \nicefrac{1}{4} & \nicefrac{3}{4} \\
	(x_1, z_1){:} & \nicefrac{1}{4} & \nicefrac{3}{4}
	\end{array}
	\]
	\medskip
	\[
	\Dal{M}{\CalZ} \Wide{=} 
	\begin{array}{ccccc}
	& x_0z_0 & x_0z_1 & x_1z_0 & x_1z_1 \\[.5em]
	x_0z_0{:} & \LMat{1.1em}{-1.8em}{2.8em}\nicefrac{1}{2} & 0 & \nicefrac{1}{2} & 0\RMat{1.6em}{-1.8em}{2.8em} \\
	x_0z_1{:} & 0& \nicefrac{1}{2}& 0& \nicefrac{1}{2} \\
	x_1z_0{:} & \nicefrac{1}{2}& 0 & \nicefrac{1}{2} & 0 \\
	x_1z_1{:} & 0 & \nicefrac{1}{2} & 0 & \nicefrac{1}{2}
	\end{array}~.
	\]
}%0629

\medskip
The definitions above show that in $\Dal{C}{\CalZ}$ the rows of the original $C$ are each repeated $2{=}\Size{\CalZ}$ times; and the subsequent update by $\Dal{M}{\CalZ}$ leaves $\CalZ$  unchanged. Observe that these definitions now account for information flows with respect to initial distributions $\Dist(\CalX{\times}\CalZ)$ where, crucially, the $\CalZ$ component is merely ``carried along''. But it captures the Dalenius effect mentioned, as we now explain.

Consider an initial distribution $\pi{\In}\Dist(\CalX{\times}\CalZ)$ such that $\pi_{x_i, z_j} = 1$ if and only if $i{=}j$, i.e.\ that $z$ is a copy of $x$'s initial value. We see that, even though $\CalZ$ is not accessed by the program at all, if ever  $y_1$ is observed then the $\CalZ$ component must certainly be $z_1$, and if $y_0$ is observed then it is $4$ times more likely to be $z_0$ than $z_1$.

Although $\CalZ$ is arbitrary, it can be shown that this Dalenius effect on any $\CalZ$ can be determined by the \HMM\ semantics \emph{specifically in the case where $\CalX{=}\CalZ$} as just above. That is, we do not have to consider ``all $\CalZ$'s'', which would be impractical. Note the construction of a fully compositional semantics for programs with hidden states is requires further extensive conceptual and technical work which we have developed elsewhere~\cite{McIver:2017}.
%By projecting onto the $\CalZ$ component, $\HMMsem{\Dal{C}{\CalZ}}{\Dal{M}{\CalZ}}$ as a whole acts as a pure channel, and if $\Size{\CalZ}{\geq}\Size{\CalX}$ then the component matrices $C$ and $M$ can be completely recovered from observations made only on the composition $\HMMsem{\Dal{C}{\CalZ}}{\Dal{M}{\CalZ}}$.
%++\Cf{Do you mean $\HMMsem{\Dal{C}{\CalX}}{\Dal{M}{\CalX}}$? I don't really understand this last paragraph. {\Ax If it makes it clearer to have $\CalX$ instead of {\CalZ} then that is fine. The important part is that the size of this extra space is large enough.}}

\section{Related work}\label{s1418}
%++\Cf{Do we need to refer to \Cite{Kalnishkan}?}
There is great diversity in approaches to information flow in (probabilistic) programs, which we have surveyed in our own earlier work \cite{mcivermeinicke10a,McIver:12,McIver:2014aa,Alvim:2014aa}. Here we have concentrated on general techniques for semantic constructions, in particular those based on monads, duality and refinement.

Refinement of probabilistic programs appeared in \cite{Jones:89} where evaluations were used to construct a powerdomain for probabilistic but possibly non-terminating computations; this was extended to include demonic choice in the discrete case in \cite{Morgan:96d,McIver:05a}, and was significantly generalised in \cite{Tix:09}. Our ``uncertainty refinement" that combines information flow with functional properties first appeared for information flow in straight-line programs in \cite{mcivermeinicke10a}, was extended to general measure spaces \cite{McIver:12} and appeared independently for the specific case of channels \cite{Alvim:2012aa}. Whereas Jones and Plotkin began with an underlying partial order over which to construct a probability space, our uncertainty-refinement order begins ``one level up'', using hyper-distributions $\Dist^2\CalX$ to encode an ``attack model'' that accounts for information flow.

Doberkat defines \emph{stochastic relations} that correspond to forward-semantic functions of type $\CalX{\Fun}\Dist\CalX$ for Markov processes: these are what we generalise by going ``one level up''. The converse of those stochastic relations \cite{Doberkat:2003aa} might improve the presentation of our \Def{d1426HB}, where a hyper is extracted from a channel and a prior, i.e.\ from a joint distribution.

Dual models for program semantics include \cite{Dijkstra:76}, then for probabilistic programs \cite{Kozen:83,Jones:90} in the purely probabilistic case. Subsequently \cite{Morgan:96d} added demonic choice. And \cite{Tix:09,G-Larrecq:2007} study dual models for probability and nondeterminism using a version of Riesz's representation theorem.
%++\Cf{Is this true for both \cite{Tix:09,G-Larrecq:2007}?}

In particular, Goubault-Larrecq's approach \cite{G-Larrecq:2007} to combining probability and nondeterminism differs from our earlier work \cite{Morgan:96d}. It uses general denotations for probabilistic programs in which nondeterminism is introduced at the level of measures (by weakening the modularity law) rather than as healthy sets of measures \cite{Morgan:96d,McIver:05a,Tix:09}. That leads naturally to a backward semantics of probabilistic demonic programs because nondeterminism is captured within integration. There is thus a strong analogy between our \UM-transformers and Goubault-Larrecq's ``previsions'' because both are continuous functionals that act on some set of tests (bounded continuous functions). The main difference is that our \UM-transformers are specifically tailored to capture security semantics, which is what leads to concavity on our set of uncertainty measures. Notice moreover that Goubault-Larrecq encounters a difficulty similar to our composition of \HMM's, that the decomposition $\HMMsem{C}{M}$ (resp.\ collinearity) is not preserved by Giry composition. Indeed, both difficulties are resolved by working in a larger space, namely, the space of abstract \HMM's (resp.\ not-necessarily-collinear continuous previsions).

In \cite{PanangadenDPP14}  a dual model for Markov processes is used to prove properties about  approximations of finite behaviours, and in \cite{GretzKM14} it is shown how expectation transformers  relate to explicit program models described by Markov processes. 

Recently Jacobs and Hasuo have explored a general categorical construction of a backward transformer semantics from a forward monadic model of probabilistic computations (discrete, continuous and quantum) \cite{Jacobs:2013,Hasuo:2014}. Their construction uses measurability as the underlying feature of ``predicates", while the stronger condition of continuity is crucial for our uncertainty measures. It would be interesting to see whether an instantiation of that categorical derivation can provide more structure for what we have done.

The concave functions advocated here for analysing information-flow properties have appeared in \cite{McIver:12,Alvim:2012aa} and have been identified in \cite{Kifer:2010} as an ingredient in privacy analysis.

\section{Conclusions and prospects}\label{s1323}

{%1357

}%1357

\begin{Figure}{f1843}{Relationship between the semantic spaces.}
	\includegraphics[scale=0.8]{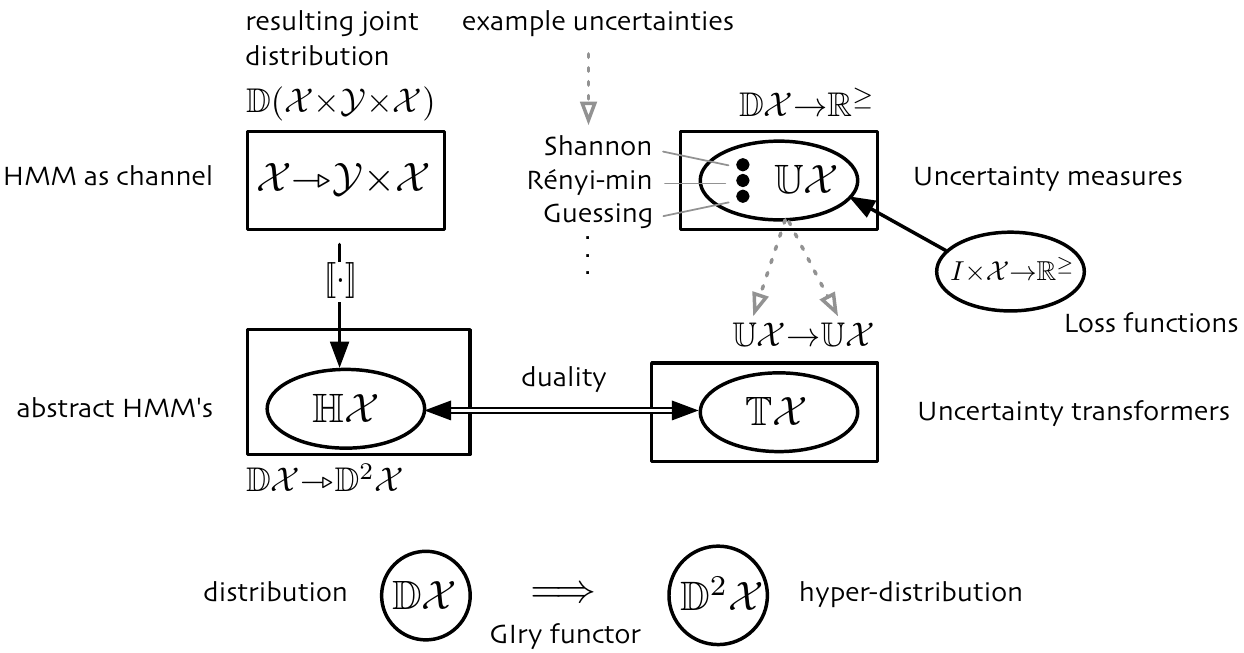}
\end{Figure}

Our principal objective was to provide an abstract setting for \HMM's based on well understood principles of semantic spaces. We did that using Giry's general monadic framework applied at the level of $\Dist\CalX$ (rather than $\CalX$); the resulting structures include a refinement order which is sensitive to both functional \emph{and} information-flow properties, and they lead to a dual, transformer space supported by theorems demonstrating the duality. \Fig{f1843} summarises the results:
\begin{itemize}

\item Top-left \Fig{f1843} shows the three-way joint distribution $\Dist(\CalX{\times}\CalY{\times}\CalX)$ produced by an \HMM\ applied to a prior of type $\Dist\CalX$: recall \Sec{s1403B}.

\item At bottom-left we formulate \emph{abstract \HMM's} over a state $\CalX$, a type $\HS\CalX$, as a monadic model for \HMM's over $\CalX$, and give their characteristic properties: recall \S\ref{s0913},\ref{s1432},\ref{s0842}.

\item At top-right we have \emph{uncertainty measures} $\Unc\CalX$ as a generalisation of diverse entropies (top centre), and we gave their characteristic properties: recall \Sec{s0731}.\,%

\item We showed (centre right) that uncertainty measures \textbf{have a complete representation} as \emph{loss functions}; recall \Sec{s1636}.

\item We gave a dual, uncertainty-transformer semantics $\TT\CalX$ of $\HS\CalX$, stating its characteristic properties (bottom right) and proved that they enable the duality with $\AHS$ (centre): recall \Thm{t1005} in \Sec{s1031}.

\item We showed how all of that is an instance of the general Giry monad as a computation, of which (finite) \HMM's use a discrete portion (bottom centre): recall \Sec{s1250}.

\item We explained how the ``Dalenius effect'' is manifested as a compositional issue in this framework, and suggested how it can be treated: recall \Sec{s0917}.

\item We stated and proved \Thm{t1005}, which we believe 
is a significant new result, in particular
its assumptions and proof.
\end{itemize}

More abstractly (recall \Sec{s0429}), we aimed to profit by joining two ideas: the established use of \HMM's as descriptions of probabilistic mechanisms having hidden state, and the established use of monads for modelling computations. Our novel use of $\Dist\CalX$ in the monad, rather than the state $\CalX$ itself, is the principal innovation that allowed this; and the synthesised hyper-distribution space that results leads to other advantages (the two $\dagger$'s below).

An immediate benefit accrues because, in monad-enabled programming languages, probabilistic-programming packages can be built very quickly and e.g.\ \cite{Erwig:06} is just one of many examples. Indeed the translation into real programs is almost elementary because of the powerful and general structures available: the Haskell prototype independently verifies the examples in Figs.~\ref{f0544m}--\ref{f0544cmnu}. (See \App{a1409} for an overview.)

More importantly, any monad brings with it both general equational properties and specific properties applying to the monad in question (such as those in \cite{Giry:81}). These conceptual tools allow reasoning about the structures modelled (\HMM's in this case) in ways that would be obscured by their more direct operational representation (e.g.\ as matrices).

\begin{itemize}
  \item
The other advantages of hypers are several: one is that they abstract from differences between entropies in a way that allows all of the entropies to be used uniformly. For example, a hyper contains all the information necessary to calculate the information leakage of a particular program fragment (typically, in the security literature, a pure channel \Sec{s1407c}), as shown in \cite{McIver:2014aa}, and furthermore the Kantorovich-metric structure of $\Dist\CalX$ we used earlier for channels \cite{Alvim:2014aa} now carries over to \HMM's.

\item
Another advantage of hypers is that their partial-order enables semantics for ``looping \HMM's'' in the standard way (least fixed-point) for computer science, rather than a direct ad-hoc definition based on matrices. Indeed a typical use of \HMM's is to run a single \HMM-step (\Sec{s1403B}) repeatedly and then to make statistical deductions about its hidden features: sophisticated mathematical tools are available for this special case \cite{Jurafsky:00}. Via abstract \HMM's we can however, in principle, handle complex, heterogeneous systems beyond (what amounts to, in the special case just above) a single loop containing just a single statement.
\end{itemize}

Our more concrete aim (again \Sec{s0429}) was to allow source-level reasoning about probabilistic programs with hidden state. Historically \emph{at the source level} this works best with backwards reasoning based on predicates (or similar) that can be embedded between program statements rather than forwards reasoning which, here, would be calculations using $\AHSpc$ directly.

Here our ``predicates'' are \UM's, which in this paper however are mathematical objects unsuitable for embedding directly in program texts (see \App{a0751}, last paragraph) As remarked in \Sec{s0534}, however, any \UM\ can be expressed as $U_l$ for some loss-function $l$ which function --crucially-- is indeed an expression based on program variables \cite{Chatzikokolakis:2014aa}. The added complexity introduced by the hidden state is that the program-logic based on that observation must represent the index-set ($I$) of the loss function; that would most likely be done by adding a special-purpose quantifier (since the loss-function index must be a bound variable within the assertion, not appearing in the program proper).

Exploiting this opportunity for a source-level quantitative logic of probabilistic hidden state is planned for future work.

\subsubsection*{Acknowledgements}
We're grateful for advice from Franck van Breugel, James Worrell, Tom Schrijvers and other members of IFIP WG 2.1, and for inspiration and insight from the INRIA Princess team. 
We acknowledge support from the Australian Research Council's grant DP120101413 and the INRIA {\'e}quipe associ{\'e}e Princess; and Morgan acknowledges the support of Data61.

%\newpage
\bibliographystyle{IEEEtran}
\bibliography{LiCS15} % Extracted what we used from probsNew into LiCS15 so that we could fiddle.

\newpage
\appendix

\section{Summary of notations \hfill{ }}\label{s1241}
\Cf{Check the page numbers ascend.}
These entries list in first-use order the points at which notation is introduced during the exposition: a detailed explanation of each is given there.

{%1010 To encapsulate macro definitions.
	\newcommand\Item[3] {\>\small#2\>\parbox[t]{24em}{\small#3}\`\small p.\pageref{#1}\\}
	\renewcommand\Space {~\\[-.5em]}
	\begin{tabbing}
		\hspace{0em}\=\hspace{8em}\=\kill
		%%%
		\Item{g1320}{$-^\dagger$}{Kleisli extension}
		\Item{g1054}{$f.x$ vs.\ $f(x)$}{Function application is ``$.$'', i.e. a dot.}
		\Item{g1042}{$\CalX{\MFun}\CalY$}{Type of a matrix.}\Space
		\Item{g1036}{$C_{x,y},C_{-,y}$ etc.}{Elements of vectors and matrixes by index; whole rows/columns.}\Space
		\Item{g1043}{$\VType{\CalX}$}{Type of vector.}\Space
		\Item{g1034}{$(\MMult)$}{Matrix multiplication: vectors automatically taken as row- or column- for conformity.}\Space
		\Item{g1052}{$(:)$ vs.\ $(\in)$}{Declaration vs.\ property.}
		\Item{g1022}{u.c.\ Roman letter}{Matrices: $C$ for channels; $M$ for transformers; $H$ for \HMM's}\Space
		\Item{g1024}{$\CalX$}{Finite set of states.}
		\Item{g1024}{$\CalY$}{Finite set of observations.}\Space
		\Item{g1026}{l.c.\ Greek letter}{Vectors, usually distributions over $\CalX$: $\pi$ for priors; $\rho$ for posteriors; $\delta$ for others.}\Space
		\Item{g1038}{$\SSum()$}{Weight (sum of elements) of vector or matrix.}
		\Item{g1039}{$\ChApp{\pi}{C}$}{Channel applied to prior.}
		\Item{g1049}{$\Nrm{-}$}{Normalisation of distribution.}
		\Item{g2002}{\emph{similar}}{wrt.\ columns of joint matrix.}
		\Item{g1056}{$y_{1,2}$ vs.\ $y_1,y_2$}{Former abbreviates latter.}
		\Item{g1058}{$\Dist$}{Discrete-distribution type constructor, a functor.}
		\Item{g1058}{$\Dist^2$}{Distribution-of-distributions.}
		\Item{g1058}{hyper}{Abbreviation of ``hyper-distribution''.}
		\Item{g1121}{inner}{Element of a hyper's base type.}
		\Item{g1121}{outer}{Distribution of a hyper over its inners.}
		\Item{g1102}{$\Hyp{\cdot}$}{Semantic function for \HMM's.}
		\Item{g1401}{$\SubDist$}{Sub-distribution.}
		\Item{g1401}{$\SubHyp$}{Sub-hyper.}
		\Item{g1011}{$x\PC{p}x'$}{The two-point distribution ``$x$ with probability $p$ and $x'$ with probability $1{-}p$''.}\Space
		\Item{g1212}{$\Point{\cdot}$}{Point distribution.}
		\Item{g1404}{$\SubPt{-}$}{Sub-point distribution.}
		\Item{g1041}{$\Supp{\delta}$}{The support of a distribution.}
		%\Item{g1103}{$J^\downarrow$}{\Cx Reduced matrix.}\Space
		\Item{g1109}{$\Avg$}{Average (of hyper); multiply in monad.}
		\Item{g1140}{$\Delta$}{Upper-case Greek for hypers.}
		\Item{g0928}{channel}{The emission part of an \HMM-step.}
		\Item{g0928}{markov}{The transition part of an \HMM-step.}
		\Item{g1117}{$\CM{C}{M},\CM{C}{},\CM{}{M}$}{One-step \HMM\ defined by channel and markov.}
		\Item{g1124}{$\NullChannel$}{Channel that releases no information.}
		\Item{g1123}{$\ID$}{Identity (Markov) transform.}
%		\Item{g1130}{``inner'' distributions}{Posteriors to which hyper assigns probabilities.}\Space
%		\Item{g1130}{``outer'' probabilities}{Probabilities assigned by hyper to posteriors.}\Space
		\Item{g0924}{\AtP}{Notation for specific hyper-distributions}\Space
		\Item{g1134}{$({;})$}{Sequential composition of \HMM's.}
		\Item{g1136}{$(\ChPar{\cdot}{\cdot})$}{Parallel composition of channels.}
%		\Item{g1226}{$H$ vs.\ $h$}{Matrix- vs.\ abstract \HMM's.}
		\Item{g1108}{$\Dist f$}{Push-forward of $f$.}
		\Item{g1143}{$(\Ref)$}{Refinement relation between hypers.}
		\Item{g1142}{$\DELTA$}{A distribution of hypers.}
		\Item{g1659}{$\Expt{\delta}{f}$}{Expected value.}
		\Item{g1210}{$\Pi,\mu,\eta$}{Giry/Lawvere functor etc.}
		\Item{g1214}{$\cal B$}{Metric-monad functor.}
		\Item{g1504}{$\KDist$}{Kantorovich metric.}
		\Item{g1233}{$d_1$}{Discrete metric.}
		%\Item{g1234}{$d_M$}{Manhattan metric.}
%		\Item{g1234}{$d_{TV}$}{Total variation metric.}
%		\Item{g1230}{$\MDist{-}{-}$}{Manhattan distance.}
		\Item{g1621}{$(\WS{p})$}{Weighted sum of distributions.}
		\Item{g1628}{$\AHS$}{Abstract \HMM's on $\CalX$.}
		\Item{g1147}{$\Unc\CalX$, \UM's}{Uncertainty measures on $\CalX$.}
		\Item{g1151}{$\Wp{h}$}{Uncertainty transformer (determined by $h$.)}
		\Item{g1254}{$\TT$}{The type of uncertainty transformers}
		\Item{g1827}{$\FunctOne$}{The everywhere-one function.}
		\Item{g1156}{$U_l$}{Uncertainty measure defined by loss-function $l$.}
		\Item{g1235}{$\Dal{H}{\CalZ}$}{Dalenius $\cal Z$-extension (of \HMM\ $H$).}
		\Item{g1146}{$(\Comp)$}{Functional composition.}
%		\Item{g1105}{$p(x),p(X|y)$ etc.}{Conventional notations for joint distributions.}
		\Item{g1613}{$\Cont\CalX$}{Continuous functions from $\Dist\CalX$ to $\Real$.}
		\Item{g1152}{$\UMet{\cdot}{\cdot}$}{Uniform metric on $\Unc\CalX$ and $\Cont\CalX$.}
		\Item{g1127}{$(\lambda \cdots)$}{Lambda abstraction.}
		\Item{g0939}{$\LossApp{\pi}{l}$}{$\pi$-skewed loss function.}
		
		%\Item{g1235}{$\Dal{C}{\CalZ}$}{Dalenius-expanded channel}
		%\Item{g1012}{$(\cdot)^T$}{matrix transpose.}
		%\Item{g1133}{\C{???}}{???}
		%\Item{g1149}{$\Expt{\Delta}{u}$}{Expected value (of $u$ on $\Delta$); in the continuous case equiv.\ $\int u \Dx{\Delta}$.}\Space
		%%%
		%\Item{g1111}{$\Cx \AHS$}{What goes here?}
	\end{tabbing}
}%1010
%\end{document}
\section{Characterisation of pure channels and pure markovs \AppFrom{\Sec{s1243}}}\label{a1105}
\Cf{Move this material into the main body?}

\medskip
\medskip
\subsection{Pure abstract markovs}\label{s1537}
Since a pure markov reveals nothing, a pure abstract markov $h{\In}\AHS$ should produce only point hypers, i.e.\ have for all $\pi{\In}\Dist\CalX$ that $h.\pi=\Point{\rho}$ for some $\rho$ (depending on $\pi$).

\label{g1146}From that we can deduce that for any pure abstract markov $h$ the effect of $\Avg\Comp h$ (on some $\pi$) is matrix multiplication by some $M$ (independent of $\pi$). That is, for any $0{\leq}p{\leq}1$ we have
\begin{equation}\label{e1217}
(\Avg\Comp h).(\pi_1\WS{p}\pi_2)
\Wide{=}
(\Avg\Comp h).\pi_1\,\WS{p}\,(\Avg\Comp h).\pi_2~,
\end{equation}
which property characterises matrix multiplication. This is because $h.(\pi_1\WS{p}\pi_2){=}\Point{\rho}$ and $h.\pi_{1,2}{=}\Point{\rho_{1,2}}$ resp.\ for some $\rho,\rho_{1,2}$, together with \Lem{l1433}, gives
\[
\Point{\rho_1}\WS{p}\Point{\rho_2}
\Wide{\Ref}
\Point{\rho} ~,
\]
and the only way that can hold is if $\rho=\rho_1\WS{p}\rho_2$, which is precisely the claim made at \Eqn{e1217} just above.

\medskip
\subsection{Pure abstract channels}\label{s0906}
A pure channel is one that releases information about the distribution on $\CalX$ but does not change it: one can think of the transformation part as the identity matrix. Thus \Eqn{e1217} above suggests that we should have that $\Avg\Comp h$ is the identity for a pure channel, i.e.\ that $\Avg.(h.\pi){=}\pi$. This is necessary, but turns out not to be sufficient: we explore a fuller characterisation of channels later (\App{a1632}).

\section{Equivalent presentations of refinement: \Lem{l0827} \AppFrom{\Sec{s1432}}}\label{s0851}

\Lem{l0827} concerned two definitions of uncertainty refinement, showing them to be equivalent: one was formulated for joint distributions (defined at \Eqn{e1005} within the lemma), suitable for discrete reasoning; and the other was formulated for hypers (\Def{d1449}), suitable for extension to more general reasoning (e.g.\ proper measures). We sketch the proof of that equivalence in \App{s0851A} immediately below.

In this section however we present an example, two hypers $\Delta_{S,I}$ shown to satisfy $\Delta_S{\Ref}\Delta_I$ in both presentations (\Def{d1449}, \Lem{l0827}), with an explanation of how to move from one presentation to the other.

As in \Eqn{e0852n} of \Sec{s1407c}, we use the following notation for discrete distributions where specific values in the support are named: we write
\begin{equation}\label{e0902}
\begin{array}{l@{\,}l@{\,}l}
x_1 &\AtP& p_1 \\
x_2 &\AtP& p_2 \\
\multicolumn{3}{l}{\textit{etc...}}
\end{array}
\end{equation}
If these are laid out horizontally, we enclose them in double set-brackets $\DDist{\cdots} $ separated by commas: thus $\DDist{H\AtP\nicefrac{2}{3},T\AtP\nicefrac{1}{3}}$ describes a coin twice as likely to give heads as tails. If the double brackets are used without probabilities (and thus also without $\AtP$'s) then the intended distribution is uniform, so that $\DDist{H,T}$ describes a fair coin; a convenient special case of that is $\DDist{H}$ for the point distribution on $H$, the coin that gives heads every time.\,%
\footnote{In the semantic space we write $\Point{x}$ for that: here we are syntactic.}

Let $\CalX$ be the set $\DSet{H,T}$ of coin-flip results. We choose our two hypers as follows, presenting them as at \Eqn{e0902}:
\[
\begin{array}{rcl}
\Delta_S &=& \left[
\begin{array}{l@{\hspace{1em}}l@{\,}l}
H\PC{{\nicefrac{2}{3}}}T &\AtP& \nicefrac{1}{2} \\ % 0913: Dunno why nicefrac needs extra {}.
H\PC{{\nicefrac{1}{3}}}T &\AtP& \nicefrac{1}{2}
\end{array}
\right] \\[3ex]
\Delta_I &=& \left[
\begin{array}{l@{\hspace{1em}}l@{\,}l}
H\PC{{\nicefrac{2}{3}}}T &\AtP& \nicefrac{1}{3} \\ % 0913.
H\PC{{\nicefrac{1}{2}}}T &\AtP& \nicefrac{1}{3} \\
H\PC{{\nicefrac{1}{3}}}T &\AtP& \nicefrac{1}{3}~.
\end{array}
\right]
\end{array}
\]

The first hyper $\Delta_S$ represents choosing fairly between two biased coins and having the chosen one secretly flipped: we know which coin was flipped, but we are not allowed to see the outcome of the flip. In $\Delta_I$ however we choose fairly between \emph{three} coins: the two biased coins from before, and a fair one. Again the chosen one is secretly flipped; again we are not allowed to see the outcome.
%\Cf{I wonder whether it would be better to see the flip outcome and to try to guess which coin produced it. That's more the ``channel'' point of view: the input is ``which coin'' and the channel reveals the flip. On the other hand, in an \HMM\ we really might be better going this ``other way around'' because the currently selected $H,T$ is the hidden state we are interested in. Channels don't have state.
%
%In both cases the guessing strategy is very simple; but for $S$ your chance of guessing right is $2/3$ while for $I$ it has dropped to $11/18$ (from $12/18$). This is (Bayes) vulnerability going down. Entropies however should go up: for guessing entropy we move from $4/3$ for $S$ to $25/18>24/18$ for $I$; for Shannon entropy we move from $0.64$ to $0.76$.}

We argue that in any reasonable measure of secrecy, it should in the second case $\Delta_I$ be harder to guess which of $H,T$ resulted from the flip than in the first case $\Delta_S$. And it is precisely that non-specific ``in \emph{any} reasonable measure'' that uncertainty refinement $\Delta_S{\Ref}\Delta_I$ attempts to capture.\,%
\footnote{Furthermore, the powerful ``Coriaceous'' completeness property (\Lem{l0755}) shows the dual result: if some $\Delta_S,\Delta_I$ are \emph{not} in the refinement relation, that is $\Delta_S{\NotRef}\Delta_I$, then there is \emph{guaranteed} to be a uncertainty measure wrt.\ to which $\Delta_I$ is \emph{not} more secure than $\Delta_S$.}

In this case, and informally speaking, $\Delta_I$ is more secure than $\Delta_S$ because there is now a third possible case that acts as a linear combination of the existing two. That is, some of the separation between the inners $H\PC{{\nicefrac{2}{3}}}T$ and $H\PC{{\nicefrac{1}{3}}}T$ in the support of $\Delta_S$ has been merged together to become a single inner $H\PC{{\nicefrac{1}{2}}}T$ in the support of $\Delta_I$ --- and what makes the observer more uncertain is that he doesn't know how to pull that single inner apart again.

Two (reduced) joint matrices $J_{S,I}$ that give $\Delta_{S,I}$ resp.\ are
\[
\begin{array}{rcl}
J_S &=& 
\begin{array}{r@{\hspace{2em}}cc}
& a & b \\[.2em]
H{:} & \LMat{.7em}{-.5em}{1.5em}\nicefrac{1}{3} & \nicefrac{1}{6}\RMat{.9em}{-.51em}{1.5em} \\
T{:} & \nicefrac{1}{6} & \nicefrac{1}{3}
\end{array} \\[4ex]
J_I &=& 
\begin{array}{r@{\hspace{2em}}ccc}
& c & d & e \\[.2em]
H{:} & \LMat{.7em}{-.5em}{1.5em}\nicefrac{2}{9} & \nicefrac{1}{6} & \nicefrac{1}{6}\RMat{.9em}{-.51em}{1.5em} \\
T{:} & \nicefrac{1}{9} & \nicefrac{1}{6} & \nicefrac{2}{9} 
\end{array}
\end{array}
\]
where the observation spaces are $\CalY_S{=}\DSet{a,b}$ and $\CalY_I{=}\DSet{c,d,e}$ respectively. (the column names are arbitrary.) Now the refinement matrix that establishes (according to \Lem{l0827}) that $\Delta_S{\Ref}\Delta_I$ is $R{\In}\CalY_S{\MFun}\CalY_I$ given by
\vspace{-1em}
\[
\begin{array}{rcl}
R &=& 
\begin{array}{r@{\hspace{2em}}ccc}
& c & d & e \\[.2em]
a{:} & \LMat{.7em}{-.5em}{1.5em}\nicefrac{2}{3} & \nicefrac{1}{3} & 0\RMat{1.5em}{-.51em}{1.5em} \\
b{:} & 0 & \nicefrac{1}{3} & \nicefrac{2}{3} 
\end{array}
\end{array}
\]
which, read columnwise, says in its column $c$ that to make Column $c$ of $J_I$ you take $\nicefrac{2}{3}$ of Column $a$ of $J_S$ and none of Column $b$ of $J_S$. The middle column $d$ of $R$ is where the actual refinement lies, that Column $d$ of $J_I$ is made by adding $\nicefrac{1}{6}$ of each of Columns $a,b$ of $J_S$ together. This is where $J_I$ (equiv.\ $\Delta_I$) reveals less than $J_S$ (equiv.\ $\Delta_S$) does about the distribution on $\CalX{=}\DSet{H,T}$. And, as the lemma suggests, we indeed have $J_S{\MMult}R=J_I$.

The alternative, more abstract presentation of this is in terms of \Def{d1449}, i.e.\ where the $\DELTA$ we are looking for, that establishes $\Delta_S{\Ref}\Delta_I$ at the hyper-level directly, can be given as (the denotation of) a joint distribution $J{\In}\Dist\CalX{\MFun}\CalY_I$ itself: we will have $\DELTA\Defs\Hyp{J}$ which, because $J$'s source type is $\Dist\CalX$, will have type $\Dist^2(\Dist\CalX) = \Dist^3\CalX$ as we expect from $\Hyp{\cdot}$. The rows of $J$ will be labelled by the support of $\Delta_S$, i.e.\ it will have only two rows so that we have
\vspace{-1em}
\begin{equation}\label{e1238}
\begin{array}{rcl}
J &=& 
\begin{array}{r@{\hspace{2em}}ccc}
& c & d & e \\[.2em]
H\PC{{\nicefrac{2}{3}}}T: & \LMat{.7em}{-.5em}{1.5em}\nicefrac{1}{3} & \nicefrac{1}{6} & 0\RMat{1.5em}{-.51em}{1.5em} \\
H\PC{{\nicefrac{1}{3}}}T: & 0 & \nicefrac{1}{6} & \nicefrac{1}{3} 
\end{array}~.
\end{array}
\end{equation}

If on the other hand we were to write $\DELTA{=}\Hyp{J}$ as a hyper directly (performing the various
normalisations etc.)\ we would have
\[
\begin{array}{rcl}
\DELTA &=& \left[
\begin{array}{l@{\hspace{1em}}l@{\,}l}
\Point{~H\PC{{\nicefrac{2}{3}}}T~} &\AtP& \nicefrac{1}{3} \\ % 0913.
(H\PC{{\nicefrac{2}{3}}}T) \PC{{\nicefrac{1}{2}}} (H\PC{{\nicefrac{1}{3}}}T)
&\AtP& \nicefrac{1}{3} \\
\Point{~H\PC{{\nicefrac{1}{3}}}T~} &\AtP& \nicefrac{1}{3}
\end{array}
\right]
\end{array} ~,
\]
with each inner here corresponding to a row of \Eqn{e1238}.

Now $\Avg.\DELTA$ is given by the calculation
\begin{Reason}
	\Step{}{
		&\Point{~H\PC{{\nicefrac{2}{3}}}T~}{\times}\nicefrac{1}{3} \\
		+&(H\PC{{\nicefrac{2}{3}}}T) \PC{{\nicefrac{1}{2}}} (H\PC{{\nicefrac{1}{3}}}T){\times}\nicefrac{1}{3} \\
		+&\Point{~H\PC{{\nicefrac{1}{3}}}T~}{\times}\nicefrac{1}{3}
	}
	\Space
	\Step{$=$}{
		H\PC{{\nicefrac{2}{3}}}T) \PC{{\nicefrac{1}{2}}} (H\PC{{\nicefrac{1}{3}}}T)
	}
	\Step{$=$}{
		\Delta_S~.
	}
\end{Reason}
This can also be seen (indeed is easier to see) if we simply take the left-marginal of $J$, for which you add the columns together: you get
\[
\begin{array}{rcl}
\begin{array}{r@{\hspace{2em}}c}
& c{+}d{+}e \\
&= 1 \\[.6em]
H\PC{{\nicefrac{2}{3}}}T: & \LMat{.7em}{-.5em}{1.5em}\nicefrac{1}{2}\RMat{1.5em}{-.51em}{1.5em} \\
H\PC{{\nicefrac{1}{3}}}T: & \nicefrac{1}{2} 
\end{array}
\end{array}~,
\]
which is again $\Delta_S$.

For the other direction we obtain $(\Dist\Avg).\DELTA$ by $\Avg$'ing each inner of $\DELTA$ while preserving the (outer) probabilities.
\footnote{Recall that the inners of $\DELTA$ are themselves hypers, which is why they can be $\Avg$'d.}
That gives
\[
(\Dist\Avg).\DELTA \Wide{=}
\left[
\begin{array}{l@{\hspace{1em}}l@{\,}l}
H\PC{{\nicefrac{2}{3}}}T &\AtP& \nicefrac{1}{3} \\ % 0913.
H\PC{{\nicefrac{1}{2}}}T &\AtP& \nicefrac{1}{3} \\
H\PC{{\nicefrac{1}{3}}}T &\AtP& \nicefrac{1}{3}
\end{array}
\right]~,
\]
because
\begin{eqnarray}
\Avg.\Point{~H\PC{{\nicefrac{2}{3}}}T~} &=& H\PC{{\nicefrac{2}{3}}}T \nonumber\\
\Avg.(H\PC{{\nicefrac{2}{3}}}T) \PC{{\nicefrac{1}{2}}} (H\PC{{\nicefrac{1}{3}}}T) &=& H\PC{{\nicefrac{1}{2}}}T \nonumber\\
\Avg.\Point{~H\PC{{\nicefrac{1}{3}}}T~} &=& H\PC{{\nicefrac{1}{3}}}T \nonumber
\end{eqnarray}
And so that the remaining question is ``How do we get such a $\DELTA$ from a given $R$\/?\,"

Remember that the support of $\Delta_S$ is $\DSet{H\PC{{\nicefrac{2}{3}}}T,H\PC{{\nicefrac{1}{3}}}T}$. Make a distribution $\pi_S$ by mapping those (inner) distributions of $\Delta_S$ onto the labels in $\CalY_S$ associated uniquely with them in $J_S$. (The association is unique because $J_S$ is reduced.) That gives us that $\pi_S$ is of type $\Dist\CalY_S$ and has value $a\PC{{\nicefrac{1}{2}}}b$.

Now form the joint-distribution matrix $\ChApp{\pi_S}{R}$, i.e.
\[
\ChApp{
	\begin{array}{rcc}
	a{:} & \LMat{.7em}{-.5em}{1.5em}\nicefrac{1}{2}\RMat{1.5em}{-.51em}{1.5em} \\
	b{:} & \nicefrac{1}{2}
	\end{array}
	\hspace{1em}}{\hspace{.7em}\raisebox{.6em}{$%1227
		\begin{array}{ccc}
		c & d & e \\[.3em]
		\LMat{.7em}{-.5em}{1.5em}\nicefrac{2}{3} & \nicefrac{2}{3} & 0\RMat{1.5em}{-.51em}{1.5em} \\
		0 & \nicefrac{1}{3} & \nicefrac{2}{3}
		\end{array}
		$}%1227
}
\hspace{.7em}=
\raisebox{.6em}{$%1227A
	\begin{array}{rccc}
	& c & d & e \\[.2em]
	a{:} & \LMat{.7em}{-.5em}{1.5em}\nicefrac{1}{3} & \nicefrac{1}{6} & 0\RMat{1.5em}{-.51em}{1.5em} \\
	b{:} & 0 & \nicefrac{1}{6} & \nicefrac{1}{3} 
	\end{array}
	$}%1227A
\]
which (like $R$ itself) is of type $\CalY_S{\MFun}\CalY_I$. (But note that $R$ is a channel matrix, whereas $\ChApp{\pi_S}{R}$ is a joint-distribution matrix.)

Now use the relabelling in the reverse direction on the rows of the joint distribution above (as ``new row-labels'' at right above) to get a matrix with the same contents but now of type $\Dist\CalX{\MFun}\CalY_I$. It is
\[
\begin{array}{r@{\hspace{2em}}ccc}
& c & d & e \\[.2em]
H\PC{{\nicefrac{2}{3}}}T: & \LMat{.7em}{-.5em}{1.5em}\nicefrac{1}{3} & \nicefrac{1}{6} & 0\RMat{1.5em}{-.51em}{1.5em} \\
H\PC{{\nicefrac{1}{3}}}T: & 0 & \nicefrac{1}{6} & \nicefrac{1}{3} 
\end{array}
\]
which is exactly the $J$ we had at \Eqn{e1238} above, and as above we get $\DELTA$ via $\DELTA{=}\Hyp{J}$.

Thus in this example we have illustrated how one might move between the two equivalent definitions of refinement. Each one has a witness: in the hyper-formulation it is the distribution on hypers $\DELTA$; and in the matrix formulation is is a post-processing ``refinement matrix'' $R$. The sketch proof (\App{s0851A}) shows how to obtain each from the other in general.

\section{Monadic vs.\ matrix presentations of refinement \AppFrom{\Sec{s1432}}}\label{s0851A}
In \App{s0851} we gave an example of the two equivalent presentations of refinement; here we give a proof (sketch) that it can always be done.

\begin{ReLemma}{Refinement of joint-distribution matrices}{l0827}
	Let $J_S{\In}\CalX'{\MFun}\CalY_S$ and $J_I{\In}\CalX'{\MFun}\CalY_I$ be joint-distribution matrices, both of them \emph{reduced} in the sense of \Def{d1426HB}, such that $\Hyp{J_{S,I}}{=}\Delta_{S,I}$ resp.
	In this section only we use $\CalX'$ as a reminder that the \emph{input} side of these $J$'s, their row-indices, is actually the \emph{output} side of the \HMM's from which they are derived, i.e.\ that $J_{x',y}=\sum_x H_{x,y,x'}$ as in \Def{d1022}.
	
	We prove the equivalence
	\[
	\Delta_S \Ref \Delta_I
	\WideRM{iff}
	J_S\MMult R = J_I \textrm{\quad for some $R$}
	\]
	where $R$ is a stochastic \emph{refinement matrix} of type $\CalY_S{\MFun}\CalY_I$ (i.e.\ such that $\sum R_{y,-}=1$ for each $y{\In}\CalY_S$).
	\Proof
	\newcommand\OneOne {\stackrel{\mbox{\rm\scriptsize1-1}}{\longleftrightarrow}}
	First we note that for any reduced joint distribution matrix $J{\In}\CalZ{\MFun}\CalZ'$ there is a one-one correspondence between $J$'s column labels, i.e.\ elements of $\CalZ'$, and the support of the hyper $\Delta{=}\Hyp{J}$ that $J$ defines: it is the function $j{\In}\CalZ'\OneOne\Supp{\Delta}$ from \Def{d1426HB}, injective into $\Dist\CalZ$ because $J$ is reduced. We write $(\OneOne)$ to emphasise our one-one use of it below.
	
	\textbf{$R$ makes $\DELTA$}:\quad
	We show first that for $J_{S,I},\Delta_{S,I}$ and $R$ as above we can construct a suitable $\DELTA$. Let the relabelling associated with $J_S$ be $j_S{\In}Y_S\OneOne\Supp{\Delta_S}$.
	Relabel $\Delta_S$ so that it is a distribution of support $\Dist Y_S$, so that we can use \Def{d1426HB} to define $\DELTA\Defs\Hyp{\ChApp{\Delta_S}{R}}$, noting that the types of (relabelled) $\Delta_S{\in}\Dist\CalY_S$ and of $R{\in}\CalY_S{\MFun}\CalY_I$ are precisely what \Def{d1426HB} requires to produce a result of type $\Dist^2\CalY_S$. Now relabel this (back again) to make an element of $\Dist^2\Supp{\Delta_S}$, that is of $\Dist^3\CalX$ because $\Supp{\Delta_S}{\subseteq}\Dist\CalX$.
	
	We have $\Delta_S{=}\Avg.\DELTA$ immediately, from the remark following \Def{d1416CB}.
	
	For $(\Dist\Avg).\DELTA{=}\Delta_I$ we first calculate
	\begin{Reason}
		\Step{}{
			(\Dist\Avg).\Hyp{\ChApp{\Delta_S}{R}}
		}
		\StepR{$=$}{Set $\CalD\Defs\Supp{\Delta_S}$ in \Lem{l0759} below}{
			\Hyp{M{\MMult}(\ChApp{\Delta_S}{R})}~,\\
			\quad \textrm{where $M_{x,\rho}\Defs\rho.x$ for $x{\In}\CalX$ and $\rho{\In}\Supp{\Delta_S}$.}
		}
	\end{Reason}
	Now for arbitrary $x{\In}\CalX$ and $y_I{\In}\CalY_I$ we continue
	\begin{Reason}
		\Step{}{
			(M{\MMult}(\ChApp{\Delta_S}{R}))_{x,y_I}
		}
		\Step{$=$}{
			\sum_{\rho{\In}\CalD}M_{x,\rho}\,(\Delta_S)_{\rho}\,R_{\rho,y_I}
		}
		\StepR{$=$}{Defn.\ $M$; $\CalD{=}\Supp{\Delta_S}$}{
			\sum_{\rho{\In}\Supp{\Delta_S}}\rho.x\,(\Delta_S)_{\rho}\,R_{\rho,y_I}
		}
		\StepR{$=$}{$\Supp{\Delta_S}{=}\CalY_S$; $\Delta_S{=}\Hyp{J_S}$}{
			\sum_{y_S{\In}\CalY_S}(J_S)_{x,y_S}\,R_{\rho,y_I}
		}
		\StepR{$=$}{$J_I{=}J_S{\MMult}R$}{
			J_I~,
		}
	\end{Reason}
	whence $(\Dist\Avg).\DELTA = \Hyp{J_I} = \Delta_I$ as required.
	
	\textbf{$\DELTA$ makes $R$}:\quad
	To show that from $\DELTA$ we can construct a suitable $R$, we do similar calculations to the above, but in the reverse direction.
	%\Cf{Maybe should fill this in somewhat. But I feel there must be a \emph{much} easier way of doing this, so I think I won't spend more time now on it. With luck, the example will be enough for reviewers, and they won't want to read this proof carefully anyway.}
\end{ReLemma}

\begin{Lemma}{Technical lemma}{l0759}
	Let $\CalD{\subseteq}\Dist\CalX$ be some finite set of distributions on $\CalX$, and let $J{\In}\CalD{\MFun}\CalY$ be a joint-distribution matrix between (those) \emph{distributions} on $\CalX$ and some observation space $\CalY$. Then $\Dist\Avg.\Hyp{J}=\Hyp{M{\MMult}J}$, where $M{\In}\CalX{\MFun}\CalD$ is defined $M_{x,\delta}\Defs \delta.x$ for $x{\In}\CalX$ and $\delta{\In}\CalD$.
	\Proof
	%Arithmetic, and careful treatment of indices.
	%\Cf{Need a better proof than this\ldots}
	Let us match inners (and associated weight) of $\Hyp{M{\MMult}J}$ with that of $\Dist\Avg.\Hyp{J}$. These are finitely supported distributions so the following sums are all finite.
	
	Let $y{\In}\CalY$. On the one hand, the $y$-inner $\delta$ of $\Hyp{M{\MMult}J}$ satisfies, for every $x{\In}\CalX$,
\[
\delta.x = \frac{(M\MMult J)_{x,y}}{\sum_x (M\MMult J)_{x,y}} = \frac{\sum_\rho M_{x,\rho}J_{\rho,y}}{\sum_x\sum_\rho M_{x,\rho} J_{\rho,y}} = \frac{\sum_\rho \rho.xJ_{\rho,y}}{\sum_\rho J_{\rho,y}}
\]
and this inner has weight $\sum_\rho J_{\rho,y}$.

On the other hand, the $y$-inner $\Delta$ of $\Hyp{J}$ satisfies, for every $\rho$,
\[
			\Delta.\rho = 
			\frac{J_{\rho,y}}{\sum_\rho J_{\rho,y}}~.
\]
This inner has weight $\sum_\rho J_{\rho,y}$. Applying the $\Avg$, we get 
\[\Avg.\Delta.x = \sum_{\rho}\rho.x\Delta.\rho = \frac{\sum_\rho \rho.xJ_{\rho,y}}{\sum_\rho J_{\rho,y}}\]
Since $\Dist\Avg$ simply distributes through the inners of $\Hyp{J}$, we deduce that $\Dist\Avg.\Hyp{J}$ has the exact same inners as $\Hyp{M\MMult J}$ with the exact same weights. That is, the two discrete hyper-distributions are equal.
\end{Lemma}

\section{Properties of the refinement order $(\Ref)$ \hfill{ }}\label{a1006}

\subsection{Abstract \HMM's are $(\Ref)$-monotonic}\label{a1113}\AppFrom{\Sec{s0842A}}\\
Super-linearity (\Lem{l1433}) is equivalently $(\Ref)$-monotonicity of the Kleisli-extension $h^\dagger$ of any $h{\In}\AHS$; that is, it is equivalent to the more general $\Delta_1{\Ref}\Delta_2 \Imp h^\dagger.\Delta_1\Ref h^\dagger.\Delta_2$.  Assuming $(\Ref)$-monotonicity and recalling that $\Point{\cdot}$ is the point distribution,
we have trivially the inequality $\Point{\pi_1} \WS{p} \Point{\pi_2}\Ref\Point{\pi_1 \WS{p} \pi_2}$ and so

\begin{Reason}
	\Step{}{
		h.\pi_1\,\WS{p}\,h.\pi_2
	}
	\StepR{$=$}{defn.~$h^\dagger$}{
		h^\dagger.\Point{\pi_1}\,\WS{p}\,h^\dagger.\Point{\pi_2}
	}
	\StepR{$=$}{$h^\dagger$ linear}{
		h^\dagger.(\Point{\pi_1}\WS{p}\Point{\pi_2})
	}
	\Space
	\StepR{$\Ref$}{$\Point{\pi_1}\WS{p}\Point{\pi_1}\Ref\Point{\pi_1\WS{p}\pi_2}$;\\
		assumption that $h^\dagger$ is monotonic}{
		h^\dagger.\Point{\pi_1\WS{p}\pi_2}
	}
	\Space
	\StepR{$=$}{defn.~$h^\dagger$}{
		h.(\pi_1\WS{p}\pi_2) ~.
	}
\end{Reason}

For the other direction (sketch), in the discrete case we note that a proof of $\Delta_1{\Ref}\Delta_2$ can be broken down into a succession of column-merges (in the matrix representation), each of them being of the form ``replace $\Point{\pi_1}\WS{p}\Point{\pi_2}$ by $\Point{\pi_1\WS{p}\pi_2}$''. 

\medskip
\subsection{Composition of abstract \HMM's respects the refinement order}\label{a1025}\AppFrom{\Sec{s0842A}}\\
We show that sequential composition of abstract \HMM's respects the refinement order $(\Ref)$ on both sides, i.e.\ that for $h,h_{1,2}{\In}\AHS$ we have both
\begin{eqnarray}
&& h_1\Ref h \Wide{\Imp} h_1;h_2\Ref h;h_2 \label{e1427A} \\
& \textrm{and\quad} & h_2\Ref h \Wide{\Imp} h_1;h_2\Ref h_1;h~. \label{e1427B}
\end{eqnarray}

%\Tf{The following proofs may be redundante (as Carroll mentions below) and may generate confusion (circularity?). Maybe we should put \Sec{a1306} and \Sec{a1139} before these proofs and then use them to prove \Eqn{e1427A} and \Eqn{e1427B} for HMM's. \Cx I would like to keep them in their present order, because that matches their occurrence in the main text. But I have referred forward (as you can see), and dropped the earlier proofs. Thanks for the suggestion!}

Although this can be argued directly in terms of abstract \HMM's, it is easier if we use the \UM's defined later (\Sec{s1244A}). For \Eqn{e1427A} we have
\begin{Reason}
	\Step{}{
		h_1\Ref h
	}
	\StepR{iff}{\App{a1139} just below}{
		\Wp{h_1}\leq\Wp{h}
	}
	\Step{implies}{
		~\\\Wp{h_1}\Comp\Wp{h_2}~\leq~\Wp{h}\Comp\Wp{h_2}
	}
	\StepR{iff}{\Cor{c1020} in \App{a1625}}{
		\Wp{(h_1;h_2)}\leq\Wp{(h_1;h)}
	}
	\StepR{iff}{\App{a1139}}{
		h_1;h_2\Ref h_1;h ~.
	}
\end{Reason}

And for \Eqn{e1427B} we have 
\begin{Reason}
	\Step{}{
		h_2\Ref h
	}
	\StepR{iff}{\App{a1139} just below}{
		\Wp{h_2}\leq\Wp{h}
	}
	\Space
	\StepR{implies}{$\Wp{h_1}$ is $(\leq)$-monotonic, \Lem{l0846A}(\ref{i1243})}{
		~\\\Wp{h_1}\Comp\Wp{h_2}~\leq~\Wp{h_1}\Comp\Wp{h}
	}
	\Space
	\StepR{iff}{\Cor{c1020} in \App{a1625}}{
		\Wp{(h_1;h_2)}\leq\Wp{(h_1;h)}
	}
	\StepR{iff}{\App{a1139}}{
		h_1;h_2\Ref h_1;h ~.
	}
\end{Reason}
\medskip
\subsection{Refinement of transformers}\label{a1139}\AppFrom{\Sec{s1408}}\\
Here we prove the correspondence between the forwards- and the backwards manifestations of refinement $(\Ref)$, i.e.\ that we have
\[
h_1\Ref h_2
\WideRM{iff}
\Wp{h_1} \leq  \Wp{h_2} ~,
\]
where on the \rhs we have extended $(\leq)$ pointwise, i.e.\ meaning $\Wp{h_1}.u.\pi\leq\Wp{h_1}.u.\pi$ for all $u{\In}\Unc\CalX$ and $\pi{\In}\Dist\CalX$. We reason
\begin{Reason}
	\Step{}{
		h_1\Ref h_2
	}
	\WideStepR{iff}{pointwise extension $(\Ref)$}{
		h_1.\pi \Ref h_2.\pi \textrm{\quad for all $\pi{\In}\Dist\CalX$}
	}
	\Space
	\WideStepR{iff}{\Lem{l0755}, soundness and completeness}{
		\Expt{h_1.\pi}{u} \leq \Expt{h_2.\pi}{u}
		\quad\begin{tabular}[t]{l}
			for all $\pi{\In}\Dist\CalX$ 
			and all $u{\In}\Unc\CalX$
		\end{tabular}
	}
	\Space
	\WideStepR{iff}{defn.\ $\Wp{()}$}{
		\Wp{h_1}.u.\pi \leq  \Wp{h_2}.u.\pi
		\quad\begin{tabular}[t]{l}
			for all $\pi{\In}\Dist\CalX$ \\
			and for all $u{\In}\Unc\CalX$
		\end{tabular}
	}
	\Space
	\StepR{iff}{pointwise extension}{
		\Wp{h_1} \leq  \Wp{h_2} ~.
	}
\end{Reason}

\medskip
\subsection{Composition respects transformer refinement}\label{a1306} \AppFrom{\Sec{s1408}}\\
For $t_{1,2}{\In}\Unc\CalX$ we have defined $t_1{\Ref}t_2$ to be simply that $t_1.u{\leq}t_2.u$ for all $u{\In}\Unc\CalX$. Here we show that functional composition of transformers respects that refinement order $(\Ref)$ on both sides, i.e.\ that for $t,t_{1,2}{\In}\TT\CalX$ we have both $t_1{\Ref}t \Imp t_1{\Comp}t_2\Ref t{\Comp}t_2$ and $t_2\Ref t \Imp t_1{\Comp}t_2\Ref t_1{\Comp}t$.

In fact it is trivial from the property (imposed by $\TT\CalX$) that transformers are $(\leq)$-monotonic, that is \Lem{l0846A}(\ref{i1243}).
%}%1029

\subsection{Soundness and completeness: \Lem{l0755}}\label{a0810}\AppFrom{\Sec{s1244A}}\\
We mention soundness and completeness in this paper because it provides an important justification for our definition and use of the general uncertainty measures and, in particular, their transformers.

The \underline{soundness} part of \Lem{l0755} is related to the Data-Processing Inequality, the \textit{DPI} \cite{Cover:2006aa} , which concerns two channels $C{\In}\CalX{\MFun}\CalY$ and $R{\In}\CalY{\MFun}\CalZ$. (Note that the channel $R$ here takes the \emph{observations} $\CalY$ of Channel $C$ as its input. Our \HMM's do not take observations as input.)

Informally stated, the \emph{cascade} of $C$ and $R$ is the channel given by the matrix multiplication $C{\MMult}R$, and the \textit{DPI} states that the information leakage from $C{\MMult}R$ cannot be more than the leakage from $C$ alone: adding another child to the game ``Chinese Whispers'' cannot make the eventual output less ridiculous.

We call this \emph{soundness} because it states that a no-less-secure hyper wrt.\ our uncertainty refinement order indeed cannot be less uncertain when tested with \emph{any} uncertainty measure. This result is proved in in \cite{McIver:2014aa,Alvim:2012aa}.

The \underline{com}p\underline{leteness} part of \Lem{l0755} is related to the ``Coriaceous Conjecture'' partially proved in \cite{Alvim:2012aa}, which became the 
Coriaceous Property (\textit{CP}) in \cite{McIver:2014aa} when its proof, for channels, was presented in complete form based on McIver's earlier, complete proof in \cite{mcivermeinicke10a} for hypers.\,%
\footnote{\Label{n0840}Geoffrey Smith has since told us that it follows from a result of Blackwell \cite{Blackwell:1951}.}
In \cite{McIver:2014aa} terms, the \textit{CP} is that if there \emph{is no} $R$ such that $C_1{\MMult}R{=}C_2$ then there is a ``gain function'' (for us here, a loss function, which determines a special form of uncertainty measure in our terms) that is witness to the non-existence of such an $R$. The importance of the \textit{CP} for quantitative information-flow security was explained in \cite{Alvim:2012aa}, and it was proved there to hold for many interesting special cases of $C_{1,2}$. But not for all of them.

The \textit{CP} was proved to extend beyond the discrete case, to proper measure spaces, in \cite{McIver:12}.

\section{Proof of \Lem{l1719A} \AppFrom{\Sec{s1031}}}\label{a1613}

This technical lemma assures the well definedness of our dual space: we have defined our uncertainty measures $\Unc\CalX$ as functions in $\Dist\CalX{\Fun}\NNReal$ with certain properties; and we have stated that $\Wp{h}.u$ is also an uncertainty measure. Thus we must show that $\Wp{h}.u\in\Dist\CalX{\Fun}\NNReal$ and that it satisfies the properties for membership of $\Unc\CalX$.

\begin{ReLemma}{Well-definedness of \Def{d1701A}}{l1719A}
	If $h{\In}\HS\CalX$ is an abstract \HMM\ and $u{\In}\Unc\CalX$ is a \UM, then $\Wp{h}.u$ is in $\Unc\CalX$.
	\Proof 
	Since $h{\in}\HS\CalX$, we have $h{\In}\AHSpc$ satisfying Lems.~\ref{l1432},\ref{l1433}. We must show for any $u{\In}\Unc\CalX$ and $\delta{\In}\Dist\CalX$ that $\delta{\mapsto}\Expt{h.\delta}{u}$ is in $\Unc\CalX$, i.e.\ that it is in $\Dist\CalX\Fun\NNReal$, is concave and is continuous (\Def{d1709A}).
	
	Membership of $\delta{\mapsto}\Expt{h.\delta}{u}$ in $\Dist\CalX\Fun\NNReal$ is trivial.
	
	\smallskip
	\underline{For concavit}y:\quad Because $u{\in}\Unc\CalX$ we know it is itself concave; and we have that $h$ satisfies the properties in \Def{d0943}. We now reason
	\begin{Reason}
		\Step{}{
			\Wp{h}.u.(\pi_1~\WS{p}~\pi_2)
		}
		\StepR{$=$}{\Def{d1701A}}{
			\Expt{h.(\pi_1\WS{p}\pi_2)} u
		}
		\Space
		\WideStepR{$\geq$}{
			\Def{d0943}, hence $(h.\pi_1)\WS{p}(h.\pi_2) \Ref h.(\pi_1\WS{p}\pi_2)$ \\
			$u$ concave, hence \Lem{l1228} (non-strict) applies
		}{
			~\\\Expt{(h.\pi_1)\WS{p}(h.\pi_2)} u
		}
		%\WideStepR{$\geq$}{
		% \Lem{l1433}: $u$ concave hence $\Delta{\mapsto}\Expt{\Delta}{u}$ monotonic\\
		% \Lem{l1228}: $(h.\pi_1)\WS{p}(h.\pi_2) \Ref h.(\pi_1\WS{p}\pi_2)$}{
		% ~\\\Expt{(h.\pi_1)\WS{p}(h.\pi_2)} u
		%}
		\Space
		\StepR{$=$}{$\Expt{}{}$ is linear}{
			\Expt{h.\pi_1}{u}~~\WS{p}~~\Expt{h.\pi_2}{u}
		}
		\StepR{$=$}{\Def{d1701A}}{
			\Wp{h}.u.\pi_1~\WS{p}~\Wp{h}.u.\pi_2 ~,
		}
	\end{Reason}
	as required.
	%Note that the appeal to \Lem{l1433} was where we used the hypothesis that $h$ is an abstract \HMM\ (rather than merely of type $\AHSpc$).
	
	\smallskip
	\underline{For continuit}y:\quad
	We must show that $\Wp{h}.u$ is continuous, given that both $u,h$ are themselves continuous.
%	Because $h$ itself is continuous, we need only show that in general the function $\Delta{\mapsto}\Expt{\Delta}{u}$ is continuous wrt.\ Kantorovich on the left, in $\Delta$ for fixed continuous $u$. This is proved in \Lem{l1527A} just below. It in turn requires several supporting theorems, given in \App{a1328} immediately following.
Because $h$ itself is continuous, we need only show that in general the function $\Delta{\mapsto}\Expt{\Delta}{u}$ is continuous wrt.\ Kantorovich on the left, in $\Delta$ for fixed continuous $u$.  This follows from the fact that $\Dist\CalX$ is a compact metric space so that the Kantorovich metric metrizes the weak topology on $\Dist^2\CalX$. That is, $\Expt{\Delta_n}{u}$ converges to $\Expt{\Delta}{u}$ for every continuous function $u{\In}\Dist\CalX{\to}\NNReal$ iff $\Delta_n$ converges to $\Delta$ wrt. Kantorovich metric.
\end{ReLemma}

\section{Extension of transformers \AppFrom{\Sec{s1408}}}\label{s1805}
The core ingredient in the proof of this theorem is the Riesz Representation Theorem for linear functionals (linear maps from a normed vector space to $\Real$). A difficulty however originates from the fact that the representation theorem is stated on the space of all continuous functions $\Cont\CalX$ (defined below), but our linear function $t$ is defined only from the subspace $\Unc\CalX$ to itself. 

\begin{Definition}{Space of continuous functions}{d1302}\label{g1613}
	We define $\Cont\CalX$ to be the set of all \emph{continuous functions} from $\Dist\CalX$ (with the Kantorovich metric) to $\Real$ (with the ordinary metric). This set is endowed with the \emph{uniform metric} $\UMet{\cdot}{\cdot}$,  \label{g1152} defined
	\begin{equation}\label{e1634}
	\UMet{u_1}{u_2} \Wide{\Defs} \sup_{\delta{\In}\Dist\CalX}|u_1.\delta-u_2.\delta|~,
	\end{equation}
	that turns $\Cont\CalX$ into a complete metric space.
\end{Definition}

Yet $\Unc\CalX$ is a sub-metric space of $\Cont\CalX$ under the uniform metric $\UMet{\cdot}{\cdot}$. More importantly, we prove that the vector space generated by $\Unc\CalX$ is dense in $\Cont\CalX$. (See \Lem{l1643} and \Fig{f0904}.) This is essential to ensure that if $t$ extends to a continuous linear function over $\Cont\CalX$, then such an extension is necessarily unique. We will show in \Thm{l1701} that such an extension always exists.

\begin{Figure}{f0904}{The sum of convex (red, upper) and concave (blue, lower) functions gives a zig-zag (black, middle).}
	\includegraphics[scale=0.3]{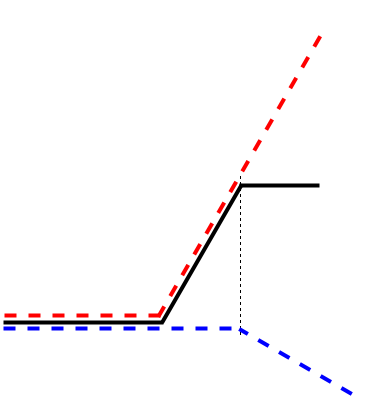}
	\captionsetup{singlelinecheck=off,font=footnotesize}
	\caption*{Every continuous piecewise-linear function can be expressed as the sum of finitely many convex and concave functions using the construction shown in this figure. This provides a geometrical view of the Concave Density \Lem{l1643}. By summing up concave and convex functions, we get non concave (resp. convex) functions. In fact, the density theorem shows that any arbitrarily shaped function can be approximated using finite sums and products of  scaled concave functions.}
	\captionsetup{singlelinecheck=on,font=normalsize}
\end{Figure}

\begin{Lemma}{Concave density}{l1643}
	The vector space generated by $\Unc\CalX$ is dense in $\Cont\CalX$ \wrt $\UMet{\cdot}{\cdot}$.
	\Proof
	This result essentially follows from \cite[Pro.~2.2]{Borwein:2011}. We give the proof here for completeness. 
	
	Let $\Span{\Unc\CalX}$ be the set of functions that can be written as the difference of two positive concave functions from $\Dist\CalX$ to $\Real$. Then $\Span{\Unc\CalX}$ coincides with the real vector space generated by $\Unc\CalX$ (by grouping positively and negatively weighted components). Equivalently, every function in $\Span{\Unc\CalX}$ is the difference of two positive continuous convex functions: if $f = u_1-u_2$ for $u_{1,2}{\In}\Unc\CalX$, then
	\[
	f = (-u_2-c) - (-u_1 - c)
	\]
	where $c = \min(\inf_{\delta{\In}\Dist\CalX}-u_1.\delta, \inf_{\delta{\In}\Dist\CalX}-u_2.\delta)$. The constant $c$ is finite because $\Dist\CalX$ is compact. The functions $-u_{1,2}-c$ are positive, continuous and convex functions.
	
	Now let us apply the Stone-Weierstrass Density Theorem \cite[Thm.~5]{Stone:1948} on $\Span{\Unc\CalX}$ which is a subset of $\Cont\CalX$. 
	
	To do that we need first to show that $\Span{\Unc\CalX}$ is an algebra (i.e. has a zero and unit, is closed under scalar multiplication and addition and pointwise multiplication of $f$'s). In addition $\Span{\Unc\CalX}$ must ``vanish nowhere" on $\Dist\CalX$ and ``separate points". (See below for explanations of those properties.)
	
	\medskip
	
	\underline{$\Span{\Unc\CalX}$ is an algebra:}
	Since $\Span{\Unc\CalX}$ is a vector space, the constant functions $\mathbf{0},\FunctOne$ (identically $0$ and $1$ resp.) and the functions $cf,f+g$ are in $\Span{\Unc\CalX}$ for every $c{\In}\Real$ and $f,g{\In}\Span{\Unc\CalX}$.
	
	Let $f,g{\In}\Span{\Unc\CalX}$ be such that $f=u_1{-}u_2$ and $g = v_1{-}v_2$ where $u_{1,2},v_{1,2}$ are positive continuous convex functions. Notice that 
	\[
	f^2 \Wide{=} 2(u_1^2 + u_2^2) - (u_1 + u_2)^2
	\]
	where $u_{1,2}^2$ and $(u_1{+}u_2)^2$ are positive convex functions (because the square of a non-negative convex function is convex). That is, we have $f^2\in\Span{\Unc\CalX}$. Now
	\[
	fg \Wide{=} (f+g)^2 - (f^2 + g^2)
	\]
	and thus $fg{\in}\Span{\Unc\CalX}$, because we have just shown that all of ${(f+g)^2},f^2,g^2$ are in $\Span{\Unc\CalX}$.
	
	\medskip
	
	\underline{$\Span{\Unc\CalX}$ vanishes nowhere:} We must show that for each $\delta{\In}\Dist\CalX$ there is some
	$f{\In}\Span{\Unc\CalX}$ such that $f\delta \neq 0$. But this is immediate since  $\FunctOne.\delta\neq 0$ for every $\delta{\In}\Dist\CalX$ and $\FunctOne{\in}\Span{\Unc\CalX}$.
	
	\medskip
	
	\underline{$\Span{\Unc\CalX}$ separates points:} We must show that for every pair $\delta \neq \delta'$ there is some $f \in \Span{\Unc\CalX}$ such that $f.\delta \neq f.\delta'$. We argue as follows.
	
	Given $\delta{\In}\Dist\CalX$ (fixed) and define $f^{\delta}.\delta'{\Defs}\KMet{\delta'}{\delta}$ for  $\delta'{\In}\Dist\CalX$. Observe that for $\delta'\neq \delta$ we have that $0 = f^\delta.\delta < \KMet{\delta'}{\delta} = f^\delta.\delta'$. Thus it suffices to show that $f^\delta{\in}\Span{\Unc\CalX}$, and  we continue as follows.
	
	For every $\delta_{1,2}{\In}\Dist\CalX$, we have
	\begin{Reason}
		\Step{}{
			f^\delta.(\delta_1\WS{p}\delta_2)
		}
		\StepR{$=$}{$d_K$ is Kantorovich distance; Definition of $f^\delta$}{
			\KMet{(\delta_1\WS{p}\delta_2)}{\delta}
		}
		\StepR{$\leq$}{$\KMet{\delta'}{\delta}$ is convex, for fixed $\delta$}{
			\KMet{\delta_1}{\delta}\WS{p}\KMet{\delta_2}{\delta} 
		}
		\StepR{$=$}{Definition of $f^\delta$}{
			f^\delta.\delta_1\WS{p}f^\delta.\delta_2
		}
	\end{Reason}
	That is $\FunctOne-f^\delta\in\Unc\CalX$ and thus $\Span{\Unc\CalX}$ separates points. 
	
	By the Stone-Weierstrass Theorem \cite[Thm.~5]{Stone:1948}, we have $\Span{\Unc\CalX}$ is dense in $\Cont\CalX$.
\end{Lemma}

The extension of a transformer $t{\In}\Unc\CalX{\to}\Unc\CalX$ to a continuous linear function from $\Cont\CalX$ to itself is done in two stages. Firstly, $t$ is extended \emph{linearly} to a continuous linear function $t'{\In}\Span{\Unc\CalX}{\to}\Cont\CalX$. This step is justified in \Thm{l1701}. Secondly, $t'$ is extended \emph{continuously} to a continuous linear function $\tilde{t}{\In}\Cont\CalX{\to}\Cont\CalX$. This step uses the density proven in \Lem{l1643} and is shown in \Lem{t1642} below.

\begin{Lemma}{Extension from $\Span{\Unc\CalX}$ to $\Cont\CalX$}{t1642}
	Every continuous linear function $t$ from $\Span{\Unc\CalX}$ to $\Cont\CalX$ extends uniquely to a continuous linear function $\tilde{t}$ from $\Cont\CalX$ to itself. 
	\Proof
	This result follows from the Continuous Linear Extension Theorem  \cite[Ch.~4 Thm.~10.1]{Loomis:1967}. 
	
	All we need to show is that the (Cauchy) completion of $\Span{\Unc\CalX}$ is $\Cont\CalX$, which follows from the fact that $\Span{\Unc\CalX}$ is dense in $\Cont\CalX$ (\Lem{l1643}) and that $\Cont\CalX$ is a complete normed vector space when endowed with the uniform norm $\UNorm{f}{\Defs}\UMet{f}{\mathbf{0}}$. 
\end{Lemma}

\begin{Theorem}{Extension from $\Unc\CalX$ to $\Cont\CalX$}{l1701}
	Every transformer extends uniquely to a positive continuous linear function from $\Cont\CalX$ to itself.
	
	\Proof
	Let $t{\In}\TT\CalX$ be a transformer. It suffices to prove that $t$ has a positive continuous extension $t'$ on the sub-vector space $\Span{\Unc\CalX}$. If such a $t'$ exists then a unique extension $\tilde{t}{\In}\CT$, which is positive 
	\footnote{For the positiveness of the continuous extension, if $f$ is a positive continuous function that is the uniform limit of a sequence of $f_n$'s in $\Span{\Unc\CalX}$, then the sequence of positive continuous functions $\max(\mathbf{0},f_n){\in}\Span{\Unc\CalX}$ also converges to $f$ \wrt the uniform metric. The reason is $|f.\delta - \max(0,f_n.\delta)|\leq |f.\delta - f_n.\delta|$, for every $\delta{\In}\Dist\CalX$ and positive $f$. Thus $t.f$ has to be positive.}
	and continuous, can be deduced using \Lem{t1642}.

	Let $f{\In}\Span{\Unc\CalX}$, there exists $u_{1,2}\in\Unc\CalX$ such that $f{=}u_1{-}u_2$. We define $t'.f{=}t.u_1 - t.u_2$.
	
	\underline{$t'$ is well-defined}:\quad 
	We must show that $t'.f$ is independent of how $f$ is written as the difference of two uncertainty measures. Firstly, notice that if $u_1{-}u_2\in\Unc\CalX$ for some $u_{1,2}\in\Unc\CalX$ then $t.(u_1{-}u_2) = t.u_1 - t.u_2$.
	%++\Tf{Otherwise, there exists a $\delta{\In}\Dist\CalX$ such that $t.(u_1-u_2).\delta \neq t.u_1.\delta - t.u_2.\delta$ and thus 
	%++\[
	%++	t.u_1.\delta = t.(u_1-u_2).\delta + t.u_2.\delta\neq t.u_1.\delta - t.u_2.\delta + t.u_2.\delta = t.u_1.\delta
	%++\]
	%++which is a contradiction. This wasn't obvious to me, but was to Annabelle. So, I added it here, just in case I forget again.}
	Secondly, let $f = u_1 - u_2 = v_1 - v_2$. Then $(u_1+v_2) - (u_2 + v_1) = \FunctZero$, which is in $\Unc\CalX$. Therefore, we have $t.(u_1+v_2) - t.(u_2 + v_1) = \FunctZero$, and that implies $t.u_1-t.u_2 = t.v_1 - t.v_2$ by linearity of $t$.
	
	\underline{$t'$ is linear and unique}:\quad Linearity is clear and it implies uniqueness of the extension $t'$ over $\Span{\Unc\CalX}$.
	
	\underline{$t'$ is $1$-Lipschitz}: \quad Let $f,g{\In}\Span{\Unc\CalX}$ be such that we have $f = u_1{-}u_2$ and $g = v_1{-}v_2$. Then 
	\begin{Reason}
		\Step{}{
			\UMet{t'.f}{t'.g}
		}
		\StepR{$=$}{Definition of $t'$}{
			\UMet{(t.u_1 - t.u_2)}{(t.v_1-t.v_2)}
		}
		\StepR{$=$}{$t$ is linear, $u_i+v_j{\in}\Unc\CalX$}{
			\UMet{t.(u_1 + v_2)}{t.(v_1+u_2)}
		}
		\StepR{$\leq$}{$t$ is $1$-Lipschitz}{
			\UMet{(u_1 + v_2)}{(v_1+u_2)}
		}
		\StepR{$=$}{Definition of $f,g$}{
			\UMet{f}{g}~.
		}
	\end{Reason}
	Therefore, $t'$ is also continuous.
	
	\underline{$t'$ is positive}: (i.e.\ it maps non-negative functions to non-negative functions). This follows from monotonicity of $t$.
	
	By \Lem{t1642}, the extension $t'$ further extends into a continuous positive linear function $\tilde{t}{\In}\Cont\CalX{\Fun}\Cont\CalX$ with $\tilde{t}.u = t.u$ for every $u{\In}\Unc\CalX$.
\end{Theorem}

\section{Proof of \Cor{c1020} \AppFrom{\Sec{s1408}}}\label{a1625}
This proof is made easier by operating in a slightly more general space than $\AHS$, i.e.\ the measurable subset of $\AHSpc$, not taking advantage of the stronger conditions that characterise $\AHS$ within it. In this section only we write $\WpE{}$ for the function defined as at \Def{d1701A} but over the larger space.
\begin{Lemma}{Transformer composition}{l1448}
	For any (measurable) $h_{1,2}{\In}\AHSpc$ we have that $\WpE{(h_1;h_2)} = \WpE{h_1}{\Comp}\WpE{h_2}$.
	\Proof
	\label{g1127}
	\begin{Reason}
		\Step{}{
			\WpE{(h_1;h_2)}.u.\pi
		}
		\StepR{$=$}{\Def{d1701A} extended to $\WpE{()}$}{
			\Expt{(h_1;h_2).\pi}{u}
		}
		\StepR{$=$}{$h_1;h_2$ is Kleisli composition}{
			\Expt{\Avg.(\Dist h_2.(h_1.\pi))} u
		}
		\Space
		\StepR{$=$}{$\Expt{\Avg.\DELTA} u = \Expt{\DELTA} (\LAbs{\Delta}{\Expt{\Delta}u})$ from ($\dag$) below \\ $\lambda$ is lambda-abstraction}{
			\Expt{\Dist h_2.(h_1.\pi)} (\LAbs{\Delta}{\Expt{\Delta}{u}})
		}
		\Space
		\StepR{$=$}{$\Expt{\Dist h_2.\Delta} F = \Expt{\Delta}{(F{\Comp}h_2)}$ from ($\ddag$) below}{
			\Expt{h_1.\pi} ((\LAbs{\Delta}{\Expt\Delta u})\Comp h_2)
		}
		\StepR{$=$}{make $\pi'$ explicit}{
			\Expt{h_1.\pi}{(\LAbs{\pi'}{((\LAbs{\Delta}{\Expt{\Delta}{u}})}\Comp h_2).\pi')}
		}
		\StepR{$=$}{$\Delta\Defs h_2.\pi'$}{
			\Expt{h_1.\pi}{(\LAbs{\pi'}{\Expt{h_2.\pi'}{u}})}
		}
		\StepR{$=$}{\Def{d1701A}}{
			\WpE{h_1}.(\LAbs{\pi'}{\Expt{h_2.\pi'} u}).\pi
		}
		\StepR{$=$}{\Def{d1701A}}{
			\WpE{h_1}.(\WpE{h_2}.u).\pi~.
		}
	\end{Reason}
	The identities $(\dag)$ and $(\ddag)$ were proven by Giry (\cite[Sec.~3 p.70]{Giry:81}). In $(\ddag)$, $F$ maps every hyper $\Delta$ to $\Expt{\Delta}{u}$.
\end{Lemma}

%\medskip
Remarkably, it is quite easy to show that $\Wp{()}$ is an injection over all of $\AHSpc$.
\begin{Lemma}{$\Wp{()}$ is an injection on $\AHSpc$}{l0857}
	If $\Wp{h_1}{=}\Wp{h_2}$ for some (measurable) $h_{1,2}{\In}\AHSpc$, then $h_1{=}h_2$.
	\Proof
	We reason 
	\begin{Reason}
		\Step{}{
			h_1\neq h_2
		}
		\StepR{$\Imp$}{for some $\pi{\In}\Dist\CalX$}{
			h_1.\pi\neq h_2.\pi
		}
		\StepR{$\Imp$}{\WLOG; $(\Ref)$-antisymmetry from \Sec{s1432}}{
			h_1.\pi\NotRef h_2.\pi
		}
		\StepR{$\Imp$}{\Lem{l0755} completeness (\emph{Coriaceous}),for some $u{\In}\Unc\CalX$}{
			\Expt{h_2.\pi}{u}< \Expt{h_1.\pi}{u}
		}
		\StepR{$\Iff$}{\Def{d1701A}}{
			\Wp{h_2}.u.\pi< \Wp{h_1}.u.\pi 
		}
		\Step{$\Imp$}{
			\Wp{h_1}\neq\Wp{h_2}~.
		}
	\end{Reason}
\end{Lemma}

Our next step is to use \Thm{t1005} to show that indeed $h_1;h_2{\in}\AHS$, so that $\WpE{}$ can be replaced by $\Wp{()}$ in \Lem{l1448} just above. We have
\begin{Lemma}{$\AHS$ closed under composition}{l1619}
	For $h_{1,2}{\In}\AHS$ we have $h_1;h_2{\in}\AHS$.
	\Proof
	If $h_{1,2}{\In}\AHS$ then $\Wp{h_{1,2}}{\in}\TT\CalX$ from Lems.~\ref{l0846A},\ref{l0846B}; and since those properties are closed under composition, we have that $\Wp{h_1}{\Comp}\Wp{h_2}{\in}\TT\CalX$ as well.
	
	From \Thm{t1005} there is then a unique $h{\In}\AHS$ such that $\Wp{h}=\Wp{h_1}{\Comp}\Wp{h_2}$; but examination of \Lem{l0857} shows membership of $\AHS$ is not necessary for that uniqueness: it applies to the whole of (measurable) $\AHSpc$. That is, there no other measurable $h$ in all of $\AHSpc$ such that $\WpE{h}=t$.
	
	From \Lem{l1448} we know that $\WpE{(h_1;h_2)}=t$, and so we must have $h_1;h_2=h\in\AHS$.
\end{Lemma}

Thus we can conclude
\begin{ReCorollary}{Transformer composition}{c1020}
	For any $h_{1,2}{\In}\AHS$ we have that also $h_1;h_2{\in}\AHS$, and furthermore $\Wp{(h_1;h_2)} = \Wp{h_1}{\Comp}\Wp{h_2}$.
	\Proof Lemmas \ref{l1448},\ref{l1619} just above.
\end{ReCorollary}

%\Cf{Removed\begin{quote}It remains to show that $h_1;h_2$ satisfies Lems.~\ref{l1432},\ref{l1433}. Since $h_{1,2}{\in}\AHS$, we know also that $\Wp{h_{1,2}}{\in}\C{\TT\CalX}$ and that they satisfy Lems.~\ref{l0846A},\ref{l0846B}. Since those properties are preserved by composition, we have that $t{=}\Wp{h_1}{\Comp}\Wp{h_2}$ also satisfies these lemmas. By \Thm{t1005}, there exists a unique $h{\In}\AHS$ such that $t = \Wp{h}$.\end{quote}}
%Since $\Wp{(h_1;h_2)}$ preserves continuity and concavity, \Lem{l0857} implies that $h{=}h_1;h_2$. Therefore, $h_1;h_2{\in}\AHS$ as required.
%\Cf{Why do we refer to continuity and concavity here?}

\section{Calculation of $\Wp{\HMMone{P}}$ \AppFrom{\Sec{s1224}}}\label{a0751}

In \Sec{s1224} a sample analysis was done on a very small program to show how, if the post-uncertainty is fixed, a pre-uncertainty can be calculated once and for all; and that then that pre-uncertainty can be used to investigate the security implications of a number of different priors, without having to re-analyse the program for each one.

Here we give the calculations for $\Wp{(\cdot)}$ in \Sec{s1224}. We note below however that ideally the pre-uncertainty would be calculated by \emph{source-level} reasoning; but that is not what we do here. (See also our ``more concrete aim'' in \Sec{s1323} concerning source-level reasoning.)

Let $P$ be the program set out in \Fig{f0544cm} (and also \Fig{f0544cmnu} from \App{a0407}). As usual for weakest preconditions, we work from post- to pre-. Let $u$ be the \UM\ from \Sec{s0903}, reflecting the circumstances of an attacker whose principal concern is whether the two bits of \XS\ are the same. 

Beginning with the second statement, since with transformers we work from the back towards the front, we expect informally that $\Wp{\HMMone{\texttt{\XS:= \XS$\PC{{\nicefrac{1}{2}}}$-\XS}}}.u$ is just $u$ again --- since the assignment does not affect $\XS[0]{=}\XS[1]$, whichever branch is taken. Calculation confirms that: for arbitrary $\pi$ we have
\begin{Reason}
	\Step{}{
		\Wp{\HMMone{\texttt{\XS:= \XS$\PC{{\nicefrac{1}{2}}}$-\XS}}}.u.\pi
	}
	\Space
	\StepR{$=$}{semantics of \texttt{\XS:= \XS$\PC{{\nicefrac{1}{2}}}$-\XS}}{
		\Expt{
			\mbox{\small$\begin{array}{r@{~}c@{~}l}
				[(& (\pi_{00}{+}\pi_{11})/2,(\pi_{01}{+}\pi_{10})/2, \\
				& (\pi_{10}{+}\pi_{01})/2,(\pi_{11}{+}\pi_{00})/2 & )]
				\end{array}$}
		}{u}
	}
	\Space
	\StepR{$=$}{expectation over point hyper}{
		u.
		\begin{array}[t]{r@{~}c@{~}l}
			(& (\pi_{00}{+}\pi_{11})/2,(\pi_{01}{+}\pi_{10})/2, \\
			& (\pi_{10}{+}\pi_{01})/2,(\pi_{11}{+}\pi_{00})/2 & )
		\end{array}
	}
	\Space
	\StepR{$=$}{definition $u$ from \Sec{s1224}}{
		& (\pi_{00}{+}\pi_{11})/2 + (\pi_{11}{+}\pi_{00})/2 \\
		\min &
		(\pi_{01}{+}\pi_{10})/2 + (\pi_{10}{+}\pi_{01})/2
	}
	\Space
	\Step{$=$}{
		(\pi_{00}{+}\pi_{11})
		~\min~
		(\pi_{01}{+}\pi_{10})
	}
	\StepR{$=$}{definition $u$ again}{
		u~,
	}
\end{Reason}
as we expected.

Continuing towards the front of the program we now calculate again for arbitrary $\pi$, but from just above able to use the same $u$ that we started with, that
\begin{Reason}
	\Step{}{
		\Wp{\HMMone{\texttt{\Leak\ $\XS[0]\,\PC{{\nicefrac{1}{2}}}\,\XS[1]$}}}.u.\pi
	}
	\Space
	\StepR{$=$}{
		semantics of \texttt{print $\XS[0]\,\PC{{\nicefrac{1}{2}}}\,\XS[1]$} \\
		define $s_0\Defs\pi_{00}{+}(\pi_{01}{+}\pi_{10})/2$ \\
		$s_1\Defs(\pi_{01}{+}\pi_{10})/2+\pi_{11}$
	}{\Expt{
			\mbox{\small$\begin{array}{r@{~}c@{~}l}
				& \CVec{\pi_{00}/s_0,\pi_{01}/2s_0,\pi_{10}/2s_0,0} \\
				\PC{{s_0}} & \CVec{0,\pi_{01}/2s_1,\pi_{10}/2s_1,\pi_{11}/s_1} \\
				\end{array}$}
		}{u}
	}
	\Space
	\StepR{$=$}{$\Expt{}{}$ linear, applied to two-point hyper (\Def{d1954})}{
		& u.\CVec{\pi_{00}/s_0,\pi_{01}/2s_0,\pi_{10}/2s_0,0} \\
		\PC{{s_0}} &
		u.\CVec{0,\pi_{01}/2s_1,\pi_{10}/2s_1,\pi_{11}/s_1}
	}
	\Space
	\StepR{$=$}{definition $u$\\from previous calculation}{
		\pi_{00}\min(\pi_{01}{+}\pi_{10})/2 + (\pi_{01}{+}\pi_{10})/2\min\pi_{11}~,
	}
\end{Reason}
as claimed in \Sec{s1224}.

We stress that calculating $\Wp{()}$ this way for any but the smallest programs is \emph{not practical at all}. For a practical calculus, instead the formulation of uncertainties as loss functions would be used to write them as expressions at the source level, i.e.\ over program variables, and then using formal manipulations in a quantitative program logic (extending e.g.\ \cite{Kozen:83,Morgan:96d}).

The issue of source-level reasoning is discussed further in the conclusion \Sec{s1323}.

\section{Using loss functions to characterise pure channels \hfill{ }}\label{s1211}\label{a1632}

With uncertainty transformers, we can be more precise about the properties satisfied by pure-(abstract) channel \HMM's specifically. As with markovs the mechanism by which information is released is independent of the (probability) values associated with the prior; in fact  it only depends on the underlying state value, that is $\CalX$. This property can be described neatly in terms of a ``multiplicative property" on transformers which, in addition, provides a 
characterisation of transformers which correspond to channels.
%We give details in \App{a1632}.
%In this section we show how the \emph{multiplicative property} sketched in \Sec{s1211} characterises pure channels.
We begin with a motivating example.

%\begin{window}[4,r,\raisebox{0pt}[1.4em][0pt]{\(%
% ~~~C^\pi~\Defs~~
%  \left(\begin{array}{cc}
%   \pi_0 & \pi_1\\
%   0 & 1\\
%  \end{array}\right)~.
%\)},{}]
Take $\CalX{=}\DSet{0,1}$. It's easy to construct an $h{\In}\AHS$ with the property that for all $\pi{\In}\Dist\CalX$ we have $\Avg.(h.\pi)=\pi$, which is to say that its markov is the identity, but it is still not a pure channel: we  simply ``cheat'' by using a different channel for each prior. Take for example the $\pi$-indexed channels given by the matrix
\[
C^\pi\Wide{\Defs}
\left(\begin{array}{cc}
\pi_0 & \pi_1\\
0 & 1\\
\end{array}\right)~.
\]

The function defined $f.\pi\Defs \AbCh{\ChApp{\pi}{C^\pi}}$ does not satisfy $f{=}\HMMsem{C}{}$ for any \emph{single} fixed $C$, and this example provides the insight for characterising pure channels: they have a simple multiplicative property, which we express using loss functions as follows.
%\end{window}
%\Cf{But is $f$ in $\AHS$ --- does it satisfy super-linearity?}

\begin{Definition}{Multiplicativity of transformers}{d1442}
	\label{g0939}
	For loss-function $l{\In}I{\Fun}\CalX{\Fun}\NNReal$ and $\pi{\In}\Dist\CalX$ define a $\pi$-skewed loss function $(\LossApp{\pi}{l}).i.x\Defs l.i.x{\times}\pi.x$.
	%\Tf{This needs to be uniform \wrt the footnote of \Def{d1124B} where the ordinary multiplication is denoted by juxtaposition instead of $\times$. I tried to write it as $(l.i.x)(\pi.x)$ but it looks strange.\CBar}
	We then say that transformer $t{\In}\TT\CalX$
	%\Cf{Was $\UT$}
	is \emph{multiplicative} if for any $\pi_{1,2}{\In}\Dist\CalX$  and loss function $l$ we have $t.(U_{\LossApp{\pi_1}{l}}).\pi_2 = t.(U_{\LossApp{\pi_2}{l}}).\pi_1$.\,%
	\footnote{This notation is by analogy with $\ChApp{\pi}{C}$ that ``multiplies $\pi$ in'' from the $\CalX$ side of a matrix; in $\LossApp{\pi}{C}$ the $\pi$ is multiplied in from the other side.}
\end{Definition}

%\Tf{I've changed the reference to \Thm{t1005} as Thm\{t1253\} disappeared. We need to make sure it's right. Also, the exact properties for $\Wp h$ are now given in \Lem{l1719A}, \Lem{l0846A} and \Lem{l0846B}. {\Ax I have checked that the properties hold for this $\Wp{h}$ definition since they are all determined by a fixed $\delta$. Tahiry --- could you check too? 
%\Tx I've checked continuity of $f$ (which was a $h$ in previous versions) explicitly with respect to Kantos on $\Dist\CalX$ and $\Dist^2\CalX$. Super-linearity is not trivial!\TBar }}

\begin{Lemma}{Channels are multiplicative}{l1453}
	Let $C{\In} \CalX{\MFun}\CalY$ be a channel matrix. Then $\Wp{\HMMsem{C}{}}$ is multiplicative.
	\Proof This follows because the identity transformer is multiplicative, i.e. $(U_{\LossApp{\pi_1}{l}}).\pi_2 = (U_{\LossApp{\pi_2}{l}}).\pi_1$, and that $\Wp{()}$ applied to a pure channel maps any given loss function to a sum of loss functions ``scaled" by the columns.% (details in the appendix at \Lem{l1045}).
	%\Cf{This lemma is now undefined. What do you want to do?}
	%\Cf{This \Lem{l1045} is currently in the Green Room. Should we move it to an appendix? In any case, check this ``immediacy'' claim.
	%{\Ax It seems to be immediate -- \Lem{l1045} simply works out wp's for various special cases, including channels.  
	%}}
\end{Lemma}

The following fact shows that this multiplicative property in fact characterises channels.
\begin{Lemma}{}{t2117}
	Let $f{\In}\AHSpc$ be such that $f.\pi$ has finite support for every $\pi{\In}\Dist\CalX$; assume it satisfies the pure-channel property from \Sec{s0906}; and assume that $\Wp{f}$ is multiplicative as just above.
	Then there is some set of observations $\CalY$ and channel $C{\In} \CalX{\MFun}\CalY$ such that $f=\HMMsem{C}{}$.
	\Proof
	Let $N$ be the size of $\CalX$ and let $\upsilon$ be the uniform distribution on $\CalX$.\,%
	\footnote{It is \emph{upsilon} for ``uniform''.}
	Define $\Delta \Defs f.\upsilon$ and let $\CalY$ be the support of $\Delta$, a finite set of distributions that will be used as column indices. Then define $C{\In}\CalX{\MFun}\CalY$ by
	\[
	C_{x,y} \Wide{\Defs} N\times\Delta.y.x ~,
	\]
	so that $f.\upsilon = \Delta =\HMMsem{C}{}.\upsilon$.
	%++\Cf{Explain better why this is.}
	We now show that in fact $f.\pi = \HMMsem{C}{}.\pi$ for all $\pi{\In}\Dist\CalX$.
	
	We have for any loss function $l$ that
	\begin{Reason}
		\Step{}{
			\Expt{f.\pi}{U_l}%.(f.\pi)
		}
		\Step{$=$}{
			\Wp{f}.U_l.\pi
		}
		\StepR{$=$}{define $l'\Defs N{\times}l$}{
			\Wp{f}.U_{\LossApp{\upsilon}{l'}}.\pi
		}
		\StepR{$=$}{assumption $\Wp{f}$ multiplicative}{
			\Wp{f}.U_{\LossApp{\pi}{l'}}.\upsilon
		}
		\Step{$=$}{
			\Expt{f.\upsilon}{(U_{\LossApp{\pi}{l'}})}%.(f.\upsilon)
		}
		\StepR{$=$}{defn.~$C$}{
			\Expt{\HMMsem{C}{}.\upsilon}{(U_{\LossApp{\pi}{l'}})}%.(\HMMsem{C}{}.\upsilon)
		}
		\StepR{$=$}{reverse steps above;\\ $\Wp{\HMMsem{C}{}}$ multiplicative}{
			\Expt{\HMMsem{C}{}.\pi}{U_l}~,%.(\HMMsem{C}{}.\pi) ~,
		}
	\end{Reason}
	so $f.\pi{=}\HMMsem{C_\Delta}{}.\pi$ since hypers are determined by loss functions \cite{mcivermeinicke10a,McIver:2014aa}, thus $f{=}\HMMsem{C_\Delta}{}$ because $\pi$ was arbitrary.
\end{Lemma}
%++\Cf{It might be clearer to have a lemma that says that for any $\Delta=f.\pi$ with $\pi=\Avg.\Delta$ there is a channel $C^\Delta$ such that also $\Delta=\HMMsem{C}{}.\pi$. The trouble is (as the example shows) that it is not the same $C^\Delta$ for different $\pi$'s. Still it would make the proof more direct: just take $C^{f.\upsilon}$.}

%\EndDocument
\end{document}
%%%%%
\cleardoublepage %%%%%%%%%%%%% Material here will probably be moved into the main text %%%%%%%%%%%%%
\hrule
~\\\begin{center}\Huge Green Room
\end{center}~\\
\hrule

\cleardoublepage% Temporary.

\section{Temporarily in Green Room}
\subsection{\HMM's and the Dalenius Desideratum}\label{ss0534}
{\Ax%0535
{\Xx%0712
	We now sketch the argument to show  that if $\CalZ{=}\CalX$ (more generally if $\Size{\CalZ}{\geq}\Size{\CalX}$) we have captured the Dalenius effect in $\HMMsem{\Dal{C}{\CalZ}}{\Dal{M}{\CalZ}}$.
	\Cf{I think we should change this to an ``it can be shown'', and refer to a plausibility argument in the appendix. For me, there are too many loose ends here. We could refer to Chen, actually.}

	For $\delta{\In} \Dist\CalZ$ define $\hat{\delta}{\In}\Dist(\CalX{\times}\CalZ)$ by $\hat{\delta}.x.z= \textit{$\delta.z$ if $z{=}x$ else 0}$. (Recall we are assuming $\CalZ{=}\CalX$.) Similarly for loss function $l{\In} I {\MFun}\CalZ$,
	\Cf{Should we type $l$ this way earlier, too?}
	define $\hat{l}{\In} I {\MFun}\CalX{\times}\CalZ$ as $\hat{l}.x.z = l.z$. Finally define transformer $t{\In}\Unc\CalZ {\Fun} \Unc\CalZ$ so that $t.U_l.\delta \Defs \Wp{\HMMsem{\Dal{C}{\CalZ}}{\Dal{M}{\CalZ}}}.U_{\hat{l}}.\hat{\delta}$. 
	
	Observe first, that $\hat{\delta}$ effectively makes a copy of $\CalZ$ in its first component, so that the correlation between $\CalX$ and $\CalZ$ is exact.  Second,  the loss function $U_{\hat{l}}$ does not depend at all on the $\CalX$ component of $\CalX{\times}\CalZ$. Third, the transformer $t$ is defined by making observations on ${\HMMsem{\Dal{C}{\CalZ}}{\Dal{M}{\CalZ}}}$, and projecting onto the $\CalZ$ component.
	
	\Cf{A technicality: we are using that any $u$ can be written as $U_l$.}
	Using these definitions we have the following facts.
	\Cf{For which we need that $t$ is actually a transformer, i.e.\ satisfies the healthiness conditions we impose on them. Is that obvious? Actually \Thm{t2117} starts from a forward $f$, not a backwards $t$. So we have to assume there's such an $f$, as well?}
	
	\begin{enumerate}
		\item $t$ is multiplicative:
		\begin{Reason}
			\Step{}
			{t.U_{\LossApp{\pi_1}{l}}.\pi_2} 
			\Step{$=$}
			{\Wp{\HMMsem{\Dal{C}{\CalZ}}{\Dal{M}{\CalZ}}.U_{\hat{\LossApp{\pi_1}{l}}}.\hat{\pi_2}}}
			\WideStepR{$=$}{\Thm{t1008}, $\Dal{M}{\CalZ}$ does not update $\CalZ$}
			{\Wp{\HMMsem{\Dal{C}{\CalZ}}{}.U_{\hat{\LossApp{\pi_1}{l}}}.\hat{\pi_2}}}
			\WideStepR{$=$}{\Lem{l1453} and $\Dal{C}{\CalZ}$ is a channel}
			{\Wp{\HMMsem{\Dal{C}{\CalZ}}{}.U_{\hat{\LossApp{\pi_2}{l}}}.\hat{\pi_1}}}
			\StepR{$=$}{As above}
			{\Wp{\HMMsem{\Dal{C}{\CalZ}}{\Dal{M}{\CalZ}}.U_{\hat{\LossApp{\pi_2}{l}}}.\hat{\pi_1}}}
			\Step{$=$}
			{t.U_{\LossApp{\pi_2}{l}}.\pi_1}~.
		\end{Reason}
		
		\item Since $t$ is multiplicative, by \Thm{t2117} there must be some channel matrix $K{\In}\CalZ{\MFun}\CalY$ corresponding to $t$.
		
		\item In fact $K$ corresponds exactly to the original channel $C$ as follows:  for any loss function $l$, and prior $\delta{\In}\Dist\CalZ$, we have that $\Wp{\HMMsem{{K}}{}}.U_l.\delta = \Wp{\HMMsem{{C}}{}}.U_{l\circ f}.\delta'$, where $\delta'.x = \delta.(f.x)$, and $f$ is an injective function from $\CalX$ to $\CalZ$.
		
	\end{enumerate}
}%0712
}%0535

\subsection{$\Wp{\cdot}$ examples}\label{s1933}
\Af{In these examples below there is a horrible use of notations already set up.
In particular see for instance $(\ChApp{C_{-,y}}{l^T})^T$. This is actually
just the $y$'th column of C being multiplied row-wise into $l$. It's used
in all of these examples and the last section and looks awful with all the transposes
cluttering everything up.
Can we have a notation such as $\pi \star l$ for this? I can put a little section in the appendix
setting out the very simple arithmetic operations needed to sort out these proofs. \CBar
}
In this section we illustrate our $\Wp{\cdot}$ definitions on \HMM's. 

\begin{Lemma}{}{l1045}
Let $l$ be a loss function of type $I{\Fun}\CalX{\Fun}\NNReal$, $C{\In} \CalX{\MFun}\CalY$ a channel matrix and  $M{\In} \CalX{\MFun}\CalX$ a Markov matrix. Then
\begin{enumerate}
	\item $\Wp{\HMMsem{}{M}}.U_l = U_{l{\MMult}M^T}$
	\item $\Wp{\HMMsem{C}{}}.U_l = \sum_y U_{(\ChApp{C_{-,y}}{l^T})^T}$.
	\item $\Wp{\HMMsem{C}{M}}.U_l = \sum_y U_{M{\MMult}(\ChApp{C_{-,y}}{l^T})^T}$.
\end{enumerate}
\Proof We reason as follows for a pure Markov matrix:
\begin{Reason}
	\Step{}
	{\Wp{\HMMsem{}{M}}.U_l.\pi}
	\Step{$=$}
	{U_l.(\HMMsem{}{M}.\pi)}
	\StepR{$=$}{\Def{d1022}}
	{U_l.(\pi{\MMult}M)}
	\Step{$=$}
	{\min_{i{\In} I}~\Expt{(\pi{\MMult}M)}{l.i}}
	\Step{$=$}
	{\min_{i{\In} I}~\sum_{x{\In}\CalX}(\pi{\MMult}M)_x \cdot l_{ix} }
	\Step{$=$}
	{\min_{i{\In} I}~\sum_{x{\In}\CalX} \sum_{x'{\In}\CalX} \pi_{x'}  M_{x'x}  l_{ix} }
	\Step{$=$}
	{\min_{i{\In} I}~\sum_{x'{\In}\CalX} \pi_{x'} \sum_{x{\In}\CalX}  M_{x'x} l_{ix} }
	\Step{$=$}
	{\min_{i{\In} I}~\sum_{x'{\In}\CalX} \pi_{x'} \sum_{x{\In}\CalX}   l_{ix} M^T_{xx'} }
	\Step{$=$}
	{U_{l{\MMult}M^T}.\pi~.}
\end{Reason}
Similarly, for a pure channel we reason
\begin{Reason}
	\Step{}
	{\Wp{\HMMsem{C}{}}.U_l.\pi}
	\Step{$=$}
	{\Expt{\HMMsem{C}{}.\pi}{U_l}}
	\Step{$=$}
	{\sum_{y{\In} \CalY}\min_{i{\In} I} \sum_{x{\In} \CalX}l_{ix}C_{xy}\pi_x}
	\Step{$=$}
	{\sum_{y{\In} \CalY}\min_{i{\In} I} \sum_{x{\In} \CalX}\pi_x (\ChApp{C_{-,y}}{l^T})_{xi}}
	\Step{$=$}
	{\sum_{y{\In} \CalY} U_{{(\ChApp{C_{-,y}}{l^T})^T}}.\pi~.}
\end{Reason}

\Af{Can we write $(\ChApp{\pi}{l^T})^T$ using some kind of shorthand so we don't use the transposes all the time?
	I'd like to be able to use commutativity rules, but with this notation it gets pretty messy... Oh ... I already put in a plea for this :-) \CBar}

Finally we observe by \Thm{t1008} that since $\HMMsem{C}{M} = \HMMsem{C}{};\HMMsem{}{M}$ we have that 
$\Wp{\HMMsem{C}{M}}.U_l = \Wp{\HMMsem{C}{}}.(\Wp\HMMsem{}{M}.U_l)$, and the result follows directly from (1) and (2).
\Cfe[../Carroll/WPseqComp.tex]{Have we shown that sequential composition is functional composition for Wp? {\Ax Good point. I added a place for that at \Thm{c1020}. Is that the right place to mention it? \Tx I copied Carroll's proof from the Endnote there.}\CBar}
\end{Lemma}

\section{\Cx Dalenius doodlings}\label{s1321}
Here's some thoughts and possible notations for our final ``Dalenius'' section, where we expand the state space etc.
{\newenvironment{Item}[1] {\item\textbf{#1:}\begin{enumerate}}{\end{enumerate}}%0757
\begin{enumerate}
	\begin{Item}{How to motivate it}
		\item The Dalenius problem is well known for channels \Cite{Dwork etc}; it's not a problem \emph{we} have introduced.
		\item In fact it's a contribution (of ours) to point out that the problem occurs with \HMM's as well.
		\item The problem is in essence that there can be ``third party''
		\footnote{First party is the secret (holder of $\CalX$); second party is the adversary (who observes $\CalY$); and third party is arbitrary (holder of some $\CalZ$ possibly correlated with $\CalX$).}
		correlations of the input $\CalX$ with some unknown (to us) $\CalZ$, and leaks inherent in an \HMM\ in $\MH$ can leak information about $\CalZ$ as well, if there is a known \emph{a priori} correlation between $\CalX$ and $\CalZ$.
	\end{Item}
	
	\begin{Item}{How to formalise it}
		\item Given is an \HMM\ $H{\In}\MH$ and third-party data $\CalZ$. The effect of $H$ on some prior $\pi{\In}\Dist\CalX$ is given by the joint distribution $J$ from \Def{d1022}.
		\item We intend informally that what we will write as $\Dal{H}{\CalZ}$ of type $(\CalX{\times}\CalZ)\MFun\CalY{\times}(\CalX{\times}\CalZ)$ should act on priors $\Dist(\CalX{\times}\CalZ)$ (rather than on $\CalX$ alone) but that the $\CalZ$-component should be ``left entirely alone'', that is neither read nor written by the mechanism that $\Dal{H}{\CalZ}$ describes.
		\item\label{i0827} Thus we define
		\begin{equation}\label{e0835}
		(\Dal{H}{\CalZ})_{x,z,y,x',z'}
		\Wide{\Defs}
		\begin{array}[t]{l@{\quad}p{10em}}
		H_{x,y,x'} & if $z'{=}z$ \\
		0 & otherwise,
		\end{array}
		\end{equation}
		so that the conditional probabilities for $y,x'$ depend only on $H$ and the original $x$ ($z$ not read), and that $z'$ always equals $z$ ($z$ not written).
	\end{Item}
	
	\begin{Item}{Good behaviour under composition}
		\item We have for $H^{1,2}{\In}\CalX\MFun\CalY^{1,2}{\times}\CalX$ resp.\ that
		\begin{Reason}
			\Step{}{
				(\,\Dal{(H^1;H^2)}{\CalZ}\,)_{x,z,y_1,y_2,x',z'}
			}
			\Space
			\StepR{$=$}{\Eqn{e0835}}{
				\begin{array}[t]{l@{\quad}p{10em}}
					(H^1;H^2)_{x,y_1,y_2,x'} & if $z'{=}z$ \\
					0 & otherwise
				\end{array}
			}
			\Space
			\Step{$=$}{
				\begin{array}[t]{l@{\quad}p{10em}}
					\sum_{x''}H^1_{x,y_1,x''}H^2_{x'',y_2,x'} & if $z'{=}z$ \\
					0 & otherwise
				\end{array}
			}
		\end{Reason}
		and, from the other direction, that
		\begin{Reason}
			\Step{}{
				(\,\Dal{H^1}{\CalZ};\Dal{H^2}{\CalZ}\,)_{x,z,y_1,y_2,x',z'}
			}
			\Step{$=$}{
				\sum_{x'',z''}{(\Dal{H^1}{\CalZ})}_{x,z,y_1,x'',z''}{(\Dal{H^2}{\CalZ})}_{x'',z'',y_2,x',z'}
			}
			\Space
			\StepR{$=$}{one-point rule;\\\Eqn{e0835}}{
				\begin{array}[t]{l@{\quad}p{10em}}
					\sum_{x''}H^1_{x,y_1,x''}H^2_{x'',y_2,x'} & if $z'{=}z$ \\
					0 & otherwise,
				\end{array}
			}
		\end{Reason}
		so that indeed $\Dal{(H^1;H^2)}{\CalZ}=\Dal{H^1}{\CalZ};\Dal{H^2}{\CalZ}$.
	\end{Item}
	
	\begin{Item}{How to use it}
		\item We know from our simple ``bad guy'' example
		\footnote{\Label{n1035}If $C^1$ ``is'' $\textbf{reveal}~0$ and $C^2$ is $\textbf{reveal}~x$ and $M$ is $x\Gets0$, then $\HMMsem{C^1}{M}{=}\HMMsem{C^2}{M}$ --- but obviously $C^1{\neq}C^2$. The issue is that the two \HMM's set (final) $x$ to the same value, and reveal the same (nothing) about that \emph{final} value of $x$: it is certain to be 0 in both cases.\par But they do reveal different things about the initial value of $x$, from which $\HMMone{\cdot}$ has abstracted. Having abstracted in that way, it is not possible to analyse any Dalenius correlation.}
		that we cannot in general recover $C$ from $\HMMsem{C}{M}$. (We can recover $M$ however.)
		
		\item In contrast, we \emph{can} recover $C$ from $\HMMone{\Dal{\CM{C}{M}}{\CalX}}$ where, in this case, we are taking the third-party ``Dalenius data'' to be $\CalX$'s initial state, that is $\CalZ{=}\CalX$.
		
		\item Let $\upsilon{\In}\Dist(\CalX^2)$ be uniform on the initial state $x$ and correlate to the third-party $x_0$ exactly, thus given by $\upsilon_{x,x_0}\Defs \textrm{$\nicefrac{1}{\Size{\CalX}}$ if $x{=}x_0$ else 0}$. Then $C$ can be recovered from $\HMMone{\Dal{\CM{C}{M}}{\CalX}}.\upsilon$.
	\end{Item}
	
	\begin{Item}{A nice theorem}%\label{i1620}
		\item\label{i1509} In principle we have to consider \emph{all} possible third-party Dalenius $\CalZ$'s if we want to use $\HMMone{\cdot}$ to capture the effect of some \HMM\ $H$ on any one them: i.e.\ we must consider all possible $\Dal{H}{\CalZ}$'s. But in fact just $\CalZ{=}\CalX$ is enough: that is
		\begin{quote}
			If $\HMMone{\Dal{H^1}{\CalX}}{=}\HMMone{\Dal{H^2}{\CalX}}$ then
			\begin{tabular}[t]{@{}l}
				$\HMMone{\Dal{H^1}{\CalZ}}=\HMMone{\Dal{H^2}{\CalZ}}$ \\
				for all $\CalZ$~,
			\end{tabular} 
		\end{quote}
		i.e.\ that $\HMMone{\Dal{H}{\CalZ}}$ for arbitrary $\CalZ$ is determined by just $\HMMone{\Dal{H}{\CalX}}$ in particular. A sketch proof is given just below at \ref{i0928}).
		\item\label{i0928} Sketch proof of \ref{i1509}): We start by supposing that $h{\In}\Dist(\CalX^2){\Fun}\Dist^2(\CalX^2)$ is $\HMMone{\Dal{H}{\CalX}}$ for some $H{\In}\MH$ and that we are given $\pi{\In}\Dist(\CalX{\times}\CalZ)$. We must express $\HMMone{\Dal{H}{\CalZ}}.\pi$ in terms of $h$ and $\pi$.
		\begin{enumerate}
			\item\label{i1612} Write joint distribution $\pi$ as $\ChApp{\delta}{D}$ for $\delta{\In}\Dist\CalX$ and Dalenius channel $D{\In}\CalX{\MFun}\CalZ$.
			\item Construct distribution $\delta^0{\In}\Dist(\CalX^2)$ such that $\delta^0_{x,x0}\Defs\textrm{$\delta.x$ if $x{=}x_0$ else 0}$, so that $\delta^0$ is the same as $\delta$ except that $x$ has been ``duplicated into $x_0$''.
			\item Construct $\Delta{\In}\Dist^2(\CalX^2)$ so that $\Delta\Defs h.\delta^0$.
			\item\label{i1611} Note that we used $D$ to ``take a copy'' of the initial correlation between each $x$ and $\CalZ$ separately: our job now is to ``paste'' that correlation back onto the $(x,x_0)$-points in the final $\Delta$. For each $(x,x_0)$ in the double-support
			\footnote{i.e.\ the support of the support}
			of $\Delta$ the $x$ is the output value, and its associated $x_0$ records ``where that $x$ came from''.
			\item For \ref{i1611}) we note that each inner $\rho$ of $\Delta$ is of type $\Dist(\CalX^2)$, but can equivalently be considered to be of type $\CalX'{\MFun}\CalX_0$;
			\footnote{Really the type is $\CalX{\MFun}\CalX$; the $\CalX'$ and $\CalX_0$ is to help remember which $\CalX$ is representing what.}
			and we  recall the ``copied'' $D{\In}\CalX_0{\MFun}\CalZ$ from \ref{i1612}) above. The matrix multiplication $\rho{\MMult}D$ is then of type $\CalX'{\MFun}\CalZ$ which we can cast back again to type $\Dist(\CalX{\times}\CalZ)$.
			\item Thus we have $\HMMone{\Dal{H}{\CalZ}}.\pi= \Dist({\MMult}D)\Delta$ where $D,\Delta$ are as above, achieving our objective of deriving $\HMMone{\Dal{H}{\CalZ}}$ from $\HMMone{\Dal{H}{\CalX}}$.
		\end{enumerate}
	\end{Item}
	
	\begin{Item}{An example}
		\item\Cfe[../Carroll/DaleniusExample.txt]{Here is an example of \ref{i1620}).}
	\end{Item}
	
	\begin{Item}{How to write it}
		\item Rather than explicitly extend an $H{\In}\CalX{\Fun}\CalY{\times}\CalX$ with a Dalenius $\CalZ$, and then apply the semantic function $\HMMone{\cdot}$, we can just define an ``enhanced'' semantic function $\HMMoneDal{H}{\CalZ}\Defs\HMMone{\Dal{H}{\CalZ}}$ that acts as if that extension had been done.
		\item Because for $H{\In}\CalX{\Fun}\CalY{\times}\CalX$ we are assured by \ref{i1620}) above that $\HMMoneDal{H}{\CalX}$ is sufficiently discriminating for all $\CalZ$, we can define $\HMMoneD{\cdot}\Defs\HMMoneDal{\cdot}{\CalX}$, i.e.\ with $\CalX$ assumed if no Dalenius is explicitly given.
		\item Thus if $\CalX$ is fixed our ``Dalenius aware'' semantics is --finally-- given by the function
		\[
		\HMMoneD{\cdot}{\In} \Dist(\CalX^2) \Fun \Dist^2(\CalX^2)~.
		\]
		Since it is an instance of our earlier $\HMMone{\cdot}$, merely using $\CalX^2$ rather than $\CalX$ on its own, it satisfies all the properties of the simpler ``Dalenius ignorant'' treatment.
	\end{Item}
	\Cf{Refer to Chen.}
\end{enumerate}
}%0757

\section{\Ax Applications}
\Cf{I gave this a section number: don't know why it didn't have one.}
Hidden Markov Processes are a generic model applicable in scenarios where insufficient information is available to infer likely behaviour. Two well-known  predictors of behaviour are called "smoothing" and "filtering" and we show how standard algorithms for estimating these predictions can be described in the logical framework presented here.

\subsection*{Filtering} 
Let $H$ be an HMM with hidden variable $x$ with value drawn from $\CalX$. The filtering problem asks for the computation of the most likely value of $x$ given all the observations up to some time $t$. Let $\pi$ be an initial prior distribution for $x$'s initial value, then the filtering problem asks for the conditional probability $H^t.\pi.\pi'.\{{\sf x'}\}/H^t.\pi.\pi'$, where $\pi'$ 
corresponds to the overall observations at time $t$ and ${\sf x'}$ is a value drawn from ${\CalX}$.

The ``forward algorithm" \Cite{} keeps a record of the hyper distribution, updating it at each stage, thus using the following property:

\begin{equation}\label{e1058}
H^{t+1}.\pi ~~ = ~~ ((H^t);H).\pi~.
\end{equation}

\subsection*{Smoothing}
The smoothing problem is similar to filtering: given some $k \leq t$ it asks for the most likely value of the hidden $x$ at time $k$, taking into account all observations up until time $t$., i.e. $(H^k; y:= x; H^{t-k}).\pi.\pi'.\{{y=\sf x'}\}/H^t.\pi.\pi'$~, where $y$ is a fresh hidden variable set at time $k$ and unchanged thereafter. As for the forward algorithm, the filtering algorithm can be calculated as follows:

\begin{equation}\label{e1117}
H^k;y:= x; H^{t-k+1}.\pi ~~ = ~~ (H^k;y:= x; H^{t-k}); H;.\pi~.
\end{equation}

\section{\Tx Tahiry's stuff}\label{sec:tahiry}

{\Xx %1653
\subsection{Related Work}
The main contributions of this paper are the monadic presentation and a duality theorem for Abstract HMMs. Note that we do not claim the originality of using monad structures to express models of computation with quantitative properties. 

We do not claim the originality of using backward semantics by means of transformers in this paper. In fact, such a technique is traced back to Dijkstra~\cite{Dijkstra:76}, and later Kozen~\cite{Kozen:83}, Jones~\Cite{Jones' Thesis} and McIver\&Morgan~\Cite{MM04}, where they described how transformers provide a surprisingly expressive dual to the popular forward relational semantics for programs. More recently, Panangaden et al~\Cite{Panangaden09} extended such technique for Markov processes among other properties, such as a converging approximation with finite behaviours. Our results establish such a duality for Hidden Markov Models which are the primary abstractions of computations for quantitative information flow.

\textbf{Approximating Markov Processes By Averaging
	ICALP 2009 , Panangaden, Plotkin et al:} This paper is about the dual view of Markov processes as (measurable) function transformers. It is mainly written from a measure theoretic point of view of measure theory (e.g. tests are general measurable functions), so no continuity/concavity or specific properties for tests. The main result is Thm. 4 which ensures the convergence of finite approximants to a minimal bisimulation equivalent Markov chain. \textbf{Main similarity is that we are using HMM which are extension of the LMP they're are using, so if we restrict to pure transitions, we should get (a fragment) of their model back} \Tf{Maybe only a fragment because they also consider sub-probability measures too.}.

\textbf{1) Kozen's paper on wp for deterministic programs ---
	2) Also Claire Jones' thesis for the same --- 3) McIver and Morgan - Partial Correctness for Probabilistic programs for
	domain theory, wp and so forth. You can also cite our book}:  these works provided the first probabilistic interpretation of predicate transformers for programs.

\textbf{McIver, Meinicke and Morgan LiCS12 for setting up the general domain equations
	and the Coriaceous result for the general measure space.}: this is where the general properties of $\AHS$ are explored and referred from.

\textbf{McIver, Morgan and Smith, POST14 for abstract channels}: re-read this.

\textbf{1) Giry's paper on monadic stuff --- 2) van Breughel's paper on Kantorovich stuff}: These contains most of the categorical approach background.
}%1653

\subsection{Channels are $4$-Lipschitz}\label{sec:k-lippie}
Recall that the Kantorovich distance $\KMet{\delta}{\delta'}$ between two distributions over $\CalX$ could be explicitly computed using the Manhattan metric. That is, 
\[
\KMet{\delta}{\delta'}\Wide{=}\frac{1}{2}\MDist{\delta}{\delta'} \Wide{=} \frac{1}{2}\sum_{x\in\CalX}|\delta.x-\delta'.x|~.
\]
\begin{definition}
Two distributions $\delta,\delta'\in\Dist\CalX$ are \emph{simply distant} at $x,x'{\In}\CalX$ if there exists a subdistribution $\alpha$ on $\CalX$ and $\varepsilon>0$ such that $\delta = \alpha + \varepsilon\{x\}$ and $\delta'=\alpha + \varepsilon\{x'\}$. 
\end{definition}
If $\delta$ and $\delta'$ are simply distant then $\KMet{\delta}{\delta'} = \varepsilon$.
\begin{Lemma}{Simply distant sequence}{l1056}
Let $\delta,\delta'\in\Dist\CalX$, there exists a sequence $\delta_0=\delta,\delta_1,\dots,\delta_r=\delta'$ such that $\delta_i$ and $\delta_{i+1}$ are simply distant, for $i=0,\dots r-1$ and 
\[
\KMet{\delta}{\delta'} = \sum_{i=0}^{r-1}\KMet{\delta_i}{\delta_{i+1}}
\]
\Proof
See Annabelle's notes on 3-Lippi.
\end{Lemma}

In the reminder of this section, $C = \{C_{x,y}\}_{(x,y)\in\CalX\times\CalY}$ is a channel matrix. When needed, we assume an enumeration of $\CalX$ (resp. $\CalY$) as $\{1,2,\dots, m\}$ (resp. $\{1,2,\dots,n\}$).

\begin{Lemma}{EMS for simply distant priors}{lem:strategy}
Let $\pi,\pi'$ be two simply distant priors at $x,x'$ such that $\HMMsem{C}{}.\pi = \sum_{y\in\CalY}\lambda_y\delta_y$ and $\HMMsem{C}{}.\pi' = \sum_{y\in\CalY}\lambda_y'\delta_y'$. Then, the matrix 
$$
\left(\begin{array}{ccccc}
\lambda_1'&0&0&\cdots &\varepsilon(C_{x,1} - C_{x',1})\\
0&\lambda_2'&0&\cdots &\varepsilon(C_{x,2} - C_{x',2})\\
0&0&\lambda_3'&\cdots &\varepsilon(C_{x,3} - C_{x',3})\\
\vdots &\vdots &\vdots &\ddots &\vdots \\
0&0&0&\cdots&\lambda_n\\
\end{array}\right)
$$
is an earth moving strategy from hyper $\HMMsem{C}{}.\pi$ to $\HMMsem{C}{}.\pi'$.

\Proof%\begin{proof}
Let $y\in\CalY$, it follows from the expression 
$$\lambda_y = (\alpha.x + \varepsilon)C_{x,y} + \alpha.x'C_{x',y} + \sum_{z\in\CalX\setminus\{x,x'\}}\alpha.zC_{z,y}$$
and 
$$\lambda_y' = \alpha.xC_{x,y} +( \alpha.x' + \varepsilon)C_{x',y} + \sum_{z\in\CalX\setminus\{x,x'\}}\alpha.zC_{z,y}$$
that such a matrix has row (resp. column) marginal $\HMMsem{C}{}.\pi$ (resp. $\HMMsem{C}{}.\pi'$).
\end{Lemma}
Notice that if some entry, say $C_{x,2}- C_{x',2}$, is negative; then
we use the matrix
$$
\left(\begin{array}{ccccc}
\lambda_1'&0&0&\cdots &\varepsilon(C_{x,1} - C_{x',1})\\
0&\lambda_2'&0&\cdots &0\\
0&0&\lambda_3'&\cdots &\varepsilon(C_{x,3} - C_{x',3})\\
\vdots &\vdots &\vdots &\ddots &\vdots \\
0&\varepsilon(C_{x',2} - C_{x,2})&0&\cdots&\gamma\\
\end{array}\right)
$$
where $\gamma$ is some real number such that 
$\gamma + \varepsilon(C_{x',2}- C_{x,2}) = \lambda_n$ and 
$\gamma + \varepsilon(C_{x,1} - C_{x',1}) + \varepsilon(C_{x,3} - C_{x',3}) + \cdots = \lambda'_n$.

\begin{Lemma}{Technical lemma}{l2139}
Let $\pi,\pi'$ be simply distant at $x,x'{\In}\CalX$ such that $\HMMsem{C}{}.\pi = \sum_{y\in\CalY}^n\lambda_y\delta_y$ and $\HMMsem{C}{}.\pi' = \sum_{y\in\CalY}^n\lambda_y'\delta_y'$. Then, for every fixed $y\in\CalY$:
$$\KMet{\delta_y}{\delta_y'}\leq \frac{\max(C_{x,y},C_{x',y})}{\max(\lambda_y,\lambda_y')}\varepsilon.$$
\Proof
Let $\pi$ and $\pi'$ be simply distant prior at $x,x'$ and $y\in\CalY$. We have 
\begin{footnotesize}
	\begin{align*}
		\KMet{\delta_y}{\delta_y'} & = \frac{\left|\lambda_y'(\alpha.x+\varepsilon) - \lambda_y\alpha.x\right|C_{x,y}}{2\lambda_y\lambda_y'}\\
		&\qquad + \frac{\left|\lambda_y'\alpha.x'-\lambda_y(\alpha.x'+\varepsilon)\right|C_{x',y}}{2\lambda_y\lambda_y'}\\
		&\qquad + \sum_{z\in\CalX\setminus\{x,x'\}}\frac{\left|\lambda_y'-\lambda_y\right|\alpha.zC_{z,y}}{2\lambda_y\lambda_y'}\\
	\end{align*}
\end{footnotesize}

Using the explicit expression of $\lambda_y$ and $\lambda_y'$ in the proof of Lemma~\ref{lem:strategy}, we obtain 
\begin{footnotesize}
	\begin{align*}
		|\lambda_y'(\alpha.x + \varepsilon) - \lambda_y\alpha.x|& = \varepsilon\left[(\alpha.x + \alpha.x' + \varepsilon)C_{x',y} + \sum_{z\in\CalX\setminus\{x,x'\}}\alpha.zC_{z,y}\right]
	\end{align*}
\end{footnotesize}
Similarly,
\begin{footnotesize}
	\begin{align*}
		|\lambda_y'\alpha.x' - \lambda_y(\alpha.x'+\varepsilon)| & = \varepsilon\left[(\alpha.x+\alpha.x' + \varepsilon)C_{x,y}\right. \\
		& + \left.\sum_{z\in\CalX\setminus\{x,x'\}}\alpha.zC_{z,y}\right]
	\end{align*}
\end{footnotesize}
and $|\lambda_y' - \lambda_y| = \varepsilon|C_{x,y} - C_{x',y}|$. If $C_{x,y}\geq C_{x',y}$, then
\begin{footnotesize}
	\begin{align*}
		\KMet{\delta_y}{\delta_y'} & = \frac{\varepsilon\left[(\alpha.x + \alpha.x' + \varepsilon)C_{x',y} + \sum_{z\in\CalX\setminus\{x,x'\}}\alpha.zC_{z,y}\right]C_{x,y}}{2\lambda_y\lambda'_y}\\
		&\qquad + \frac{ \varepsilon\left[(\alpha.x+\alpha.x' + \varepsilon)C_{x,y} + \sum_{z\in\CalX\setminus\{x,x'\}}\alpha.zC_{z,y}\right]C_{x',y}}{2\lambda_y\lambda'_y}\\
		&\qquad + \frac{\varepsilon\left[(C_{x,y} - C_{x',y})\sum_{z\in\CalX\setminus\{x,x'\}}\alpha.zC_{z,y}\right]}{2\lambda_y\lambda'_y}\\
		& = \frac{\varepsilon}{\lambda_y\lambda_y'}\left((\alpha.x + \alpha.x' + \varepsilon)C_{x',y} + \sum_{z\in\CalX\setminus\{x,x'\}}\alpha.zC_{z,y}\right)C_{x,y}
	\end{align*} 
\end{footnotesize}
Since $C_{x',y}\leq C_{x,y}$, we have 
\begin{footnotesize}
	\begin{align*}
		(\alpha.x + \alpha.x' + \varepsilon)C_{x',y} + \sum_{z\in\CalX\setminus\{x,x'\}}\alpha.zC_{z,y}\leq\min(\lambda_y,\lambda_y')
	\end{align*}
\end{footnotesize} 
and therefore
\begin{align*}
	\KMet{\delta_y}{\delta_y'}\leq\frac{C_{x,y}\varepsilon}{\max(\lambda_y,\lambda_y')}
\end{align*}
The case $C_{x,y}\leq C_{x',y}$ gives the upper bound 
\begin{align*}
	\KMet{\delta_y}{\delta_y'}\leq\frac{C_{x',y}\varepsilon}{\max(\lambda_y,\lambda_y')}
\end{align*}
which entails the claimed result.
\end{Lemma}

\begin{Theorem}{Channels are Lipschitz}{t1120}
Given a channel $C$, the map $\HMMsem{C}{}{\In}\Dist\CalX{\Fun}\Dist^2\CalX$ is $(2+\eta)$-Lipschitz, where $\eta$ is the maximal Euclidean distance between pairs of rows of $C$ (up to the last column)

\Proof%\begin{proof}
Let $C$ be a channel matrix and $\pi,\pi'$ be two simply distant priors at $x,x'$. By Lemma~\ref{lem:strategy}, we have 
\begin{footnotesize}
	\begin{align*}
		\KMet{\HMMsem{C}{}.\pi}{\HMMsem{C}{}.\pi'}
		& \leq \sum_{y\in\CalY\setminus\{n\}}\lambda_y'\KMet{\delta_y}{\delta_y'} + \lambda_n\KMet{\delta_n}{\delta_n'} \\ 
		& + \varepsilon \sum_{y\in\CalY\setminus\{n\}}|C_{x,y} - C_{x',y}|~.
	\end{align*}
\end{footnotesize}
By Lemma~\ref{l2139}, 
\[
\max(\lambda_y,\lambda_y')\times\KMet{\delta_y}{\delta_y'}\Wide{\leq}\max(C_{x,y},C_{x',y})\varepsilon~.
\]
Since, for every $z\in\CalX$, $\sum_{y\in\CalY}C_{z,y} = 1$, we have $\sum_{y\in\CalY}\max(C_{x,y},C_{x',y})\leq 2$. Hence, we obtain 
\[
\KMet{C.\pi}{C.\pi'}\Wide{\leq}(2+\eta_{xx'})\varepsilon \Wide{=} (2+\eta_{xx'})\KMet{\pi}{\pi'}
\]
where $\eta_{xx'} = \sum_{y\in\CalY\setminus\{n\}}|C_{x,y} - C_{x',y}|$.

In the general case of non-simply distant priors, we quantify the indices of $\eta_{xx'}$ to obtain a general upper bound, i.e., let 
\begin{align*}
	\eta = \max_{x,x'}\sum_{y\in\CalY\setminus\{n\}}|C_{x,y} - C_{x',y}|
\end{align*}
By \Lem{l1056}, there exists a sequence $\pi_0=\pi,\pi_1,\dots,\pi_r=\pi'$ such that $\pi_i$ and $\pi_{i+1}$ are simply distant. We have: 
\begin{Reason}
	\Step{}{
		\KMet{\HMMsem{C}{}.\pi}{\HMMsem{C}{}.\pi'}
	}
	\StepR{$\leq$}{Triangular inequality}{
		\sum_{i=0}^{r-1}\KMet{\HMMsem{C}{}.\pi_i}{\HMMsem{C}{}.\pi_{i+1}}
	}
	\StepR{$\leq$}{$\pi_{i,i+1}$ are simply distant}{
		\sum_{i=0}^{r-1}\eta\KMet{\pi_{i}}{\pi_{i+1}}
	}
	\StepR{$=$}{\Lem{l1056}}{
		\eta\KMet{\pi}{\pi'}
	}
\end{Reason}
\end{Theorem}

Notice that in general $0\leq \eta \leq 2$. The case $\eta=2$ happen only where the last column of $C$ has $0$ on both the first and second rows. But \Thm{t1120} still holds if we permute the columns, therefore, $\eta$ can lowered to
\begin{align*}
\max_{x,x'}(2\KMet{C_{x,-}}{C_{x',-}} - \max_y|C_{x,y} - C_{x',y}|)
\end{align*}
where $\KMet{C_{x,-}}{C_{x',-}}$ is the Kantorovich distance between the row distributions of the channel matrix $C$ associated to $x$ and $x'$. This implies that in the deterministic case, $\eta{=} 2 {-} 1 {=} 1$ and we deduce that deterministic channels are $3$-Lipschitz.

\subsection{A Markov update is $1$-Lipschitz}
A Markov update $M$ is of type $\Dist \CalX{\Fun}\Dist \CalX$. To obtain a program, we apply the unit on $M$ that is $\{M\}.\delta = \{M.\delta\}$ 
\Tf{Change to the proper notation for the unit of the Giry monad.}
where $\{M\}:\Dist \CalX{\Fun}\Dist^2S$. Concretely, $M$ is a transition matrix and $M.\delta$ is the application of that matrix on the prior $\delta$. Hence $M$ is a linear, thus Lipschitz function. Moreover, since $\KMet{\{\delta\}}{\{\delta'\}} = \KMet{\delta}{\delta'}$, we also deduce that $\{M\}$ is Lipschitz with the same Lipschitz constant as $M$.

\begin{Lemma}{Markov updates are Lipschitz}{l1457}
Markov updates are $1$-Lipschitz.
\Proof
All Markov updates $M$ are constructed from a map $K:\CalX{\Fun}\Dist \CalX$ such that 
$$M = \{\Avg.\Dist.K\}$$
where $(\Dist,\Avg,\{\ \})$ is the Giry monad. Since $\CalX$ is endowed with the discrete metric and $\Dist \CalX$ has the Kantorovitch lifting of that metric, we deduce that $K$ is $1$-Lipschitz. Therefore, $M$ is $1$-Lipschitz.
\end{Lemma}

\subsection{Counterexample: channels are not $1$-Lipschitz}
I'll be constructing an example of a channel that is not $1$-Lipschitz here.

%%%%%%%%%%%%%End of Tahiry's new Green Room

\section{\Cx Assorted definitions}
\Cf{This is an unordered collection of things we might need.}

\begin{Definition}{Average of hyper}{d1208}
\Cf{Put something here: be sure to point out it's \emph{multiply} from the monad.}
\end{Definition}

\begin{Definition}{Point distribution}{d1417P}
For $z$ in $\CalZ$ the point distribution $\Point{z}$ is the distribution in $\Dist\CalZ$ assigning probability 1 to $z$ and probability 0 to all other elements of $\CalZ$.\end{Definition}

\begin{Definition}{Support of a distribution}{zd1510}
\Cf{Don't seem to need this.}
Given discrete distribution $\delta{\In}\Dist\CalZ$, we write $\Supp{\delta}$ for the \emph{support} of $\delta$, the set of elements $z{\In}\CalZ$ for which $\delta.z$, the probability assigned by $\delta$ to $z$, is not zero. Obviously $\delta{\in}\Dist\CalZ$ implies $\Supp{\delta}{\subseteq}\CalZ$; if in fact $\Supp{\delta}{=}\CalZ$ then we say that $\delta$ is \emph{full support}.
\end{Definition}
Note that $\Supp{\Point{z}}{=}\DSet{z}$.

\begin{Definition}{Two-point distribution}{zd1954}
\Cf{Maybe we do need this: see \Fn{n1413.}}
For $x_{1,2}{\In}\CalX$ we write $x_1\PC{p} x_2$ for the distribution in $\Dist\CalX$ assigning probability $p$ to $x_1$ and probability $1{-}p$ to $x_2$.
\end{Definition}
Note that $x_1\PC{1} x_2 = \Point{x_1}$ and $x_1\PC{0} x_2 = \Point{x_2}$ and, if $0{<}p{<}1$, then $\Supp{x_1\PC{p} x_2}{=}\DSet{x_1,x_2}$.

\begin{Definition}{Weighted sum}{zd1955}
For $\delta_{1,2}{\In}\Dist\CalX$ we write $\delta_1\WS{p} \delta_2$ for the weighted sum of the two distributions, so that $(\delta_1\WS{p} \delta_2).x = p\delta_1.x+(1{-}p)\delta_2.x$.
\end{Definition}
Note that taking Defs.~\ref{d1954},\ref{d1955} and \ref{d1417P} together gives us two ways of writing a two-point distribution, since $x_1\PC{p} x_2 = \Point{x_1}\WS{p} \Point{x_2}$.

As stated in \Def{d1030}, an uncertainty measure can be applied to a hyper, in which case it is implicitly lifted via expected value. In the case that the uncertainty measure is generated by a loss function $h$ (as explained in \Def{d1321A}), there is a simple way to apply the uncertainty $L_h$ to joint distributions directly, i.e.\ without converting them first to hypers.

\begin{Lemma}{Loss-function uncertainty applied to joint distribution}{l1041}\\
\Cf{Don't seem to need this.}
For loss function $u$ extend the uncertainty $L_u$ to subdistributions via the original formula: that is for $\delta{\In}\SubDist\CalX$ define $L_u.\delta\Defs\min_{i\in I}~\sum_x\delta.x\times h.i.x$.
\Cf{There might be a case for a more general requirement on uncertainties when they are extended to subdistributions. We require that they be concave, at the moment, but (initially) we don't apply them to subdistributions. So, instead, require them to be \emph{sublinear} over $\SubDist\CalX$. (Or should that be superlinear? Remember what Plotkin said.) That is they \emph{scale} and they \emph{subdistribute} addition. If we did that, then \Lem{l1041} would apply to all uncertainties, not just those generated from loss functions.}
Then provided $J$ has no all-zero column, we have $L_u.\Hyp{J} = \sum_y L_u.J_{-,y}$.
\Proof
We calculate
\begin{Reason}
	\Step{}{
		L_u.\Hyp{J}
	}
	\StepR{$=$}{\Eqn{e0958}}{
		L_u.(\sum_{y{\In}\CalY}~\SubPt{J_{-,y}})
	}
	\StepR{$=$}{\Def{d1417}}{
		L_u.(\sum_{y{\In}\CalY}~\sum J_{-,y}\times\Point{\Nrm.J_{-,y}})
	}
	\StepR{$=$}{linearity lifted $L_u$}{
		\sum_{y{\In}\CalY}~\sum J_{-,y}\times L_u.(\Nrm.J_{-,y})
	}
	\WideStepR{$=$}{scaling unlifted $L_u$}{
		\sum_{y{\In}\CalY}~(\,\sum J_{-,y}\times (L_u.J_{-,y})/\sum J_{-,y}\,)
	}
	\Step{$=$}{
		\sum_{y{\In}\CalY}~L_u.J_{-,y} ~.
	}
\end{Reason}
\end{Lemma}

Note that we have trivially from \Def{d1124} that
\[
\ChApp{(\delta_1\WS{p}\delta_2)}{C} \Wide{=}
(\ChApp{\delta_1}{C}) ~\WS{p}~ (\ChApp{\delta_2}{C}) ~.
\]

By analogy with using semantic brackets $\Hyp{\cdot}$ for the meaning of a joint distribution (as a hyper), we use semantic brackets for the meaning of a channel (matrix), as an abstract channel (function).

%\end{document} % Temporary
%%%%% Begin Sargasso %%%%%
\cleardoublepage %%%%%%%%%%%%% Material below here will probably be thrown away %%%%%%%%%%%%%
\hrule
~\\\begin{center}\Huge Sargasso
%\Cf{Material in here is (probably) not going to make it into this paper,
%or has been superseded.
%But we keep it here so that we can easily find it if needed
%(i.e.\ avoiding the need to search emails or earlier svn versions.}
\end{center}~\\
\hrule
{\Xx%0818 To localise grey, and disabling of \label and \ref.
\renewcommand\label[1] {}
\renewcommand\ref[1] {$\emptyset$}
\bigskip
\newcommand\NS {\bigskip\noindent---\\} % Next Sargasso

\NS
\subsection{Example program with non-uniform prior \AppFrom{\Sec{s0825}}}\label{a0407}

In \Fig{f0544cm} we gave a small program, on a two-bit state space, that illustrated an information leak (via a \verb+print+ statement) followed by a probabilistic update. In the case of a uniform input (prior distribution), the program's output turned out to have the same distribution as had the input: that is, although information had been leaked by the \verb+print+, that leak was rendered useless, made out-of-date by the subsequent probabilistic update.

The intuitive explanation for this was that e.g.\ a printed \verb+0+ would tell us that \emph{at that point} the state could not be \verb+11+ (as in \Fig{f0544c} from \Sec{s1407c}). But the subsequent update could convert a \verb+11+ into a \verb+00+, and so the final state could after all be \verb+00+.

We then cautioned that the conclusion that the program was (wrt.\ the final state) ``leak free'' was unjustified in general, since with a different prior there could well be proper leakage.

Here we give an example showing that to be so. In \Fig{f0544cmnu} the initial hyper is ``skewed'', that is
\begin{figure}
	{\tt\small
		\begin{tabbing}
			// \textit{\XS\ is set uniformly from \{01,10,11\}.} \\
			print xs[0] $\PC{{\nicefrac{1}{2}}}$ xs[1]~; \\
			xs:= xs $\PC{{\nicefrac{1}{2}}}$ -xs \\
	\end{tabbing}}
	\small This system is as in \Fig{f0544cm} except that the prior initial distribution differs: at least one bit of \XS\ is known to be 1.
	\caption{Simple-channel program excluding \XS=00 initially}\label{f0544cmnu}
\end{figure}
\begin{equation}\label{e0651a}
\begin{array}{r@{\hspace{3em}}l@{\,}l}
\CVec{0,\nicefrac{1}{3},\nicefrac{1}{3},\nicefrac{1}{3}} &\AtP& 1~,
\end{array}
\end{equation}
so that with certainty ($\AtP1$) it is known that the initial distribution is $\CVec{0,\nicefrac{1}{3},\nicefrac{1}{3},\nicefrac{1}{3}}$. Via the first statement \texttt{print xs[0] $\PC{{\nicefrac{1}{2}}}$ xs[1]} an attacker will with probability $\nicefrac{1}{3}$ (resp.\ $\nicefrac{2}{3}$) observe \verb+0+ (resp.\ \verb+1+) and revise his belief of \XS's distribution as in the first (resp.\ second) row here:
\[
\begin{array}{r@{\hspace{3em}}l@{\,}l}
\CVec{0,\nicefrac{1}{2},\nicefrac{1}{2},0} &\AtP& \nicefrac{1}{3} \\
\CVec{0,\nicefrac{1}{4},\nicefrac{1}{4},\nicefrac{1}{2}} &\AtP& \nicefrac{2}{3}
\end{array}
\]
And after the second statement \texttt{xs:= xs $\PC{{\nicefrac{1}{2}}}$ -xs} the hyper for the \emph{current} (and final) distribution of $\XS$ will have become
\begin{equation}\label{e0651b}
\begin{array}{r@{\hspace{3em}}l@{\,}l}
\CVec{0,\nicefrac{1}{2},\nicefrac{1}{2},0} &\AtP& \nicefrac{1}{3} \\
\CVec{\nicefrac{1}{4},\nicefrac{1}{4},\nicefrac{1}{4},\nicefrac{1}{4}} &\AtP& \nicefrac{2}{3} ~,
\end{array}
\end{equation}
where in the $\nicefrac{1}{3}$-case he is better off finally than initially (since he knows \XS\ cannot be 00 or 11), but in the other case he is worse off (since \XS=00 has become possible). Thus if the attacker's choice is either to guess \XS's initial value or to run the program and guess \XS's final value, he can use these hypers to help make up his mind depending on his own criteria for the utility of his planned theft, the social context in which he is operating.\,%
\footnote{Compare a thief's two alternatives for stealing a credit card: \emph{``Steal it now, since the wallet is just sitting there.''}\ and \emph{``Steal it after the card is used at an ATM where he can see some digit of the PIN, but there is a risk his victim will notice and choose a new PIN.''}}

For example, the Shannon entropy of \XS\ is initially $\lg(3){\sim}1.6$, but finally it is conditionally $\nicefrac{1}{3}{\Atimes}1+\nicefrac{2}{3}{\Atimes}2=\nicefrac{2}{3}{>}1.6$: if the attacker is using Shannon entropy to make his decision, he should act sooner rather than later.

On the other hand, the one-guess probability (R{\'e}nyi min-entropy) of \XS\ is initially $\nicefrac{1}{3}$; and finally it is the same, at $\nicefrac{1}{3}{\Atimes}\nicefrac{1}{2}+\nicefrac{2}{3}{\Atimes}\nicefrac{1}{4}=\nicefrac{1}{3}$. 
If the attacker is using this criterion, it does not matter when he acts.

In either case, the hypers \Eqn{e0651a} and \Eqn{e0651b} contain all the information necessary for his decision: the bit-values printed are themselves not important \emph{for his decision}, which is why we can quotient our semantics by abstracting from them. (He does however need those values when he \emph{makes} his attack, if he decides ``later''.)

These calculations are confirmed in \App{a1409}.

\NS
Having reached \Sec{s1008} we have two formulations of \HMM's: the original, matrix style (\Sec{s1403B}); and the ``abstract'' hyper-distribution style (\Sec{s0913}), and there is a semantic function $\HMMone{\cdot}$ that takes the first to the second (\Sec{s0913}). Each has its own definition of a sequential composition operator (Secs.~\ref{s1243},\ref{s1105}). It's a routine but inescapable obligation to show that the semantic function commutes with the two definitions of that composition.

\begin{ReTheorem}{Composition faithfully denoted}{t1008}\\
	Let $H^{1,2}{\In}\MH$ be \HMM's. Then
	\[
	\HMMone{H^1;H^2} \Wide{=} \HMMone{H^1};\HMMone{H^2}~,
	\]
	where \Eqn{e0946} is used for $({;})$ on the left and \Def{d1210} for $({;})$ on the right.
	\Proof
	\label{g1105}We use conventional $p(\cdot,\cdot,\cdots)$-style joint-distribution notations for this.
	
	Write $p_1(x,y_1,x'')$ for the probability $H^1_{x,y_1,x''}$ that the joint distribution $H^1$ assigns to the event ``$x$ is input, and $y,x''$ are output''; write $p_2(x'',y_2,x')$ similarly for $H^2$; and write $p(x,y_1,y_2,x')$ for the (matrix) composition $H^1;H^2$ as a whole.
	
	We follow the usual notational conventions for these $p$'s wrt.\ marginal- and conditional distributions, e.g.\ that $p_1(y)$ is the marginal probability of the event $Y{=}y$ induced by $H_1$ and that $p_1(x|y_1,x'')$ is the conditional probability of $X{=}x$ given $Y_1{=}y_1\land X''{=}x''$. By e.g.\ $p_1(Y)$ we mean the $y$-marginal distribution as a whole, and by $p_1(X|y_1)$ we mean the \emph{a-posteriori} distribution on the input induced by having observed some $y_1$ (and having marginalised by summing over all $x''$).
	
	We avoid potential divisions by zero (i.e.\ due to a $\CalY$ marginal's being zero at some $y$) by using \WLOG\ reduced matrices (\Def{d1426HB}).
	
	Now fix arbitrary $\pi{\In}\VType{\CalX}$. The first step $\HMMone{H^1}.\pi$ produces hyper $p_1(X''|y_1)\AtP p_1(y_1)$ with $y_1$ varying over $\CalY$, where $p_1(x,y_1,x'')=\pi_x H^1_{x,y,x''}$.\,%
	\footnote{Note that the omitted $x$ in $p_1(X''|y_1)$ indicates an implicit $\sum_x$, following the usual conventions for the $p()$ notations.}
	By ``$y_1$ varying over $\CalY$'' we mean e.g.\ a distribution presented in the style of \Eqn{e0852n} for all $y_1$ in $\CalY$.
	
	And from an arbitrary $\pi''$, the second step would produce $p_2(X'|y_2)\AtP p_2(y_2)$ with $y_2$ (also) varying over $\CalY$, and $p_2(x'',y_2,x')=\pi''_{x''}H^2_{x'',y_2,x'}$. To carry out the Kleisli composition, we must let the second-step prior $\pi''$ range over the inners $p_1(X''|y_1)$ from the first step.
	
	Thus letting $\pi''$ be some $p_1(X''|y_1)$ from the first step, we get from \Def{d1022} that
	\begin{Reason}
		\Step{}{
			{p_2}_{y_1}(x'',y_2,x')
		}
		\Step{$=$}{
			\pi''_{x''}H^2_{x'',y_2,x'}
		}
		\StepR{$=$}{set $\pi''\Defs p_1(X''|y_1)$}{
			p_1(X''|y_1)_{x''}H^2_{x'',y_2,x'}
		}
		\Step{$=$}{
			p_1(x''|y_1)H^2_{x'',y_2,x'} ~,
		}
	\end{Reason}
	where the $y_1$-subscript in the ${p_2}_{y_1}$, which we are defining on the left, captures its dependence on $p_2$. 
	
	With that, we have that Kleilsi composition $\HMMone{H^1};\HMMone{H^2}$ applied to $\pi$ is
	\begin{equation}\label{e1342}
	{p_2}_{y_1}(X'|y_2) \quad\AtP~p_1(y_1){p_2}_{y_1}(y_2) ~,
	\end{equation}
	with $y_{1,2}$ varying over $\CalY^2$.
	
	What we would like to know is whether this Kleisli-generated \Eqn{e1342} is the same hyper
	\begin{equation}\label{e1211}
	p(X'|y_1,y_2) \quad\AtP~ p(y_1,y_2)~,
	\end{equation}
	with $y_{1,2}$ varying over $\CalY^2$, that would result from \Def{d1022} applied to $\ChApp{\pi}{(H^1;H^2)}$ directly.
	Thus we now prove for all $y_{1,2}$ the two equalities
	\begin{eqnarray}
	\textrm{\small \underline{Kleisli-generated \Eqn{e1342}}} && \textrm{\small\underline{\Def{d1022} generated \Eqn{e1211}}} \nonumber\\
	p_1(y_1){p_2}_{y_1}(y_2) &=& p(y_1,y_2) \label{e1051a} \\
	{p_2}_{y_1}(X'|y_2) &=& p(X'|y_1,y_2) \label{e1051b}
	\end{eqnarray}
	that together will establish the equality of the two hypers. Beginning with \Eqn{e1051a} we calculate
	\begin{Reason}
		\Step{}{
			p_1(y_1){p_2}_{y_1}(y_2)
		}
		\Step{$=$}{
			p_1(y_1)\sum_{x'}{p_2}_{y_1}(y_2,x')
		}
		\Step{$=$}{
			\sum_{x'}p_1(y_1){p_2}_{y_1}(y_2,x')
		}
		\StepR{$=$}{\Lem{l1625} below}{
			\sum_{x'}p(y_1,y_2,x')
		}
		\Step{$=$}{
			p(y_1,y_2)~.
		}
	\end{Reason}
	
	For \Eqn{e1051b} we prove that ${p_2}_{y_1}(x'|y_2) = p(x'|y_1,y_2)$ for arbitrary $x'$, calculating
	\begin{Reason}
		\Step{}{
			{p_2}_{y_1}(x'|y_2)
		}
		\Step{$=$}{
			{p_2}_{y_1}(y_2,x')/{p_2}_{y_1}(y_2)
		}
		\StepR{$=$}{\Lem{l1625}}{
			p(y_1,y_2,x')/p_1(y_1)~/{p_2}_{y_1}(y_2)
		}
		\StepR{$=$}{\Eqn{e1051a}}{
			p(y_1,y_2,x')/p(y_1,y_2)
		}
		\Step{$=$}{
			p(x'|y_1,y_2)~.
		}
	\end{Reason}
	And with those two equalities we have our result.
\end{ReTheorem}

The following technical lemma, used in the proof of \Thm{t1008}, simplifies the relationship between $p_1,p_2$ and $p$:
\begin{Lemma}{Technical lemma}{l1625}
	Continuing with the notations of above, we show that
	\[
	p_1(y_1){p_2}_{y_1}(y_2,x')
	\Wide{=}
	p(y_1,y_2,x') ~.
	\]
	\Proof
	\begin{Reason}
		\Step{}{
			p_1(y_1){p_2}_{y_1}(y_2,x')
		}
		\Step{$=$}{
			p_1(y_1)\sum_{x''}{p_2}_{y_1}(x'',y_2,x')
		}
		\StepR{$=$}{calculation above}{
			p_1(y_1)\sum_{x''}p_1(x''|y_1)H^2_{x'',y_2,x'}
		}
		\Step{$=$}{
			\sum_{x''}p_1(y_1,x'')H^2_{x'',y_2,x'}
		}
		\Step{$=$}{
			\sum_{x,x''}p_1(x,y_1,x'')H^2_{x'',y_2,x'}
		}
		\Step{$=$}{
			\sum_{x,x''}\pi_x H^1_{x,y_1,x''}H^2_{x'',y_2,x'}
		}
		\Step{$=$}{
			\sum_{x}\pi_x\sum_{x''}H^1_{x,y_1,x''}H^2_{x'',y_2,x'}
		}
		\Step{$=$}{
			\sum_{x}p(x,y_1,y_2,x')
		}
		\Step{$=$}{
			p(y_1,y_2,x')~.
		}
	\end{Reason}
\end{Lemma}

\NS
%\begin{abstract}
Security leaks via (information theoretic) channels on the one hand and, on the other, via (non-interference style) observations of imperative probabilistic programs with hidden state, are in fact each special cases of Hidden Markov Models (\HMM's). In the mathematics literature, \HMM's are typically described via three-way joint probability distributions over initial hidden state, observations and final (still hidden) state.

Here we provide a contrasting, computer-science account of ``abstract \HMM's'' as program semantics, computations in the Giry monad, with the novelty that instead of acting on a program-state $\CalX$ they act on distributions $\Dist\CalX$ directly, so that the monadic type of probabilistic computations \emph{with} hidden state $\CalX$ becomes $\AHSpc$ rather than the $\CalX{\Fun}\Dist\CalX$ typical for probabilistic programs when state is not hidden.

Characteristic of this approach are: identification of the common properties shared by many extant, diverse entropy measures; definition of a ``more secure than'' partial order in a space $\Dist^2\CalX$ of ``hyper-distributions''; soundness- and completeness theorems that relate the security order to testing via (generalised) entropies; and formulation of a dual ``predicate-transformer style'' post-entropy to pre-entropy description of \HMM\ computations. Further, technical contributions include significant theorems about the properties of the semantic spaces and the relationships between them.
%\end{abstract}

\NS
%\section{Introduction}\label{s1656}
%\subsection{Setting and overview}

Probabilisitic sequential programs with hidden state are examples of Hidden Markov Models, or \HMM's,\,%
\footnote{We use apostrophe uniformly for suffixes of acronyms.}
whose typical formulation is a joint probability distribution over initial state, observations, and final state.
% (\Sec{s1403B}).
Probabilistic channels are a special case, in which the initial- and final state are equal: information flows from initial state to observations.% (\Sec{s1407c}).

We propose an alternative to that typical \HMM\ formulation: a representation, based on the Giry monad, that we believe is more suitable for program semantics. (The resulting benefits we hope for are explained just below in \Sec{s0429} and elaborated in the conclusion \Sec{s1323}.) Indeed the monadic view of \emph{probabilistic} computations in particular is well established \cite{Moggi:89,Giry:81}: the type of computations on a set $\CalX$ is $\CalX{\Fun}\Dist\CalX$ where type-constructor $\Dist$ makes distributions on its base type $\CalX$; the Kleisli extension is then of type $\Dist\CalX{\Fun}\Dist\CalX$, representing action of multiplying an initial-state-distribution vector by a Markov-matrix. % to give a final-state-distribution vector. 
But that does not account for hidden state and information flow.

Our point of departure is to include hidden state by beginning with $\Dist\CalX$ as the basis space (rather than simply $\CalX$): the computation type we obtain is ``one level up'', of type $\Dist\CalX{\Fun}\Dist^2\CalX$, the extension is $\Dist^2\CalX{\Fun}\Dist^2\CalX$; and we call the double-distribution type ``hyper-distributions''.

In earlier work we have used hyper-distributions to establish compositionality results \cite{mcivermeinicke10a}, to explore the effect of including demonic nondeterminism \cite{McIver:12} and, separately, to generalise the information-flow treatment of probabilistic channels \cite{McIver:2014aa,Alvim:2014aa}. A second common theme  has been the generalisation of entropies (such as Shannon) to a more abstract setting where only their essential properties are preserved \cite{mcivermeinicke10a,Alvim:2012aa,McIver:12,Alvim:2014aa}. Here we bring all those separate strands together and take a significant further step.

Dual to the ``forward'' representation of a computation, from some initial- to final structure, there is often a ``backward'' representation from some final- to initial test. %(A conspicuous early example is weakest preconditions \cite{Dijkstra:76}: the forwards view of a nondeterministic program is $S{\Fun}\Pow S_\bot$, and the dual, backwards view is $\Pow S{\Fun}\Pow S$ where in the backwards case $\Pow S$ represents a test of whether a particular state $s$ satisfies a pre- or post-condition.)

Our further step is to show that there is a dual backwards view for \HMM's as well. We call it ``uncertainty transformers'', functions that transform ``post- uncertainty measures'' into ``pre- uncertainty measures''. We and others have argued that specific entropies (e.g.\ Shannon) are too limited for security work generally \cite{Smith:2009aa}, and focus here on the essential properties such a measure should have: continuity and concavity. That view is supported by  powerful theorems that such a generalisation supports, and a methodological criterion that uncertainty measures can capture contexts in a way that individual styles of entropy cannot.

A second further step is to extend our recent treatment \cite{Alvim:2014aa} of the \emph{Dalenius Desideratum}, the ``collateral'' leakage of information due to unknown correlations with third-party data, from merely channels (a ``read only'' scenario  \cite{Dalenius:1977aa,Dwork:2006aa}, such as access to a statistical database) to programs that might alter the database (thus ``read/write'' as well). The concerns relate also to compositionality, where the third-party data is represented by variables in scope of a program fragment but to which it does not explicitly refer \cite{mcivermeinicke10a}.

To make the presentation accessible to the broader security community, we do not begin from the Giry monad: rather we first work in elementary terms. In \Sec{s1250} we introduce the more general perspective which will then be seen to have informed our earlier constructions.

\NS
{%1247
	%\subsection{Principal contributions and aims of the paper: summary
	%\protect\Cf{Add specific references to sections, lemmas and theorems only once we have agreed on this text.}
	%}\label{s0429}
	%\Cf{We must be very clear here what is new \emph{in this paper}, since a great deal of what we present here is a collecting-together of results we (and our colleagues) have already published in other places. However worthy that is, it might not in itself be enough for this venue, and so our substantial new contributions here have to be highlighted. It looks to me like our new material begins in \Sec{s0842}.}
	
	\noindent Our \textbf{principal contributions} are these:
	\begin{Figure}{f1843}{Relationship between the semantic spaces}
		\vspace{24ex}
		%\ImageInText{0pt}{1.7em}{0.6}{Spaces.pdf} %fonts fixed
		%\ImageInText{0pt}{1.7em}{0.6}{Carroll/Spaces.pdf} %fonts fixed
		\ImageInText{0pt}{1.7em}{0.6}{Spaces.pdf} %fonts fixed
		
		%\caption{Relationship between the semantic spaces}
	\end{Figure}
	
	\noindent--- Noting that \HMM's generalise both information channels and probabilistic programs, we introduce a \underline{novel monadic model} of ``abstract'' \HMM's (\Sec{s0913}), functions from distributions to distributions-of-distributions, to serve as denotations for probabilistic programs with hidden state. Composition of abstract \HMM's is the Kleisli composition of the distribution monad.
	
	\noindent--- We explain how the partial \underline{securit}y-\underline{increasin}g \underline{order} on distributions of distributions, the so called \emph{hyper-distributions}, %, together with a soundness-and-completeness result that relates that order to a testing order, 
	promotes to a novel order for \HMM's.
	
	\noindent--- We formulate a ``backwards", transformer model, dual to the ``forward'' monadic semantics; the duality requires a non-trivial extension of the Riesz representation theorem, applying it to hyper-distributions. The transformer model is based on  \underline{uncertaint}y \underline{measures} on distributions: these generalise diverse popular entropies, and our approach exposes the relationship between the concavity of entropies as functions on $\Dist\CalX$ and the logical structure of \HMM's. 
	
	%\noindent--- We introduce \underline{uncertaint}y\underline{ measures} on distributions as a generalisation the variety of currently popular entropies, and use them to construct a novel uncertainty-transformer semantics, dual to the ``forward'' Kleisli-composition semantics for \HMM's.
	
	\noindent--- We formulate  \underline{characteristic} ``\underline{healthiness}'' p\underline{ro}p\underline{erties} of both the forward- and dual, backward semantics for \HMM's.
	
	The \HMM-as-monad, the uncertainty transformers, and their healthiness conditions are completely novel (we believe); although the partial order and the hyper-distributions have appeared earlier, they are here consolidated and related to common notions of entropy. 
	
	\noindent Our \textbf{principal aims} in this paper are these:
	
	\noindent--- (More abstract) We want to identify and explain the connection between \HMM's as typically presented in mathematics (joint distributions and matrices), and the monadic structures typical of computer science for modelling denotationally, in this case, the semantics of quantitative information flow for sequential programs. The theorems establishing healthiness conditions, and the forwards/backwards duality, are essential parts of that connection.
	
	\noindent--- (More concrete) We want to provide the basis for a source-level reasoning method, analogous to Hoare logic or weakest preconditions, for quantitative non-interference in sequential programs. For this, the dual, transformer semantics for \HMM's seems to be a necessary first step, together with the representation theorem
	\Cf{Which one?}
	which relates security goals  to the behaviour of a program.
	% together with the representation theorem for uncertainties based on ``loss functions'' that allows them to be expressed over program variables.
	
	In the conclusion \Sec{s1323} we discuss \textbf{the benefits} we expect achieving these aims would bring.
	%}%1247
	
	\NS
	This is the abstract we submitted.
	\begin{quote}
		We give a monad-inspired presentation of Hidden Markov Models (\HMM's) which is a precursor to a programming logic for source-level reasoning about quantitative non-interference in sequential probabilistic programs (of which probabilistic channels are a special case). Based on the Giry monad, but beginning with distributions $\Dist\CalX$ over state $\CalX$ (rather than $\CalX$ itself), our computations have type $\AHSpc$. We believe monads better suit program semantics than the explicit joint distributions typically used for \HMM's. 
		
		The approach encourages a ``predicate-transformer style'' post- to pre-semantics, dual to $\AHSpc$, of type $\Unc\CalX{\Fun}\Unc\CalX$ where $\Unc\CalX$ generalises entropies (e.g.\ Shannon) in $\Dist\CalX{\Fun}\NNReal$. For quantitative information flow, this is new.
		
		Our technical contributions, besides formulating these structures, is to identify and prove important theorems about how the structures relate to each other.
		
		[``abstracted abstract'', 125 words]
	\end{quote}
	
	\NS
	\C{\sf If you want to revise the abstract, please put a whole new one (use cut-and-paste) in your colour at the end, so that the versions are chronological and the colour identifies who's responsible for each.}
	
	We provide an abstract account of Hidden Markov Models and Channels as used
	for Quantitative Information Flow. The novelty is to use the Giry Monad on a space of
	distributions $\Dist\CalX$, so that HMM steps and channels are continuous mappings $\Dist\CalX \Fun \Dist^2\CalX$. 
	The Giry Functor $\Dist$ plays a crucial role in providing enough structure to enable distinctions between
	prior and posterior probability distributions required for an information-flow analysis.
	
	We also study the dual property-based ``transformer" model, using the concave functions on
	$\Dist\CalX$ to formulate quantitative properties, and identified in previous work \cite{McIver:12,McIver:2014aa}
	as characterising a robust ``security refinement" order. We prove the equivalence between 
	the transformer model and the model of continuous mappings $\Dist\CalX \Fun \Dist^2\CalX$, and characterise when
	those mappings are HMM steps or Channels. \\ {}[128 words]
	
	\NS
	We embed a semantics for probabilistic programs, with hidden state, within a probabilistic monad; the aim is to take advantage of general monadic structures and operations in order to reason about quantitative information flow.
	
	The monadic view of probabilistic computations is well established \cite{Moggi:89,Giry:81}: the type of computations on a set $\CalX$ is $\CalX{\Fun}\Dist\CalX$ where type-constructor $\Dist$ makes distributions on its base type $\CalX$; the Kleisli extension is then of type $\Dist\CalX{\Fun}\Dist\CalX$, representing action of multiplying an initial-state-distribution vector by a Markov-matrix to give a final-state-distribution vector.
	
	By taking $\Dist\CalX$ as the state space (rather than simply $\CalX$) we obtain in the same monad a model $\Dist\CalX{\Fun}\Dist^2\CalX$, one level up, that we show is suitable for quantitative information flow: the Kleisli extension is then of type $\Dist^2\CalX{\Fun}\Dist^2\CalX$, representing the action of a Hidden Markov Model (\HMM) on ``hyper-distributions'', that is distributions of distributions.
	
	Furthermore, just as there is a dual ``backwards'' view for nondeterministic programs, the weakest preconditions \cite{Dijkstra:76}, and a similar construction for deterministic probabilistic programs \cite{Kozen:83}, and again for probabilistic/demonic programs \cite{Morgan:96d,McIver:05a}, we show here that there is a dual backwards view for \HMM's.
	\footnote{We use apostrophe uniformly for suffixes of acronyms.}
	We call them ``uncertainty transformers''.
	
	We also justify the introduction of ``uncertainty measures'' on distributions, the functions that the uncertainty transformers transform, as a generalisation of entropies. We and others have argued that specific entropies (like e.g.\ Shannon) are too limited for security work generally \cite{Smith:2009aa}, and propose here the essential properties are simply continuity and concavity. That view is supported by several powerful theorems that such a generalisation supports, and a methodological criterion that uncertainty measures are adaptable enough to capture contexts in a way that individual entropies are not.
	
	Finally, we show how our constructions allow reasoning about the \emph{Dalenius Desideratum} that concerns the ``collateral'' leakage of information due to pre-existing correlations of the program state with other data that the program does not access, and of which the program designer might have been unaware. It applies not only to statistical databases  \cite{Dwork:2006aa}, but also to compositionality of semantics \cite{mcivermeinicke10a}.
	
	To make the presentation accessible to the security community, we first work in elementary terms, i.e.\ without monads explicitly; in \Sec{s1250} we introduce the latter, more general perspective which informed our constructions.
	
	\NS
	Channels and hidden-state programs are separate special cases of Hidden Markov Models (\HMM's), typically described via three-way joint probability distributions.
	
	Our novel, program-semantics approach to \HMM's instead uses the Giry monad on distributions $\Dist\CalX$ over a state $\CalX$: computations become $\AHSpc$ rather than the usual $\CalX{\Fun}\Dist\CalX$.
	
	This accommodates unification of many diverse entropy measures, allows a security-based partial order over ``hyper-distributions'' $\Dist^2\CalX$ with soundness- and completeness theorems relating that order to testing via (generalised) entropies, and permits a dual ``predicate-transformer style'' post-entropy to pre-entropy semantics for \HMM's. \\ {}[``abstracted abstract'', 101 words]
	
	\NS
	%{\Cx%1247s
	\subsection*{Principal contributions and aims of the paper}
	
	\noindent Our \textbf{principal contributions} are these:
	
	\noindent--- We note that \HMM's model both information flow via channels and probabilistic sequential computation on a hidden state, and provide a \underline{novel monadic model} of ``abstract'' \HMM's as functions from distributions to distributions-of-distributions, so called \emph{hyper-distributions}, in which their composition is (the usual) Kleisli composition of a monad.
	
	\noindent--- We identify, motivate and explain a partial \underline{securit}y-\underline{increasin}g \underline{order} on hyper-distributions, together with a soundness-and-completeness result that relates that order to a testing order.
	
	\noindent--- We introduce \underline{uncertaint}y\underline{ measures} on distributions as a generalisation the variety of currently popular entropies, and use them to construct a novel uncertainty-transformer semantics, dual to the ``forward'' Kleisli-composition semantics for \HMM's.
	
	\noindent--- We formulate and prove \underline{characteristic} ``\underline{healthiness}'' p\underline{ro}p\underline{erties} of both the forward- and dual, backward semantics for \HMM's.
	
	The \HMM-as-monad, the uncertainty transformers, and their healthiness conditions are completely novel (we believe); the partial order and the hyperdistributions have appeared, partially, elsewhere but are here consolidated. 
	
	\noindent Our \textbf{principal aims} in this paper are these:
	
	\noindent--- (More abstract) We want to identify and explain the smooth connection between \HMM's as typically presented in the mathematics literature(joint distributions and matrices) and the monadic structures for modelling denotationally, in the computer-science style, the semantics of quantitative information flow for sequential programs. For these, the significant theorems establishing healthiness conditions are essential.
	
	\noindent--- (More concrete) We want to provide the basis for a source-level reasoning method, analogous to Hoare logic or weakest preconditions, for quantitative non-interference in sequential programs. For this, the novel dual ``backwards'' semantics for \HMM's, seems to be essential, together with a representation theorem for uncertainties based on ``loss functions''.
	
	In the conclusion \Sec{s1323} we discuss \textbf{the benefits} we expect achieving these aims would bring.
	%}%1247s
	
	\NS
	\section*{\HMM's and the Dalenius Desideratum
		\protect{\Af{Integrate \Sec{s1321} with this.}}}\label{s0917}
	\Cf{SeeZ \Fn{n1230}.}
	Our presentation provides a generic framework for describing the behaviour of programs which combine both a Markov update and
	a release of information and, as such, are examples of \HMM's. It concentrates on the transformation of initial- to final states, and about release of information concerning that final state.
	
	There are important areas however in which leakage from the initial state is important, even if that state is overwritten by the markov part of the \HMM. The issue is that what the initial state \emph{was} might reveal information about what some other correlated state still \emph{is}, or might independently have become, even if that other state is not mentioned in the program at all.
	
	The semantics given here does not, so far, preserve that initial information leak (if it exists). For example given a markov $M$ such that $M_{x,x'} = 1$ if and only if $x'{=}x_0$ for some fixed state $x_0{\In} \CalX$, we have $\HMMsem{C}{M} = \HMMsem{}{M}$ for any channel $C$ --- that is, any leaking by $C$ of the initial state is not captured in the semantics.
	
	However the release of initial information can be important \Cite{Dwork}. For example, statistical databases can reveal information about seemingly unrelated data if there is some interesting, but undocumented correlation between it and the data stored in the database. That information release persists even if the database is subsequently updated.
	
	The presentation of \HMM's as matrices (i.e.\ the original) preserves information about the initial state. Up to this point, however, we have deliberately abstracted from that. In this section we will show how to adapt our framework to account for explicit leakage from the initial state when we need to.
	
	Let $C{\In}\CalX{\MFun}\CalY$ be a channel and let $M{\In}\CalX{\MFun}\CalX$ be a markov. Let $\CalZ$ be a fresh type. \label{g1235}We write $\Dal{C}{\CalZ}$ for the expanded channel, of type $\CalX{\times}\CalZ\MFun\CalY$, defined
	$(\Dal{C}{\CalZ})_{x,z,y}\Defs~C_{x,y}$~.
	Similarly we write $\Dal{M}{\CalZ}{\In}\CalX{\times}\CalZ\MFun\CalX{\times}\CalZ$ for the markov $(\Dal{M}{\CalZ})_{x,z,x',z'}\Defs~M_{x,x'}$ if $z{=}z'$ otherwise $0$~.
	The important aspect of these definitions is that for any initial distribution $\pi{\In}\Dist(\CalX{\times}\CalZ)$, the joint distribution obtained by either $\ChApp{\pi}{(\Dal{C}{\CalZ})}$ or $\ChApp{\pi}{(\Dal{M}{\CalZ})}$ does not access the $\CalZ$ component.
	
	Here is an example. Consider the channel and markov
	%\LMat{.7em}{-.5em}{1.5em}\nicefrac{1}{3} & \nicefrac{1}{6}\RMat{.9em}{-.51em}{1.5em}
	\[
	C~=~ 
	\begin{array}{ccc}
	& y_0 & y_1 \\
	x_0{:} & \LMat{.7em}{-.5em}{1.5em}\nicefrac{1}{2} & \nicefrac{1}{2}\RMat{.9em}{-.51em}{1.5em} \\
	x_1{:} & \nicefrac{1}{3} & \nicefrac{2}{3}
	\end{array}
	\hspace{3em}
	M~=~ 
	\begin{array}{ccc}
	& x_0 & x_1 \\
	x_0{:} & \LMat{.7em}{-.5em}{1.5em}1 & 0\RMat{.9em}{-.51em}{1.5em} \\
	x_1{:} & \nicefrac{1}{4} & \nicefrac{3}{4}
	\end{array}
	\]
	
	If we take $\CalZ \Defs \{z_0, z_1\}$, the channel $\Dal{C}{\CalZ}$ is
	\[
	\Dal{C}{\CalZ}\Wide{=} 
	\begin{array}{c@{\quad\quad}cc}
	& y_0 & y_1 \\
	(x_0, z_0){:} & \LMat{.7em}{-1.8em}{2.8em}\nicefrac{1}{2} & \nicefrac{1}{2}\RMat{.9em}{-1.8em}{2.8em} \\
	(x_0, z_1){:} & \nicefrac{1}{2} & \nicefrac{1}{2} \\
	(x_1, z_0){:} & \nicefrac{1}{3} & \nicefrac{2}{3} \\
	(x_1, z_1){:} & \nicefrac{1}{3} & \nicefrac{2}{3}
	\end{array}
	\]
	and similarly the markov $\Dal{C}{\CalZ}$ is
	\[
	\Dal{M}{\CalZ} \Wide{=} 
	\begin{array}{ccccc}
	& x_0z_0 & x_0z_1 & x_1z_0 & x_1z_1 \\
	x_0z_0{:} & \LMat{.7em}{-1.8em}{2.8em}1 & 0 & 0 & 0\RMat{.9em}{-1.8em}{2.8em} \\
	x_0z_1{:} & 0& 1& 0& 0 \\
	x_1z_0{:} & \nicefrac{1}{4}& 0 & \nicefrac{3}{4} & 0 \\
	x_1z_1{:} & 0 & \nicefrac{1}{4} & 0 & \nicefrac{3}{4}
	\end{array}~.
	\]
	
	The definitions above show that for  $\Dal{C}{\CalZ}{\In} \CalX{\times}\CalZ\MFun\CalX{\times}\CalZ$ the
	rows of the original channel matrix are each repeated $|\CalZ|$ times; moreover a subsequent update of the state in $\CalX$ by
	$\Dal{M}{\CalZ}$ will leave the $\CalZ$ part of the state unchanged. In fact we can recover the all the original ingredients from the
	\emph{properties} of the semantic function $\HMMsem{\Dal{C}{\CalZ}}{\Dal{M}{\CalZ}}$, rather than the syntactic description of
	the component matrices, provided that $|\CalX| \leq |\CalZ|$. We sketch how to do this next.
	\Cf{Appendix? Haven't read this yet.}
	
	For $\delta{\In} \Dist\CalZ$, let $f{\In} \CalX {\Fun}\CalZ$ be a (1-1) function, and let $\hat{\delta}{\In} \CalX{\times}\CalZ$ be defined so that $\hat{\delta}.x.z= \delta.z$ if $z{=}f.x$, and zero otherwise. Similarly for loss function $l{\In} I {\MFun}\CalZ$, define $\hat{l}{\In} I {\MFun}\CalX{\times}\CalZ$ as $\hat{l}.x.z = l.z$. Finally define transformer $t{\In}\Unc\CalZ {\Fun} \Unc\CalZ$ by $t.U_l.\delta \Defs \Wp{\HMMsem{\Dal{C}{\CalZ}}{\Dal{M}{\CalZ}}}.U_{\hat{l}}.\hat{\delta}$. Using these definitions we have the following facts.
	
	\begin{enumerate}
		\item $t$ is multiplicative:
		\begin{Reason}
			\Step{}
			{t.U_{(\ChApp{\pi_1}{l^T})^T}.\pi_2}
			\Step{$=$}
			{\Wp{\HMMsem{\Dal{C}{\CalZ}}{\Dal{M}{\CalZ}}.U_{\hat{(\ChApp{\pi_1}{l^T})^T}}.\hat{\pi_2}}}
			\WideStepR{$=$}{\Thm{t1008}, $\Dal{M}{\CalZ}$ does not update $\CalZ$}
			{\Wp{\HMMsem{\Dal{C}{\CalZ}}{}.U_{\hat{(\ChApp{\pi_1}{l^T})^T}}.\hat{\pi_2}}}
			\WideStepR{$=$}{\Lem{l1453} and $\Dal{C}{\CalZ}$ is a channel}
			{\Wp{\HMMsem{\Dal{C}{\CalZ}}{}.U_{\hat{(\ChApp{\pi_2}{l^T})^T}}.\hat{\pi_1}}}
			\StepR{$=$}{As above}
			{\Wp{\HMMsem{\Dal{C}{\CalZ}}{\Dal{M}{\CalZ}}.U_{\hat{(\ChApp{\pi_2}{l^T})^T}}.\hat{\pi_1}}}
			\Step{$=$}
			{t.U_{(\ChApp{\pi_2}{l^T})^T}.\pi_1}~.
		\end{Reason}
		
		\item Since $t$ is multiplicative, by \Thm{t2117} there must be some channel matrix $K{\In}\CalZ{\MFun}\CalY$ corresponding to $t$.
		
		\item In fact $K$ corresponds exactly to the original channel $C$ as follows:  for any loss function $l$, and prior $\delta{\In}\Dist\CalZ$, we have that $\Wp{\HMMsem{{K}}{}}.U_l.\delta = \Wp{\HMMsem{{C}}{}}.U_{l\circ f}.\delta'$, where $\delta'.x = \delta.(f.x)$.
		
	\end{enumerate}
	
	\NS
	\section{Further healthiness conditions for \HMM's}\label{s1211}
	{\Cx%1216
		With the uncertainty transformers of \Sec{s1636}, we can be more precise about the healthiness conditions satisfied specifically by pure-channel \HMM's. (Our characterisation of pure channels in \Sec{s1407c} was quite loose.)
		
		Recall that \Cite{McIver and Morgan, Kozen, Panangaden, Plotkin} have used a dual semantics to characterise probabilistic program models containing features that augment basic probabilistic transitions. The hypers combine information flow with probabilistic updates, and then a simple characterisation based on refinement characterises pure transitions (\Sec{s1537}). In this section we show, using the dual transformer model, that we can characterise pure channels as well, and that a combination of the forwards- and backwards approaches allows us to characterise single-step \HMM's. \Af{Find the correct terminology. \Cx For what?}
	}%1216
	
	Fundamental for our earlier \Thm{t1005} (\Sec{s1408}) was that the transformers are {\Tx  1-Lipschitz \sout{continuous} and linear \sout{additive}} on $\Unc\CalX$. 
	\Tf{I've changed the reference to \Thm{t1005} as Thm\{t1253\} disappeared. We need to make sure it's right. Also, the exact properties for $\Wp h$ are now given in \Lem{l1719A}, \Lem{l0846A} and \Lem{l0846B}. {\Ax I have checked that the properties hold for this $\Wp{h}$ definition since they are all determined by a fixed $\delta$. Tahiry --- could you check too?\TBar }}
	However there remain many  functions $h$ in $\AHSpc$ that are not determined by some fixed underlying stochastic matrix as channels are (\Sec{s0918}). Take $\CalX{=}\DSet{x_1,x_2}$ so that any $\pi{\In}\Dist\CalX$ is determined by $\pi_{x_1}$ alone, and for each such $\pi$ define a (channel) matrix
	\Tf{In the definition of $C^\pi$, shouldn't $\pi_{x_i}$ be $\pi.x_i$?{\Ax We are using $\pi_x$ for $\pi$ evaluated at $x$ I think? I've just corrected it in the matrix.} \Cx Actually we are using $\pi_x$ --- I'll fix all that stuff, don't worry.}
	\[
	C^\pi \Wide{\Defs} \left(
	\begin{array}{cc}
	\pi_{x_1} & \pi_{x_2}\\
	0 & 1\\
	\end{array}\right)~,
	\]
	together with a function $h{\In}\Dist\CalX {\Fun} \Dist^2 \CalX$ defined so that $h.\pi \Defs \HMMsem{C^\pi~}{}.\pi$. 
	The resulting transformer $\Wp{h}$ satisfies the conditions for \Thm{t1005} but, because we ``cheat'' by adapting the channel to the \emph{distribution} of the input, there is no \emph{fixed} underlying stochastic matrix that determines $h$ for all $\pi{\In}\Dist\CalX$ (and therefore not for $\Wp{h}$ either). 
	This example provides the insight for channels' distinguishing feature:  in  the transformer representation, channels have a simple multiplicative property. 
	
	\begin{Definition}{Multiplicative}{d1442}
		We say that a transformer $t{\In}\C{\TT\CalX}$ is \emph{multiplicative} if for any $\pi_{1,2}{\In}\Dist\CalX$  and loss function $l$
		we have $t.U_{(\ChApp{\pi_1}{l^T})^T}.\pi_2 = t.U_{(\ChApp{\pi_2}{l^T})^T}.\pi_1$.
		\Cf{I'll see whether this expression can be simplified. {\Ax Great -- also in the previous section.}}
	\end{Definition}
	
	\begin{Lemma}{Channels are multiplicative}{l1453}
		Let $C{\In} \CalX{\MFun}\CalY$ be a channel matrix. Then $\Wp{\HMMsem{C}{}}$ is multiplicative.
		\Cf{Be careful to be clear about \emph{what} is being given these healthiness conditions. For example, this section is supposed to be about HC's for abstract \HMM's. But this lemma states an HC for a transformer, not an \HMM.}
		\Proof This is immediate from \Lem{l1045} and that $U_{(\ChApp{\pi_1}{l^T})^T}.\pi_2 = U_{(\ChApp{\pi_2}{l^T})^T}.\pi_1$~.
	\end{Lemma}
	
	The next result shows that the multiplicative property in fact characterises channels.
	
	\begin{Theorem}{}{t2117}
		Let $h{\In} \AHSpc$ be such that $h.\pi$ has finite support for every $\pi{\In}\Dist\CalX$. In addition let $h{\Ref}I$
		\Cf{Is $I$ the denotation of \textbf{skip}? {\Ax Yes.} \CBar}
		and $\Wp{h}$ be multiplicative.
		Then there is some channel $C{\In} \CalX{\MFun}\CalY$ such that $h= \HMMsem{C}{}$.
		\Proof
		\Cf{Is this proof a candidate for the appendix? {\Ax Possibly --- it depends what sort of proofs we want to have in the main text. 
				This proof demonstrates the utility of the transformer view so I vote that it stays ---  unless there are other proofs we prefer that do it.} \CBar I have not checked the proof yet.}
		Let $\Delta \Defs h.\upsilon$, where $\upsilon$ is the uniform distribution over $\Dist\CalX$. We define $C_\Delta{\In}\CalX{\MFun}\CalY$ as follows.  Since $h.\upsilon$ has finite support, we can assume  a finite labelling of the support $\CalY \Defs \{y_1, \dots y_k\}$. 
		
		Now define $(C_\Delta)_{x,y} \Defs \Delta.y.x/(\Avg\Delta.x)$; observe that since $h.\upsilon \Ref I.\upsilon$ we must have $\Avg\Delta.x = \upsilon.x > 0$. We now show that $h = \HMMsem{C_\Delta}{}$.
		
		Given any loss function $l'$, we write it as $l' = (\ChApp{\upsilon}{l^T})^T$, where $l.i.x = |\CalX|\cdot l'.i.x$. We now reason:
		\begin{Reason}
			\Step{}
			{\Wp{h}.U_{l'}.\delta}
			\StepR{$=$}{Re-write $l'$ as above}
			{\Wp{h}.U_{(\ChApp{\upsilon}{l^T})^T}.\delta}
			\StepR{$=$}{$\Wp{h}$ is multiplicative}
			{\Wp{h}.U_{(\ChApp{\delta}{l^T})^T}.\upsilon}
			\StepR{$=$}{$h.\upsilon = \HMMsem{C_\Delta}{}.\upsilon$}
			{\Wp{\HMMsem{C_\Delta}{}}.U_{(\ChApp{\delta}{l^T})^T}.\upsilon}
			\WideStepR{$=$}{$C_\Delta$ is a channel matrix; \Lem{l1453}}
			{\Wp{\HMMsem{C_\Delta}{}}.U_{(\ChApp{\upsilon}{l^T})^T}.\delta}
			\StepR{$=$}{Re-write $l'$ as above}
			{\Wp{\HMMsem{C_\Delta}{}}.U_{l'}.\delta~.}
		\end{Reason}
		Finally we deduce that $h =\HMMsem{C_\Delta}{}$, since hypers are determined by loss functions \cite{mcivermeinicke10a,McIver:2014aa}.
		\Tf{Is this a consequence of Kostas's representation of concave functions (or density of finite loss functions) or an independent argument? I know from Coriaceous that expected values of uncertainty measures define hypers. So by that density, expected values of loss functions also define hypers. {\Ax No -- this is simply referring to the Coriaceous i.e. that the refinement defined directly on hypers is equivalent to
				the order defined on tests. Basically the Coriaceous says that we only need to consider tests defined by loss functions because the two
				orders are the same.} \TBar}
	\end{Theorem}
	
	\NS
	\section{A $\Wp{\cdot}$ example for \Fig{f0544cm}}
	
	We can use our $\Wp{\cdot}$ semantics to answer questions about \Fig{f0544cm} over all priors. For example suppose we wish to know whether, in the final state, $\XS[0]$ and $\XS[1]$ are the same or different. Thus we want to know, in the final state, whether $(\XS[0]=\XS[1])$ or $(\XS[0]\neq\XS[1])$, taking into account the observations and what we can infer from the program text. We use $(\XS[0]=\XS[1])$ and $(\XS[0]\neq\XS[1])$ to form a loss function $U_l$ where $l$ corresponds to the labels defined by the two predicates.  for For shorthand we write
	
	\[
	(\XS[0]=\XS[1]) \AND (\XS[0]\neq\XS[1])~,
	\]
	for the associated the loss function  defined $(\XS[0]=\XS[1]) \AND (\XS[0]\neq\XS[1]).\pi = (a+d) \Min (b+c)$, for prior $\pi$, and $\pi.(00) = a, \pi.(01)= b, \pi.(10)= c, \pi.(11) = d$.
	
	Writing $P$ for the two lines of code in \Fig{f0544cm}, and $\pi$ for the prior we have that
	
	\[
	\Wp{P}.\pi \Wide{=} a~\Min (b+c)/2 + d ~\Min (b+c)/2~.
	\] 
	
	The summands correspond to the two observations during program execution; note that the subsequent Markov state transition does not
	change whether the $\XS[0]$ and $\XS[1]$ are the same or different, however some information has been released to improve the observer's ability to guess some properties about the the final value of the secret, in particular whether the bits are the same or different.
	
	Thus for a prior where $b+c= 1/2$, and therefore where $a+d = 1/2$ his initial chance of guessing whether the two bits are the same or different is $1/2$. However after running the program his best guess now depends on the relative values of $a$ and $d$. For example if $a=1/2$ and and $d=0$ his guess (for loss) is now $1/4$ since if he observes $0$ then he knows that the state originally is most likely to have been $(00)$ and therefore his strategy (to lose) is to guess that
	$\XS[0]\neq\XS[1]$; similarly if he observes $1$ he then he chooses (to lose) $\XS[0]=\XS[1]$, as he knows for sure that $\XS[0]\neq\XS[1]$.
	
	\subsection{$\Wp{()}$ calculation details}
	
	Let $P$ be the program set out in \Fig{f0544cm}, and let $\pi$ be the prior with $\pi.(00) = a, \pi.(01)= b, \pi.(10)= c, \pi.(11) = d$.
	
	\begin{Reason}
		\Step{}
		{\Wp{P}.(\XS[0]=\XS[1]) \AND (\XS[0]\neq\XS[1])}
		\WideStepR{$=$}{\textit {xs:= xs $\PC{{\nicefrac{1}{2}}}$ -xs} leaves postcondition invariant}
		{\Wp{\textit{(print xs[0] $\PC{{\nicefrac{1}{2}}}$ xs[1]})}.(\XS[0]=\XS[1]) \AND (\XS[0]\neq\XS[1])}
		\WideStepR{$=$}{Instantiate observations}
		{(0=\XS[0]=\XS[1]) \AND (\XS[0]\neq\XS[1])/2 ~~+}
		\Step{}
		{ ~~~~(1=\XS[0]=\XS[1]) \AND (\XS[0]\neq\XS[1])/2~.}
	\end{Reason}
	
	We can now evaluate this expression for $\pi$ as above, to see that 
	\[
	\Wp{P}.\pi \Wide{=} a~\Min (b+c)/2 + d ~\Min (b+c)/2~.
	\]
	
	\NS
	For example, in \Fig{f0544cmnu} the initial hyper is
	\begin{figure}
		{\tt\small
			\begin{tabbing}
				// \textit{\XS\ is set uniformly from \{01,10,11\}.} \\
				print xs[0] $\PC{{\nicefrac{1}{2}}}$ xs[1]~; \\
				xs:= xs $\PC{{\nicefrac{1}{2}}}$ -xs \\
		\end{tabbing}}
		\small This system is a in \Fig{f0544cm} except that the prior initial distribution differs: at least one bit of \XS is known to be 1.
		\caption{Simple-channel program excluding \XS=00 initially}\label{f0544cmnu}
	\end{figure}
	\begin{equation}\label{ze0651a}
	\begin{array}{r@{\hspace{3em}}l@{\,}l}
	\CVec{0,\nicefrac{1}{3},\nicefrac{1}{3},\nicefrac{1}{3}} &\AtP& 1~,
	\end{array}
	\end{equation}
	that is with certainty ($\AtP1$) it is known that the initial distribution is $\CVec{0,\nicefrac{1}{3},\nicefrac{1}{3},\nicefrac{1}{3}}$. Via \texttt{print xs[0] $\PC{{\nicefrac{1}{2}}}$ xs[1]} an attacker will wprob\Cf{Define this abbreviation.} $\nicefrac{1}{3}$ (resp.\ $\nicefrac{2}{3}$) have observed 0 (resp.\ 1) and revised his belief of \XS's distribution as in the first (resp.\ second) row here:
	\[
	\begin{array}{r@{\hspace{3em}}l@{\,}l}
	\CVec{0,\nicefrac{1}{2},\nicefrac{1}{2},0} &\AtP& \nicefrac{1}{3} \\
	\CVec{0,\nicefrac{1}{4},\nicefrac{1}{4},\nicefrac{1}{2}} &\AtP& \nicefrac{2}{3}
	\end{array}
	\]
	And after \texttt{xs:= xs $\PC{{\nicefrac{1}{2}}}$ -xs} the hyper for the \emph{current} (and final) distribution of $\XS$ will have become
	\begin{equation}\label{ze0651b}
	\begin{array}{r@{\hspace{3em}}l@{\,}l}
	\CVec{0,\nicefrac{1}{2},\nicefrac{1}{2},0} &\AtP& \nicefrac{1}{3} \\
	\CVec{\nicefrac{1}{4},\nicefrac{1}{4},\nicefrac{1}{4},\nicefrac{1}{4}} &\AtP& \nicefrac{2}{3} ~,
	\end{array}
	\end{equation}
	where in the $\nicefrac{1}{3}$-case he is better off finally than initially (since he knows \XS\ cannot be 00 or 11), but in the other case he is worse off (since \XS=00 has become possible). Thus if the attacker's choice is either to guess \XS's initial value or to run the program and guess \XS's final value, he can use these hypers to help make up his mind.
	\footnote{Compare a thief's two alternatives for stealing a credit card: \emph{``Steal it now, since the wallet is just sitting there.''}\ and \emph{``Steal it after the card is used at an ATM where he can see some digit of the PIN, but there is a risk his victim will notice and choose a new PIN.''}}
	The Shannon entropy of \XS\ is initially $\lg(3){\sim}1.6$, but finally it is conditionally $\nicefrac{1}{3}{\times}1+\nicefrac{2}{3}{\times}2=\nicefrac{2}{3}{>}1.6$: if the attacker is using Shannon entropy to make his decision, he should act sooner rather than later. On the other hand, the one-guess probability of \XS\ is initially $\nicefrac{1}{3}$; and finally it is the same, at $\nicefrac{1}{3}{\times}\nicefrac{1}{2}+\nicefrac{2}{3}{\times}\nicefrac{1}{4}=\nicefrac{1}{3}$. If the attacker is using this criterion, it does not matter when he acts. In either case, the hypers \Eqn{e0651a} and \Eqn{e0651b} contain all the information necessary for his decision: the bit-values printed are themselves not important \emph{for his decision}, which is why we can quotient our semantics by abstracting from them. (He does however need those values when he \emph{makes} his attack, if he decides ``later''.)
	
	\NS
	\subsection{Examples}*\label{s1418}
	\Cf{Compare \Sec{s1418n}.}
	We examine an example \HMM-step. Set the state-space $\CalX{=}\{0,1,2\}$ and observation-space $\CalY{=}\{0,1\}$; let channel $C$ reveal the parity of $x$ with probability $\nicefrac{2}{3}$ and the opposite of that parity with probability \nicefrac{1}{3}; and let markov $M$ choose final $x'$ uniformly so that $x{\leq}x'{\leq}2$, i.e.\ it cannot decrease its input. Assume the input distribution $\pi$ is uniform. The matrices $C,M$ and vector $\pi$ are therefore
	
	{\small%1423
		\begin{equation}\label{e1425}\small
		C = \left(
		\begin{array}{cc}
		\nicefrac{2}{3} & \nicefrac{1}{3} \\
		\nicefrac{1}{3} & \nicefrac{2}{3} \\
		\nicefrac{2}{3} & \nicefrac{1}{3}
		\end{array}
		\right)
		\hspace{0.5em}
		M = \left(
		\begin{array}{ccc}
		\nicefrac{1}{3} & \nicefrac{1}{3} & \nicefrac{1}{3} \\
		0               & \nicefrac{1}{2} & \nicefrac{1}{2} \\
		0               & 0               & 1
		\end{array}
		\right)
		\hspace{0.5em}
		\pi = \left(
		\begin{array}{c}
		\nicefrac{1}{3} \\
		\nicefrac{1}{3} \\
		\nicefrac{1}{3}
		\end{array}
		\right)
		\end{equation}
	}%1423
	
	We look first at the abstract channel $\AbCh{C}$ applied to $\pi$. From Defs.~\ref{d1124B},\ref{d1416CB} the joint distribution matrix $\ChApp{\pi}{C}$ is 
	\begin{equation}\label{e0837}
	\ChApp{\pi}{C} \Wide{=} 
	\left(
	\begin{array}{cc}
	\nicefrac{2}{9} & \nicefrac{1}{9} \\
	\nicefrac{1}{9} & \nicefrac{2}{9} \\
	\nicefrac{2}{9} & \nicefrac{1}{9}
	\end{array}
	\right)~,
	\end{equation}
	giving the $y$-marginal as the column sums $\RVec{\nicefrac{5}{9},\nicefrac{4}{9}}$ with corresponding $\CalX$-posteriors as normalised columns $\CVec{\nicefrac{2}{5},\nicefrac{1}{5},\nicefrac{2}{5}}$ and $\CVec{\nicefrac{1}{4},\nicefrac{1}{2},\nicefrac{1}{4}}$. Thus the abstract channel $\AbCh{C}$ takes incoming (uniform) prior $\CVec{\nicefrac{1}{3},\nicefrac{1}{3},\nicefrac{1}{3}}$ to the two-point-supported hyper that assigns probability $\nicefrac{5}{9}$ to the posterior $\CVec{\nicefrac{2}{5},\nicefrac{1}{5},\nicefrac{2}{5}}$ and $\nicefrac{4}{9}$ to the posterior $\CVec{\nicefrac{1}{4},\nicefrac{1}{2},\nicefrac{1}{4}}$. In tabular form we write that hyper, a discrete distribution (on distributions), as
	\begin{equation}\label{e0852}
	\begin{array}{r@{\hspace{3em}}l@{\,}l}
	\textrm{``inner'' distributions} & \multicolumn{2}{@{}l}{\textrm{``outer'' distribution}} \\
	\CVec{\nicefrac{2}{5},\nicefrac{1}{5},\nicefrac{2}{5}} &\AtP& \nicefrac{5}{9} \\
	\CVec{\nicefrac{1}{4},\nicefrac{1}{2},\nicefrac{1}{4}} &\AtP& \nicefrac{4}{9}~,
	\end{array}
	\end{equation}
	where in general $z_1\AtP p_1,~z_2 \AtP p_2,\cdots$ is the discrete distribution that assigns probability $p_1$ to $z_1$ etc. In \Eqn{e0852} the values $z_1,z_2,\cdots$ are themselves (inner, posterior) distributions.
	
	\label{g1130}We now include the markov $M$, so that we find $\HMMsem{C}{M}.\pi$ in the case where both $C$ \emph{and} $M$ have operated. Recalling \Def{d1728} we can calculate in general
	\begin{Reason}
		\Step{}{
			J_{x',y}
		}
		\Step{$=$}{
			\sum_x\pi_xH_{x,y,x'}
		}
		\StepR{$=$}{Special case that $H$\\is a single step $\CM{C}{M}$}{
			\sum_x\pi_xC_{x,y}M_{x,x'}
		}
		\Step{$=$}{
			\sum_x(\ChApp{\pi}{C})_{x,y}M_{x,x'}
		}
		\Step{$=$}{
			\sum_x(\ChApp{\pi}{C})^T_{y,x}M_{x,x'}
		}
		\Step{$=$}{
			((\ChApp{\pi}{C})^T{\MMult}M)_{y,x'} ~,
		}
		\Step{$=$}{
			((\ChApp{\pi}{C})^T{\MMult}M)^T_{x',y} ~,
		}
		\Step{$=$}{
			(M^T{\MMult}(\ChApp{\pi}{C}))_{x',y} ~,
		}
	\end{Reason}
	\label{g1012}where we have written $(\cdot)^T$ for matrix transpose. From \Eqn{e0837} that gives for this example that $\HMMsem{C}{M}.\pi$ is
	\[
	\left(    
	\begin{array}{ccc}
	\nicefrac{1}{3} & 0               & 0 \\
	\nicefrac{1}{3} & \nicefrac{1}{2} & 0 \\
	\nicefrac{1}{3} & \nicefrac{1}{2} & 1
	\end{array}
	\right)
	\MMult
	\left(
	\begin{array}{cc}
	\nicefrac{2}{9} & \nicefrac{1}{9} \\
	\nicefrac{1}{9} & \nicefrac{2}{9} \\
	\nicefrac{2}{9} & \nicefrac{1}{9}
	\end{array}
	\right)
	=
	\left(
	\begin{array}{cc}
	\nicefrac{2}{27} & \nicefrac{1}{27} \\
	\nicefrac{7}{54} & \nicefrac{4}{27} \\
	\nicefrac{19}{54} & \nicefrac{7}{27}
	\end{array}
	\right)
	\]
	which, written as at \Eqn{e0852}, is
	\begin{equation}\label{e0852A}
	\begin{array}{l@{\hspace{3em}}l@{\,}l}
	\CVec{\nicefrac{2}{15},\nicefrac{7}{30},\nicefrac{19}{30}} &\AtP& \nicefrac{5}{9} \\
	\CVec{\nicefrac{1}{12},\nicefrac{1}{3},\nicefrac{7}{12}} &\AtP& \nicefrac{4}{9}~,
	\end{array}
	\end{equation}
	\label{g1132}Note that the ``outer'', that is $y$-marginal probabilities are again $\RVec{\nicefrac{5}{9},\nicefrac{4}{9}}$, i.e.\ the same as at \Eqn{e0852} and unaffected by $M$. \label{g1133}The new ``inner'', that is output-posterior distributions are obtained by left-multiplying the inners for $\ChApp{\pi}{C}$ by $M^T$.
	\footnote{Doing it that way, i.e.\ on the inners of $\AbCh{\ChApp{\pi}{C}}$ after normalisation, is easier arithmetic; but the formula $M^T{\MMult}(\ChApp{\pi}{C})$ is easier to state.}
	
	What \Eqn{e0852A} tell us is that our \emph{a-posteriori} belief of the distribution on $\CalX$ of the \emph{output} of $\CM{C}{M}$, applied to a uniformly distributed input, will with probability $\nicefrac{5}{9}$ be $\CVec{\nicefrac{2}{15},\nicefrac{7}{30},\nicefrac{19}{30}}$ and with probability $\nicefrac{4}{9}$ be $\CVec{\nicefrac{1}{12},\nicefrac{1}{3},\nicefrac{7}{12}}$.
	
	Our ``a priori'' estimate of the output distribution is calculated by applying the markov to the (known) prior distribution on the input: that is, it is ``prior'' in the sense that it is our before-execution belief of what the final distribution will be. Our after-execution estimate of the final distribution is determined by the observations emitted by the \HMM's channel component, which we know only once the \HMM\ has run: therefore it is called ``a posteriori.''
	\footnote{This is consistent with the use of these terms for channels, because for a pure channel the final- and initial distributions on the state are the same: we use the channel's emissions (output) to revise our belief about both.}
	
	\NS
	\Cf{%1035
		\label{n1031} An arbitrary $H{\In}\MH$ can be decomposed into $H_1;H_2$ with $H_1{=}\CM{}{M_1}$ and $H_2{=}\CM{C_2}{M_2}$ where
		\begin{itemize}
			\item $M_1{\In}\MH$ takes initial $\CalX$ into an expanded state-space $\CalY{\times}\CalX$; with the default channel, thus $H_1$ releases no information,
			\item $C_2{\In}\CalY{\times}\CalX{\MFun}\CalY$ releases the $\CalY$-component and
			\item $M_2{\In}\CalY{\times}\CalX{\MFun}\CalX$ contracts the state-space back to just $\CalX$, projecting onto the $\CalX$-component and discarding the $\CalY$ component.
		\end{itemize}
	}%1035
	
	\NS
	\Cfe[../Carroll/HMMsteps1.tex]{
		This \Lem{l1719A} is a crucial result, that $\Wp{()}$ is well defined, and we have previously worked on it for single \HMM\ steps, i.e.\ that for any channel $C$ and Markov $M$ we have that $\Wp{\CM{C}{M}}.u$ is concave and continuous if $u$ itself is. This we can do (right?)}
	\Cfe[../Carroll/AKMstuff.tex]{There was some material here (from Annabelle I think) about gain-functions distinguishing hypers. We need to put it somewhere, but I'm not sure it belongs here. So for now I've put it in an endnote. Can we Sargasso it? {\Ax Yes.}}
	
	\NS
	Once we figure out precisely what the conditions on $\UT$ are in \Thm{t1005}, we will give that subset of $\UT$ a name, like $\TS$. The overall idea is that
	\begin{itemize}
		\item ``Syntactic'' \HMM's are matrices in $\MH$ for some $\CalY$. But possibly not \emph{all} matrices in that set can be expressed as the sequential composition of primitive \HMM's like $\CM{C}{M}$ for $C{{\In}\CalX{\MFun}\CalY}$ and $M{{\In}\CalX{\MFun}\CalX}$.\par\emph{We don't know at the moment how to characterise that subset of matrices}; and we don't have time to find out, for this paper. \textbf{Actually, we do: see \Fn{n1031}.}
		
		\item Denotational, that is $\HMMone{\cdot}$-images of elements of $\MH$, whether ``real'' \HMM's or not, are in $\AHSpc$. But not all elements of $\AHSpc$ are such images, although we know that the ones that are do satisfy \Lem{l1432} and \Lem{l1433}.
		
		\item The subset of $\AHSpc$ satisfying \Lem{l1432} and \Lem{l1433} we call $\AHS$. That is $\HMMone{\MH}{\subseteq}\AHS$, so to speak. They are our \emph{abstract \HMM's}. (What we \emph{don't} know is whether in fact $\HMMone{\MH}{=}\AHS$.)
		
		\item In similar fashion, \Lem{l1719A} gives $\Wp{(\AHS)}{\subseteq}\UT$. But not all transformers in $\UT$ are necessarily in $\Wp{(\AHS)}$. \Thm{t1005} is supposed to characterise $\Wp{(\AHS)}$ within $\UT$, and that subset of $\UT$ will be called $\TS$. The theorem would then effectively be \emph{Every $t{\In}\TS$ is $\Wp{h}$ for some $h{\In}\AHS$}, i.e.\ that $\TS{\subseteq}\Wp{(\AHS)}$.
		
		\item To finish it off, we'd need also to show that \emph{Every $\Wp{h}$ for $h{\In}\AHS$ is in $\TS$}, that is $\Wp{(\AHS){\subseteq}\TS}$ whence in fact $\Wp{(\AHS){=}\TS}$. We don't have that at the moment: or do we?
		See \Fn{n1130}.
		
		\item The Dijkstra-analogy, on which this is based, is
		\begin{itemize}
			\item Every strict relation in $S_\bot{\Fun}S_\bot^\uparrow$ is taken by $\Wp{()}$ to a strict and conjunctive transformer, and
			\item Every strict and conjunctive transformer is the $\Wp{()}$-image of some strict relation in $S_\bot{\Fun}S_\bot^\uparrow$.
		\end{itemize}
	\end{itemize}
	
	\NS
	\Cfe[../Carroll/HP.tex]{Some thoughts on connecting all this to \cite{Morgan:96d}.}
	
	\NS
	\Tf{\label{n1841}It's needed because the linearity of $t:\Cont\CalX{\Fun}\Cont\CalX$ is not enough to write $t = \Wp h$ for some $h$. To use the Riesz representation theorem, we need the continuity of $t$. Moreover $\Integral{f}{\Delta}$ is necessarily continuous in $f$.
		\Ax\ Does linearity of $t$ imply that it is continuous (using compactness and boundedness and things like that...)? \Cx\ Yes, I think it should. Think about simplifying \Lem{l0846B} accordingly. Furthermore, doesn't it follow immediately the continuity of $\Expt{-}$?
		\Tx I don't think linearity implies continuity in general (unless we use some weird maths without the axiom of choice). In fact, as soon as our vector space has infinitely many linearly independent family of vectors, then there always are discontinuous linear maps (see http://en.wikipedia.org/wiki/Discontinuous\_linear\_map)\CBar {\Ax Good. Let's keep continuity.}
		\Tx This comment can be archived/removed?\CBar
	}
	
	\NS
	\Cfe[../Carroll/HMMsteps.tex]{So we've shown that $\CM{C}{};\CM{}{M}=\CM{C}{M}$. But what about the other order, i.e.\ as $\CM{}{M};\CM{C}{}$?}
	
	\NS
	\Cfe[../Carroll/NoMCexample.txt]{Annabelle and I did come calculations to do with reducing \HMM's to a general form $\HMMone{M};\HMMone{C}$. The conclusion seems to be that you can't in general.}
	
	\NS
	\Cfe[../Carroll/Uniform.tex]{Here is some information about the uniform metric.}
	
	\NS
	{%1124
		\subsection*{Properties preserved by $\Wp h$}
		This section shows that some properties of $h{\In}\AHS$ are straightforwardly" translated into properties of $\Wp h$.
		
		\textbf{Preservation of Lipschitzness}
		
		Let $\Lip_p\CalX$, for $p{\In}\Real$, be the set of $p$-Lipschitz functions from $\Dist\CalX$ to $\Real$. The set of all Lipschitz functions is written $\Lip\CalX$.
		\begin{Definition}{Preservation of $\Lip\CalX$}{d1429}
			A (extended) transformer $t:\Cont \CalX\to \Cont \CalX$ \textit{strongly preserves} $\Lip \CalX$ if for every $p\in\NNReal$, there exists $q\in\NNReal$ such that $t.\Lip_p\CalX\subseteq \Lip_q \CalX$.
		\end{Definition}
		
		It follows from \Def{d1429} that $t$ strongly preserves $\Lip \CalX$ iff $t.\Lip_1\CalX\subseteq\Lip_p\CalX$ for some $p\in\NNReal$. The forward implication is clear. As for the backward one, let $p\in\NNReal$ such that $t.\Lip_1\CalX\subseteq\Lip_p\CalX$ and $f\in \Lip_q\CalX$, for some $q\in\NNReal$. Then $\frac{1}{q}f\in \Lip_1\CalX$ and since $t$ is linear, $t.f = q\ \!t.(\frac{1}{p}f)\in \Lip_{pq}\CalX$.
		
		\begin{Lemma}{Characterisation of Lipschitzness}{t1443}
			An abstract HMM $h{\In}\AHS$ is $p$-Lipschitz, for some $p$, iff $\Wp h$ strongly preserves $\Lip \CalX$.
			\Proof
			($\Rightarrow$) Let $h$ be $p$-Lipschitz and $f\in \Lip_1\CalX$.
			Let $\delta_{1,2}\in\Dist \CalX$, we have
			\[
			|\Wp h.f.\delta_1 - \Wp h.f.\delta_2| \Wide{\leq} \KMet{h.\delta_1}{h.\delta_2} \leq p\KMet{\delta_1}{\delta_2}
			\]
			That is, $\Wp h.\Lip_1\CalX\subseteq \Lip_p\CalX$.
			
			($\Leftarrow$) Assume that $\Wp h$ strongly preserves $\Lip \CalX$ such that $\Wp h.\Lip_1\CalX\subseteq \Lip_p\CalX$, for some $p{\In}\NNReal$. We have 
			\begin{align*}
				\KMet{h.\delta}{h.\delta'} & = \sup_{f\in\Lip_1 \CalX}\left|\int_{h.\delta}f - \int_{h.\delta'}f\right|\\
				&\leq \sup_{f\in\Lip_1 \CalX}\left|\Wp h.f.\delta - \Wp h.f.\delta'\right|\\
				&\leq p\KMet{\delta}{\delta'} 
			\end{align*}
			That is, $h$ is $p$-Lipschitz.
		\end{Lemma}
		
		\begin{Theorem}{Lipschitz Abstract HMM}{l1255}
			Let $t:\Cont \CalX{\Fun}\Cont \CalX$ be an extended transformer that strongly preserves $\Lip \CalX$. There exists a $p$-Lipschitz abstract HMM $h$ such that $t = \Wp h$.
			
			\Proof
			Follows from \Thm{t1005} and \Thm{t1443}.
		\end{Theorem}
		
		\textbf{Preservation of Concavity}
		\begin{Definition}{Concave}{d1448}
			A function $h{\In}\Dist\CalX{\Fun}\Dist^2\CalX$ is \textit{concave} if 
			\[
			h.\pi_1 \WS{p} h.\pi_2 \Wide{\Ref} h.(\pi_1 \WS{p} \pi_2)~.
			\]
			where $\Ref$ is the refinement order on $\Dist\CalX{\Fun}\Dist^2\CalX$.
		\end{Definition}
		
		\begin{Theorem}{Characterisation of concavity}{t1528}
			A denotation $h{\In}\Dist\CalX\to\Dist^2\CalX$ is concave iff $\Wp h$ preserves concave functions.
			\Proof
			The forward implication is given by \Lem{l1719A}.
			
			Conversely, let $\pi_1,\pi_2{\In}\Dist\CalX$ and $\Wp P$ be preserving concave tests.
			\begin{Reason}
				\Step{}{
					h.\pi_1 \WS{p} h.\pi_2 \Wide{\Ref} h.(\pi_1 \WS{p} \pi_2)~
				}
				\StepR{$\Leftarrow$}{Definition of $\Ref$,}{}
				\Step{}{
					\forall u{\In}\Unc\CalX\ \left[\Wp h.u.\pi_1\WS{p}\Wp h.u.\pi_2\right]\leq \Wp h.u.(\pi_1 \WS{p} \pi_2)
				}
				\StepR{$\Leftarrow$}{$\Wp h$ preserves concave tests.}{}
				\Step{}{
					True
				}
			\end{Reason}
			This proves that $h$ is indeed concave.
		\end{Theorem}
	}%1124
	
	\NS
	\section{Equivalence between the two presentations}\label{s1505}
	
	{%2140
		\Cf{%1015
			This is Tahiry' Riesz theorem, originally from \Sec{s1408}. Will it become a technical lemma used in the proof of \Thm{t1005}?
			\begin{quote}
				\begin{Theorem}{Riesz representation for transformers}{t1253}
					Let $t{\In}\UT$ be additive and continuous on $\Unc\CalX$. Then there exists a continuous function $h{\In}\Dist\CalX \rightarrow \Dist^2 \CalX$, such
					that $t = \Wp{h}$.
				\end{Theorem}
			\end{quote}
			\Tx This is a duplication of \Thm{t1005} which is proven directly, i.e., the proof starts with a transformer (type $\Unc\CalX{\Fun}\Unc\CalX$), extends it to get a type $\Cont\CalX{\Fun}\Cont\CalX$, then use previous proofs.
		}%1015
		
		\Cf{%1019
			Does the $h$  satisfy \Lem{l1433} as well? \Ax At the moment, possibly not. I think there must be additional conditions on $t$ (like concave preserving) for that. \Cx However it works out, this theorem should state that the $h$ you get satisfies \Lem{l1432} and \Lem{l1433}.
			\Tx Yes it does, see \Sec{sec:tahiry} \Thm{t1528}.
		}%1019
		
		The Giry construction and uncertainty transformers of the previous sections are linked closely by the $\Wp$ correspondence of \Def{d1701A}.  In this section, we consider a more general set of transformers: we only require them to map $\Cont\CalX$ (the set of continuous functions from $\Dist\CalX$ to $\Real$) into itself. That is, for every $P{\In}\Dist\CalX\to\Dist^2\CalX$, $\Wp P{\In}\Dist\CalX\to\Dist^2\CalX$. The results of this section can be trivially applied to \Def{d1701A}. 
		
		{\Xx%0826
			We are mainly interested in finding what properties of transformers are essential in characterising the image of $\Wp$. Firstly, the map $\Wp P$ is \emph{linear}, i.e. 
			\[
			\Wp P.(af + bg) \Wide{=} a\Wp P.f + b\Wp P.g
			\] 
			where $a,b$ are real numbers and $f,g{\In}\Cont\CalX$. 
			\Tf{In previous sections, Carroll used $u$ for uncertainty measures (continuous+concave). Shall I use the same notation for continuous functions instead of the $f,g$'s? \Cx I would stick with $f,g$ if they are not necessarily concave.}
			Secondly, it is \emph{positive}: for every test $f\geq 0$, $\Wp P.f\geq 0$ holds. Thirdly, we show in \Thm{t1053} that it is \emph{well-defined}, i.e., $\Wp P.f$ is indeed a continuous function from $\Dist\CalX$ to $\Real$. Fourthly, if we endow the space $\Cont\CalX$ with the uniform metric 
			\begin{equation}\label{e1025}
			\UMet{f}{g} \Wide{=} \sup_{x\in\CalX}|f.x-g.x|~, % Space before punctuation here. (CCM)
			\end{equation}
			then $\Wp P$ is also \emph{continuous} from $\Cont\CalX$ to $\Cont\CalX$. Lastly, $\Wp p$ is \emph{total}, i.e., $\Wp P.\FunctOne = \FunctOne$ ($\FunctOne$ is the constant function that evaluates to $1$ everywhere on $\Dist\CalX$). 
		}%0826
		
		These five properties provide the characterisation we seek. Given a continuous linear transformer $t:\Cont\CalX{\Fun}\Cont\CalX$ such that $t.\FunctOne = \FunctOne$, we show the existence of a continuous function $P:\AHSpc$ such that $\Wp P = t$ (\Thm{t1102}).
		
		\subsection*{$\Wp$ is well-behaved}
		Let us firstly show that $\Wp h$ satisfies these properties.
		{\Xx%0854
			\begin{Theorem}{$\Wp h$ is linear continuous}{t1053}
				Let $h{\In}\AHSpc$ be a continuous function. Then $\Wp h$ is well-defined, linear, continuous and total.
				\Proof
				It is clear that $\Wp h$ is linear, positive and total. The remaining properties are proved in \Lem{l1527A} and \Lem{l1530}.
			\end{Theorem}
			\begin{Lemma}{$\Wp h$ is continuous}{l1530X}
				Let $h{\In}\AHSpc$. Then $\Wp h{\In}\Cont\CalX{\Fun}\Cont\CalX$ is continuous with respect to the uniform topology.
				\Cf{\label{n1130}You were explaining (at lunch) why we need this; but I'm not sure I understood. Could you say why? 
					\Tx It's needed because the linearity of $t:\Cont\CalX{\Fun}\Cont\CalX$ is not enough to write $t = \Wp h$ for some $h$. To use the Riesz representation theorem, we need the continuity of $t$. Moreover $\Integral{f}{\Delta}$ is necessarily continuous in $f$.
					\Ax Does linearity of $t$ imply that it is continuous (using compactness and boundedness and things like that...)?
					\Tx I don't think linearity imply continuity in general (unless we use some weird maths without the axiom of choice). In fact, as soon as our vector space has infinitely many linearly independent family of vectors, then there always are discontinuous linear maps (see http://en.wikipedia.org/wiki/Discontinuous\_linear\_map)
				}
				\Proof
				Let $f,g{\In}\Cont\CalX$; then we have % Let..., then... is ungrammatical (in English). CCM
				\begin{Reason}
					\Step{}{
						\UMet{\Wp h.f}{\Wp h.g}
					}
					\StepR{$=$}{Definition of $\UNorm{\cdot}$}{
						\sup_{\delta\in\Dist\CalX} \left|\Integral{f}{h.\delta} - \Integral{g}{h.\delta}\right|
					}
					\StepR{$\leq$}{$\left|\Integral{(f-g)}{h.\Delta}\right|\leq\Integral{|f-g|}{h.\Delta}$}{
						\sup_{\delta\in\Dist\CalX} \Integral{|f-g|}{h.\delta}
					}
					\StepR{$\leq$}{$|f-g|\leq\UMet{f}{g}$}{
						\sup_{\delta\in\Dist\CalX} \Integral{\UMet{f}{g}}{h.\delta} 
					}
					\StepR{$=$}{$\Integral{\FunctOne}{h.\delta} = 1$}{
						\UMet{f}{g} ~.
					}
					
				\end{Reason}
				Hence $\Wp h$ is $1$-Lipschitz, thus continuous on $\Cont\CalX$. 
			\end{Lemma}
		}%0854
		
		\subsection*{Linear and continuous transformers yield $\AHSpc$}
		Given a continuous function $h{\In}\AHSpc$, the transformer $\Wp h$ transforms tests linearly and continuously. The converse of \Thm{t1053} also holds, that is, a linear and continuous transformer $t$ is the $\Wp$ of some $h$ (\Thm{t1102}). 
		
		{\Xx%0940
			Let $t$ be a transformer and $\pi{\In}\Dist\CalX$. The \emph{$\pi$-projection} 
			\Tf{Fiber may be more precise!}
			of $t$ maps $f{\In}\Dist\CalX$ to $t.f.\pi\in\Real$ linearly. Projections are linear forms (real valued) and we will rely on the fact that integrations are in one-to-one correspondence with continuous linear forms~\Cite{Riesz}.

			\begin{Lemma}{Every $\pi$-projection is continuous}{l1102}
				If $t{\In}\Cont \CalX{\Fun} \Cont \CalX$ is a continuous linear map, then for each $\pi{\In}\Dist \CalX$, $f\mapsto t.f.\pi$ is a continuous linear form on $\Cont \CalX$. 
				\Proof
				Let $f,g{\In}\Cont \CalX$, we have
				\begin{Reason}
					\Step{}{
						|t.f.\pi- t.g.\pi| &\leq \UMet{t.f}{t.g}
					}
					\StepR{$\leq$}{Definition of $\UNorm{\ }$}{
						\UMet{t.f}{t.g}
					}
					\StepR{$=$}{$t$ is linear}{
						\UNorm{t.(f - g)}
					}
					\StepR{$\leq$}{$t$ is continuous}{
						\sup_{\UNorm{u}=1}{|t.u|}\UMet{f}{g}
					}
				\end{Reason}
				The quantity $\sup_{\UNorm{u}=1}{|t.u|}$ is finite because $t$ is linear continuous~\Cite{Functional Analysis Ref.}. Hence, the $\pi$-projection is Lipschitz and thus continuous.
			\end{Lemma}
		}%0940
		
		{\Xx%1004
			We establish a technical lemma that ensures the existence of a $1$-Lipschitz function which maximises the total variation in the definition of the Kantorovich metric on $\Dist\CalX^2$. It is also used to deduce uniform convergence from weak convergence in the proof of~\Thm{t1053}.
			
			A family $\CalA$ of tests is said to be \emph{uniformly bounded} if there exists a real number $M$ such that, for every $f{\In} \CalA$ and $\pi{\In}\Dist\CalX$, $|f.\pi|\leq M$. The family $\CalA$ is \emph{equicontinuous at $x{\In} X$}  if for each $\varepsilon >0$ there exists $\eta>0$ such that for every $f\in \CalA$ and $x'{\In} X$, we have $|f(x) - f(x')|\leq \varepsilon$. The family $\CalA$ is \emph{equicontinuous} if it is equicontinuous everywhere. Intuitively, all functions in an equicontinuous family have the same ``degree of continuity."
			
			\begin{Lemma}{Technical lemma}{l1224}
				Let $\pi{\In}\Dist \CalX$. The set 
				$$\Lip_1^\pi \CalX = \{f{\In}\Lip_1\CalX\ |\ f.\pi = 0\}$$
				is equicontinuous and uniformly bounded.
				
				\Proof
				Let $f{\In}\Lip_1^\pi \CalX$. For every $\delta{\In}\Dist \CalX$, 
				$$|f.\delta| = |f.\delta - f.\pi|\leq |\delta - \pi|\leq 1,$$
				that is, $\Lip_1^\alpha\CalX$ is uniformly bounded.
				
				The equicontinuity follows from Lipschitzness. Let $\delta{\In}\Dist \CalX$ and $\varepsilon>0$. For every $f{\In}\Lip_1^\pi\CalX$ and $\delta'{\In}\Dist \CalX$ such that $|\delta-\delta'|<\eta = \varepsilon$, we have $|f.\delta-f.\delta'|\leq |\delta-\delta'|<\varepsilon.$
			\end{Lemma}
			
			Since the set $\Lip_1^\pi\CalX$ is closed (\wrt $\UNorm{\ }$), the Arzela-Ascoli Theorem~\Cite{Arzela-Ascoli} implies that $\Lip_1^\pi$ is compact.
		}%1004
		
		\begin{Theorem}{Characterisation of $\Wp$}{t1102} Let $t{\In}\Cont \CalX{\Fun} \Cont \CalX$ be linear, continuous, positive and total. There exists a \emph{unique} $h{\In}\Dist \CalX{\Fun} \Dist^2 \CalX$, continuous, such that $t = \Wp h$.
			\Proof 
			{\Xx%1102
				Let $\pi{\In}\Dist\CalX$. By \Lem{l1102}, the $\pi$-projection of $t$ is a continuous linear form on $\Cont \CalX$. Therefore, there exists a unique regular Borel measure $\Delta_\pi{\In}\Dist^2\CalX$~\footnote{Every finite Borel measure on a metric space is regular~\Cite{Parthasarathy} Theorem \T{number}.}  such that $t.f.\pi = \int_{\Delta_\pi} f$, for every $f{\In}\Cont\CalX$~\Cite{Riesz}. Define $h$ to be the function that associates $\Delta_\pi$ to each $\pi{\in}\Dist\CalX$.
				
				$h.\pi$ is a probability measure for every $\pi$. In fact, it is a positive measure because the $t$ is positive 
				\Tf{Given an open set $O$, $h.\pi.O = \sup\{t.f.O\ |\ 0\leq f\leq 1\wedge f\in\Dist\CalX\wedge\mathrm{supp}(f)\subseteq O\}$. Does this need to be stated explicitly, to convince the reader that $h.\pi$ is indeed positive.}
				and total because
				\[
				\Integral{\FunctOne}{h.\pi} \Wide{=} t.\FunctOne.\pi \Wide{=} \FunctOne.\pi \Wide{=} 1~.
				\]
				
				It remains to show that $h$ is continuous with respect to the Kantorovich metric on $\Dist\CalX$ and $\Dist^2\CalX$. Let $\pi_n$ be a sequence of distributions converging to $\pi{\In}\Dist\CalX$. For every fixed $f{\In}\Cont \CalX$, $\Integral{f}{h.\pi_n} = t.f.\pi_n$ converges to $t.f.\pi = \int_{h.\pi}f$ because $t.f$ is continuous. That is, $h.\pi_n$ converges weakly to $h.\pi$. 
				\Tf{The definition of weak convergence could be found in~\Cite{Parthasarathy}.}
				Let us show that that convergence is uniform in $f$ on the set of $1$-Lipschitz functions using Ranga's result~\Cite{Ranga}. 
				
				Let $\delta:\Dist\CalX$. We define $\Lip_1^\delta\CalX$ to be the set of $1$-Lipschitz functions from $\Dist\CalX$ to $\Real$ that evaluates to $0$ at $\delta$. We have 
				\begin{equation}\label{z_e1230}
				\KMet{h.\pi_n}{h.\pi} = \sup_{f\in\Lip_1^\delta\CalX}\left|\Integral{f}{h.\pi_n}-\Integral{f}{h.\pi}\right|	
				\end{equation}
				because $\KMet{}{}$ is invariant by translation. By \Lem{l1224}, $\Lip_1^\delta \CalX$ is equicontinuous and uniformly bounded. Since the metric space $\Dist\CalX$ is separable (it is compact), we can apply Ranga's result (\Cite{Ranga} Theorem 3.1) which says that the \rhs of \Eqn{e1230} is very close to $0$ for $n$ sufficiently large. Therefore, $h$ is sequentially continuous, thus continuous.
			}%1102
		\end{Theorem}
	}%2140

	\NS
	{\Xx %1651
		The Uniform Limit Theorem
		A sequence of functions $F_n{\In}\Dist^2\CalX\to \Real$ converges uniformly to $F{\In}\Dist^2\CalX\to\Real$ iff
		
		\[
		\lim_{n\to\infty}\sup_{\Delta{\In}\Dist^2\CalX}\left|F_n.\Delta-F.\Delta\right| \Wide{=} 0
		\]
		
		The convergence is called uniform because the rate of convergence does not depend on $\Delta{\In}\Dist^2\CalX$.
		\begin{Theorem}{Uniform Limit \Cite{Real Analysis Ref.}}{t2143}
			If $F_n$ converges uniformly to $F$ and all $F_n$ are continuous then the limit $F$ is continuous.
		\end{Theorem}
	} %1651
	
	\NS
	\section{Draft proof of \Thm{t1005}}\label{s0920}
	\begin{ReTheorem}{Characterisation of transformers}{t1005}
		Let $t{\In}\UT$ satisfy the following properties:
		\Cf{Move the definition of these properties to \Sec{s0935}.}
		\Cf{\label{n1553}I will try writing this wrt.\ $\Unc\CalX$ rather than $\Cont\CalX$, to see what happens. \textbf{I do understand however} (see \Fn{n1208}) that (later) in order to use Riesz we might have to frame this within $\Cont\CalX$ instead of (the smaller) $\Unc\CalX$. In that case our condition on $u$ in \Sec{s0935} will have to change. Nevertheless the \emph{statement} of \Thm{t1005} should (in my opinion) not refer to $\Cont\CalX$.
			\Tx Indeed, at the moment, my proof is based on $\Cont\CalX$. Notice however that $\Unc\CalX$ generates a vector subspace of $\Cont\CalX$. Therefore, by the Hahn-Banach Extension Theorem (http://en.wikipedia.org/wiki/Hahn-Banach\_theorem), $t{\In}\Unc\CalX\to\Unc\CalX$ extends into a (positive?) continuous linear map $t'{\In}\Cont\CalX\to\Cont\CalX$. This extension is not unique though (unless $\Unc\CalX$ is dense in $\Cont\CalX$, which I think isn't the case). So, the constructed $h$ will depend on which extension we choose. All $h$'s will however give the same result when applied against a given uncertainty measure. \TBar
		}
		\begin{enumerate}
			\item It is \emph{linear}, so that for $a_{1,2}{\In}\NNReal$ and $u_{1,2}{\In}\Unc\CalX$ we have
			\[
			t.(a_1u_1+a_2u_2) \Wide{=} a_1t.u_1 + a_2t.u_2~.
			\]
			\item It is \emph{positive}, so that $t.u.\delta{\geq}0$ for every $u{\In}\Unc\CalX$ and $\delta{\In}\Dist\CalX$.
			\Cf{I'm using $\pi$ for priors and $\rho$ for posteriors and $\delta$ for distributions that could be either. State this convention in the intro to the whole paper.}
			\item It is \emph{continuous} wrt.\ the uniform metric $\UMet{\cdot}{\cdot}$, defined
			\[
			\UMet{u_1}{u_2} \Wide{\Defs} \T{\sup_{\delta\in\Dist\CalX}|u_1.\delta-u_2.\delta|} ~.
			\]
			\Cf{I will leave this for now: but this uniform metric is topologically equivalent to the Kantorovich, on this space. Could we use Kanto istead? 
				\Tx I think the sup is on $\Dist\CalX$, not $\CalX$.
			}
			\item It is \emph{total}, so that $t.\FunctOne{=}\FunctOne$ where $\FunctOne.\delta\Defs1$ for all $\delta{\In}\Dist\CalX$.
		\end{enumerate}
		
		Then there is a function $h{\In}\AHS$ such that $t = \Wp{h}$.
		\Proof
		We give the proof further below, after some technical lemmas.
	\end{ReTheorem}
	
	{%0940
		Let $t{\In}\UT$ be a transformer and take $\delta{\In}\Dist\CalX$. Then the \emph{$\delta$-projection} 
		\Tf{Fiber might be more precise!}
		of $t$ maps $u{\In}\Unc\CalX$
		\Cf{Originally $f{\In}\Dist\CalX$ here. Should it have been $f{\In}\Cont\CalX$? Anyway, I have put it into $\Unc\CalX$ to see what happens. 
			\Tx Yes, $f$ should have been in $\Cont\CalX$.
		}
		to $t.u.\delta\in\Real$ linearly. Projections are linear forms (real valued) and we will rely on the fact that integrations are in one-to-one correspondence with continuous linear forms~\Cite{Riesz}.
		
		\begin{Lemma}{Every $\delta$-projection is continuous}{l1102}
			If $t{\In}\UT$ is continuous and linear, then for each $\delta{\In}\Dist\CalX$ we have that $u\mapsto t.u.\delta$ is a continuous linear form on $\Unc\CalX$. 
			\Proof
			Take $u_{1,2}{\In}\Unc\CalX$; then we have
			\begin{Reason}
				\Step{}{
					|t.u_1.\delta{-}t.u_2.\delta|
				}
				\StepR{$\leq$}{Definition of $\UNorm{\ }$}{
					\UMet{t.u_1}{t.u_2}
				}
				\StepR{$=$}{$t$ is linear}{
					\UNorm{t.(u_1{-}u_2)}
				}
				\StepR{$\leq$}{$t$ is continuous}{
					\sup_{u\,{\rm st.}\,\UNorm{u}=1}|t.u|\times\UMet{u_1}{u_2}
				}
			\end{Reason}
			The quantity $	\sup_{u\,{\rm st.}\,\UNorm{u}=1}|t.u|$ is finite because $t$ is linear and continuous~\Cite{Functional Analysis Ref., maybe Rudin91}. Hence the $\delta$-projection is Lipschitz and thus continuous.
		\end{Lemma}
	}%0940
	
	{%1005
		Next we establish a technical lemma that ensures the existence of a $1$-Lipschitz function which maximises the total variation in the definition of the Kantorovich metric on $\Dist^2\CalX$. It is also used to deduce uniform convergence from weak convergence in the proof of~\Thm{t1053}.
		
		\C{Let a \emph{test} be a function in $\Dist\CalX{\Fun}\Real$.}
		\Cf{Added this. Could they be uncertainty measures?
			\Tx Not really (unless we use Kostas' metric and define the Kantorovich metric to be the sup of total variations on concave, 1-Lipschitz functions), the family here is a translated set of 1-Lipschitz functions used in the definition of the Kantorovich metric.
		}
		A family $\CalA$ of tests is said to be \emph{uniformly bounded} if there exists a real number $M$ such that for every $u{\In}\CalA$ and $\delta{\In}\Dist\CalX$ we have $|u.\delta|\leq M$. The family $\CalA$ is said to be \emph{equicontinuous at $\delta{\In}\Dist\CalX$} if for each $\varepsilon{>}0$ there exists $\eta{>}0$ such that for every $u{\In}\CalA$ and $\delta'{\In}\Dist\CalX$ \C{with $|\delta{-}\delta'|{<}\eta$} we have $|u.\delta{-}u.\delta'|{\leq}\varepsilon$.
		The family $\CalA$ is (simply) \emph{equicontinuous} if it is equicontinuous at every $\delta$. Intuitively, all functions in an equicontinuous family have the same ``degree of continuity".
		
		\begin{Lemma}{Technical lemma}{l1224}
			For any $\delta{\In}\Dist\CalX$ the \C{family of tests}
			\[
			\Lip_1^\delta \CalX \Wide{\Defs} \{\T{u{\In}\Unc\CalX}\ |\ u.\delta = 0\textrm{ and } u \textrm{ 1-Lipschitz}\}
			\]
			is equicontinuous and uniformly bounded.
			\Proof
			Take arbitrary $u{\In}\Lip_1^\delta \CalX$. For every $\delta'{\In}\Dist \CalX$ we have
			\[
			|u.\delta'|~=~|u.\delta'-0|~=~|u.\delta'{-}u.\delta|~\leq ~|\delta' {- }\delta|~\leq~1,~
			\]
			that is, $\Lip_1^\delta\CalX$ is uniformly bounded.
			
			The equicontinuity also follows from Lipschitzness. Take $\delta{\In}\Dist \CalX$ and $\varepsilon{>}0$. For every $u{\In}\Lip_1^\delta\CalX$ and $\delta'{\In}\Dist \CalX$ such that $|\delta-\delta'|<\eta = \varepsilon$, we have $|u.\delta-u.\delta'|\leq |\delta-\delta'|<\varepsilon.$
		\end{Lemma}
	}%1005
	
	We can now give the proof of \Thm{t1005}:
	\Cf{Refer to \Lem{l0903}.}
	\begin{ReTheorem}{Characterisation of transformers}{t1005}
		Under the assumptions above, there exists a \T{\emph{unique}}
		\Tf{This uniqueness may be a problem if $t$ is of type $\Unc\CalX\to\Unc\CalX$.}
		$h{\In}\AHSpc$ satisfying the conditions of \Sec{s0935} such that $t{=}\Wp h$.
		\Proof
		Take $\delta{\In}\Dist\CalX$. By \Lem{l1102}, the $\delta$-projection of $t$ is a continuous linear form on $\Unc\CalX$. Therefore, there exists a unique regular Borel measure $\Delta_\delta{\In}\Dist^2\CalX$
		\footnote{Every finite Borel measure on a metric space is regular~\Cite{Parthasarathy} Theorem \T{number}.}
		such that $t.u.\delta = \Expt{\Delta_\delta}{u}$, for every $u{\In}\Unc\CalX$~\Cite{Riesz}.
		\Cf{\label{n1208}This use of Riesz is presumably where we need to be using $\Cont\CalX$ rather than $\Unc\CalX$?
			\Tx Yes, but there is a work-around. See \Fn{n1553}.
		}
		Define $h.\delta\Defs\Delta_\delta$ for each $\delta{\In}\Dist\CalX$. \C{We now check that $h$ has the required properties.}
		
		First, we have that $h.\delta$ is a probability measure for every $\delta$. In fact it is a positive measure, because the $t$ is positive;
		\Tf{Given an open set $O$, $h.\pi.O = \sup\{t.f.O\ |\ 0\leq f\leq 1\wedge f\in\Dist\CalX\wedge\mathrm{supp}(f)\subseteq O\}$. Does this need to be stated explicitly, to convince the reader that $h.\pi$ is indeed positive.}
		and it is total because
		\[
		\Expt{h.\delta}{\FunctOne} \Wide{=} t.\FunctOne.\pi \Wide{=} \FunctOne.\pi \Wide{=} 1~.
		\]
		
		It remains to show
		that $h$ has the properties demanded by Lemmas \ref{l1432},\ref{l1433}. For \Lem{l1432}, we let $\delta_n$ be a sequence of distributions in $\Dist\CalX$ converging to $\delta{\In}\Dist\CalX$. \T{It suffices to show that the limit of  $\KMet{h.\delta_n}{h.\delta}$ is $0$.}
		
		For every fixed $u{\In}\Unc \CalX$ we have that $\Expt{h.\delta_n}{f} = t.u.\delta_n$ converges to $t.u.\delta = \Expt{f}{h.\delta}$ because $t.u$ is continuous. That is, $h.\delta_n$ converges weakly to $h.\delta$. 
		Let us show that that convergence is uniform in $u$ on the set of $1$-Lipschitz functions using Ranga's result~\Cite{Ranga}. 
		
		Take $\delta'{\In}\Dist\CalX$, and \T{recall that $\Lip_1^{\delta'}\CalX$ is the set of $1$-Lipschitz functions from $\Dist\CalX$ to $\Real$ that evaluate to $0$ at $\delta'$}.
		\Cf{We did something similar above: is this a repetition?
			\Tx It was a redefinition. Maybe that extra sentence can be removed as the definition of $\Lip_1^{\delta'}\CalX$ will be very close to this proof.
		}
		We have 
		\begin{equation}\label{e1230}
		\KMet{h.\delta_n}{h.\delta}
		\Wide{=}
		\sup_{u\in\Lip_1^{\delta'}\CalX}\left|\Expt{h.\delta_n}{u}-\Expt{h.\delta}{u}\right|	
		\end{equation}
		because $\KMet{}{}$ is invariant by translation. By \Lem{l1224} we know that $\Lip_1^{\delta'} \CalX$ is equicontinuous and uniformly bounded. Since the metric space $\Dist\CalX$ is separable (it is compact), we can apply Ranga's result (\Cite{Ranga} Theorem 3.1) which says that the \rhs of \Eqn{e1230} is very close to $0$ for $n$ sufficiently large. Therefore, thus continuous.
		\Cf{What's \emph{sequentially} continuous?
			\Tx removed it and added a sentence stating that all we need is the convergence of $h.\delta_n$ to $h.\delta$ \wrt the Kantorovich metric. Preservation of limits of sequences is sufficient for continuity in metric spaces.
		}
		
		\Cf{This proof taken from Tahiry's \Thm{t1528}.} For \Lem{l1433}
		{%1232
			suppose that $t{=}\Wp{h}$ is in $\UT$ and take arbitrary $\delta_{1,2}{\In}\Dist\CalX$ Then we reason
			\begin{Reason}
				\Step{}{
					h.\delta_1 \WS{p} h.\delta_2 ~\Ref~ h.(\delta_1 \WS{p} \delta_2)
				}
				\StepR{if}{for all $u{\In}\Unc\CalX$\\ \Lem{l0755}}{
					\Expt{(h.\delta_1 \WS{p} h.\delta_2)}{u} ~\leq~ \Expt{h.(\delta_1 \WS{p} \delta_2)} u
				}
				\StepR{if}{Defn.\ \Wp{()}}{
					& \Wp{h}.u.\delta_1\WS{p}\Wp{h}.u.\delta_2 \\\leq & \Wp{h}.u.(\delta_1 \WS{p} \delta_2)
				}
				\StepR{if}{Defn.~$\Unc\CalX$}{
					\Wp{h}.u \in \Unc\CalX ~,
				}
			\end{Reason}
			which was our assumption.
		}%1232
	\end{ReTheorem}
	
	\NS
	\section{\Ax Transformers etc.\protect{\Af{Being re-worked in \Sec{s0917}.}}}\label{s1244}
	
	\begin{itemize}
		\item Definition of the HMMs as transformers by defining Wp
		
		\item Observe that we can apply the Riesz representation theorem to show that there is an underlying hyper distribution in the case that
		the functional is linear.
		
		\item An example to show that linearity is not sufficient to characterise Channels (and therefore HMMs as well);
		
		\item Introduce the "multiplicative" property
		
		\item Show that this plus linearity characterises channels
		
	\end{itemize}
	
	\begin{Definition}{Disorder tests}{d1709}
		Continuous concave functions
	\end{Definition}
	
	\begin{Definition}{Transformers}{d1701}
		Let $P : \Dist \CalX \rightarrow \Dist^2\CalX$ be a function. Let $t$ be a real-valued function 
		$\Dist\CalX \rightarrow \Real$. Define $\Wp{P}.t.\delta\in \Real$ as follows:
		
		\[
		\Wp{P}.t.\delta \Defs \int_{P.\delta} t~.
		\]
		We write $ \int_{P.\delta} t$ as $\Expt{{P.\delta}}{t}$.
	\end{Definition}
	
	\Af{Need to connect this Wp definition to Giry composition and list the properties we get "for free".}
	
	\begin{Lemma}{Disorders preserved}{l1719}
		If $P$ is a channel and $t\in Disroders$ then $\Wp{P}.t \in Disorders$.
		Furthermore $\Wp{P}.t \leq t$.
		
		\Proof
		For each $\delta_1, \delta_2$ and $0\leq p\leq 1$, there is some channel $B$ such that: 
		\[
		[\delta_1 \PC{p} \delta_2, B]  = [\delta_1] \PC{p} [\delta_2]~,
		\]
		and therefore that $\Wp{B}.g.(\delta_1 \PC{p} \delta_2) = g.\delta_1 \PC{p} g.\delta_2$, for any function $g: \Dist\CalX \rightarrow \Real$.
		\Af{Fact in the appendix.}
		
		We can now prove that $\Wp{C}.t$ is concave.
		\Cf{Should be writing $\delta_1\WS{p}\delta_2$, not $\delta_1\PC{p}\delta_2$.}
		\begin{Reason}
			\Step{}
			{\Wp{C}.t .(\delta_1 \PC{p} \delta_2)}
			
			\StepR{$\leq$} { $B; C \Ref C $} 
			{\Wp{(B; C)}.t .(\delta_1 \PC{p} \delta_2)}
			
			\StepR{$=$} {Giry composition}
			{\Wp{B}(\Wp{C}t).(\delta_1\PC{p}\delta_2)}
			
			\StepR{$=$}{Definition of B}
			{\Wp{C}.t.\delta_1  \PC{p}  \Wp{C}.t.\delta_2}
		\end{Reason}
		As required.
		
	\end{Lemma}
	
	\begin{Lemma}{}{l1628}
		If $P \Ref \Skip$ then $\Avg P.\delta = \delta$. \Af{We need to define refinement\ldots\ \Cx More generally if $P_1\Ref P_2$ then $\Avg.(P_1.\delta)=\Avg.(P_2.\delta)$ for all $\delta$. That will of course be $\delta$ itself if $P_{1,2}$ are channels; but if $P$ isn't a channel we can have $\Avg.(P.\delta){\neq}\delta$.}
	\end{Lemma}
	
	\subsection*{Gain functions}
	
	An interesting class of Disorder tests was introduced elsewhere \cite{Alvim:2012aa}.
	
	\begin{Definition}{Gain functions}{d1321}
		A \emph{gain function} is a Disorder test $V_g{\In}\Dist \CalX \rightarrow \Real$, where $g$ is a matrix $\CalY \times \CalX$ of nonnegative values. Given $\delta\in \Dist\CalX$, define $V_g[\delta]$ as follows:
		\[
		V_g[\delta] \Wide{\Defs} \min_{y\in \CalY} (g.y)\cdot \delta~,
		\]  
		where $f \cdot w \Defs \sum_{x\in\CalX}f.x \times w.x$~.\Af{These are actually co-gain functions.}
	\end{Definition}
	
	It was shown elsewhere \cite{McIver:2014aa} that any hypers $\Delta_{1,2}$ are either equal, or can be distinguished by a gain function, i.e. there is some
	gain function $V_g$ such that $\Expt{{\Delta_1}}{V_g} \neq \Expt{{\Delta_2}}{V_g}$.  This allows us to use gain functions to distinguish between functions using \Def{d1701}.
	
	\begin{Lemma}{Gain functions distinguish}{l1341}
		Given functions $P_{1,2}{\In}\Dist \CalX \rightarrow \Dist^2\CalX$, we have that $P_1 = P_2$ if and only if $\Wp{P_1}.V_g = \Wp{P_2}.V_g$ for all gain functions $V_g$.
	\end{Lemma}
	
	Gain functions satisfy the following.
	
	\begin{Lemma}{Gain function arithmetic}{l1347}
		Let $V_g$ be a gain function and $\alpha$ a function $\CalX \rightarrow\Real$. \Af{Might have to be non-negative reals.}
		Define $V_{g\star \alpha}$ as the gain function determined by $(g\star\alpha).y.x \Defs 
		g.y.x \times \alpha.x$. Then for any $\alpha_{1,2} \in \Dist\CalX$ we have $V_{g\star{\alpha_1} }[\alpha_2] =  V_{g\star{\alpha_2} }[\alpha_1]$.
		\Proof
		
		Follows immediately from \Def{d1321}.
	\end{Lemma}
	
	A similar property to \Lem{l1347} holds for channels as well.
	
	\begin{Lemma}{Channel properties}{l1402}
		Let $C$ be a channel. The following properties hold.
		
		Let $V_g$ a gain function and $\delta_{1,2} \in \Dist\CalX$. Then
		\[
		\Wp{C}.V_{g\star{\delta_1}}.\delta_2 \Wide{=} \Wp{C}.V_{g\star{\delta_2}}.\delta_1~.
		\]

		Observe that $\Wp{C}.V_g.\delta \leq V_g.\delta$ follows from \Cite{LiCS12}.
		
		Write the columns of $C$ as $C[z]$, where $z$ ranges over some finite index set $\CalZ$. Then
		\[
		\Wp{C}.V_g.\delta \Wide{=} \sum_{z\in \CalZ} V_{g\star{C[z]}}.\delta~.
		\]
		The result now follows easily since $\alpha\star\beta = \beta\star\alpha$, and \Lem{l1347}.
	\end{Lemma}
	
	It turns out that this important property is what distinguishes channels amongst all functions $\Dist\CalX \rightarrow \Dist^2\CalX$.
	
	\begin{Definition}{Multiplicative}{d1425}
		Let $P{\In} \Dist\CalX \rightarrow \Dist^2\CalX$. We say that $P$ is \emph{multiplicative} provided that
		\[
		\Wp{P}.V_{g\star{\delta_1}}.\delta_2 \Wide{=} \Wp{P}.V_{g\star{\delta_2}}.\delta_1~,
		\]
		for all $\delta_{1, 2} \in \Dist\CalX$ and gain functions $V_g$.
		\Cf{I \emph{think} $g$ is the gain function, and $V_g$ is the vulnerability (for us, uncertainty) derived from it. To avoid confusion, we should probably write $U_g$ or maybe $\underline{V}_g$.}
	\end{Definition}
	
	We are almost ready to prove that multiplicative functions correspond to channels; we
	will do so by showing that if $P$ is multiplicative then $P.\delta$ is determined by $P.u$, where
	$u$ is the uniform distribution. To show further that there is a channel $C_P$ such that $P.\delta = C_P.\delta$
	for all $\delta$ we show how hypers and channels are in 1-1 correspondence.
	
	\begin{Definition}{hyper to Channel}{d1539}
		Let $\Delta \in \Dist^2\CalX$ have finite support $\CalY \subseteq \Dist\CalX$. Further let $\Avg.\Delta.x > 0$ for all $x \in \CalX$. We define a channel $C_\Delta \in \CalX \times \CalY \rightarrow \Real$
		\Cf{See however \Note{n1016}.}
		\[
		C_\Delta.x.y \Wide{\Defs} (\Delta.y\times y.x)/\Avg.\Delta.x~.
		\]
	\end{Definition}
	
	\Def{d1539} is well-defined, since each row sums to $1$:
	\[
	\sum_{y\in\CalY} C_\Delta.x.y = \sum_{y\in\CalY} (\Delta.y\times y.x)/\Avg.\Delta.x = \Avg.\Delta.x/\Avg.\Delta.x ~.
	\]
	
	We have the following properties of $C_\Delta$.
	
	\begin{Lemma}{}{l1626}
		\begin{enumerate}
			\item $C_\Delta.(\Avg.\Delta) = \Delta~.$
		\end{enumerate}
	\end{Lemma}
	
	We can now characterise channels as multiplicative functions which are anti-refinements of $\Skip$.
	
	\begin{Theorem}{Multiplicative determines channels}{t1556}
		Let $P{\In} \Dist\CalX \rightarrow \Dist^2\CalX$. If $P$ is multiplicative and $P \Ref \Skip$
		then there exists a channel $C$ such that $P.\delta = C.\delta$ for all $\delta\in \Dist\CalX$.
		\Proof
		
		It is sufficient to show that $P= C_{P.u}$, where $u$ is the uniform distribution. We reason as follows for gain function $V_g$ and $\delta\in \Dist\CalX$.
		\begin{Reason}
			\Step{}
			{\Wp{P}.V_{g\star u}.\delta}
			\StepR{$=$}{$P$ multiplicative; \Def{d1425}}
			{\Wp{P}.V_{g\star \delta}.u}
			\StepR{$=$}{\Lem{l1626} $C_{P.u}.u = P.u$; \Lem{l1628}}
			{\Wp{C_{P.u}}.V_{g \star \delta}.u}
			\StepR{$=$}{$C_{P.u}$ is a channel; \Lem{l1402}}
			{\Wp{C_{P.u}}V_{g \star u}.\delta~.}
		\end{Reason}
		Since hypers are determined by gain functions,
		\Cf{Note that $g \star u$ ranges over all gain functions if $g$ itself does.}
		the result now follows.
	\end{Theorem}
	
	Note that multiplicative \C{is necessary: it} does not follow from $P\Ref \Skip$, as the following example shows:
	
	For example, consider the map
	\begin{equation}\label{e1956}
	\begin{array}{ccc}
	\left(
	\begin{array}{c}
	p\\
	(1{-}p)\\
	\end{array}\right)&
	\mapsto& 
	\left(
	\begin{array}{cc}
	p^2 & p(1{-}p)\\
	0 & (1{-}p)\\
	\end{array}\right)
	\end{array}
	\end{equation}
	The matrix on the right-hand side has entries summing to $1$. It represents a hyper distribution with support size $2$ as follows: each column can be normalised to recover the priors in the support, and each such prior occurs with probability the sum of the entries in its column. For instance in \Eqn{e1956} the priors in the support are $(1, 0)^T$ occurring with probability $p^2$, and $(p/(1{+}p),   1/(1{+}p))^T$ occurring with probability $1{-}p^2$. Viewed in this light, \Eqn{e1956} is a deterministic program, as it is of the correct type, and can be shown to be continuous and refined by $\Skip$. 
	\Af{Continuity follows by direct calculation, because for any $0 \leq p, q, k \leq 1$ we have that $p^2 \PC{k} q^2 \geq (p \PC{k} q)^2$.} 
	\Af{Is it the case that multiplicative implies $P \Ref \Skip$ ?}
	\Af{It is not a channel because there is no constant matrix which delivers the correct output hypers.}

	%%%%%%%%%%%%%Annabelle's new Green Room section about HMMs AGR824
	\section{\Ax Characterising \HMM's\protect{\Af{Being re-worked in \Sec{s0917}.}}}\label{s0921}
	
	HMMs are traditionally described as a process with two  kinds of ``output". The first is related to information flow concerning the initial state,
	and the second is a Markov update of the state. The overall combination of the initial information flow followed by the Markov update suggests a model of HMM as a function of type $\Dist \CalX \rightarrow \Dist^2 \CalX$, however this loses the initial information flow 
	which can be important when considering compositional properties%
	
	For example using a naive model for HMMs as a channel (Giry-) composed with a Markov update does not preserve the information about the initial state. (This is an important aspect of HMMs.) For example:
	\[
	{\textit reveal}(h); h:=0 ~~=~~ h:=0
	\]
	in such a model, so that the fact that the initial reveal of the initial value of $h$ is not preserved (in this naive model), renders a non-compositional semantics in contexts where there is a non-trivial initial correlation with a (fresh) hidden variable $h'$.
	\Af{Shall we go into this?}

	\subsection*{Modelling correlations in channels and HMMs}
	
	Let $C$ be a channel of type $\Dist\CalX \rightarrow \Dist^2\CalX$. We show how $C$ can in fact be regarded as a channel 
	$\Dist(\CalX{\times}\CalY) \rightarrow \Dist^2(\CalX{\times}\CalY)$ over the set of joint distributions $\Dist(\CalX{\times}\CalY)$.
	
	\begin{Definition}{Dalenius lifting}{d0813}
		Let $C$ be a channel of type $\Dist\CalX \rightarrow \Dist^2\CalX$ and $\CalY$ a finite type (state space). Define $\LEntan{C}{\CalY}: \Dist(\CalX{\times}\CalY) \rightarrow \Dist^2(\CalX{\times}\CalY)$ via the transformer semantics:
		
		\[
		\Wp\LEntan{C}{\CalY}.V_g.\delta~~\Defs~~ \sum_i V_{g\star C[i]}.\delta~,
		\]
		where $g\star C[i](w, x, y) \Defs g(w,x,y)\times C[i](x)$~, and $\delta \in \Dist(\CalX{\times}\CalY)$.
	\end{Definition}
	
	The Dalenius lifting uses the correlation between $\CalX$ and $\CalY$ together with the information about $\CalX$ released by the channel $C$ in order to make deductions about the information released by $\CalY$.  \Def{d0813} clearly defines a channel by theorem~\ref{t1556}; more concretely we can compute the actual channel matrix in the following example. Consider the channel:
	\[
	C \Wide{=} 
	\begin{array}{c}
	x_0\\
	x_1
	\end{array}
	\left( \begin{array}{cc}
	1/2 & 1/2\\
	1/3 & 2/3
	\end{array}\right)
	\]
	
	If we take $\CalY \Defs \{y_0, y_1\}$, then the channel $\LEntan{C}{\CalY}$ becomes:
	
	\[
	\LEntan{C}{\CalY} \Wide{=} 
	\begin{array}{c}
	(x_0, y_0)\\
	(x_0, y_1)\\
	(x_1, y_0)\\
	(x_1, y_1)\\
	\end{array}
	\left( \begin{array}{cc}
	1/2 & 1/2\\
	1/2 & 1/2\\
	1/3 & 2/3\\
	1/3 & 2/3
	\end{array}\right)
	\]

	In fact if we
	only look at the information flow with respect to $\CalY$, then we can describe the information released as a channel
	which depends on $C$ and the correlation with $\CalX$.
	
	Let $\alpha : \CalY \rightarrow \Dist \CalX$ and $\pi \in \Dist \CalY$. We write $\LEntan{\pi}{\alpha}$ for the joint distribution in $\Dist (\CalX \times \CalY)$ defined:
	\[
	\LEntan{\pi}{\alpha}(x, y) \Defs \pi.y\times\alpha.y.x~.
	\]
	In fact this makes $\pi$ the  marginal distribution on $\CalY$ relative to $\LEntan{\pi}{\alpha}$. Moreover, any joint distribution on $\CalX \times \CalY$ can be decomposed uniquely in this way.
	Using these conventions we can define a channel which leaks information about $\CalY$.
	
	\begin{Definition}{Dalenius channel}{d0841}
		Let $C:\Dist\CalX \rightarrow \Dist^2\CalX$ be a channel on $\CalX$ and $\alpha : \CalY \rightarrow \Dist \CalX$ define a correlation map taking $\pi{\In} \Dist\CalY$ to $\LEntan{\pi}{\alpha} \in \Dist(\CalX{\times}\CalY)$.  We define a channel $C_\alpha:\Dist\CalY \rightarrow \Dist^2\CalY$ by
		\[
		\Wp C_\alpha.V_g.\pi \Wide{\Defs} \Wp(\LEntan{C}{\CalY}).V_{\REntan{g}{\CalX}}.(\LEntan{\pi}{\alpha})~,
		\]
		where $V_g$ is a gain function on $\CalY$ and ${\REntan{g}{\CalX}}(w,x, y) \Defs g(w, y)$.
	\end{Definition}
	
	Observe that \Def{d0841} does indeed define a channel on $\CalY$; this follows from \Def{d0813} which implies that $\Wp C_\alpha.V_{g\star \pi_1}.\pi_2 = \Wp C_\alpha.V_{g\star\pi_2}.\pi_1$, and thus by  theorem \ref{t1556} that $C_\alpha$ is a channel.
	
	Applying these definitions to our example above and using $\alpha(y_0) \Defs (1/4, 3/4)$ and  $\alpha(y_1) \Defs (1, 0)$, we see that $C_\alpha$ is produced by taking the $\alpha(y_0)$-weighted average of the rows $(x_0, y_0)$ and $(x_1, y_0)$ and the $\alpha(y_1)$-weighted average of the rows $(x_0, y_1)$ and $(x_1, y_1)$ to produce:
	
	\[
	C_\alpha \Wide{=} 
	\begin{array}{c}
	y_0\\
	y_1
	\end{array}
	\left( \begin{array}{cc}
	3/8 & 5/8\\
	1/2 & 1/2
	\end{array}\right)
	\]
	
	An important special case of this is where the correlation is exact: define $\iota$ to be the identity mapping, so that $\LEntan{C}{\CalX}$ preserves a copy of the information flow wrt the initial value. In this case the construction $C_\iota$ simply selects one each of the original rows of $C$, so that $C_\iota$ is equivalent (as an abstract channel) to the original $C$.
	
	Similarly we can promote a Markov matrix $M{\In}\Dist\CalX \rightarrow \Dist\CalX$ to $\LEntan{M}{\CalY}: \Dist (\CalX\times \CalY) \rightarrow \Dist (\CalX\times \CalY)$.
	
	\[
	\LEntan{M}{\CalY}(x,y).(x',y') \Wide{\Defs} M.x.x' ~\textit{iff}~ y=y'~.
	\]
	
	We can now define formally an HMM.
	
	\begin{Definition}{HMM}{d1330}
		Given a channel $C{\In}\Dist\CalX \rightarrow \Dist^2\CalX$ and a Markov matrix $M{\In}\Dist\CalX \rightarrow \Dist\CalX$, an HMM has type
		$\Dist\CalX^2 \rightarrow \Dist^2\CalX^2$
		is constructed as:
		\[
		\LEntan{C}{\CalX} ; [\LEntan{M}{\CalX}]~,
		\]
		where $[\LEntan{M}{\CalX}]$ is $\Pnt \circ \LEntan{M}{\CalX}$ and $\Pnt$ is the Giry Monadic function which makes the point distribution from its argument. \footnote{For clarity we often give the type of an HMM to be $\Dist(\CalX{\times}\CalY) \rightarrow \Dist^2(\CalX{\times}\CalY)$. }
	\end{Definition}
	
	Given $\delta{\In}\Dist(\CalX \times \CalY)$ we define the right and left marginals: $\LeftMar{\delta}{\In}\Dist\CalX$ is $\LeftMar{\delta}.x \Defs \sum_{y\in\CalY}\delta.x.y$, and $\RightMar{\delta}.y \Defs \sum_{x\in\CalX}\delta.x.y$. We have the following properties of HMMs.
	
	\begin{Lemma}{HMM properties}{l1608}
		Let $H: \Dist(\CalX {\times} \CalY) \rightarrow \Dist^2(\CalX {\times} \CalY) $ be an HMM.
		\Af{Here I find it somewhat confusing to have the type as $\Dist \CalX^2 \rightarrow \Dist^2 \CalX^2$ and so have used $\CalY$ for the second $\CalX$. What do you(s) think?}
		
		The following properties hold.
		
		\begin{enumerate}
			\item For fixed $\beta: \CalY \rightarrow \Dist\CalX$, and any $\delta \in \Dist Y$, the mapping $C_{\beta}.\delta \Defs  (\Dist\RightMar{\cdot})H.\REntan{\beta}{\delta}$ is a channel $\Dist \CalY \rightarrow\Dist^2 \CalY$. \Af{Need to have two functions: one for $\beta$ and one for $\alpha$...}
			
			\item For any $\pi{\In}\Dist\CalX$  and any $\alpha{\In} \CalX \rightarrow \Dist\CalY$, the mapping $M_{\CalX}.\pi \Defs   \Avg\circ (\Dist\RightMar{\cdot})H.\LEntan{\pi}{\alpha}$  of type $\Dist \CalX \rightarrow\Dist \CalX$ is a Markov process.
			
			\item When $\CalY$ is identified with $\CalX$,  and $H=  \LEntan{C'}{\CalX} ; [\LEntan{M'}{\CalX}]$  for some channel $C'$ and markov matrix $M'$, then $C_{\iota}$ as constructed in (1) above is equivalent to $C'$ and $M$ as constructed at (2) above is equivalent to $M'$, so that $H$ itself is equivalent to $\LEntan{C_{\iota}}{\CalX} ; [\LEntan{M}{\CalX}]$.
		\end{enumerate}
		\Proof
		(1) follows because
		\begin{Reason}
			\Step{}
			{\Wp C_\beta.\delta}
			\Step{$=$}
			{\int_{ (\Dist\RightMar{\cdot})H.\REntan{\beta}{\delta}} V_g}
			\Step{$=$}
			{\int_{ H.\REntan{\beta}{\delta}} V_{\REntan{g}{\CalX}}}
			\Step{$=$}
			{\Wp H.V_{\REntan{g}{\CalX}}. \REntan{\beta}{\delta}}
			\StepR{$=$}{$H= [\LEntan{C}{\CalY}];[\LEntan{M}{\CalY}]$}
			{\Wp([\LEntan{C}{\CalY}];[\LEntan{M}{\CalY}]).V_{\REntan{g}{\CalX}}.\REntan{\beta}{\delta}}
			\StepR{$=$}{$g$ does not mention $\CalX$}
			{\Wp [\LEntan{C}{\CalY}].V_{\REntan{g}{\CalX}}.\REntan{\beta}{\delta}~.}
		\end{Reason}
		The result follows from the comment after \Def{d0841}.
		
		(2) and (3) are now obvious, aren't they?
	\end{Lemma}
	The property \Lem{l1608}(1) says that for a fixed set of conditional distributions defined by $\beta$, the HMM $H$ acts as a channel on
	the right marginal distributions of $\REntan{\beta}{\delta}$. In particular, when we take $\beta$ to be $\iota$, such that $\REntan{\iota}{\delta}$ represents the distribution such that the correlation between $\CalX$ and $\CalY$ is exact, then the channel defined is exactly the information leaked about the initial value of $x$. Next, the property \Lem{l1608}(2) says that if we only consider the average change of state of the variable $x$,  without considering information flow,  then we can associate it with a fixed Markov Process. Finally \Lem{l1608}(3) says that $H$ is determined by  the information flow of the initial value of $x$, put together with the average change of state.
	
	\NS
	\subsection*{Metrics based on specialised sets of tests}\label{s2238}
	
	\begin{Definition}{Ascoli sets}{d1434}
		A family $\CalA$ of tests is said to be \emph{uniformly bounded} if there exists a real number $M$ such that, for every $f\in \CalA$ and $\pi\in \Dist\CalX$, $|f.\pi|\leq M$. The family $\CalA$ is \emph{equicontinuous at $x\in X$}  if for each $\varepsilon >0$ there exists $\eta>0$ such that for every $f\in \CalA$ and $x'\in X$, we have $|f(x) - f(x')|\leq \varepsilon$. The family $\CalA$ is \emph{equicontinuous} if it is equicontinuous everywhere. A family that is uniformly bounded and equicontinuous is called \emph{Ascoli}~\Cite{Ascoli}.
	\end{Definition}
	
	The first example of a family that is Ascoli is the singleton set of test $\{f\}$ (or any finite set of tests). Such a family is particularly useful if computing the leakage with respect to a particular uncertainty measure is required (e.g. $f$ abstracts a strategy).
	
	The set of $1$-Lipschitz tests that vanishes at a given distribution $\pi$ is another example of family that is Ascoli.
	
	\begin{Lemma}{$\Lip_1^\pi \CalX$ is Ascoli}{l1224}
		Let $\pi:\Dist \CalX$. The set 
		$$\Lip_1^\pi \CalX = \{f\in\Lip_1\CalX\ |\ f.\pi = 0\}$$
		is equicontinuous and uniformly bounded.
		
		\Proof
		Let $f\in\Lip_1^\pi \CalX$. For every $\delta:\Dist \CalX$, 
		$$|f.\delta| = |f.\delta - f.\pi|\leq |\delta - \pi|\leq 1,$$
		that is, $\Lip_1^\alpha\CalX$ is uniformly bounded.
		
		The equicontinuity follows from Lipschitzness. Let $\delta\in\Dist \CalX$ and $\varepsilon>0$. For every $f\in \Lip_1^\pi\CalX$ and $\delta':\Dist \CalX$ such that $|\delta-\delta'|<\eta = \varepsilon$, we have $|f.\delta-f.\delta'|\leq |\delta-\delta'|<\varepsilon.$
	\end{Lemma} 
	
	\NS
	\subsection*{Proof of \Thm{t1008} from \Sec{s1008}}\label{a1008}
	\begin{ReTheorem}{Composition faithfully denoted}{t1008}\\
		Let $H^{1,2}{\In}\MH$ be \HMM's. Then
		\[
		\HMMone{H^1;H^2} \Wide{=} \HMMone{H^1};\HMMone{H^2}~.
		\]
		\Proof
		\Cf{Check this proof for possible divisions by zero.}
		We use the notations of \Sec{s1008}. Write $p_1(x,y_1,x'')$ for the probability $H^1_{x,y_1,x''}$ that the joint distribution $H^1$ assigns to the event ``$x$ is input, and $y,x''$ are output''; write $p2(x'',y_2,x')$ similarly for $H^2$; and write $p(x,y_1,y_2,x')$ for the (syntactic) composition $H^1;H^2$ as a whole.
		
		We follow the usual notational conventions for these $p$'s wrt.\ marginal- and conditional distributions, e.g.\ that $p_1(y)$ is the marginal probability of $Y{=}y$ induced by $H_1$ and that $p_1(x|y_1,x'')$ is the conditional probability of $X{=}x$ given $Y_1{=}y_1\land X''{=}x''$. Further, by e.g.\ $p_1(Y)$ we mean the $y$-marginal distribution as a whole, and by $p_1(X|y_1)$ we mean the \emph{a-posteriori} distribution on the input induced by having observed some $y_1$ (and having marginalised by summing over all $x''$).
		
		Now let $H^i{=}(C^i,M^i)$ and start from some arbitrary $\pi$ to apply the Kleisli composition. The first step $\HMMone{H^1}$ produces hyper $p_1(X''|y_1)\AtP p_1(y_1)$ with $y_1$ varying over $\CalY$, where from \Def{d1022} we have $p_1(x,y_1,x'')=\pi_xC^1_{x,y_1}M^1_{x,x''}$. And from an arbitrary $\pi''$, the second step would produce $p_2(X'|y_2)\AtP p_2(y_2)$ with $y_2$ (also) varying over $\CalY$, and $p_2(x'',y_2,x')=\pi''_{x''}C^2_{x'',y_2}M^2_{x'',x'}$. To carry out the Kleisli composition, we must let the second-step prior $\pi''$ range over the inners $p_1(X''|y_1)$ from the first step.
		
		Thus letting $\pi''$ be some $p_1(X''|y_1)$ from the first step, we get from \Def{d1022} that
		\begin{Reason}
			\Step{}{
				{p_2}_{y_1}(x'',y_2,x')
			}
			\Step{$=$}{
				\pi''_{x''}C^2_{x'',y_2}M^2_{x'',x'}
			}
			\StepR{$=$}{set $\pi''\Defs p_1(X''|y_1)$}{
				p_1(X''|y_1)_{x''}C^2_{x'',y_2}M^2_{x'',x'}
			}
			\Step{$=$}{
				p_1(x''|y_1)C^2_{x'',y_2}M^2_{x'',x'} ~,
			}
		\end{Reason}
		where the $y_1$-subscript in the ${p_2}_{y_1}$, which we are defining on the left, captures its dependence on $p_2$. 
		
		With that, we have that Kleilsi composition $\HMMone{H^1};\HMMone{H^2}$ applied to $\pi$ is
		\begin{equation}\label{z_e1342}
		{p_2}_{y_1}(X'|y_2) \quad\AtP~p_1(y_1){p_2}_{y_1}(y_2) ~,
		\end{equation}
		with $y_{1,2}$ varying over $\CalY^2$, where by ``$y_{1,2}$ varying over $\CalY^2$'' we mean e.g.\ a distribution presented in the style of \Eqn{e0852} for all $(y_1,y_2)$ in $\CalY^2$.
		
		What we would like to know is whether this Kleisli-generated \Eqn{e1342} is the same hyper
		\begin{equation}\label{z_e1211}
		p(X'|y_1,y_2) \quad\AtP~ p(y_1,y_2)~,
		\end{equation}
		with $y_{1,2}$ varying over $\CalY^2$, that would result from \Def{d1022} applied to $\ChApp{\pi}{(H^1;H^2)}$ directly.
		Thus we now prove for all $y_{1,2}$ the two equalities
		\begin{eqnarray}
		\textrm{\small \underline{Kleisli-generated \Eqn{e1342}}} && \textrm{\small\underline{\Def{d1022} generated \Eqn{e1211}}} \nonumber\\
		p_1(y_1){p_2}_{y_1}(y_2) &=& p(y_1,y_2) \label{e1051a} \\
		{p_2}_{y_1}(X'|y_2) &=& p(X'|y_1,y_2) \label{e1051b}
		\end{eqnarray}
		that together will establish the equality of the two hypers. Beginning with \Eqn{e1051a} we calculate
		\begin{Reason}
			\Step{}{
				p_1(y_1){p_2}_{y_1}(y_2)
			}
			\Step{$=$}{
				p_1(y_1){\times}\sum_{x'}{p_2}_{y_1}(y_2,x')
			}
			\Step{$=$}{
				\sum_{x'}p_1(y_1){p_2}_{y_1}(y_2,x')
			}
			\StepR{$=$}{\Lem{l1625} below}{
				\sum_{x'}p(y_1,y_2,x')
			}
			\Step{$=$}{
				p(y_1,y_2)~.
			}
		\end{Reason}
		
		For \Eqn{e1051b} we prove that ${p_2}_{y_1}(x'|y_2) = p(x'|y_1,y_2)$ for arbitrary $x'$, calculating
		\begin{Reason}
			\Step{}{
				{p_2}_{y_1}(x'|y_2)
			}
			\Step{$=$}{
				{p_2}_{y_1}(y_2,x')/{p_2}_{y_1}(y_2)
			}
			\StepR{$=$}{\Lem{l1625}}{
				p(y_1,y_2,x')/p_1(y_1)~/{p_2}_{y_1}(y_2)
			}
			\StepR{$=$}{\Eqn{e1051a}}{
				p(y_1,y_2,x')/p(y_1,y_2)
			}
			\Step{$=$}{
				p(x'|y_1,y_2)~.
			}
		\end{Reason}
		And with those two equalities we have our result.
	\end{ReTheorem}
	
	The following technical lemma, used in the proof of \Thm{t1008}, simplifies the relationship between $p_1,p_2$ and $p$:
	\begin{Lemma}{Technical lemma}{l1625}
		Continuing with the notations of \Sec{s1008}, we calculate
		\[
		p_1(y_1){p_2}_{y_1}(y_2,x')
		\Wide{=}
		p(y_1,y_2,x') ~.
		\]
		\Proof
		\begin{Reason}
			\Step{}{
				p_1(y_1){p_2}_{y_1}(y_2,x')
			}
			\Step{$=$}{
				p_1(y_1)\sum_{x''}{p_2}_{y_1}(x'',y_2,x')
			}
			\StepR{$=$}{calculation above}{
				p_1(y_1)\times\\\quad\sum_{x''}p_1(x''|y_1)C^2_{x'',y_2}M^2_{x'',x'}
			}
			\Step{$=$}{
				\sum_{x''}p_1(y_1,x'')C^2_{x'',y_2}M^2_{x'',x'}
			}
			\Step{$=$}{
				\sum_{x,x''}p_1(x,y_1,x'')C^2_{x'',y_2}M^2_{x'',x'}
			}
			\Step{$=$}{
				\sum_{x,x''}\pi_xC^1_{x,y_1}M^1_{x,x''}C^2_{x'',y_2}M^2_{x'',x'}
			}
			\Step{$=$}{
				\sum_{x}\pi_x\sum_{x''}C^1_{x,y_1}M^1_{x,x''}C^2_{x'',y_2}M^2_{x'',x'}
			}
			\Step{$=$}{
				\sum_{x}p(x,y_1,y_2,x')
			}
			\Step{$=$}{
				p(y_1,y_2,x')~.
			}
		\end{Reason}
	\end{Lemma}
	
	\NS
	\begin{Lemma}{$\Wp$'s are well defined}{l1527}
		Let $P:\AHSpc$ and $f:\Cont\CalX$ be continuous functions. Then $\Wp P.f$ is continuous.
		\Proof % This is how the IEEE wants us to do it. (CCM)
		The proof relies on the fact that the function $F.\Delta=\Integral{f}{\Delta}$, which maps hypers to real numbers, is continuous under the Kantorovich metric on hypers. In fact, if that function is continuous then $\Wp P.f = F\circ P$ is continuous because $F$ and $P$ are continuous.
		
		Let $\Delta,\Delta':\Dist^2\CalX$ be two hypers.
		\Tf{where is the notation for hypers introduced?}
		We make first the stronger assumption that
		$f$ is a $k$-Lipschitz function, for some $k{>}0$. We have 
		\begin{Reason}
			\Step{}{ % Indent interiors of \Step etc. (CCM)
				\left|\Integral{f}{\Delta} - \Integral{f}{\Delta'}\right|
			}
			\StepR{$=$}{$\frac{1}{k}f$ is $1$-Lipschitz}{
				k\left|\Integral{\frac{1}{k}f}{\Delta} - \Integral{\frac{1}{k}f}{\Delta'}\right|
			}
			\StepR{$\leq$}{Definition of $\KMet{\cdot}{\cdot}$}{
				k\KMet{\Delta}{\Delta'} ~, % Finish calculations with punctuation, if appropriate. (CCM)
			}
		\end{Reason}
		\Cf{I guess $\KMet{\cdot}{\cdot}$ is the K-metric. We need to make it look prettier. But also we need to remember here that the spaces are 1-bounded, i.e.\ that the K-distance is never more than 1.}
		Hence the map $F$ is $k$-Lipschitz in this case, so continuous. 
		
		Now, let $f\in\Cont\CalX$ be an arbitrary continuous function, i.e.\ no longer necessarily Lipschitz. By Theorem~\Cite{Lipschitz Density Theorem}, there exists a sequence $f_n$ of Lipschitz functions such that
		\Cf{$\UMet{}{}$ defined somewhere? \Tx Equation \Eqn{e1025} \CBar}
		\[
		\lim_{n{\Fun}\infty}\UMet{f}{f_n}\Wide{=} 0~.
		\]
		The sequence $f_n$ generates a sequence $F_n:\Dist^2\CalX{\Fun}\Real$ defined by $F_n.\Delta = \Integral{f_n}{\Delta}$. Each function $F_n$ is continuous because $f_n$ is $k$-Lipschitz for some $k$ (depending on $n$). For every $m:\Natural$ and $\delta:\Dist S$, $|f_m.\delta| \leq \sup_n \UNorm{f_n}$. The \rhs is finite because: 
		
		\begin{itemize}
			\item[-] $\UNorm{f_n}$ converges to $\UNorm{f}$ (continuous image of a convergent sequence),
			\Cf{And $\UMet{}{}$ is continuous?
				\Tx this follows from a general property of metrics as we discussed previously. Given a metric $d$ and a fixed point $x_0$, we have 
				\[
				|d(x,x_0) - d(x',x_0)|\leq d(x,x')
				\]
				for all $x,x'$. So $f\mapsto \UMet{f}{0}$ is $1$-Lipschitz.\CBar
			}
			\item[-] and so there exists $N\in \Natural$ such that $|\UNorm{f_k} - \UNorm{f}|\leq 1$ for every $k\geq N$. That is, $\sup_n\UNorm{f_n} \leq \max(\max_{n\leq N} \UNorm{f_n},\UNorm{f}+1)$.
		\end{itemize}
		\Cf{Paused here for now.}
		
		Therefore, by the Dominated Convergence Theorem~\Cite{DCT}, $f$ is $\Delta$-integrable and the sequence $F_n$ converges to $F$ such that $F.\Delta = \int_{\Delta}f$, for every $\Delta\in\Dist^2S$ (this convergence is pointwise). Let us show that $f$ is necessarily continuous by showing that $F_n$ converges uniformly to $F$. 
		
		Let $\varepsilon>0$. For every $\Delta\in\Dist^2\CalX$; we have 
		\begin{Reason}
			\Step{}{
				|F_n.\Delta - F.\Delta|
			}
			\StepR{$=$}{Definition of $F$ and $F_n$}{
				\left|\Integral{f_n}{\Delta} - \Integral{f}{\Delta}\right|
			}
			\StepR{$\leq$}{$\left|\Integral{g}{\Delta}\right|\leq\Integral{|g|}{\Delta}$}{
				\Integral{\left|f_n - f\right|}{\Delta}
			}
			\StepR{$\leq$}{$|f_n-f|\leq\UMet{f_n}{f}$ and $\Integral{\FunctOne}{\Delta} = 1$}{
				\UMet{f_n}{f} ~.
			}
		\end{Reason}
		Since $\lim_{n{\Fun}\infty}\UMet{f_n}{f} = 0$, there exists $N\in\Natural$ (that depends on $\varepsilon$ only) such that $|F_n.\Delta-F.\Delta| < \varepsilon$ for every $n\geq N$. By the Uniform Limit Theorem~\Cite{ULT}, the uniform limit of a sequence of continuous functions is continuous, that is, $F$ is continuous.
	\end{Lemma}
	
	\NS
	\begin{Definition}{Denotation of \HMM}{d1022}
		\Cf{This definition shouldn't be in terms of $C,M$ --- just use $H$.}
		Let $H\Defs\CM{C}{M}$ be an \HMM\ with with $C{\In}\CalX{\MFun}\CalY$ and $M{\In}\CalX{\MFun}\CalX$. Its denotation, of type $\AHSpc$, is called an \emph{abstract \HMM} and is defined for $\pi{\In}\Dist\CalX$ by
		\[
		\HMMone{H}.\pi
		\Wide{=}
		\HMMsem{C}{M}.\pi
		\Wide{\Defs}
		\Hyp{\ChApp{\pi}{J}}~,
		\]
		where the joint distribution matrix $J{\In}\CalX{\MFun}\CalY$ is given by $J_{x',y}\,\Defs\sum_x H_{x,y,x'}=\sum_x\pi_xC_{x,y}M_{x,x'}$.
		\footnote{\Label{n1736}%1736
			Recalling $H_{x,y,x'}\Defs C_{x,y}M_{x,x'}$ we can calculate
			\begin{Reason}
				\Step{}{
					J_{x',y}
				}
				\Step{$=$}{
					\sum_x\pi_xH_{x,y,x'}
				}
				\Step{$=$}{
					\sum_x\pi_xC_{x,y}M_{x,x'}
				}
				\Step{$=$}{
					\sum_x(\ChApp{\pi}{C})_{x,y}M_{x,x'}
				}
				\Step{$=$}{
					\sum_x(\ChApp{\pi}{C})^T_{y,x}M_{x,x'}
				}
				\Step{$=$}{
					((\ChApp{\pi}{C})^T{\MMult}M)_{y,x'} ~,
				}
				\Step{$=$}{
					((\ChApp{\pi}{C})^T{\MMult}M)^T_{x',y} ~,
				}
			\end{Reason}
			whence we have the useful fact that $J=M^T{\MMult}(\ChApp{\pi}{C})$.
		}%1736
	\end{Definition}
	
	\NS
	\section{\Cx The structure of hyper-space}\label{s1432G}
	\Cf{This to go at \Sec{s1432} above.}
	
	The space $\Dist^2\CalX$, i.e.\ of hyper-distributions on $\CalX$, has a number of properties that have been explored previously \Cite{ICALP10,LiCS12,MSCS14}. The important property for us here is that $\Dist^2\CalX$ has a partial order $(\Ref)$ of ``uncertainty refinement''
	\footnote{It's called ``entropy refinement'' in \Cite{MSCS Def 5.1} and ``secure refinement'' in \Cite{LiCS12 VI-A} and \Cite{ICALP10 \S6.1}.}
	such that for two hypers, a specification and implementation $\Delta_{S,I}{\In}\Dist^2\CalX$, their being related $\Delta_S{\Ref}\Delta_I$ means that, in a sense we make precise below,
	\Cf{Probably in the section on uncertainties.}
	the implementation (-representing) $\Delta_I$ never releases more information that the specification $\Delta_S$ does.
	We repeat the definition of refinement here:
	\begin{Definition}{Uncertainty refinement}{d1449}
		Let $\Delta_{S,I}{\In}\Dist^2\CalX$ be two hypers on $\CalX$. We say that $\Delta_S$ is refined by $\Delta_I$ just when there is a distribution $\DELTA{\In}\Dist^3\CalX$, that is a distribution of \emph{hypers}, such that
		\[
		\Delta_S ~=~ \Avg.\DELTA \WideRM{and} (\Dist\Avg).\DELTA ~=~ \Delta_I~.
		\]
	\end{Definition}
	The advantage of this abstract formulation is that it is defined on hypers directly, and can readily be generalised to the continuous case, i.e.\ when we are working with proper measures rather than discrete distributions \Cite{LiCS12,MSCS}.
	But here we remain discrete, and in that case $(\Ref)$ can also be given a syntactic characterisation \Cite{McIver:2014aa,CSF12}:
	\Cf{Mention however how in those papers it's between \emph{channels}, not JD's, and that it goes the other way, i.e.\ is \emph{anti}- (composition) refinement.}
	\begin{Lemma}{Refinement of joint-distribution matrices}{l0827}
		Let $J_S{\In}\CalX'{\MFun}\CalY_S$ and $J_I{\In}\CalX'{\MFun}\CalY_I$ be joint-distribution matrices, both of them \emph{reduced} in the sense of \Def{d1426HB}, such that $\Hyp{J_{S,I}}{=}\Delta_{S,I}$ resp.
		\Cf{I've used $\CalX'$ here as a reminder that the \emph{input} side of the $J$'s is actually the \emph{output} side of the \HMM, i.e.\ that $J_{x',y}=\sum_x H_{x,y,x'}$ as in \Def{d1022}. Should we do that throughout? Or is it more confusing than helpful?}
		Then
		\[
		\Delta_S \Ref \Delta_I
		\WideRM{iff}
		J_S\MMult R = J_I
		\]
		for some stochastic \emph{refinement matrix} $R{\In}\CalY_S{\MFun}\CalY_I$ (i.e.\ such that $\sum R_{y,-}=1$ for each $y{\In}\CalY_S$).
		\Proof
		Given in \App{s0851}.
	\end{Lemma}
	
	With \Lem{l0827} the reflexivity and transitivity of relation $(\Ref)$ is clear from elementary matrix properties.
	For antisymmetry we refer to \cite[Thm~6]{McIver:2014aa},
	\Cf{Do we need to say anything about the fact that PoST is using channels whereas we are using hypers?}
	whose supporting Lemma 1 there we repeat here in the current notation.
	\begin{Lemma}{Strict monotonicity}{l1228}
		Given are two hypers $A,B{\In}\Dist^2\CalX$ and a strictly concave function $f{\In}\Dist\CalX{\Fun}\NNReal$.
		
		If $A{\StrictRef}B$ then $\Dist f.A < \Dist f.B$.
		\Proof Proved for channels in \cite[Lem~1]{McIver:2014aa}; the proof for hypers is essentially identical.
	\end{Lemma}
	We now have antisymmetry, because $A{\Ref}B{\Ref}A$ and $A{\neq}B$ implies $A{\StrictRef}B{\StrictRef}A$ whence from from \Lem{l1228} the contradiction $\Dist f.A<\Dist f.B<\Dist f.A$ for any strictly concave $f{\In}\Dist\CalX{\Fun}\NNReal$ of our choice (such as Shannon entropy) \Cite{PoST, Thm 1}.
	
	The space $(\Dist^2\CalX,\Ref)$ has neither a least- nor a greatest element, and it is not closed under $\Ref$-sups. It can be completed however by moving to proper measures \Cite{where?}; but we do not need that here.
	
	\NS
	A pure channel is an \HMM\ whose transfer-part is missing (equiv.\ is the identity): it is wholly determined by some channel matrix $C$. The denotation $\AbCh{C}$ is given in \Def{d1416CB}.
	
	\begin{Definition}{Hyper determines channel and prior}{d1539A}
		Let hyper $\Delta{\In}\Dist^2\CalX$ have finite support some $\CalY\subseteq\Dist\CalX$; note that $\CalY$ is thus a set of (posterior) distributions. Let $\delta{\In}\CalX$ be $\Avg.\Delta$ and suppose that $\delta$ has full support, i.e.\ that $\Supp{\Delta}{=}\CalX$.
		
		We define a channel $C_\Delta{\In} \CalX{\times}\CalY\Fun\Real$ as
		\[
		C_\Delta.x.y \Wide{\Defs} \Delta.y\times y.x/\delta.x~,
		\]
		where the division on the right is well defined because by assumption $\delta.x$ is never 0. Also, the channel rows are 1-summing, because for each $x$ we have
		\[
		\sum_y (\Delta.y\times y.x/\delta.x)
		\Wide{=}
		(\sum_y \Delta.y\times y.x)/\delta.x
		\Wide{=}
		\delta.x/\delta.x ~.
		\]
		
		Under the conditions stated, we say that $\Delta$ \emph{determines} $C_\Delta,\delta$.
		\Cf{\label{n1016}But (as we discussed) there's an issue here with the fact that a given abstract channel $f$ will produce different supports for $C_{f.\delta}$, depending on the $\delta$. \Def{d1044A} is therefore not quite right.}
	\end{Definition}
	
	With \Def{d1539A} we can characterise which functions are abstract channels. We have
	\begin{Definition}{Abstract channel}{d1044A}
		An \emph{abstract channel} is a function $P$ of type $\AHSpc$ such that there is a (concrete) channel $C$ so that for all full-support priors $\delta{\In}\Dist\CalX$ we have that $P.\delta$ determines $C,\delta$.
	\end{Definition}
	A trivial consequence of \Def{d1044A} is that for any abstract channel $P$ and full support $\delta$ we have $\Avg.(P.\delta){=}\delta$.
	
	\NS
	The function $\Avg{\In}\Dist^2\CalX{\Fun}\Dist\CalX$ ``collapses'' a hyper by averaging its inner distributions according to their outer probabilities.
	\footnote{It is multiply $\mu$ from the probability monad, as explained in \Sec{???}.}
	Averaging the output hyper of an abstract \HMM\ recovers the distribution on the output state, ignoring the effect of any (possible) information leaked by a channel component. Thus $\Avg$ applied to \Eqn{e0852A} in the example of \Sec{s1418} above, i.e.\ $\HMMsem{C}{M}.\pi$, should give the output distribution produced by $M$ applied directly to $\pi^T$ which, from \Eqn{e1425}, is
	\[
	\left(
	\nicefrac{1}{3}~
	\nicefrac{1}{3}~
	\nicefrac{1}{3}
	\right)
	\left(
	\begin{array}{ccc}
	\nicefrac{1}{3} & \nicefrac{1}{3} & \nicefrac{1}{3} \\
	0               & \nicefrac{1}{2} & \nicefrac{1}{2} \\
	0               & 0               & 1
	\end{array}
	\right)
	\Wide{=}
	\left(
	\nicefrac{1}{9}~
	\nicefrac{5}{18}~
	\nicefrac{11}{18}
	\right)~.
	\]
	And the average
	\(  \nicefrac{5}{9}\RVec{\nicefrac{1}{9},\nicefrac{5}{18},\nicefrac{11}{18}}
	+ \nicefrac{4}{9}\RVec{\nicefrac{1}{12},\nicefrac{1}{3},\nicefrac{7}{12}}
	\)
	from \Eqn{e0852A} is indeed $\RVec{\nicefrac{1}{9},\nicefrac{5}{18},\nicefrac{11}{18}}$ as just above.
	
	\NS
	Let's suppose we have a prior $\pi$ and two \HMM-steps over state $\CalX$ given by $I{=}(C,M)$ and $J{=}(D,N)$, with the rows of $\pi,C,M,D,N$ indexed by $\CalX$ and the columns of $C,D$ indexed by observables $\CalY$ respectively. That is, our types are
	\[
	\begin{array}{rl}
	\pi{\In} & \VType{\CalX} \\
	C,D{\In} & \CalX{\MFun}\CalY \\
	M,N{\In} & \CalX{\MFun}\CalX~.
	\end{array}
	\]
	
	We saw the \HMM\ view of the composition $H=I;J$ at \Eqn{e0946} above.
	First we take the conventional \HMM\ view of what happens when you start with $\pi$ and then ``do'' $H$ followed by $I$. The result is a joint distribution $p(x,y,x',z,x'') = \pi_xC_{x,y}M_{x,x'}D_{x',z}N_{x',x''}$. And what we are interested in is $p(X'',(Y,Z))$ and the hyper $\sum_{y,z}~p(y,z){\times}\Point{p(X''|y,z)}$ that it produces.
	
	\NS
	But first I'll summarise the notation that the calculation above seems to have synthesised. I hope it's pretty close to what ``they'' use.
	\begin{enumerate}
		\item Vector- and matrix typing is made more concise by the use of a variant arrow ``$\MFun$''. So $\pi{\In}\VType{\CalX}$ means $\pi$ is an $\CalX$-indexed vector of reals; and $M{\In}\CalX{\MFun}\CalY$ means that $M$ is a matrix of reals, with rows indexed by $\CalX$ and columns by $\CalY$. Thus the type of $\pi\MMult M$ is $\VType{\CalY}$.
		\item Vector- and matrix indexing is done by subscripts, as in $\pi_x$ and $M_{x,y}$: we consider vectors to be degenerate matrices, as usual. For $M$ the first index $x$ is for the row, and the $y$ is for the column; for $\pi$ it's for whichever dimension is not degenerate, so that it doesn't matter whether $\pi$ is a row- or a colun vector.
		
		\item If we follow the usual practice of writing $p(\cdot,\cdot,\cdots)$ to refer to a fixed (joint) distribution matrix $J$ say over some $\CalX{\times}\CalY{\times}\cdots$ then of course $p(x,y,\cdots)$ means $J_{x,y,\cdots}$. But the $p(\cdots)$ notation has by convention some useful shortcuts, providing certain variable-naming conventions are followed.
		
		\item Suppose (for all items from here on) we always use $x$ for a value in $\CalX$ and $y$ for a value in $\CalY$ etc. And suppose (in this item only) that $p,J$ are over only $\CalX{\times}\CalY$. Then by $p(x)$ we mean $\sum_y p(x,y)$ and similarly $p(y) = \sum_y p(x,y)$ --- these are the marginal probabilities. Moreover writing $p(x,y)$ is the same as writing $p(y,x)$ because the variable names tell us which is which. (Nevertheless we'll try to write them in a consistent order.) The important thing is that the naming convention tells you which variable(s) are left out, and thus which variable(s) must be summed-over.
		
		\item The usual conditional notation applies: thus $p(x|y) = p(x,y)/p(y)$. The missing-variable convention works here too, since if our space is (now) $\CalX{\times}\CalY{\times}\CalZ$ we have for example
		\begin{Reason}
			\Step{}{
				p(x|y)
			}
			\StepR{$=$}{apply missing-$z$ convention\\\emph{before} resolving conditional}{
				\sum_z p(x,z|y)
			}
			\StepR{$=$}{defn.\ conditional}{
				\sum_z (~p(x,y,z)/p(y)~)
			}
			\Step{$=$}{
				(~\sum_z p(x,y,z)~)/p(y)
			}
			\StepR{$=$}{apply missing-$z$ convention\\\emph{after} resolving conditional}{
				p(x,y)/p(y)
			}
			\StepR{$=$}{defn.\ conditional}{
				p(x|y)~.
			}
		\end{Reason}
		
		\item To refer to the whole distribution rather than simply to a probability it assigns, we use capital Roman letters (as for random variables). Thus $p(X|y)$ is the conditional distribution on $\CalX$ induced by the single value $y$.
		\footnote{More explicit notation would be $p(X|Y{=}y)$; but this is the kind of clutter we hope to avoid.}
		Normally, all variables left of the conditional sign ``$|$'' must be the same case; if there's no ``$|$'' written, it's as if it is at the extreme right.
		
		\item When we are dealing with distributions as a whole, we can consider them to be matrices (or vectors). Thus with a space $\CalX{\times}\CalY$ for $J,p$ we can consider $p(X,Y)$ to be a matrix. Then we have $p(X,Y)_{x,y} = p(x,y)$.
		
		\item If we (abnormally) write e.g.\ $p(x,Y)$ we mean the \emph{sub-}distribution on $\CalY$ determined by $x$, which is (thus) not normalised. Normalising it would give $p(Y|x)$. That means that $p(X,Y)_{-,y}$ should be the $y$-column (unnormalised) of $p(X,Y)$ as a matrix, i.e.\ $p(X,y)$.
		\footnote{We probably won't be using these; but I'm just putting in everything I can think of.}
	\end{enumerate}
	
	\NS
	We start with a distribution $\pi$ over $\CalX$. The first \HMM-step $H^1$ produces a hyper over $\CalX$, that is an outer distribution over inner distribution(see above) the hyper $\sum_{y_1}~p_1(y_1){\times}\Point{p_1(X''|y_1)}$, where $p_1(x,y_1,x'')\Defs\pi_xC_{x,y_1}M_{x,x''}$. And from an arbitrary $\pi''$, the second step produces $\sum_{y_2}~p_2(y_2){\times}\Point{p_2(X'|y_2)}$ where $p_2(x'',y_2,x')\Defs\pi''_{x''}D_{x'',y_2}N_{x'',x'}$. To carry out the Kleisli composition, we must let the second-step prior $\pi''$ range over the inners $p_1(X''|y_1)$ from the first step.
	
	\NS
	The lifting inherent in Kleisli-composition first applies the right-hand step to each inner (i.e.\ posterior) from the left-hand step, preserving the way in which they are all combined together by weighted average. Then a ``flatten'', a.k.a.\ average, or multiply from the monad, is performed on the result. Thus, letting $\pi''$ be some $p_1(X''|y_1)$ from the first step, we get from \Eqn{e1055} that
	\begin{Reason}
		\Step{}{
			{p_2}_{y_1}(x'',y_2,x')
		}
		\Step{$=$}{
			\pi''_{x''}D_{x'',y_2}N_{x'',x'}
		}
		\StepR{$=$}{set $\pi''\Defs p_1(X''|y_1)$}{
			p_1(X''|y_1)_{x''}D_{x'',y_2}N_{x'',x'}
		}
		\Step{$=$}{
			p_1(x''|y_1)D_{x'',y_2}N_{x'',x'} ~,
		}
	\end{Reason}
	where the $y_1$-subscript in the ${p_2}_{y_1}$ which we are defining on the left captures its dependence on $p_2$. 
	Now the application of the lifted second step to the hyper produced by the first step then gives us
	\[
	\sum_{y_1}~p_1(y_1){\times}\Point{\,\sum_{y_2}~{p_2}_{y_1}(y_2){\times}\Point{{p_2}_{y_1}(X'|y_2)}\,}
	\]
	which, when flattened, becomes
	\begin{equation}%\label{e1342}
	\sum_{y_1,y_2}~p_1(y_1){p_2}_{y_1}(y_2){\times}\Point{{p_2}_{y_1}(X'|y_2)}
	\end{equation}
	That's the Kleisli composition.
	
	\NS
	\subsection*{Tying down the \HMM/hyper connection}
	\subsubsection{Justifying our interpretation}
	We won't be putting this in the paper, I think; but it has to be done somewhere.
	
	An \HMM-step is given by two (stochastic) matrices $C,M$. The matrix $C{\In}\CalX{\MFun}\CalY$ is a channel, where we use ``$\MFun$'' for the ``type'' of the matrix in the sense of taking rows to columns: actually $C$, as a function, is of type $\CalX{\times}\CalY\Fun\Real$. The $M{\In}\CalX{\MFun}\CalX$ is a Markov-transition matrix. We say that the \HMM-step, i.e.\ the pair ($C,M$) taken together,  denotes a certain hyper, and here we give our justification for that. I'm going to try to do it in what might be the ``conventional'' style.
	
	Define $H{\In}\MH$ by $H_{x,y,x'}\Defs C_{x,y}M_{x,x'}$.
	\footnote{This is parallel composition $\ChPar{C}{M}$ of $C$ and $M$.}
	This (also stochastic) matrix $H$ is our syntactic representation of the \HMM\ given by $(C,M)$ together. If we apply it to a prior $\pi{\In}\VType{\CalX}$,
	\Cf{Here I'm experimenting with $\VType{\CalX}$ as the ``type'' of a vector over $\CalX$, i.e.\ $\CalX{\Fun}\Real$. But maybe this is going to far\ldots\ Does it clarify, or clutter?\par A nice thing about it however is e.g.\ that if $\pi{\in}\VType{\CalX}$ and $M{\in}\CalX{\MFun}\CalY$ then the matrix product $\pi{\MMult}M$ is in $\VType{\CalY}$.}
	that is forming $\ChApp{\pi}{H}$, we obtain a joint distribution $p(X,Y,X')$ over $\CalX{\times}\CalY{\times}\CalX$ with probabilities $p(x,y,x')\Defs \pi_x C_{x,y} M_{x,x'}$. We can now ask ``what is the posterior distribution of the initial state $X$ given we have observed some emission $y$?'' The answer is just $p(X|y)$. And if we ask for the posterior view of the final state the answer is $p(X'|y)$.
	\footnote{\Label{n1116}More conventionally, we'd write $p(X'{=}x'|Y{=}y)$; note that a summation over the third variable $x$ is therefore implied, since it's not mentioned. What we're doing here, to avoid clutter, is assuming the ``$X'{=}$'' if we write just $x'$. We just have to be careful with the variable names. \par A second convention I'll use is to write $p(X,Y)$ etc.\ to mean the joint distribution (in this case) as a whole, and in the order shown. Then for example $p(X,Y).(x,y)=p(x,y)$.}
	Both answers use Bayes' formula, and we will concentrate on the final state, i.e.\ in conventional notation
	\footnote{In $p(y,x')$ I'm writing the variables ``in order'' just to help keep things straight.}
	\begin{equation}%\label{e1031}
	p(X'{=}x'|Y{=}y) \Wide{=} p(X{=}x'\land Y{=}y)/p(Y{=}y)~,
	\end{equation}
	or more succinctly $p(x'|y) = p(y,x')/p(y)$.
	
	To get from here to a hyper, we abstract from $Y$: that is, rather than give for each $y{\In}\CalY$ a corresponding posterior $p(X'|y)$ over the output state, we make a distribution over those posteriors (thus a hyper), so that the probability assigned to each posterior is the sum of the (marginal) probabilities of all the $y$-values that produce it. But (I am glad to say) gives the same answer. We first apply the $C$ to the $\pi$, and get a hyper from that; then we push-forward the Markov matrix $M$ over that hyper to get our final hyper. The correspondence we're looking for is that if we convert the joint distribution $p(X',Y)$ over $\CalX'{\times}\CalY$ (note the order) in \Eqn{e1031} to a hyper, then we will get the same ``our'' final hyper.
	
	Using conventional notation (and therefore differing from \Def{d1426H}), we achieve that abstraction from the $y$-values as follows: recalling \Def{d1417P} of the point-distribution constructor $\Point{\cdot}$, we see that the hyper denoted by $H$ and $\pi$, that is $\Hyp{\ChApp{\pi}{H}}$, is given by
	\begin{Definition}{Semantics of \HMM}{d1055} We define
		\begin{equation}%\label{e1055}
		\Hyp{\ChApp{\pi}{H}} \Wide{\Defs} \sum_y~p(y){\times}\Point{p(X'|y)}~,
		\end{equation}
		where $p(x,y,x')\Defs\ \pi_xH_{x,y,x'} = \pi_xC_{x,y}M_{x,x'}$.
		The once-free $y$ is now abstracted, bound by the $\sum_y$.
	\end{Definition}
	
	This expresses that the support of $\Hyp{\ChApp{\pi}{H}}$ is the set of posteriors $p(X'|y)$ obtained as $y$ varies over $Y$, and the probability assigned to each $p(X'|y)$ is just the marginal $p(y)$ of the $y$ that produced it. If two $y$'s, say $y_{1,2}$ produce the same posterior, i.e.\ have $p(X'|y_1){=}p(X'|y_2)$, then they are automatically coalesced.
	\footnote{That is, if $p(X'|y_1){=}p(X'|y_2)$ then
		\[
		\begin{array}{cl}
		& p(y_1){\times}\Point{p(X'|y_1)}+p(y_2){\times}\Point{p(X'|y_2)} \\
		= & (p(y_1)+p(y_2)){\times}\Point{p(X'|y_1)} \\
		= & (p(y_1)+p(y_2)){\times}\Point{p(X'|y_2)}~.
		\end{array}
		\]
	}
	That's ``their way''.
	
	And this is ``our way''. First we make the hyper corresponding to the joint distribution $\ChApp{\pi}{C}$. In the style just above, that would be $\sum_y~p(y){\times}\Point{p(X|y)}$. Then by applying the push-forward of the Markov matrix $M$, we in effect apply it to each of the inners $p(X|y)$, retaining the outer weighted sum; that gives $\sum_y p(y)\times\Point{\RVec{p(X|y)}\MMult M}$, where in $\RVec{p(X|y)}\MMult M$ we have converted $p(X|y)$ to a row-vector so that we can matrix-multiply it by $M$ to give another row-vector, which latter we convert back into a probability distribution, but now over $\CalX'$ (instead of $\CalX$).
	\footnote{Really need notation for that too, or maybe a clear set of conventions explaining why we can make the conversions implicitly.}
	To show that is equal to \Eqn{e1055} we calculate for arbitrary $x'$ that
	\begin{Reason}
		\Step{}{
			(\RVec{p(X|y)}\MMult M)_{x'}
		}
		\StepR{$=$}{matrix multiplication}{
			\sum_x p(x|y)M_{x,x'}
		}
		\StepR{$=$}{defn.\ conditional}{
			\sum_x p(x,y)M_{x,x'}/p(y)
		}
		\StepR{$=$}{$p(x,y)$ determined by $\pi,C$}{
			\sum_x \pi_x C_{x,y}M_{x,x'}/p(y)
		}
		\StepR{$=$}{$p(x,y,x')$ determined\\by $\pi,C,M$}{
			\sum_x p(x,y,x')/p(y)
		}
		\StepR{$=$}{$\sum_x p(x,y,x')=p(y,x')$}{
			p(y,x')/p(y)
		}
		\StepR{$=$}{defn.\ conditional}{
			p(x'|y)
		}
		\StepR{$=$}{Fn.~\ref{n1116}}{
			p(X'|y).x'~.
		}
	\end{Reason}
	Thus $\RVec{p(X|y)}\MMult M = p(X'|y)$, since $x'$ was arbitrary.
	
	\NS
	If these are laid out horizontally, we enclose them in double set-brackets $\DDist{\cdots} $ separated by commas: thus $\DDist{H\AtP\nicefrac{2}{3},T\AtP\nicefrac{1}{3}}$ describes a coin twice as likely to give heads as tails. If the double brackets are used without probabilities (and thus also without $\AtP$'s) then the intended distribution is uniform, so that $\DDist{H,T}$ describes a fair coin; a convenient special case of that is $\DDist{H}$ for the point distribution on $H$, the coin that gives heads every time.
	
	\NS
	If we apply the push forward of $\MM{M}$ to the above, we transform the inners but maintain the outer: that is, we obtain
	\begin{eqnarray}
	\MM{M}.\DDist{x_0\AtP\nicefrac{2}{5}, & x_1\AtP\nicefrac{1}{5}, & x_2\AtP\nicefrac{2}{5}} \label{e1412CB} \\
	\MM{M}.\DDist{x_0\AtP\nicefrac{1}{4}, & x_1\AtP\nicefrac{1}{2}, & x_2\AtP\nicefrac{1}{4}} \label{e1412DB}
	\end{eqnarray}
	as the result of $\Dist\MM{M}.(\AbCh{C}.\pi)$. Calculating it through for the given $M$ gives
	\begin{eqnarray}
	\DDist{x_0\AtP\nicefrac{2}{15}, & x_1\AtP\nicefrac{7}{30}, & x_2\AtP\nicefrac{19}{30}} \label{e1412E} \\
	\DDist{x_0\AtP\nicefrac{1}{12}, & x_1\AtP\nicefrac{1}{3}, & x_2\AtP\nicefrac{7}{12}} \label{e1412F}
	\end{eqnarray}
	
	\NS
	\[
	\ChApp{\pi}{\ChPar{C}{M}} \Wide{=} 
	\left(
	\begin{array}{cccccc}
	\nicefrac{2}{27} & \nicefrac{2}{27} & \nicefrac{2}{27} &
	\nicefrac{1}{27} & \nicefrac{1}{27} & \nicefrac{1}{27} \\
	0 & \nicefrac{1}{18} &\nicefrac{1}{18} &
	0 & \nicefrac{1}{9} & \nicefrac{1}{9} \\
	0 & 0 &\nicefrac{2}{9} &
	0 & 0 & \nicefrac{1}{9}
	\end{array}
	\right)~.
	\]
	
	\NS
	Write row-vectors as $\RVec{a,b,c}$ and column vectors as $\CVec{a,b,c}$.
	\Cf{We might need something like this: \Bx Similarly if $p(X)$ refers to a probability distribution on random variable $X$ over finite sample space $\CalX$, we write $\RVec{p(X)},\CVec{p(X)}$ for that distribution presented as a row,column vector respectively with indexing in the canonical order for $\CalX$. \par\Cx See \En{n0841} for why.}
	
	\NS
	so that e.g.\ a 3-row-by-2-column matrix might be $\CVec{\RVec{a,b},\RVec{c,d},\RVec{e,f}}$.
	
	\NS
	A discrete distribution over a (finite) state with some canonical order can be given either by a row- or a column vector in which the probabilities are presented in that same order.
	
	\NS
	Here is the construction applied to our examples from \Sec{s1323}. We start with uniform prior $\pi$ as before, and the effect of channel $C$ is to produce the output hyper \Eqn{e1327} whose inner distributions are
	\begin{eqnarray}
	\DDist{x_0\AtP\nicefrac{2}{5}, & x_1\AtP\nicefrac{1}{5}, & x_2\AtP\nicefrac{2}{5}} \label{e1412AB} \\
	\DDist{x_0\AtP\nicefrac{1}{4}, & x_1\AtP\nicefrac{1}{2}, & x_2\AtP\nicefrac{1}{4}} \label{e1412BB}
	\end{eqnarray}
	The outer distribution assigns probability $\nicefrac{5}{9}$ to \Eqn{e1412A} and probability $\nicefrac{4}{9}$ to \Eqn{e1412B}.
	
	\NS
	Now we take the hyper/Kleisli view. From $\pi$, the first step produces (see above) the hyper $\sum_{y_1}~p_1(y_1){\times}\Point{p_1(X''|y_1)}$, where $p_1(x,y_1,x'')\Defs\pi_xC_{x,y_1}M_{x,x''}$. And from an arbitrary $\pi''$, the second step produces $\sum_{y_2}~p_2(y_2){\times}\Point{p_2(X'|y_2)}$ where $p_2(x'',y_2,x')\Defs\pi''_{x''}D_{x'',y_2}N_{x'',x'}$. To cary out the Kleisli composition, we must let the second-step prior $\pi''$ range over the inners $p_1(X''|y_1)$ from the first step.
	
	\NS
	With that, we have that the application of the lifted second step $\Dist\HMMone{H^2}$ to the hyper produced by the first step $\HMMone{H^1}.\pi$ is
	\[
	\sum_{y_1}~p_1(y_1){\times}\Point{\,\sum_{y_2}~{p_2}_{y_1}(y_2){\times}\Point{{p_2}_{y_1}(X'|y_2)}\,}
	\]
	which, when flattened, becomes
	\begin{equation}%\label{e1342}
	\sum_{y_1,y_2}~p_1(y_1){p_2}_{y_1}(y_2){\times}\Point{{p_2}_{y_1}(X'|y_2)}
	\end{equation}
	That's the Kleisli composition.
	
	\NS
	\subsection*{The special cases of pure channels and pure transitions}
	\Cf{Probably we will only mention these, not prove them: they must be well known from the definition of ``$;$'' on \HMM's directly.}
	We ``know'' that the Kleisli composition of two channels-as-\HMM's $C,D$ is their parallel composition $C{\parallel}D$. For that, we make $M,N$ above the identity matrix and find that $I;J$ applied to $\pi$ generates the joint distribution $p(x,y,x',z,x'') = \pi_xC_{x,y}M_{x,x'}D_{x',z}N_{x',x''}$, so that $p(x,(y,z)) = \sum_{x',x''} \pi_xC_{x,y}M_{x,x'}D_{x',z}N_{x',x''}$, which reduces to $\pi_xC_{x,y}D_{x,z}$. Since $(C{\parallel}D)_{x,(y,z)} = C_{x,y}D_{x,z}$, we have our results $\HMMsem{C}{};\HMMsem{D}{}=\HMMsem{C{\parallel}D}{}$.
	
	We also know, on the other hand, that for Markov matrices the Kliesli composition reduces to just their matrix product, i.e.\ that $\HMMsem{}{M};\HMMsem{}{N}=\HMMsem{}{M{\MMult}N}$. This time it's the channels $C,D$ that are trivial, revealing nothing. The joint distribution $p(x,y,x',z,x'') = \pi_xC_{x,y}M_{x,x'}D_{x',z}N_{x',x''}$ becomes $p(x,x',x'') = \pi_xM_{x,x'}N_{x',x''}$, where we simply remove the $y,z$ because they are the unique elements for which $C_{x,y},D_{x,z}$ are not zero (and hence are one). This time we have $p(x,x'') = \sum_{x'} \pi_xM_{x,x'}N_{x',x''} = \pi_x(M{\MMult}N)_{x,x'}$.
	
	\NS
	\subsection*{Channel syntax}
	A \emph{channel} is a (stochastic) matrix of non-negative reals in which each row sums to 1; we use upper-case Roman letters for channels, like $C$. The rows are labelled with elements from a source, or input set, typically $\CalX$; and the columns are labelled with elements from a target or output set, typically $\CalY$.
	
	For matrices in general we use these notations:
	\begin{Definition}{Matrix indexing}{d1422}
		For matrix $M$ with input,output types $\CalX,\CalY$ we write $M_{x,y}$ for the element indexed by row $x$ and column $y$. The $x$-th row of $M$ is $M_{x,-}$; the $y$-th column is $M_{-,y}$.
	\end{Definition}
	
	\begin{Definition}{Weight}{d1358}
		Let $M$ be a matrix or (a special case) a vector. Then $\sum M$ is the weight of $M$, that is $\sum_{x,y} M_{x,y}$. For e.g.\ a row $M_{x,-}$, a vector, we have $\sum M_{x,-}= \sum_y M_{x,y}$.
	\end{Definition}
	
	Thus the $x$-th row of $C$ is  $C_{x,-}$; the $y$-th column is written $C_{-,y}$; and the element at row $x$ and column $y$ is $C_{x,y}$. Each row $C_{x,-}$ as a whole is a conditional probability distribution over $\CalY$ given an input value $x{\In}\CalX$. Specifically, the $y$-th element $C_{x,y}$ of $C_{x,-}$ is the probability that an input $x$ will lead to an output $y$, so for example $\sum C_{x,-}=1$.
	
	We regard channels (here) as a \emph{syntactic} way of describing a mechanism that converts an incoming, \emph{prior} distribution over $\CalX$ into an outgoing, \emph{posterior} distribution (again) over $\CalX$ that is determined by the value $y$ that the channel produced. We use lower-case Greek letters for distributions in $\Dist\CalX,\Dist\CalY$ over $\CalX,\CalY$, typically $\pi{\In}\CalX$ and $\rho{\In}\CalY$; then $\pi_x$ is the probability that $\pi$ assigns to $x$, and similarly $\rho_y$.
	
	\subsection*{Informal channel semantics}
	A prior $\pi{\In}\Dist\CalX$ and a channel $C$ of input,output types $\CalX,\CalY$ together determine a joint distribution over $\CalX{\times}\CalY$. We have
	\begin{Definition}{Channel applied to prior}{d1124}
		Given a prior $\pi{\In}\Dist\CalX$ and a channel $C{\In}(\CalX{\times}\CalY)\Fun\NNReal$ we write $\ChApp{\pi}{C}$ for the joint distribution matrix, say $J$, that results from applying the channel to the prior: thus $(\ChApp{\pi}{C})_{x,y}\Defs \delta_x C_{x,y}$.
	\end{Definition}
	Given some matrix $J$ as above we have $\sum J=1$ because $J$ denotes a distribution over $\CalX{\times}\CalY$.
	
	Use this notation for normalisation of a sub-distribution:
	\begin{Definition}{Normalisation}{d1357}
		Let $\delta$ be a subdistribution on $\CalX$ that is not everywhere zero. Then $\Nrm.\delta$ is the full distribution obtained by normalising $\delta$, so that $\Nrm.\delta.x = \delta.x/\sum\delta$.
	\end{Definition}
	
	Now suppose that $J=\ChApp{\pi}{C}$. The \emph{posterior} distribution on $\CalX$, induced by knowing the input prior $\pi$ on $\CalX$ and having observed a particular output $y$, is $\Nrm.(J_{-,y})$. This is our intended informal semantics: having a channel $C$, a prior $\pi$ we can determine the probability $\sum J_{-,y}$ of a particular $y$'s being output and, given than, the determine the posterior distribution $\Nrm.(J_{-,y})$ it induces on $\CalX$.
	
	\subsection*{Precise channel semantics: abstract channels and hyper-distributions}
	A joint-distribution matrix $J$ contains ``too much'' information if we need only the posteriors and their respective probabilities (as explained immediately above) and not, in particular, the actual value of $y$ that led to a particular posterior. This view is appropriate in security since the information leakage of a channel concerns what an adversary can discover about its input distribution (in $\Dist\CalX$), and not (usually) concerned with the specific $y$ values whose being observed led to that discovery.
	
	Thus for example if column $y$ of the channel matrix $C$ is all zero, then column $y$ of $J=\ChApp{\pi}{C}$ will also be all zero, for any prior $\pi$. Thus output $y$ will never occur (for any prior), and we might as well leave it out.
	
	Similarly, if two columns $y_{1,2}$ of $J$ are scaled versions of each other, which we call \emph{similar} (as for triangles), then we might as well coalesce the two columns (add them together), since for a given prior the same posterior will be inferred for $y_1$ as for $y_2$ --- and the probability of inferring that posterior will be the sum of the probabilities of observing $y_{1,2}$.
	
	And finally, a 1-1 renaming of the $y$-values in $C$, and hence in $J$, has no effect on the posteriors and their respective probabilities.
	
	Abstracting from the inessential information, as suggested above, we are left with a distribution of posteriors on $\CalX$: this type $\Dist^2\CalX$ we call \emph{hyper-distributions}, and it is the semantic domain for joint distributions on $\CalX{\times}\CalY$. We write ``hyper'' for short.
	
	We introduce these notations:
	\begin{Definition}{Scaling}{d1409}
		Let $\delta$ be a subdistribution (including full distributions) on $\CalX$, and let $c$ be a scaling factor in $[0,1]$. Then $c\delta$ (or $c{\times}\delta$ if we need to be explicit) is the scaled subdistribution defined $(c\delta).x = c\times \delta.x$.
	\end{Definition}
	
	\begin{Definition}{Sub-hyper}{d1353}
		Just as a subdistribution's probabilities sum to at most one, so do a sub-hyper's \emph{outer} probabilities sum to at most 1. We define the type $\SubHyp\CalX$ to be $\SubDist(\Dist\CalX)$, that is the subdistributions over the full distributions on $\CalX$.
	\end{Definition}
	
	\begin{Definition}{Sub-point hyper}{d1417}
		Let $\delta$ be a subdistribution on $\CalX$. Then $\SubPt{\delta}$ is the sub-hyper of weight $\sum\delta$ centred on the point-hyper $\Point{\Nrm.\delta}$. That is, we define $\SubPt{\delta} = \sum\delta\times\Point{\Nrm.\delta}$.
	\end{Definition}
	
	\begin{Definition}{Subdistribution}{d1401}
		A subdistribution on $\CalX$ is a function in $\CalX{\Fun}[0,1]$ with $\sum_x\delta.x\leq 1$. We write $\SubDist\CalX$ for the set of (discrete) subdistributions in $\CalX$.
	\end{Definition}
	
	\begin{Definition}{Full distribution}{d1401F}
		A full distribution on $\CalX$ is a subdistribution $\delta$ with $\sum\delta=1$. We write $\Dist\CalX$ for the set of full distributions on $\CalX$. Trivially $\Dist\CalX\subset\SubDist\CalX$.
	\end{Definition}
	
	\begin{Definition}{Point distribution}{d1417P}
		For $x$ in $\CalX$ the point distribution $\Point{x}$ is the (full) distribution in $\Dist\CalX$ assigning probability 1 to $x$ and probability 0 to all other elements of $\CalX$. For distribution $\delta$ in $\Dist\CalX$ the point hyper $\Point{\delta}$ assigns (outer) probability 1 to (inner) distribution $\delta$ and (outer) probability 0 to all other (inner) distributions.
	\end{Definition}
	
	With the above notations, we give the semantic function taking joint-distribution matrices over $\CalX,\CalY$ to hypers in $\Dist^2\CalX$. It is
	\begin{Definition}{Joint distribution denotes hyper}{d1426H}
		Let $J{\In}\Dist(\CalX{\times}\CalY)$ be presented as a 1-summing matrix with rows over $\CalX$ and columns over $\CalY$. Then $\Hyp{J}\in\Dist^2\CalX$, the hyper denoted by $J$, is given by
		\Cf{Somewhere explain the convention that we use ``semantic brackets'' to do from matrices/vectors to denotations, since the former we can consider to be syntax.}
		\[
		\Hyp{J} \Wide{\Defs} \stackrel{\begin{array}{c}
			\makebox[0pt]{\scriptsize summation of sub-point hypers for all columns}\\[1ex]
			\downarrow
			\end{array}}
		{\sum_{\substack{y{\In}\CalY\\0\neq\sum J_{-,y}}}}
		\quad\overbrace{
			\underbrace{(\sum J_{-,y})}_{\text{scaling factor}}
			\times\underbrace{\Point{\Nrm.J_{-,y}}}_{\text{point hyper}}
			)}^{\text{sub-point hyper for this $y$-column}}~.
		\]
		That is, each column $J_{-,y}$ is taken individually, normalised, and then converted to a point hyper. But then that point hyper is scaled down by $\sum J_{-,y}$ to make it a sub-point hyper. Once those $y$-indexed sub-point hypers are all added together, we get the (full) hyper corresponding to the original joint distribution $J$.
		
		Using \Def{d1417} however allows us to write more succinctly
		\begin{equation}%\label{e0958}
		\Hyp{J}\Defs~\sum_{y{\In}\CalY}~\SubPt{J_{-,y}}~.
		\end{equation}
	\end{Definition}
	
	That finally gives us the denotation of a channel, called an \emph{abstract channel} \cite{McIver:2014aa}.
	\begin{Definition}{Denotation of channel}{d1416C}
		Let $C$ be a channel matrix from input $\CalX$ to output $\CalY$. Its denotation, of type $\AHSpc$, is given by
		\[
		\AbCh{C}.\pi
		\Wide{\Defs}
		\Hyp{\ChApp{\pi}{C}}~,
		\]
		where the $\AbCh{\cdot}$ on the left is what we are defining, and the $\Hyp{\cdot}$ on the right is given by \Def{d1426H}.
	\end{Definition}
	We do not call it an abstract channel here, since we will soon see it as a special case of an abstract \HMM.
	
	\NS
	\section{Channels: syntax and semantics}\label{s1007}
	\Cf{Carroll's reworking of his own \Sec{s1244A} below.}

	\section{Markov processes: syntax and semantics}\label{s1237}
	We now treat conventional Markov processes in the same denotational style we used for channels in \Sec{s1007} just above: then we can put channels and Markov processes together within the same semantic domain, resulting in denotations for \HMM's.
	
	\subsection*{Markov-process syntax}
	A \emph{Markov process} is a (stochastic) matrix of non-negative reals in which each row sums to 1; we use upper-case Roman letters for channels, like $M$. The rows and columns are labelled with elements from a the same set,  (here) typically $\CalX$. We regard the matrix as a \emph{syntactic} way of describing a mechanism that takes  an incoming state $x{\In}\CalX$ to an outgoing state $x'{\In}\CalX$ according to the probability distribution given by $M_{x,-}$.  That is, the probability that $x$ will be taken to $x'$ is $M_{x,x'}$.
	
	\subsection*{Markov-process hyper-distribution semantics}\label{s1332}
	Given Markov matrix $M$ and incoming distribution $\pi$, the outgoing distribution $\pi'$ is (as usual) the distribution $\pi'$ given as a row-vector ${\pi'}^T$ by the matrix multiplication $\pi^T\MMult M$, where to be explicit we are writing $-^T$ for the $\CalX$-indexed row-vector determined by a distribution on $\CalX$  and $(\MMult)$ for matrix multiplication. To make an outgoing \emph{hyper} we inject $\pi'$ into $\Dist^2\CalX$ by making it a point distribution. Thus we have
	\begin{Definition}{Denotation of Markov process}{d1416M}
		Let $M$ be a Markov matrix on $\CalX$. Its denotation, of type $\AHSpc$, is given by
		\[
		\MM{M}.\pi
		\Wide{\Defs}
		\Point{\pi^T{\MMult}M}~,
		\]
		where the $\MM{\cdot}$ on the left is what we are defining, and the $\Point{\cdot}$ on the right gives the point-hyper based on the $\pi'{\In}\Dist\CalX$ such that $\pi^T{\MMult}M$ is the row-vector ${\pi'}^T$.
	\end{Definition}
	This too we will see as a special case of an abstract \HMM.
	
	\section{Examples of denotations}\label{s1323}
	\subsection*{Notations for these examples}
	
	Write row-vectors as $\RVec{a,b,c}$ and column vectors as $\CVec{a,b,c}$ so that e.g.\ a 3-row-by-2-column matrix might be $\CVec{\RVec{a,b},\RVec{c,d},\RVec{e,f}}$.
	\Cf{We might need something like this: \Bx Similarly if $p(X)$ refers to a probability distribution on random variable $X$ over finite sample space $\CalX$, we write $\RVec{p(X)},\CVec{p(X)}$ for that distribution presented as a row,column vector respectively with indexing in the canonical order for $\CalX$. \par\Cx See
		\En{n0841} for why.}
	
	\NS
	\ENSection{Tying down the \HMM/hyper connection}
	\ENSubsection{Justifying our interpretation}
	We won't be putting this in the paper, I think; but it has to be done somewhere.
	
	{\Xx%0909
		An \HMM-step is given by two (stochastic) matrices $C,M$. The matrix $C{\In}\CalX{\MFun}\CalY$ is a channel, where we use ``$\MFun$'' for the ``type'' of the matrix in the sense of taking rows to columns: actually $C$, as a function, is of type $\CalX{\times}\CalY\Fun\Real$. The $M{\In}\CalX{\MFun}\CalX$ is a Markov-transition matrix. We say that the \HMM-step, i.e.\ the pair ($C,M$) taken together,  denotes a certain hyper, and here we give our justification for that. I'm going to try to do it in what might be the ``conventional'' style.
		
		Define $H{\In}\MH$ by $H_{x,y,x'}\Defs C_{x,y}M_{x,x'}$.
		\footnote{This is parallel composition $\ChPar{C}{M}$ of $C$ and $M$.}
		This (also stochastic) matrix $H$ is our syntactic representation of the \HMM\ given by $(C,M)$ together. If we apply it to a prior $\pi{\In}\VType{\CalX}$,
		\Cf{Here I'm experimenting with $\VType{\CalX}$ as the ``type'' of a vector over $\CalX$, i.e.\ $\CalX{\Fun}\Real$. But maybe this is going to far\ldots\ Does it clarify, or clutter?\par A nice thing about it however is e.g.\ that if $\pi{\in}\VType{\CalX}$ and $M{\in}\CalX{\MFun}\CalY$ then the matrix product $\pi{\MMult}M$ is in $\VType{\CalY}$.}
		that is forming $\ChApp{\pi}{H}$, we obtain a joint distribution $p(X,Y,X')$ over $\CalX{\times}\CalY{\times}\CalX$ with probabilities $p(x,y,x')\Defs \pi_x C_{x,y} M_{x,x'}$. We can now ask ``what is the posterior distribution of the initial state $X$ given we have observed some emission $y$?'' The answer is just $p(X|y)$. And if we ask for the posterior view of the final state the answer is $p(X'|y)$.
		\footnote{\Label{n1116}More conventionally, we'd write $p(X'{=}x'|Y{=}y)$; note that a summation over the third variable $x$ is therefore implied, since it's not mentioned. What we're doing here, to avoid clutter, is assuming the ``$X'{=}$'' if we write just $x'$. We just have to be careful with the variable names. \par A second convention I'll use is to write $p(X,Y)$ etc.\ to mean the joint distribution (in this case) as a whole, and in the order shown. Then for example $p(X,Y).(x,y)=p(x,y)$.}
	}%0909
	Both answers use Bayes' formula, and we will concentrate on the final state, i.e.\ in conventional notation
	\begin{equation}%\label{e1031}
	p(X'{=}x'|Y{=}y) \Wide{=} p(X{=}x'\land Y{=}y)/p(Y{=}y)~,
	\end{equation}
	or more succinctly $p(x'|y) = p(y,x')/p(y)$.
	
	To get from here to a hyper, we abstract from $Y$: that is, rather than give for each $y{\In}\CalY$ a corresponding posterior $p(X'|y)$ over the output state, we make a distribution over those posteriors (thus a hyper), so that the probability assigned to each posterior is the sum of the (marginal) probabilities of all the $y$-values that produce it.
	
	Using conventional notation (and therefore differing from \Def{d1426H}), we achieve that abstraction from the $y$-values as follows: recalling \Def{d1417P} of the point-distribution constructor $\Point{\cdot}$, we see that the hyper denoted by $H$ and $\pi$, that is $\Hyp{\ChApp{\pi}{H}}$, is given by
	\begin{Definition}{Semantics of \HMM}{d1055} We define
		\begin{equation}%\label{e1055}
		\Hyp{\ChApp{\pi}{H}} \Wide{\Defs} \sum_y~p(y){\times}\Point{p(X'|y)}~,
		\end{equation}
		where $p(x,y,x')\Defs\ \pi_xH_{x,y,x'} = \pi_xC_{x,y}M_{x,x'}$.
		The once-free $y$ is now abstracted, bound by the $\sum_y$.
	\end{Definition}
	
	This expresses that the support of $\Hyp{\ChApp{\pi}{H}}$ is the set of posteriors $p(X'|y)$ obtained as $y$ varies over $Y$, and the probability assigned to each $p(X'|y)$ is just the marginal $p(y)$ of the $y$ that produced it. If two $y$'s, say $y_{1,2}$ produce the same posterior, i.e.\ have $p(X'|y_1){=}p(X'|y_2)$, then they are automatically coalesced.
	\footnote{That is, if $p(X'|y_1){=}p(X'|y_2)$ then
		\[
		\begin{array}{cl}
		& p(y_1){\times}\Point{p(X'|y_1)}+p(y_2){\times}\Point{p(X'|y_2)} \\
		= & (p(y_1)+p(y_2)){\times}\Point{p(X'|y_1)} \\
		= & (p(y_1)+p(y_2)){\times}\Point{p(X'|y_2)}~.
		\end{array}
		\]
	}
	
	\ENSubsection{\HMM\ composition}
	Let $H,I{\In}\MH$ be two \HMM's. Their (syntactic) composition $H;I$ is defined \Cite{someone?}
	\begin{equation}%\label{e0946}
	(H;I)_{x,(y_1,y_2),x'} \Wide{\Defs} \sum_{x''} H_{x,y_1,x''}I_{x'',y_2,x'} ~.
	\end{equation}
	It is essentially matrix multiplication in which the ``left over'' $y_1$ from the left-hand argument is put aside, not passed on to the right argument and so remains observable: only the $x'$ is passed from $H$ to $I$, and that makes the (cut-down) $H$ and $I$ conformal.
	Note how the set of observables is now $\CalY^2$; and further compositions would increase the exponent again.
	
	\ENSubsection{Proof that composition works: notations}
	A mainstay of our interpretation is that ``syntactic'' composition of \HMM's \Eqn{e0946} is mapped by $\HMMone{\cdot}$ to the Giry (i.e.\ Kleisli) ``semantic'' composition of the components' denotations. With the above as a guide, we now tackle that.
	
	But first I'll summarise the notation that the calculation above seems to have synthesised. I hope it's pretty close to what ``they'' use.
	\begin{enumerate}
		\item Vector- and matrix typing is made more concise by the use of a variant arrow ``$\MFun$''. So $\pi{\In}\VType{\CalX}$ means $\pi$ is an $\CalX$-indexed vector of reals; and $M{\In}\CalX{\MFun}\CalY$ means that $M$ is a matrix of reals, with rows indexed by $\CalX$ and columns by $\CalY$. Thus the type of $\pi\MMult M$ is $\VType{\CalY}$.
		\item Vector- and matrix indexing is done by subscripts, as in $\pi_x$ and $M_{x,y}$: we consider vectors to be degenerate matrices, as usual. For $M$ the first index $x$ is for the row, and the $y$ is for the column; for $\pi$ it's for whichever dimension is not degenerate, so that it doesn't matter whether $\pi$ is a row- or a colun vector.
		
		\item If we follow the usual practice of writing $p(\cdot,\cdot,\cdots)$ to refer to a fixed (joint) distribution matrix $J$ say over some $\CalX{\times}\CalY{\times}\cdots$ then of course $p(x,y,\cdots)$ means $J_{x,y,\cdots}$. But the $p(\cdots)$ notation has by convention some useful shortcuts, providing certain variable-naming conventions are followed.
		
		\item Suppose (for all items from here on) we always use $x$ for a value in $\CalX$ and $y$ for a value in $\CalY$ etc. And suppose (in this item only) that $p,J$ are over only $\CalX{\times}\CalY$. Then by $p(x)$ we mean $\sum_y p(x,y)$ and similarly $p(y) = \sum_y p(x,y)$ --- these are the marginal probabilities. Moreover writing $p(x,y)$ is the same as writing $p(y,x)$ because the variable names tell us which is which. (Nevertheless we'll try to write them in a consistent order.) The important thing is that the naming convention tells you which variable(s) are left out, and thus which variable(s) must be summed-over.
		
		\item The usual conditional notation applies: thus $p(x|y) = p(x,y)/p(y)$. The missing-variable convention works here too, since if our space is (now) $\CalX{\times}\CalY{\times}\CalZ$ we have for example
		\begin{Reason}
			\Step{}{
				p(x|y)
			}
			\StepR{$=$}{apply missing-$z$ convention\\\emph{before} resolving conditional}{
				\sum_z p(x,z|y)
			}
			\StepR{$=$}{defn.\ conditional}{
				\sum_z (~p(x,y,z)/p(y)~)
			}
			\Step{$=$}{
				(~\sum_z p(x,y,z)~)/p(y)
			}
			\StepR{$=$}{apply missing-$z$ convention\\\emph{after} resolving conditional}{
				p(x,y)/p(y)
			}
			\StepR{$=$}{defn.\ conditional}{
				p(x|y)~.
			}
		\end{Reason}
		
		\item To refer to the whole distribution rather than simply to a probability it assigns, we use capital Roman letters (as for random variables). Thus $p(X|y)$ is the conditional distribution on $\CalX$ induced by the single value $y$.
		\footnote{More explicit notation would be $p(X|Y{=}y)$; but this is the kind of clutter we hope to avoid.}
		Normally, all variables left of the conditional sign ``$|$'' must be the same case; if there's no ``$|$'' written, it's as if it is at the extreme right.
		
		\item When we are dealing with distributions as a whole, we can consider them to be matrices (or vectors). Thus with a space $\CalX{\times}\CalY$ for $J,p$ we can consider $p(X,Y)$ to be a matrix. Then we have $p(X,Y)_{x,y} = p(x,y)$.
		
		\item If we (abnormally) write e.g.\ $p(x,Y)$ we mean the \emph{sub-}distribution on $\CalY$ determined by $x$, which is (thus) not normalised. Normalising it would give $p(Y|x)$. That means that $p(X,Y)_{-,y}$ should be the $y$-column (unnormalised) of $p(X,Y)$ as a matrix, i.e.\ $p(X,y)$.
		\footnote{We probably won't be using these; but I'm just putting in everything I can think of.}
	\end{enumerate}
	
	\ENSubsection{Proof that composition works: calculations}
	
	Let's suppose we have a prior $\pi$ and two \HMM-steps over state $\CalX$ given by $I{=}(C,M)$ and $J{=}(D,N)$, with the rows of $\pi,C,M,D,N$ indexed by $\CalX$ and the columns of $C,D$ indexed by observables $\CalY$ respectively. That is, our types are
	\[
	\begin{array}{rl}
	\pi{\In} & \VType{\CalX} \\
	C,D{\In} & \CalX{\MFun}\CalY \\
	M,N{\In} & \CalX{\MFun}\CalX~.
	\end{array}
	\]
	
	We saw the \HMM\ view of the composition $H=I;J$ at \Eqn{e0946} above. First we take the conventional \HMM\ view of what happens when you start with $\pi$ and then ``do'' $H$ followed by $I$. The result is a joint distribution $p(x,y,x',z,x'') = \pi_xC_{x,y}M_{x,x'}D_{x',z}N_{x',x''}$. And what we are interested in is $p(X'',(Y,Z))$ and the hyper $\sum_{y,z}~p(y,z){\times}\Point{p(X''|y,z)}$ that it produces.
	Now we take the hyper/Kleisli view. From $\pi$, the first step produces (see above) the hyper $\sum_{y_1}~p_1(y_1){\times}\Point{p_1(X''|y_1)}$, where $p_1(x,y_1,x'')\Defs\pi_xC_{x,y_1}M_{x,x''}$. And from an arbitrary $\pi''$, the second step produces $\sum_{y_2}~p_2(y_2){\times}\Point{p_2(X'|y_2)}$ where $p_2(x'',y_2,x')\Defs\pi''_{x''}D_{x'',y_2}N_{x'',x'}$. To cary out the Kleisli composition, we must let the second-step prior $\pi''$ range over the inners $p_1(X''|y_1)$ from the first step.
	
	The lifting inherent in Kleisli-composition first applies the right-hand step to each inner (i.e.\ posterior) from the left-hand step, preserving the way in which they are all combined together by weighted average. Then a ``flatten'', a.k.a.\ average, or multiply from the monad, is performed on the result. Thus, letting $\pi''$ be some $p_1(X''|y_1)$ from the first step, we get from \Eqn{e1055} that
	\begin{Reason}
		\Step{}{
			{p_2}_{y_1}(x'',y_2,x')
		}
		\Step{$=$}{
			\pi''_{x''}D_{x'',y_2}N_{x'',x'}
		}
		\StepR{$=$}{set $\pi''\Defs p_1(X''|y_1)$}{
			p_1(X''|y_1)_{x''}D_{x'',y_2}N_{x'',x'}
		}
		\Step{$=$}{
			p_1(x''|y_1)D_{x'',y_2}N_{x'',x'} ~,
		}
	\end{Reason}
	where the $y_1$-subscript in the ${p_2}_{y_1}$ which we are defining on the left captures its dependence on $p_2$. 
	Now the application of the lifted second step to the hyper produced by the first step then gives us
	\[
	\sum_{y_1}~p_1(y_1){\times}\Point{\,\sum_{y_2}~{p_2}_{y_1}(y_2){\times}\Point{{p_2}_{y_1}(X'|y_2)}\,}
	\]
	which, when flattened, becomes
	\begin{equation}%\label{e1342}
	\sum_{y_1,y_2}~p_1(y_1){p_2}_{y_1}(y_2){\times}\Point{{p_2}_{y_1}(X'|y_2)}
	\end{equation}
	That's the Kleisli composition.
	
	What we would like to know is whether this Kleisli-generated \Eqn{e1342} is the same hyper $\sum_{y_1,y_2}~p(y_1,y_2){\times}\Point{p(X'|y_1,y_2)}$ that would result from \Def{d1055} applied to $\ChApp{\pi}{H}$ directly, that is when we set $p(x,(y_1,y_2),x')\Defs (\ChApp{\pi}{H})_{x,(y_1,y_2),x'}$~.
	
	We begin with a technical lemma that, based on the calculation above, simplifies the relationship between $p_1,p_2$ and $p$:
	\begin{Lemma}{Technical lemma}{l1625}
		\[
		p_1(y_1){p_2}_{y_1}(y_2,x')
		\Wide{=}
		p(y_1,y_2,x') ~.
		\]
		\Proof
		\begin{Reason}
			\Step{}{
				p_1(y_1){p_2}_{y_1}(y_2,x')
			}
			\Step{$=$}{
				p_1(y_1)\sum_{x''}{p_2}_{y_1}(x'',y_2,x')
			}
			\StepR{$=$}{calculation above}{
				p_1(y_1)\times\\\quad\sum_{x''}p_1(x''|y_1)D_{x'',y_2}N_{x'',x'}
			}
			\Step{$=$}{
				\sum_{x''}p_1(y_1,x'')D_{x'',y_2}N_{x'',x'}
			}
			\Step{$=$}{
				\sum_{x,x''}p_1(x,y_1,x'')D_{x'',y_2}N_{x'',x'}
			}
			\Step{$=$}{
				\sum_{x,x''}\pi_xC_{x,y_1}M_{x,x''}D_{x'',y_2}N_{x'',x'}
			}
			\Step{$=$}{
				\sum_{x}\pi_x\sum_{x''}C_{x,y_1}M_{x,x''}D_{x'',y_2}N_{x'',x'}
			}
			\Step{$=$}{
				\sum_{x}p(x,y_1,y_2,x')
			}
			\Step{$=$}{
				p(y_1,y_2,x')~.
			}
		\end{Reason}
	\end{Lemma}
	
	With that lemma we prove for all $y_{1,2}$ the two equalities
	\begin{eqnarray}
	\textrm{\small \underline{Kleisli-generated}} && \textrm{\small\underline{\Def{d1055} applied to $\ChApp{\pi}{H}$}} \nonumber\\
	p_1(y_1){p_2}_{y_1}(y_2) &=& p(y_1,y_2) \label{e1051a} \\
	{p_2}_{y_1}(X'|y_2) &=& p(X'|y_1,y_2) \label{e1051b}
	\end{eqnarray}
	that together will establish the equality of the two hypers. Beginning with \Eqn{e1051a} we calculate
	\begin{Reason}
		\Step{}{
			p_1(y_1){p_2}_{y_1}(y_2)
		}
		\Step{$=$}{
			p_1(y_1){\times}\sum_{x'}{p_2}_{y_1}(y_2,x')
		}
		\Step{$=$}{
			\sum_{x'}p_1(y_1){p_2}_{y_1}(y_2,x')
		}
		\StepR{$=$}{\Lem{l1625}}{
			\sum_{x'}p(y_1,y_2,x')
		}
		\Step{$=$}{
			p(y_1,y_2)~.
		}
	\end{Reason}
	
	For \Eqn{e1051b} we prove for arbitrary $x'$ that ${p_2}_{y_1}(x'|y_2) = p(x'|y_1,y_2)$, calculating
	\begin{Reason}
		\Step{}{
			{p_2}_{y_1}(x'|y_2)
		}
		\Step{$=$}{
			{p_2}_{y_1}(y_2,x')/{p_2}_{y_1}(y_2)
		}
		\StepR{$=$}{\Lem{l1625}}{
			p(y_1,y_2,x')/p_1(y_1)~/{p_2}_{y_1}(y_2)
		}
		\StepR{$=$}{\Eqn{e1051a}}{
			p(y_1,y_2,x')/p(y_1,y_2)
		}
		\Step{$=$}{
			p(x'|y_1,y_2)~.
		}
	\end{Reason}
	
	So that's it, the proof that $\HMMone{H}=\HMMone{I;J} = \HMMone{I};\HMMone{J}$, where the ``$;$'' on the left is \HMM-composition (syntactic level) and the ``$;$'' on the right is Kleisli composition (semantic level).
	
	\ENSubsection{Special cases: pure channels and pure transitions}
	\Cf{Probably we will only mention these, not prove them: they must be well known from the definition of ``$;$'' on \HMM's directly.}
	We ``know'' that the Kleisli composition of two channels-as-\HMM's $C,D$ is their parallel composition $C{\parallel}D$. For that, we make $M,N$ above the identity matrix and find that $I;J$ applied to $\pi$ generates the joint distribution $p(x,y,x',z,x'') = \pi_xC_{x,y}M_{x,x'}D_{x',z}N_{x',x''}$, so that $p(x,(y,z)) = \sum_{x',x''} \pi_xC_{x,y}M_{x,x'}D_{x',z}N_{x',x''}$, which reduces to $\pi_xC_{x,y}D_{x,z}$. Since $(C{\parallel}D)_{x,(y,z)} = C_{x,y}D_{x,z}$, we have our results $\HMMone{C};\HMMone{D}=\HMMone{C{\parallel}D}$.
	
	We also know, on the other hand, that for Markov matrices the Kliesli composition reduces to just their matrix product, i.e.\ that $\HMMone{M};\HMMone{N}=\\HMMone{M{\MMult}N}$. This time it's the channels $C,D$ that are trivial, revealing nothing. The joint distribution $p(x,y,x',z,x'') = \pi_xC_{x,y}M_{x,x'}D_{x',z}N_{x',x''}$ becomes $p(x,x',x'') = \pi_xM_{x,x'}N_{x',x''}$, where we simply remove the $y,z$ because they are the unique elements for which $C_{x,y},D_{x,z}$ are not zero (and hence are one). This time we have $p(x,x'') = \sum_{x'} \pi_xM_{x,x'}N_{x',x''} = \pi_x(M{\MMult}N)_{x,x'}$.
	
	\NS
	A discrete distribution over a (finite) state with some canonical order can be given either by a row- or a column vector in which the probabilities are presented in that same order. On the other hand, where specific values in the support are named, we write
	\begin{equation}\label{ze0902}
	\begin{array}{l@{\,}l@{\,}l}
	x_1 &\AtP& p_1 \\
	x_2 &\AtP& p_2 \\
	\multicolumn{3}{l}{\textit{etc...}}
	\end{array}
	\end{equation}
	If these are laid out horizontally, we enclose them in double set-brackets $\DDist{\cdots} $ separated by commas: thus $\DDist{H\AtP\nicefrac{2}{3},T\AtP\nicefrac{1}{3}}$ describes a coin twice as likely to give heads as tails. If the double brackets are used without probabilities (and thus also without $\AtP$'s) then the intended distribution is uniform, so that $\DDist{H,T}$ describes a fair coin; a convenient special case of that is $\DDist{H}$ for the point distribution on $H$, the coin that gives heads every time.
	\footnote{In the semantic space we write $\Point{x}$ for that: here we are syntactic.}
	
	\subsection*{Examples' setup}
	Take a three-element state-space $\CalX={0,1,2}$ and two small programs: one is an update (i.e.\ a Markov transition), and the other is a channel. The updating matrix $M$ describes changing initial $x$ by setting final $x'$ uniformly chosen from the values at least as great as $x$. The channel matrix $C$ describes revealing the parity of $x$ with probability $\nicefrac{2}{3}$ and the opposite of that parity with probability \nicefrac{1}{3}, so that the output observable set (for the channel) is $\CalY=\DSet{0,1}$. The matrices $M,C$ are therefore
	\[
	\begin{array}{lll}
	M &=& \CVec{\RVec{\nicefrac{1}{3},\nicefrac{1}{3},\nicefrac{1}{3}},
		\RVec{0,\nicefrac{1}{2},\nicefrac{1}{2}},
		\RVec{0,0,1}} \\
	C &=& \CVec{\RVec{\nicefrac{2}{3},\nicefrac{1}{3}},
		\RVec{\nicefrac{1}{3},\nicefrac{2}{3}},
		\RVec{\nicefrac{2}{3},\nicefrac{1}{3}}} ~.
	\end{array}
	\]
	
	We assume the incoming value $x$ is uniformly distributed, so that the prior $\pi$ (as a row vector of probabilities) is $\RVec{\nicefrac{1}{3},\nicefrac{1}{3},\nicefrac{1}{3}}$.
	
	\subsection*{The denotation of $M$}\label{s1340}
	The matrix multiplication $\pi^T{\MMult}M$ is
	\( \RVec{\nicefrac{1}{3},\nicefrac{1}{3},\nicefrac{1}{3}}
	\MMult
	\CVec{\RVec{\nicefrac{1}{3},\nicefrac{1}{3},\nicefrac{1}{3}},
		\RVec{0,\nicefrac{1}{2},\nicefrac{1}{2}},
		\RVec{0,0,1}}
	~=~                           
	\RVec{\nicefrac{1}{9},\nicefrac{5}{18},\nicefrac{11}{18}}
	\)~,
	making the probabilities of $x'_{0,1,2}$ in the output state $\nicefrac{1}{9},\nicefrac{5}{18},\nicefrac{11}{18}$ resp. Because $M$'s denotation has codomain $\Dist^2\CalX$, that discrete distribution is lifted to a point hyper: the actual result of $M$ (i.e.\ its denotation) applied to $\pi$ is thus $\Point{\pi}$. That is $\MM{M}.\pi = \DDist{\DDist{x_0\AtP\nicefrac{1}{9},x_1\AtP\nicefrac{5}{18},x_2\AtP\nicefrac{11}{18}}\AtP1}$.
	\footnote{The ``$\AtP1$'' is added for emphasis: it can be omitted.}
	Intuitively that means that after $M$'s application to $\pi$ we are sure, that is  $\DDist{\cdots\AtP1}$, that the final distribution of $x'$ over $\CalX$ will be $\DDist{x_0\AtP\nicefrac{1}{9},x_1\AtP\nicefrac{5}{18},x_2\AtP\nicefrac{11}{18}}$.
	
	\subsection*{The denotation of $C$}\label{s1327}
	We have the same (uniform) $\pi$, but now the channel matrix
	\( \CVec{\RVec{\nicefrac{2}{3},\nicefrac{1}{3}},
		\RVec{\nicefrac{1}{3},\nicefrac{2}{3}},
		\RVec{\nicefrac{2}{3},\nicefrac{1}{3}}}
	\). We let $\CalY=\DSet{y_0,y_1}$; the actual $y$-values are arbitrary. The joint probability matrix $J=\ChApp{\pi}{C}$, describing a distribution in $\Dist(\CalX{\times}\CalY)$, is then
	\( \CVec{\RVec{\nicefrac{2}{9},\nicefrac{1}{9}},
		\RVec{\nicefrac{1}{9},\nicefrac{2}{9}},
		\RVec{\nicefrac{2}{9},\nicefrac{1}{9}}}
	\).
	When J is converted to a hyper via \Def{d1426H} it becomes an ``outer'' distribution $\RVec{\nicefrac{5}{9},\nicefrac{4}{9}}$ over two ``inner'' distributions $\CVec{\nicefrac{2}{5},\nicefrac{1}{5},\nicefrac{2}{5}}$ and $\CVec{\nicefrac{1}{54},\nicefrac{1}{2},\nicefrac{1}{4}}$ that we present as columns because they are the normalised \emph{columns} of $J$. The hyper $\Hyp{J} = \Hyp{\ChApp{\pi}{C}}$ is thus
	\begin{equation}%\label{e1327}
	\begin{array}{llllll}
	\OProb & \OProb & x_0\AtP\nicefrac{2}{5}, & x_1\AtP\nicefrac{1}{5}, & x_2\AtP\nicefrac{2}{5} & \CProb\AtP\nicefrac{5}{9} \\
	& \OProb & x_0\AtP\nicefrac{1}{4}, & x_1\AtP\nicefrac{1}{2}, & x_2\AtP\nicefrac{1}{4} & \CProb\AtP\nicefrac{4}{9} \\
	\multicolumn{6}{l}{\CProb ~.}
	\end{array}
	\end{equation}
	
	This output hyper represents $C$'s leaking (via $\CalY$) information about prior distribution $\pi$ (on $\CalX$) when it happens to be uniform: with probability $\nicefrac{5}{9}$ the revised, posterior belief about the distribution on $\CalX$ will be $\DDist{x_0\AtP\nicefrac{2}{5}, x_1\AtP\nicefrac{1}{5}, x_2\AtP\nicefrac{2}{5}}$, and with probability $\nicefrac{4}{9}$ it will be $\DDist{\DDist{x_0\AtP\nicefrac{2}{5}, x_1\AtP\nicefrac{1}{5}, x_2\AtP\nicefrac{2}{5}}}$.
	
	\emph{Section} {\HMM's: syntax and semantics}
	
	\subsection*{Basic structure of \HMM's: syntax}\label{s1403}
	A Hidden Markov Model comprises a set $\CalX$ of states, a set $\CalY$ of observations, and two stochastic matrices $M,C$ \Cite{Jurafsky:00}: the \emph{transition} probabilities $M_{x,x'}$ give for any two states $x,x'{\In}\CalX$ the (conditional) probability that a transition will end in final state $x'$ given that it began in initial state $x$; and the \emph{emission} probabilities $C_{x,y}$ give for any state $x$ and observation $y{\In}\CalY$ the probability that $y$ will be emitted, and thus observed, given initial state $x$. Typically an \HMM\ is analysed over a number of steps $i=0,1,\cdots$ from some initial prior distribution $\pi_0$ over $\CalX$, so that a succession of states $x',x'',\cdots$ and observations $y,y'\cdots$ occurs.
	
	We illustrate a single step in \Fig{f1420}. With $\pi_0$ the distribution of incoming state $x$, the distribution $\pi_1$ of outgoing states $x'$ is the multiplication of $\pi$ as a row-vector by $M$ as a matrix, as we saw above in \Sec{s1237}. Similarly from \Sec{s1007} we recall that the distribution of observations $y$ is the $\CalY$-marginal of the joint matrix $J=\ChApp{\pi_0}{C}$. The ``hidden'' essence of the \HMM-model is that though we cannot see the incoming $x$ and outgoing $x'$ directly, still the observation $y$ tells us something about each if we do know the incoming-state distribution $\pi_0$ and the matrices $M,C$.
	\begin{Figure}{f1420}{One step of an \HMM.}\Xx
		\vspace{29ex}
		\verb+\ImageInText{0pt}{3ex}{0.4}{HMMapri_new.pdf} %fonts fixed+
		
		\bigskip
		The emission component is some channel $C$ from $\CalX$ to $\CalY$, and the transformation component is Markov matrix $M$ from $\CalX$ to itself. The incoming prior is distribution $\pi_0$ on state-space $\CalX$, transformed into $\pi_1$ on $\CalX$ by $M$. The distribution of observables $y{\In}\CalY$ is $\rho_0$. Observing a particular $y$ determines posterior revisions of both $\pi_0$ and $\pi_1$.
		%\caption{One step of an \HMM.}
	\end{Figure}
	
	The syntactic representation of the \HMM\ (step) is the two matrices that comprise it: we write $\CM{C}{M}$.
	
	\subsection*{Posterior distributions on $\CalX$: input \emph{and} output}
	We saw in \Sec{s1327} that given a prior $\pi$ on $\CalX$ and an observation $y$ produced by a known channel $C$, we can revise our belief of the distribution on $\CalX$: these posteriors are the support of the (hyper) distribution $\AbCh{C}.\pi$, its inner distributions; and the probabilities assigned to each comprise the outer distribution on those inners.
	
	However an observed $y$ induces also a revision of the output distribution that the Markov matrix produces from the input distribution. After observing a particular $y$, we should assume that the Markov transition occurred from the posterior distribution on $\CalX$ corresponding to that $y$, not the original $\pi$. We illustrate this using our examples from \Sec{s1323}.
	
	In \Eqn{e1327} of \Sec{s1327} we saw the output hyper produced by $C$ applied to prior $\pi$; and in \Sec{s1332} we saw how to apply a Markov transition to a particular prior. To apply $M$ to \emph{all} of the posteriors produced by $C$ from prior $\pi$ we use its push-forward $\Dist\MM{M}$, defined below. Thus we will write $\Dist\MM{M}.(\AbCh{C}.\pi)$, intuitively the application of $C$ to $\pi$ followed by applications of $\MM{M}$ to all the posteriors that $\AbCh{C}.\pi$ produces.
	
	Given two sets and a function between them, we define as follows the ``push forward'' of the function that acts between discrete distributions on those sets, a function of type $\Dist\CalX{\Fun}\Dist\CalX'$:
	\footnote{We are for now specialising to the discrete case. In general the push-forward $\Dist f$ of arrow $f$ is the action of functor $\Dist$ on it. }
	{\Xx%1223
		\begin{Definition}{push forward}{d1341X}
			Given two sets $\CalZ,\CalZ'$ and a function $f{\In}\CalZ{\Fun}{\CalZ'}$, we write $\Dist f$ for the \emph{push-forward} of $f$ that is of type $\Dist\CalZ{\Fun}\Dist\CalZ'$, defined as follows: for $z'{\In}\Dist\CalZ'$ and $\delta{\In}\Dist\CalZ$ we have
			\[
			(\Dist f).z' \Wide{\Defs} \sum_{\substack{z{\In}\CalX\\f.z=z'}} \delta.z ~.
			\]
			That is, the probability $\Dist f.z'$ that the push-forward $\Dist f$ assigns to arbitrary $z'{\In}\CalZ'$ is the sum of the probabilities that $\delta$ assigns to all elements $z{\In}\CalZ$ that are taken by $f$ to $z'$, i.e.\ such that $f.z=z'$.
		\end{Definition}
	}%1223
	
	Now since $\MM{M}$ is of type $\Dist\CalX{\Fun}\Dist\CalX$, the type of its push-forward $\Dist\MM{M}$ is $\Dist^2\CalX{\Fun}\Dist^2\CalX$, i.e.\ it takes hypers to hypers. In this case it takes the (outer) distribution over the posteriors (inners) that $\AbCh{C}$ produces from prior $\pi$ and ``pushes $\MM{M}$ forward'' to produce the same (outer) distributions over those posteriors transformed each-for-each by $\MM{M}$.
	
	Here is the construction applied to our examples from \Sec{s1323}. We start with uniform prior $\pi$ as before, and the effect of channel $C$ is to produce the output hyper \Eqn{e1327} whose inner distributions are
	\begin{eqnarray}
	\DDist{x_0\AtP\nicefrac{2}{5}, & x_1\AtP\nicefrac{1}{5}, & x_2\AtP\nicefrac{2}{5}} \label{e1412A} \\
	\DDist{x_0\AtP\nicefrac{1}{4}, & x_1\AtP\nicefrac{1}{2}, & x_2\AtP\nicefrac{1}{4}} \label{e1412B}
	\end{eqnarray}
	The outer distribution assigns probability $\nicefrac{5}{9}$ to \Eqn{e1412A} and probability $\nicefrac{4}{9}$ to \Eqn{e1412B}.
	
	If we apply the push forward of $\MM{M}$ to the above, we transform the inners but maintain the outer: that is, we obtain
	\begin{eqnarray}
	\MM{M}.\DDist{x_0\AtP\nicefrac{2}{5}, & x_1\AtP\nicefrac{1}{5}, & x_2\AtP\nicefrac{2}{5}} \label{e1412C} \\
	\MM{M}.\DDist{x_0\AtP\nicefrac{1}{4}, & x_1\AtP\nicefrac{1}{2}, & x_2\AtP\nicefrac{1}{4}} \label{e1412D}
	\end{eqnarray}
	as the result of $\Dist\MM{M}.(\AbCh{C}.\pi)$. Calculating it through for the given $M$ gives
	\begin{eqnarray}
	\DDist{x_0\AtP\nicefrac{2}{15}, & x_1\AtP\nicefrac{7}{30}, & x_2\AtP\nicefrac{19}{30}} \label{e1412E} \\
	\DDist{x_0\AtP\nicefrac{1}{12}, & x_1\AtP\nicefrac{1}{3}, & x_2\AtP\nicefrac{7}{12}} \label{e1412F}
	\end{eqnarray}
	
	\textbf{subsection}{\HMM\ semantics}
	Given a (syntactic) description $\CM{C}{M}$ of an \HMM, we define its denotation $\HMMone{M,C}$ to be $\Dist\MM{M}\Comp\AbCh{C}$, that is the functional composition first applying $\AbCh{C}$ and then applying $\Dist\MM{M}$.
	\begin{Definition}{\HMM\ semantics}{d1426HMM}
		Let an \HMM\ with state $\CalX$ and observables $\CalY$ be determined by a channel matrix $C$ from $\CalX$ to $\CalY$ and a Markov matrix $M$ from $\CalX$ to itself. The denotation of $\CM{C}{M}$ is of type $\AHSpc$, given by
		\[
		\HMMsem{C}{M}.\pi \Wide{\Defs} \Dist\MM{M}.(\AbCh{C}.\pi) ~,
		\]
		where the $\HMMsem{\cdot,\cdot}$ on the left is what we are defining and the $\MM{\cdot}$'s on the right are the semantics of Markov- and channel matrices.
		That is, we have $\HMMone{C}{M}=\Dist\MM{M}\Comp\AbCh{C}$.
	\end{Definition}
	
	\NS
	\begin{verbatim}
	Executing C gives the output hyper {{ <2/5,1/5,2/5> @ 5/9, 
	<1/4,1/2,1/4> @
	4/9 }} from just above.
	
	To execute M from this point, that is after C, we use the Kleisli
	construction to apply M to each of the two posterior distributions 
	<2/5,1/5,2/5>
	and <1/4,1/2,1/4> separately, then averaging the results according to 
	the
	probabilities 5/9,4/9 associated with those two posteriors individually.
	
	For the first posterior we get
	
	<2/5,1/5,2/5> * M
	=   <2/5,1/5,2/5>
	* <[1/3,1/3,1/3], [0,1/2,1/2], [0,0,1]>
	= <2/15,7/30,19/30> ,
	
	and for the second we get
	
	<1/4,1/2,1/4> * M
	=   <1/4,1/2,1/4>
	* <[1/3,1/3,1/3], [0,1/2,1/2], [0,0,1]>
	= <1/12,1/3,7/12> .
	
	The overall result of C;M beginning from pi is therefore
	
	<2/15,7/30,19/30> @ 5/9
	<1/12,1/3,7/12>   @ 4/9 .   
	\end{verbatim}
	
	\NS
	The prior distribution is $\pi_0$ and, once we have observed a 
	particular $y$, the posterior distribution on the input can be 
	calculated from $M,C,\pi_0$ via Bayes' formula in the usual way.
	
	But we concentrate instead on the output.
	
	The prior distribution of outgoing $x'$ is $X_1$ as calculated in 
	\Sec{s1403} above. Its posterior distribution is conditioned on the 
	emitted $y_1$ actually observed: it too is determined by the usual 
	Bayes formula
	\begin{equation}%\label{e1508}
	\Pr(X_1{=}x_1|Y_1{=}y_1)
	\Wide{\Defs}
	\frac{\sum_{x_0}\, \Pr(X_0{=}x_0)E(y_1|x_0)T(x_1|
		x_0)
	}{
		\sum_{x_0}\, \Pr(X_0{=}x_0)E(y_1|x_0)}~,
	\end{equation}
	that is the (joint) probability that $x_1,y_1$ both occurred divided 
	by the overall (marginal) probability that $y_1$ occurred. Thus 
	\emph{before} we observe any $y_1$ we believe the distribution of 
	outgoing $x_1$ to be $X_1$, and \emph{after} we observe $y_1$ we 
	believe that distribution to be as~\Eqn{e1508}. This view is 
	illustrated in \Fig{f1421}.
	\begin{Figure}{f1421}{A Hidden Markov Model, APost view}
		\vspace{34ex}
		%\ImageInText{0pt}{10ex}{0.4}{HMMapost_new.pdf} % fonts fixed
		
		\bigskip
		By APost we mean that the conditional distributions $
		\delta_{\{1,2,3\}}$ are deduced \emph{after} observation of the 
		emitted values $y_{\{1,2,3\}}$, and represent a revision of the APri 
		knowledge of the outgoing state as represented in the $X_1$ of 
		\Fig{f1420}.
		%\caption{A Hidden Markov Model, APost view}
	\end{Figure}
	
	\textbf{subsection}{The attacker's point of view: an equivalent 
		representation}
	Although the matrices $T,E$ determine the \HMM\ 
	completely, we suggest that from the point of view of an attacker 
	trying to determine the state of the \HMM, it would be more useful to 
	consider a different (but equivalent) formulation: the effect of one 
	step from a known initial distribution $X_0$ is a joint distribution 
	over observations in $\CalY$ and their corresponding outgoing 
	conditional distributions over $\CalX$: this structure thus comprises 
	values $\Delta$ of type $\Dist(\CalY{\times}\Dist\CalX)$, where we write $
	\Dist\CalY$ and similar for the type of \emph{discrete distributions} 
	over $\CalX$, thus one-summing functions of type $\CalX{\Fun}[0,1]$. That 
	is, each $\Delta$ gives for a pair $(y_1,\delta_1)$ in $\CalY{\times}
	\Dist\CalX$ the probability that an attacker will observe $y_1$ and will 
	conclude from it that $x_1$ has APost distribution $\delta_1$.
	
	We call such $\Delta$-values \emph{hyper-distributions}, or just 
	\emph{hypers}. Since $\Delta$ is a joint distribution (jointly over $
	\CalY$ and $\Dist\CalX$), we can speak of its left- and right-marginal 
	distributions: the left-marginal distribution $LMarg{\Delta}$ is of 
	type $\Dist\CalY$, and is in fact just $Y_1$ from above. That is, the 
	distribution $Y_1$ of emitted observations is recovered as $
	LMarg{\Delta}$.
	
	The right-marginal distribution $RMarg{\Delta}$ of the hyper is more 
	interesting: it is of type $\Dist^2\kern-.1em\CalX$ and, although it 
	\emph{averages} to the outgoing state distribution $X_1$ (in the sense 
	shown in \Fig{f1421}), most of the popular (conditional) information-
	entropy measurements are likely to decrease, becoming less than the 
	entropy of $X_1$ itself: that decrease quantifyies the ``leak'' that 
	the emissions of $Y_1$ represent. For example, the \emph{conditional 
		Shannon Entropy} of $RMarg{\Delta}$, defined $\sum_{\delta{\In}\Dist
		\CalX} RMarg{\Delta}\kern-.2em(\delta)ShE(\delta)$ over the possible 
	APost\ distributions $\delta$ is no more than $ShE(X_1)$, the 
	Shannon Entropy of the APri outgoing distribution $X_1$ itself.\,%
	\footnote{\Label{n1656}For distribution $X$ in $\Dist\CalX$ such that $
		\Pr(X{=}x_i)$ is $p_i$, the Shannon Entropy $ShE(X)$ of $X$ is given 
		by $-\sum_i\,p_i\lg p_i$. As remarked, a number of other security-
		based definitions of entropy give the same inequality \Cite{Kopf:07}.}
	
	Thus the denotional-style semantic representation we extract from 
	\HMM-theory is the \emph{hyper-distribution} of type $
	\Dist(\CalY{\times}\Dist\CalX)$, a nesting of one distribution within 
	another. As we will see, this allows us to equip the semantic space 
	with a ``refinement'' partial order; but it is \emph{security} 
	refinement, so that for hypers $\Delta_{\{0,1\}}$ one can speak of 
	whether $\Delta_0$ is more- or less secure than $\Delta_1$ or, if not, 
	whether they are perhaps simply security-incomparable.

	\NS
	\begin{verbatim}
	==5 Executing C and then M
	
	This is the most general case, where the hidden state is both leaked 
	(somewhat)
	and altered.
	
	We start with prior <1/3,1/3,1/3>, as before.
	
	Executing C gives the output hyper {{ <2/5,1/5,2/5> @ 5/9, 
	<1/4,1/2,1/4> @
	4/9 }} from just above.
	
	To execute M from this point, that is after C, we use the Kleisli
	construction to apply M to each of the two posterior distributions 
	<2/5,1/5,2/5>
	and <1/4,1/2,1/4> separately, then averaging the results according to 
	the
	probabilities 5/9,4/9 associated with those two posteriors individually.
	
	For the first posterior we get
	
	<2/5,1/5,2/5> * M
	=   <2/5,1/5,2/5>
	* <[1/3,1/3,1/3], [0,1/2,1/2], [0,0,1]>
	= <2/15,7/30,19/30> ,
	
	and for the second we get
	
	<1/4,1/2,1/4> * M
	=   <1/4,1/2,1/4>
	* <[1/3,1/3,1/3], [0,1/2,1/2], [0,0,1]>
	= <1/12,1/3,7/12> .
	
	The overall result of C;M beginning from pi is therefore
	
	<2/15,7/30,19/30> @ 5/9
	<1/12,1/3,7/12>   @ 4/9 .   
	\end{verbatim}
	
	\NS
	\subsection*{Abstract channels}
	An ``abstract channel'' is a function of type $\AHSpc$ that denotes the effect a ``concrete'' channel has on converting a given prior distribution into a distribution of posteriors (i.e.\ into a hyper). But not all functions of that type are abstract channels, as we now explain.
	
	\begin{Definition}{Hyper determines channel and prior}{d1539AZ}
		Let hyper $\Delta{\In}\Dist^2\CalX$ have finite support some $\CalY\subseteq\Dist\CalX$; note that $\CalY$ is thus a set of (posterior) distributions. Let $\delta{\In}\CalX$ be $\Avg.\Delta$ and suppose that $\delta$ has full support, i.e.\ that $\Supp{\Delta}{=}\CalX$.
		
		We define a channel $C_\Delta{\In} \CalX{\times}\CalY\Fun\Real$ as
		\[
		C_\Delta.x.y \Wide{\Defs} \Delta.y\times y.x/\delta.x~,
		\]
		where the division on the right is well defined because by assumption $\delta.x$ is never 0. Also, the channel rows are 1-summing, because for each $x$ we have
		\[
		\sum_y (\Delta.y\times y.x/\delta.x)
		\Wide{=}
		(\sum_y \Delta.y\times y.x)/\delta.x
		\Wide{=}
		\delta.x/\delta.x ~.
		\]
		
		Under the conditions stated, we say that $\Delta$ \emph{determines} $C_\Delta,\delta$.
		\Cf{\label{n1016}But (as we discussed) there's an issue here with the fact that a given abstract channel $f$ will produce different supports for $C_{f.\delta}$, depending on the $\delta$. \Def{d1044A} is therefore not quite right.}
	\end{Definition}
	
	With \Def{d1539A} we can characterise which functions are abstract channels. We have
	\begin{Definition}{Abstract channel}{d1044A}
		An \emph{abstract channel} is a function $P$ of type $\AHSpc$ such that there is a (concrete) channel $C$ so that for all full-support priors $\delta{\In}\Dist\CalX$ we have that $P.\delta$ determines $C,\delta$.
	\end{Definition}
	A trivial consequence of \Def{d1044A} is that for any abstract channel $P$ and full support $\delta$ we have $\Avg.(P.\delta){=}\delta$.

	\subsection{The semantic function from joints to hypers}\label{s1956}
	In this section we define precisely the denotation $\Hyp{J}$ in $\Dist^2\CalX$ of a joint-distribution matrix $J{\In}\CalX{\MFun}\CalY$. The principal tool for that is the ``push forward'', here in general form:
	\begin{Definition}{Push-forward of a function}{d1341B}\label{g1108}\\
		Given sets $\CalZ,\CalZ'$ and function $f{\In}\CalZ{\Fun}{\CalZ'}$, we write $\Dist f$ for the \emph{push-forward} of $f$, a ``lifted'' function of type $\Dist\CalZ{\Fun}\Dist\CalZ'$ \cite{Fremlin:00}. For $z'{\In}\CalZ'$ and $\delta{\In}\Dist\CalZ$ we have
		\footnote{Lifting, as in $\Dist f$, binds tightest: the conventional notation for $\Dist f.\delta.z'$ would be $(\Dist f)(\delta)(z')$, so that $(\Dist f)(\delta){\in}\Dist\CalZ'$.}
		\[
		\Dist f.\delta.z' \Wide{\Defs} \sum_{\substack{z{\In}\CalZ\\f.z=z'}} \delta.z ~.\quad\footnotemark
		\LiftBox\]
		\footnotetext{$\Dist f$ is the action of functor $\Dist$ on arrow $f$: see \Sec{s1250}.}%In general $f$ must be measurable.}
	\end{Definition}
	
	We now define the semantic function itself:
	\begin{Definition}{Reduced joint-distribution denotes hyper}{d1426HB}\label{g1102}
		Let $J{\In}\CalX{\MFun}\CalY$ satisfy $1=\sum J$ so that it describes a discrete joint distribution in $\Dist(\CalX{\times}\CalY)$. \label{g1103}Recalling \Sec{s1129}, define a \emph{reduced} matrix $J^\downarrow$ by (1) removing all-zero columns from $J$ and (2) adding any similar columns of $J$ together, retaining the label of only one of them. Let the remaining labels $\CalY^\downarrow{\subseteq}\CalY$ be the column-indices of this reduced matrix $J^\downarrow$.\,%
		\footnote{The reduction is analogous to reduced \emph{channels} in \cite{McIver:2014aa}. Although $J^\downarrow$ is not unique, the ambiguity does not affect $\Hyp{J}$.}
		\footnote{We thank a referee for suggesting that this definition might be simplified by using the converse of stochastic relations, as developed by Doberkat \cite{Doberkat:2003aa}. This is discussed further in \Sec{s1418}.}
		
		Now define the $\CalY^\downarrow$-marginal $\delta_y\Defs\sum J^\downarrow_{-,y}$ of $J^\downarrow$, and note from (1) just above that it is nowhere zero (on $\CalY^\downarrow$). Define function $j{\In}\CalY^\downarrow{\Fun}\Dist\CalX$ by $j.y\Defs \Nrm{J^\downarrow_{-,y}}$, i.e.\ so that $j.y{\in}\Dist\CalX$ is the posterior distribution over $\CalX$ that $J^\downarrow$ induces given $y$.\,%
		%\footnote{This is the \emph{right conditional} of $J^\downarrow$.}
		Note from (2) that $j$ is an injection, a fact we use later in \Lem{l0827}. Then $\Hyp{J}$, the hyper in $\Dist^2\CalX$ denoted by $J$ in $\CalX{\MFun}\CalY$, is given by
		\(
		\Hyp{J}\Defs(\Dist j).\delta~.
		\)
	\end{Definition}
	
	An example is given at \Sec{s1407c} below.
	
}%0818 End Sargasso

\cleardoublepage
\hrule
~\\\begin{center}\Huge Endnotes
	%\Cf{Don't edit the LaTeX stuff just below directly (except to change its format): it's populated via endnote macros
	%in the main text.}
\end{center}~\\
\hrule

\begingroup
\parindent 0pt
\parskip 2ex
\def\enotesize{\normalsize}
\def\notesname{} % Suppress endnotes title.
\theendnotes
\endgroup
\end{document}